\documentclass[twoside,b5paper,10pt,english]{book}
\usepackage{graphicx}
\graphicspath{ {figures/} }
\usepackage{array}
\usepackage{watermark}
\usepackage{fancybox}
\usepackage[dvipsnames,usenames]{color}
\usepackage{babel}
\usepackage[Lenny]{fncychap}
\usepackage[pagestyles]{titlesec}
\usepackage{bookmark}
\usepackage{afterpage}
\usepackage[b5paper,left=2.5cm,right=1.5cm,top=2.5cm]{geometry}
\usepackage[a4,center,pdflatex]{crop}
\usepackage{hyperref}
\usepackage{amsmath,amssymb,amsfonts,bm,cancel,physics,nicefrac}
\usepackage{cite}
\usepackage{pdfpages}
\usepackage[font=footnotesize,labelfont=scriptsize]{caption,subfig}
\newpagestyle{mystyle_empt}{%
\sethead[][][ ]
{}{}{}
}
\newpagestyle{mystyle}{%
\sethead[\thepage][][ Chapter\ \thechapter.  \chaptertitle]
{Section \thesection. \sectiontitle}{}{\thepage}
\setheadrule{1pt}
}
\newpagestyle{mystyle2}{%
\sethead[\thepage][][ \chaptertitle]
{\chaptertitle}{}{\thepage}
\setheadrule{1pt}
}
\newpagestyle{mystyle3}{%
\sethead[\thepage][][ Appendix\ \thechapter.  \chaptertitle]
{Appendix\ \thechapter.  \chaptertitle}{}{\thepage }
\setheadrule{1pt}
}
\newpagestyle{mystyle4}{%
\sethead[\thepage][][ Chapter\ \thechapter.  \chaptertitle]
{Chapter\ \thechapter.  \chaptertitle}{}{\thepage }
\setheadrule{1pt}
}

\usepackage{tocbibind}
\addto{\captionsenglish}{}

\begin{document}
\frontmatter
\begin{titlepage}
\begin{center}

\thiswatermark{\put(5,-650){\includegraphics[width=1.1\textwidth]{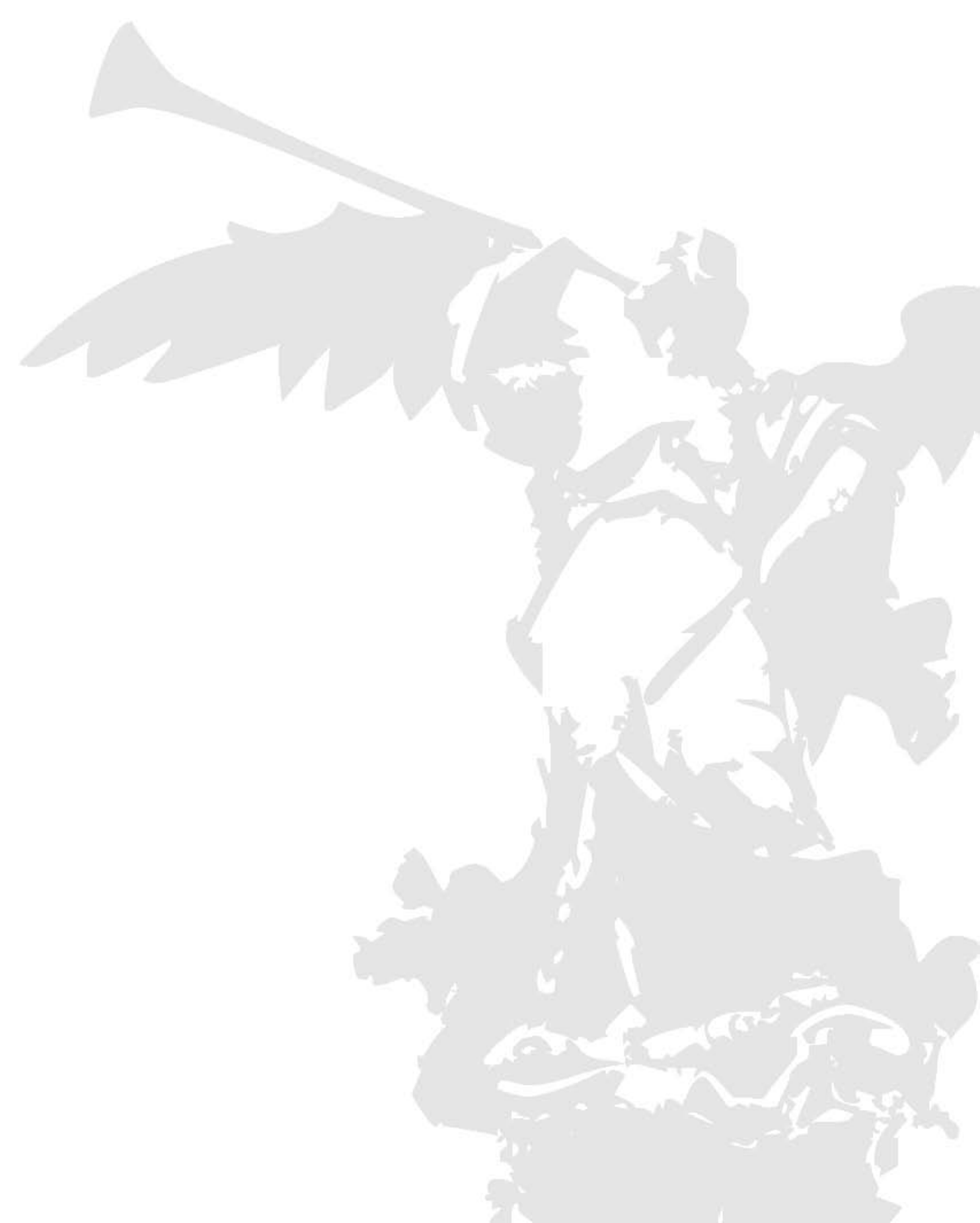}}}

\includegraphics[width=0.45\textwidth]{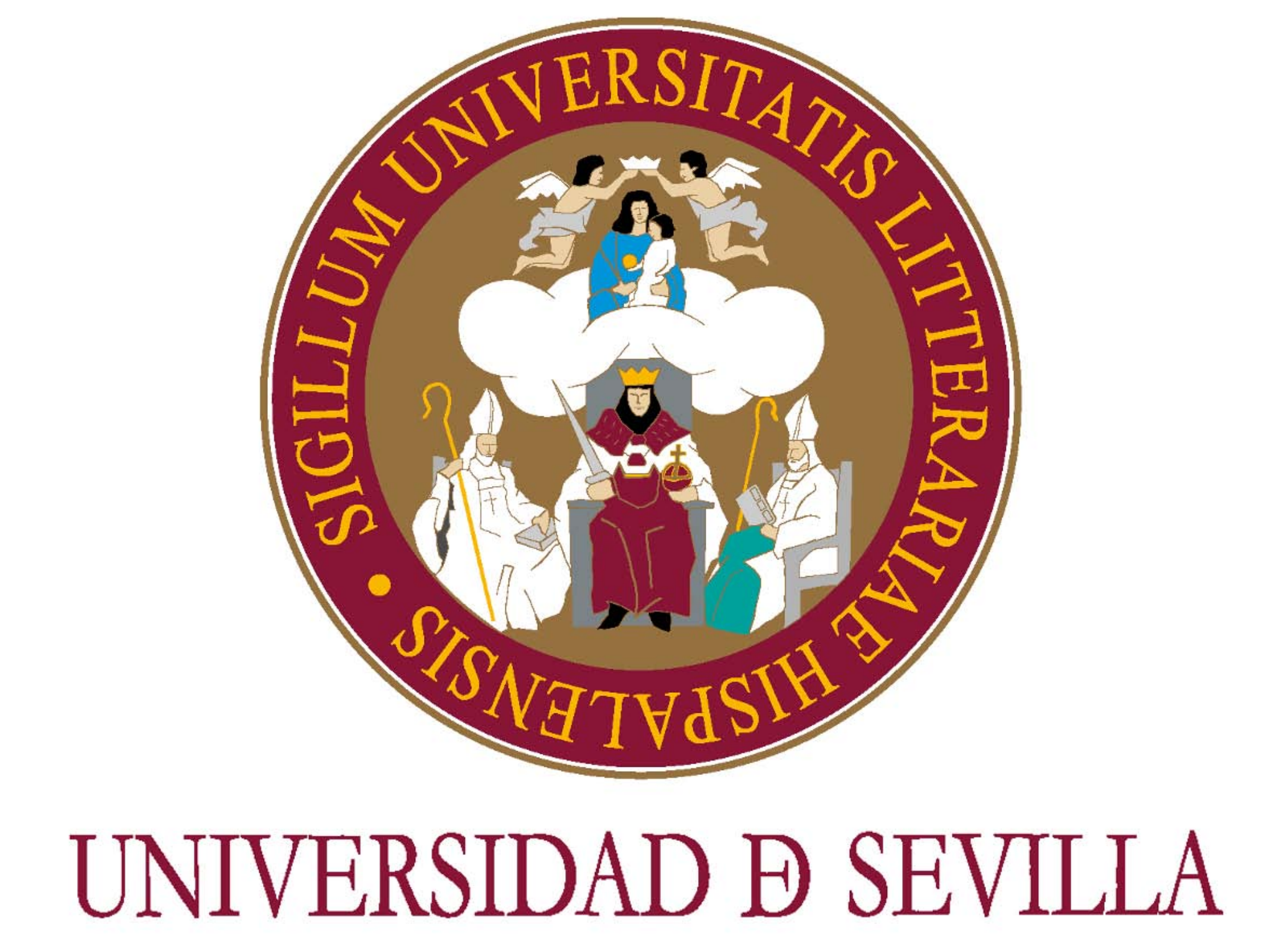}\\[1cm]

 \textsc{\Large Departamento de F\'{\i}sica At\'{o}mica \\Molecular y Nuclear}\\[1.3cm]

 \textsc{\Large Tesis Doctoral}\\[1.3cm]

 \hrulefill \\[0.6cm]
 { \huge \bfseries 
    From proteins to grains: \\ a journey through simple models
 }\\[0.6cm]

 \hrulefill \\[2.0cm]

\begin{minipage}{0.4\textwidth}
\begin{flushleft} \large
\emph{Doctorando:}\\
Carlos Alberto Plata Ramos \\
$\quad$
\\ $\quad$
\end{flushleft}
\end{minipage}
\begin{minipage}{0.4\textwidth}
\begin{flushright} \large

\emph{Director:} \\
Antonio Prados Monta\~{n}o
\emph{Tutor:} \\
Diego G\'{o}mez Garc\'{i}a
\end{flushright}

\end{minipage}
 
\vfill
 
{\large October, 2018}

\end{center}

\end{titlepage}

\thispagestyle{empty}
\pagestyle{mystyle_empt} 

\cleardoublepage

\begin{center}

\includegraphics[width=0.35\textwidth]{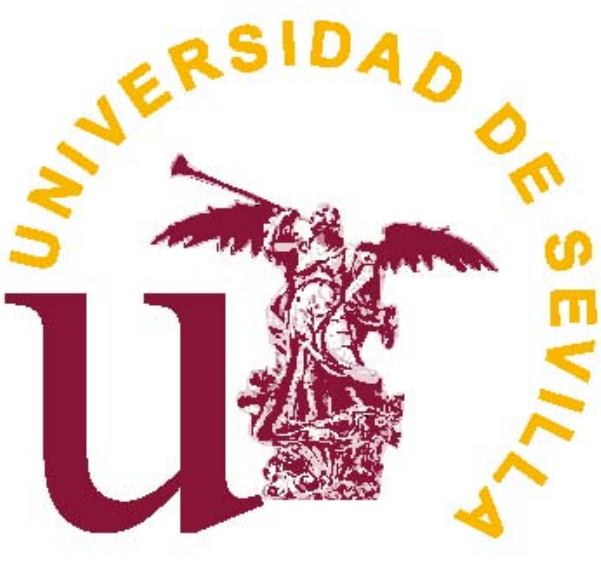}\\[0.1cm]
\hrulefill \\[0.1cm]
 \textsc{\Large Departamento de F\'{\i}sica At\'{o}mica \\Molecular y Nuclear}\\[4.3cm]
 {\Large Memoria presentada para optar al Grado \\ de Doctor por la Universidad de Sevilla por}\\
\vspace{2.5cm}

{\Large Carlos Alberto Plata Ramos}\\[0.1cm]
 \hrulefill \\[0.3cm]
{\Large V$^{\circ}$ B$^{\circ}$ del Director de Tesis} \\
\vspace{2.5cm}
{\Large Antonio Prados Monta\~no} \\[0.1cm]
 \hrulefill \\

\end{center}

\chapter*{$\quad$}
\begin{quotation}\begin{flushright}\begin{em}
No pidas tiempo al tiempo compa\~nera\\
y piensa en qu\'e gastarte todo el tiempo que nos queda\\[0.2cm]
\par\end{em}
Tino Tovar
\end{flushright}\end{quotation}

\thispagestyle{empty}
 
\chapter*{Acknowledgments / Agradecimientos}
\pagestyle{mystyle2} 
El documento que sostiene frente a usted contiene, de forma resumida, una corta vida llena de trabajo, esfuerzo e ilusi\'on. Ser\'ia una necedad pensar que dicha vida pertenece al autor en exclusiva. Por ello, sirvan estas primeras p\'aginas para agradecer de coraz\'on a todos aquellos que, de una forma u otra, pusieron su \textit{grano} en esta monta\~na de arena.

En primer lugar, tengo mucho que agradecerle a Antonio Prados, director y principal art\'ifice de las ideas presentadas en esta tesis. Ha sido un magn\'ifico gu\'ia durante este camino que comenzara, casi por sorpresa, en unas tutor\'ias sobre principios variacionales. En este negocio, tan competitivo, me ha ense\~nado con su ejemplo que la brillantez profesional no requiere de ninguna tara personal como condici\'on  necesaria. Ignorando la jerarqu\'ia, he sentido que ha escuchado y valorado mis ideas, ha confiado en mis opiniones y me he sentido apoyado por \'el en todo momento. Trabajar con Antonio es un placer del que no pretendo desprenderme y valoro su amistad como uno de los principales resultados de este trabajo.

Gracias a mi tutor, Diego G\'omez, por facilitarme, siempre de buena gana, las poco atractivas tareas bur\'ocraticas relacionadas con el doctorado. Durante esta \'epoca, son muchas las horas pasadas en la Facultad de F\'isica. Tengo que agradecer al departamento de FAMN, en su completitud, el arropo institucional y la cercan\'ia mostrada hacia mi persona. 

Haciendo zoom en lo local, merece un especial agradecimiento el \'area de F\'isica Te\'orica y, en especial, el grupo del que he formado parte estos a\~nos. Durante mis a\~nos de carrera y m\'aster, con profesionalidad y precisi\'on encomiables, Javier y M$^{\text{a}}$ Jos\'e me ense\~naron gran parte de la f\'isica estad\'istica que conozco. La p\'erdida de M$^{\text{a}}$ Jos\'e fue, sin duda, el momento m\'as doloroso que he vivido en esta facultad, el injusto truncamiento de su vida ha dejado hu\'erfanas a miles de mentes que desconocen lo que un monstruo les ha arrebatado. Desde el primer a\~no que comenzara mi labor docente, he compartido asignatura con \'Alvaro. \'El me ha ense\~nado un interesante punto de vista sobre la f\'isica m\'as sencilla. Para Maribel y Pablo s\'olo tengo buenas palabras. Les agradezco infinitamente su calidez en una profesi\'on, en ocasiones, fr\'ia. Son muchos los desayunos compartidos, y a Maribel le debo la sabia mezcla de jam\'on con roquefort que asegura una ma\~nana productiva.

During the PhD, I conducted two international research stays. The first of them was at Duke University. There, I could explore the experimental world under the supervision of Piotr Marszalek. I thank him and his group for the warm hosting, I felt like an actual member of the team. Piotr is considered a global expert in his field, even though I highlight his humility. I specially acknowledge the help and patience shown by Zack and Qing with this experimental rookie. Between both research stays, I had the oppotunity to attend the great summer school ``Fundamental Problems in Statistical Physics XIV'' in Italy. I thank teachers and students for creating such a wonderful atmosphere of comradeship.  

In my second research stay, I went to Paris under the supervision of Emmanuel Trizac. I have to say merci beaucoup to him and the whole Laboratoire de Physique Th\'eorique et Mod\`eles Statistiques. Therein, they have a nice convivial atmosphere that I enjoyed a lot. Working with Emmanuel started new projects in my career that still continue. I acknowledge him a lot his guidance and confidence. Merci en especial a In\'es, una encantadora parisina burgalesa que me abri\'o las puertas del simp\'atico n\'ucleo joven del LPTMS. I felt really comfortable among them. Sorry for not giving all the names.
 
En mi estancia parisina viv\'i en el Colegio Espa\~na. All\'i conoc\'i a un grupo extraordinario de j\'ovenes cient\'ificos de un amplio espectro de disciplinas. En tres meses formamos una peque\~na familia, compartimos penas, alegr\'ias y un mill\'on de locuras. Estoy seguro que todas ellas permanecen guardadas con cari\~no en la parcelita que todos creamos en nuestro coraz\'on de cono.
 
Los cient\'ificos, aunque a veces se olvide, somos personas y, pese a lo gratificante de nuestra tarea, no nos nutrimos \'unicamente de conocimiento. Por ello, es necesario fomentar y agradecer las acciones con las que el gobierno y diversas instituciones, p\'ublicas o privadas, apoyan y financian nuestro trabajo. En mi caso, esta tesis ha podido lle\-var\-se a cabo gracias a la financiaci\'on por parte de la Fundaci\'on C\'amara de Sevilla (01/01/2015-31/08/2015) y el Ministerio de Educaci\'on, Cultura y Deporte mediante un contrato FPU14/00241 (desde 01/09/2015). Asimismo agradezco las ayudas concedidas asociadas al contrato FPU para el desarrollo de las dos estancias de investigaci\'on que he realizado durante el periodo de formaci\'on doctoral, as\'i como el apoyo proporcionado mediante los proyectos  FIS2014-53808-P y PP2018/494 concedidos respectivamente por el Ministerio de Econom\'ia y Competitividad y la Universidad de Sevilla a trav\'es de su Plan Propio de Investigaci\'on.

Gracias Sevilla por estos 27 a\~nos maravillosos. Esta ciudad ha sido el escenario perfecto para la aventura que llega a un punto y aparte. No podr\'ia sentirme m\'as afortunado de haber nacido en la cultura de la cercan\'ia personal en la que me he criado. Pese a no tener ni caseta ni hermandad, siento que \'este es mi hogar. Si me marcho, es con el \'unico fin de regresar.

Una ciudad la hacen sus habitantes. Si mi vida en Sevilla ha sido tan buena es, sin duda, gracias a las personas de las que me he visto rodeado. Durante mi etapa predoctoral, ese c\'irculo lo ha conformado principalmente la mongolfiera assasina. Tengo mucho que agradecer a esta amalgama de singulares individuos. Gracias a Mario por su brillantez en cualquier conversaci\'on. Gracias a Carlos Alive por su carisma inigualable, no podr\'iamos haber tenido un mejor compa\~nero de expedici\'on canadiense. Gracias a Laura por su cari\~no y complicidad; y por darnos de comer siempre que surg\'ia la oportunidad, ya fuera con hambre o sin ella. Gracias a Jorge por ser la personificaci\'on de la amabilidad y el buenrollismo proclamado por Carlos. Gracias  Miguel, por ser de las pocas personas que entienden que un hueso roto no duele tanto. Gracias a Andr\'es, un ingeniero tiene mucho que aguantar de un grupo endog\'amico de f\'isicos. Gracias a todos los italianos, sevillanos de adopci\'on: Grazia, Cristina, Anna, Stefano y Alessandro. Con este \'ultimo tuve el placer de trabajar, siendo la experiencia de colaboraci\'on entre iguales m\'as gratificante de la que he disfrutado en estos a\~nos. Gracias a la sangre nueva, Llanlle y Teresa, ten\'eis una herencia maravillosa que estoy seguro que disfrutar\'eis.

Ser\'ia cruel no acordarme de mis compa\~neros del pasado. Gracias a mis medicuchas preferidas, Elena y Ana, por seguir siendo el contacto con una versi\'on m\'as joven de m\'i mismo. Gracias a mis compa\~neros de la primera generaci\'on del grado en f\'isica de la Universidad de Sevilla. Juntos, fuimos capaces de superar con \'exito cientos de obst\'aculos, y para nuestro deleite lo hicimos con el buen humor demostrado en nuestros Frankis. En especial me gustar\'ia destacar a Migue Tan, por ser mi otra mitad de un t\'andem que pasar\'a a la historia y a Maite, mi eterna compi, la amistad que nos une no sabr\'a nunca de conceptos espaciales o temporales. No dispongo de todo el espacio que cada uno se merece y son muchos los que deber\'ian ser nombrados, pero no puedo pasar por aqu\'i sin nombrar, al menos, a mis queridos: MJ, Jos\'e Alberto, Seijas, JumaX y Manu Camb\'on. 

En el camino de la educaci\'on, los compa\~neros son parte fundamental, pero $\text{?`}$qu\'e ser\'ia de este camino sin la figura del docente? Tengo much\'isimo por lo que dar las gracias a los profesores que me he encontrado a lo largo de toda mi vida. Desde educaci\'on infantil a la universidad, todos ellos me han ense\~nado algo. Ahora que, con gusto, comparto en parte su profesi\'on, intento emular a muchos de ellos. Mentir\'ia si dijera que todos fueron extraordinarios, pero incluso los malos profesores te ense\~nan pr\'acticas que evitar. Si en este punto tuviera que hacer menci\'on de alguien en especial, en positivo, le dar\'ia las gracias a Montse. Ella probablemente se marchara sin saber c\'omo una f\'isica, profesora de matem\'aticas, pod\'ia marcar tan profundamente a sus alumnos con la sencilla herramienta de una docencia exquisita y cercana.

Por \'ultimo, tengo much\'isimo que agradecerle a toda mi familia, tanto a la que comparte mi sangre como a la que no. En mi casa tuve los mejores profesores posibles. Gracias a mis padres por hacer de m\'i lo que soy, hoy me siento su obra. Jam\'as encontrar\'e la manera de devolverles todo lo bueno que, con infinito cari\~no, han puesto en m\'i. Con su ejemplo, me ense\~naron una de las lecciones m\'as presentes e importantes en mi vida: ``siempre puede encontrarse tiempo para lo que se considera importante''. Gracias a mi hermano porque, salvando las distancias, en su caminar me ha facilitado una senda que seguir con ilusi\'on. A partir de un momento, la familia se empieza a escoger, y cada d\'ia que pasa me alegro de haber escogido bien. Gracias Mercedes, por estar a mi lado y por compartirlo todo conmigo. 

En definitiva, gracias a todos, a los nombrados y a los que no, por hacer de \'esta una \textit{traves\'ia} feliz.

{\clearpage \thispagestyle{empty}}
\chapter*{List of publications}

This thesis includes, at least partially, the research contained 
in the following works:
\begin{itemize}
\item \textbf{Carlos A. Plata},  Fabio Cecconi, Mauro Chinappi, and Antonio Prados,
\textit{Understanding the dependence on the pulling speed of the unfolding 
pathway of proteins}, Journal of Statistical Mechanics P08003 (2015).  
\item Alessandro Manacorda, \textbf{Carlos A. Plata}, Antonio Lasanta, Andrea 
Puglisi, 
and Antonio Prados, \textit{Lattice models for granular-like velocity fields: 
hydrodynamic description}, Journal of Statistical Physics
 \textbf{164}, 810 (2016).
\item \textbf{Carlos A. Plata}, Alessandro Manacorda, Antonio Lasanta, Andrea 
Puglisi, 
and Antonio Prados, \textit{Lattice models for granular-like velocity fields: 
finite-size effects}, Journal of Statistical Mechanics 093203 (2016).
\item \textbf{Carlos A. Plata} and Antonio Prados, \textit{Global stability and  
$H$-theorem in
 lattice models with nonconservative interactions}, Physical Review E \textbf{95},
 052121 (2017).
\item \textbf{Carlos A. Plata} and Antonio Prados, \textit{Kovacs-like memory 
effect in 
athermal systems: linear response analysis}, Entropy \textbf{19}, 539 (2017).
\item \textbf{Carlos A. Plata} and Antonio Prados, \textit{Modelling the unfolding 
pathway of bio\-mo\-le\-cu\-les: theoretical approach and experimental prospect}. 
In Luis L. Bonilla, Efthimios Kaxiras, 
and  Roderick Melnik (editors),\textit{Coupled Mathematical Models for Physical and Biological Nanoscale Systems and Their Applications}, Springer Proceedings in Mathematics and Statistics \textbf{232}, 137 (Springer, 2018).
\item \textbf{Carlos A. Plata}, Zackary N. Scholl, Piotr E. Marszalek, Antonio 
Prados,
\textit{Relevance of the speed and direction of pulling in simple modular 
proteins}, Journal of Chemical Theory and Computation \textbf{14}, 2910 (2018).	
\end{itemize}
Other works that are not included in the thesis are:
\begin{itemize}
\item Antonio Prados and \textbf{Carlos A. Plata}, \textit{
Comment on ``Critique and correction of the currently accepted solution of the infinite spherical well in quantum mechanics'' by Huang Young-Sea and Thomann Hans-Rudolph}, Europhysics Letters \textbf{116}, 60011 (2016).
\end{itemize}

{\clearpage \thispagestyle{empty}}
\addtocontents{toc}{\protect\setcounter{tocdepth}{-1}}
\tableofcontents
\addtocontents{toc}{\protect\setcounter{tocdepth}{3}}


{\clearpage \thispagestyle{empty}}
\mainmatter

\chapter{Introduction}	
\label{ch:intro}
\newcommand{\eq}{\text{eq}}
\newcommand{\prest}{\mathcal{P}}
\newcommand{\tder}[1]{\partial_t {#1}}
\newcommand{\vv}{\bm{v}}
\newcommand{\rr}{\bm{r}}
\newcommand{\HCS}{\text{\tiny HCS}}
\newcommand{\USF}{\text{\tiny USF}}
\newcommand{\thr}{\text{th}}
\newcommand{\st}{\text{s}}
\newcommand{\av}{\text{av}}
\newcommand{\ini}{\text{ini}}

\pagestyle{mystyle}  

How...? Why...? Curiosity is a natural instinct common to the whole
mankind.  Questioning is the very first step in any intellectual
process, either science in general or physics in particular make no
exception. Statistical mechanics was born to answer a question: how is
the macroscopic world we see related to their microscopic components?
Bernoulli, Maxwell, Boltzmann, Gibbs... all of them helped to
establish the foundations of statistical mechanics.  Nevertheless, the
answer is not complete nowadays; fortunately for statistical
physicists, there is still a lot of work to do. On the one hand, the
scope is getting broader. Right now, we attempt to understand
molecular biophysics, ecology, social sciences, and much more with the
mathematical tools of statistical mechanics. On the other hand, the
``right'' theoretical framework for nonequilibrium statistical
mechanics, in contrast to its equilibrium
counterpart, is still under active development.

This thesis is devoted to the analysis, through the lens of
statistical mechanics, of two simple models motivated within two quite
different fields: biophysics and kinetic theory of granular gases.
The use of simple models for understanding, reproducing and predicting
nature is a cornerstone in physics. The goal of this kind of modeling
is to catch the essence of a complex system with the minimal,
simplest, possible ingredients.  The advantage of this approach is
twofold. First, simplicity enables a (more) rigorous mathematical treatment,
 leaving the number of necessary approximations to a
minimum. Second, the low number of ingredients allows us to isolate
the features of a system that are responsible for the emergence of a
certain behavior.

Our work is divided into two parts, corresponding to the two aforementioned
models. As stated above, we study them with the usual tools of
statistical mechanics. That means that, depending on our level of description,
our starting point is either Langevin-type equations, for the 
description of fluctuating physical quantities, or Fokker-Planck/master equations, 
for the description of probability density functions.
 On the one hand, in part \ref{part:part-bio},
 we put forward and study a elasticity model for modular proteins capable
of predicting the unfolding pathway of these macromolecules. On the other hand,
we analyze a lattice model mimicking the main features of shear modes
in granular gases in part \ref{part:granular}.

\section{Biophysics}

Biophysics is a relatively new scientific discipline. Its evident etymology
gives us a neat clue of the scope it deals with. It has to do with 
the physics of the 
biological systems. The wide range of length scales covered by biosystems makes it
natural to distinguish among several subfields within biophysics, which study
systems going from biomolecules, as DNA or RNA, to ecosystems at global scale. 

At first glance, one could argue that physics and biology seem not to
share a lot in common. In principle, physics is more conceptual and
``simplistic'', whereas biology tries to describe life in all
detail. Traditionally, this has led to two different approaches in
biophysics: the biologist's and the physicist's. In the first, biology
borrows tools from physics, either experimental or theoretical ones,
in order to analyze the biological system of interest.  In the latter,
biology provides the system to be analyzed, which is useful to
elucidate new physical phenomena.  These definitions of different
approaches stem from quite ``selfish'' standpoints and are getting
obsolete nowadays. Differences between the biologist's and the
physicist's approach have become subtler, with the borders between the
different sciences blurring more and more with time. Currently, the
most frequent view is a unified but multidisciplinary approach.

As stated above, there are several subfields within biophysics
depending on the length scale of interest. Molecular biophysics focus
on the study of biomolecules: their structure, function, and
dynamics. Two main kinds of biomolecules have been analyzed in this
context: nucleic acids (DNA and RNA) and proteins. Our understanding
of their elasticity properties is a essential step forward in our
comprehension of some of the basic mechanisms underlying how the cell
works.  Throughout the first part of this thesis, we focus on the
study of elastomechanical properties of proteins.

Proteins are, roughly, chains of amino acids linked by peptide bonds.
Amino acids are organic compounds, composed of an amine and a carboxylic acid
group, which makes any protein to have a C-terminus and a
N-terminus. There are 20 amino acids, which differ from each other in
their residue. It is the residue that gives each amino acid its
peculiarity, so to say. Some residues are polar and thus hydrophilic,
others are nonpolar and thus hydrophobic. Some of them are charged,
either positively or negatively. This is important for the
spatial arrangement of the protein, as explained below.

\begin{figure}
  \centering
 \includegraphics[width=0.75 \textwidth]{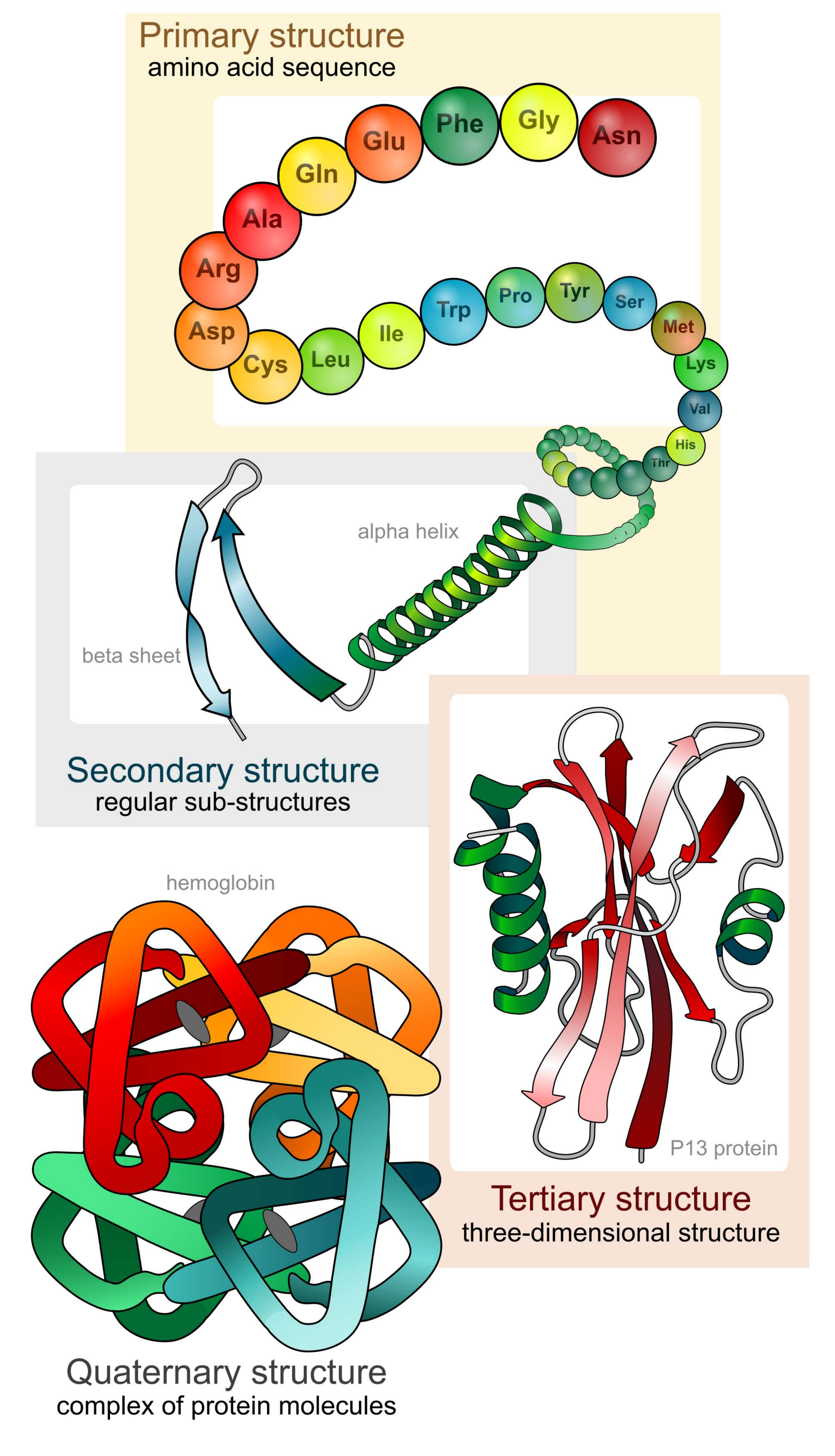}
 \caption{Visual of the different levels of description
 distinguished in the study of proteins, from primary to quaternary
 structure. In the primary structure, the different amino acids are usually
 denoted by a three letter code. Image taken from \cite{fig-structure-url}.}
  \label{ch1_fig:structures}
\end{figure}

Proteins are extraordinary complex systems, and thus they are studied
from four levels of description that are called structures. The
primary structure studies the particular sequence of amino acids: in
other words, the primary structure is determined by the ordered list
of their corresponding residues.  The secondary structure deals with the formation of stable substructures, mainly driven by
hydrogen bonding. There are two of these structures: $\alpha$-helices,
which have a coiled up shape, and $\beta$-sheets, which have a zig-zag
shape.  The tertiary structure provides the tridimensional arrangement
of the protein, which is mainly driven by the interactions between the
residues. For instance, hydrophilic residues prefer to point outwards,
closer to water, whereas hydrophobic residues prefer to point inwards,
further from water. In addition, there are also disulphur (covalent)
bonds between the thiol side chain of cysteine, van der Waals
interactions between nonpolar residues, ionic bonds between charged residues,
etc. Finally, the quaternary structure takes into account the
conformation of complex proteins comprising several polypeptide
chains. The aforementioned different levels of
structure are visualized in figure \ref{ch1_fig:structures}.

\begin{figure}
  \centering
 \includegraphics[width=0.9 \textwidth]{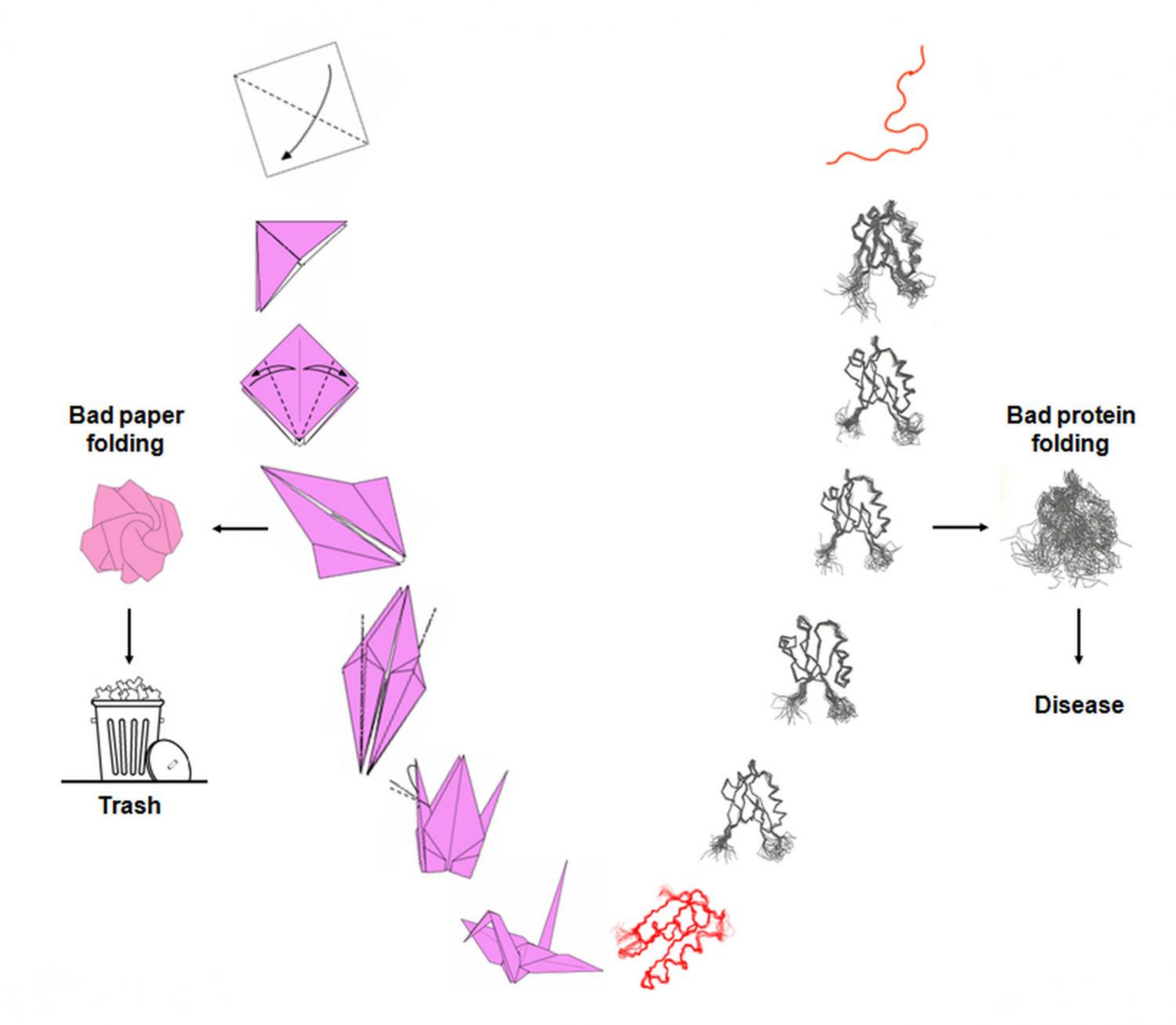}
 \caption{Origami analogy of the folding process in proteins. 
 Misfolded states are responsible for different diseases. 
 Image taken from \cite{fig-origami-url}.}
  \label{ch1_fig:origami}
\end{figure}

One of the burning issues in biophysics is the folding and unfolding
of proteins.  Why? On the one hand, most proteins in the body work
properly just in their folded state.  Nevertheless, there are
misfolded states, metastable in a physical language; proteins in these
states are responsible for some diseases as Alzheimer's, Parkinson's or
the bovine spongiform encephalopathy \cite{So03,Ha17}.  This fact can
be intuitively understood with the nice parallelism between protein
folding and origami figures depicted in figure
\ref{ch1_fig:origami}. It is when the mechanism responsible for
discarding the misfolded proteins---that is, throwing them into the
trash---does not  properly work that these diseases appear.  On the
other hand, there are also proteins with mechanical functions that
unfold during the extension of muscles. Hence, it is natural that a
huge community of biophysicists tries to improve our current
understanding of the processes of folding and unfolding
\cite{TOMyH10}.

\subsection{Single-molecule experiments}

The development of the so-called single-molecule experiments in the
last decades has triggered a whole new area of investigation on the 
elastomechanical properties of biomolecules \cite{R06,KyL10,MyD12,HyD12}. 
Up to that breakthrough, experiments were carried out in bulk. In bulk
experiments,
many particles are involved and thus the only information obtained was about
average and collective behavior.

\begin{figure}
  \centering
 \includegraphics[width=0.8 \textwidth]{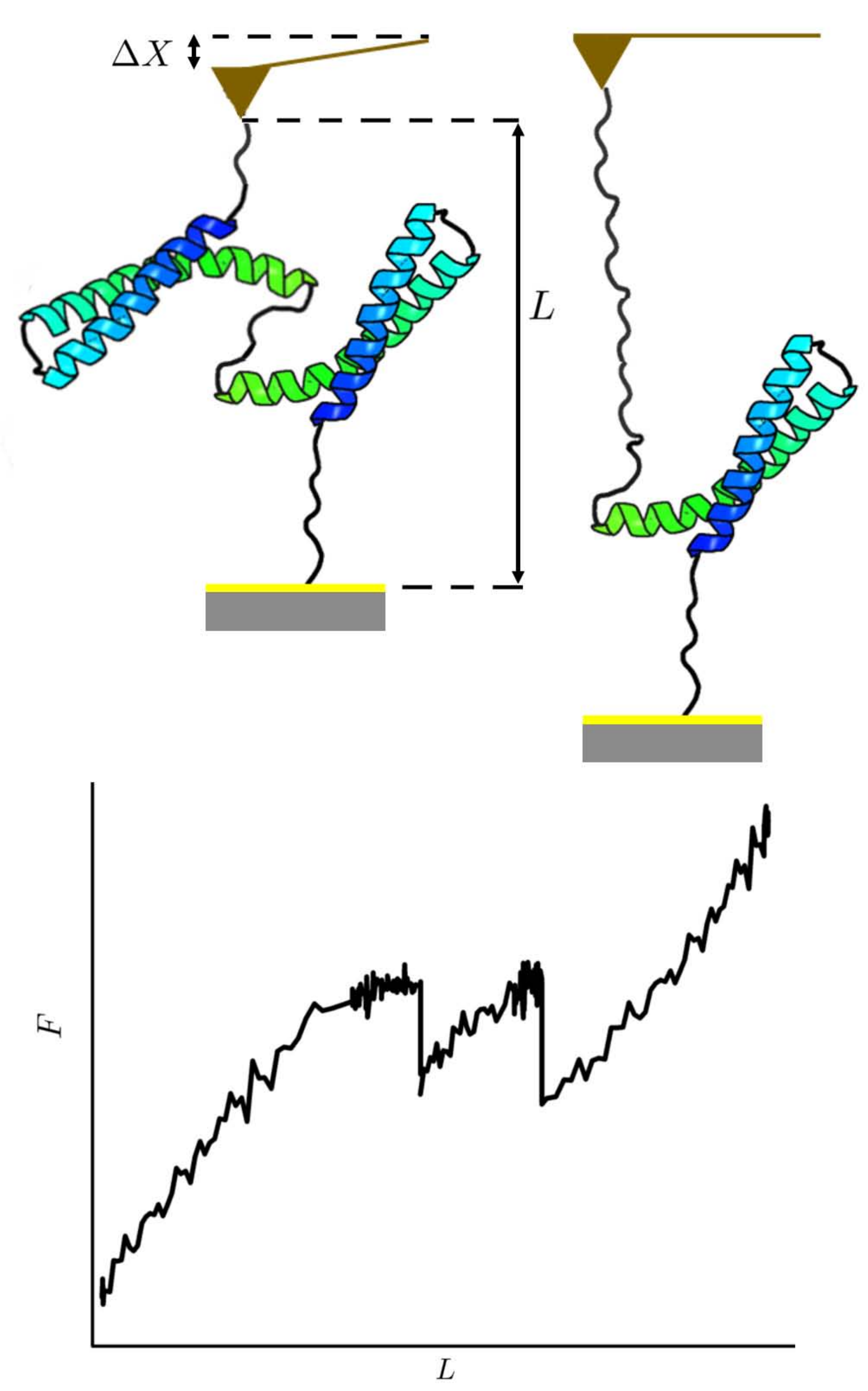}
 \caption{(Top) Sketch of the experimental setup in an AFM
   experiment with a modular biomolecule comprising two modules. 
   On the left, the position of the platform has been 
   shifted, producing an elongation  $L$
   of the molecule and bending the cantilever a magnitude
   $\Delta X$. On the right, the force is almost relaxed because of
   the unraveling of one of the modules. (Bottom) Typical force-extension
   curve output of the length control AFM experiment above. The rips in the force
    account for the unfolding of the modules.}
  \label{ch3_fig:sketch-experiment}
\end{figure}

The most used single-molecule techniques are laser optical tweezers
(LOT) and atomic force microscopy (AFM). In the LOT case, the molecule
is caught between two beads that are optically trapped by lasers. In
turn, in the AFM case, the molecule is tightened between a subtract
and the tip of a cantilever.  AFM excels because of its extensive use
and, specifically, has played a crucial role in the study of modular
proteins
\cite{COFMBCyF99,FMyF00,HDyT06}. Figure~\ref{ch3_fig:sketch-experiment}
shows a sketch of the experimental setup in a pulling experiment of a
molecule comprising two modules. The biomolecule is stretched between
the platform and the tip of the cantilever. The spring constant of the
cantilever is $k_{c}$, which is usually in the range of $10-100$ pN/nm
\cite{AyA06}.  The stretching of the molecule makes the cantilever
bend by $\Delta X$, and then the force can be recorded as
$F=k_{c}\Delta X$.  The total length of the whole system $\Delta X+L$,
is the sum of the bending of the cantilever and the molecule's
elongation.

Usually, AFM can operate in two modes depending on the control
parameter, either length or force. In length control experiments, the
position of the platform where the sample rests is controlled by a
piezoelectric material and the resulting force is measured.  In force
control experiments, the force is controlled by a feedback algorithm
and the length is recorded.  Therefore, in both modes the output of
the experiment is a force-extension curve. This force-extension curve
provides a fingerprint of the elastomechanical properties of the
molecule under study.

Here, we focus on length control experiments with ``modular
biomolecules''.  With this general terminology we allude to both
polyproteins \cite{CSMGCyB15,MyB17} (proteins comprising smaller
protein modules or domains) and structurally simpler proteins with
intermediate states stemming from the unfolding of stable
substructures named ``unfoldons'' \cite{ByR08,GMTCyC14}, see below.  The
heterogeneity of natural polyproteins makes it quite complicate to
study them.  For that reason, the generation of artificial engineered
homopolyproteins \cite{BBTBSRyC03,STyC09}, proteins composed of
identical (or very similar) repeats, has been a milestone in the
advancement of single-molecule experiments.
 
When a modular biomolecule is pulled in a length control AFM
experiment, a sawtooth pattern comes about in the force-extension
curve \cite{COFMBCyF99,FMyF00,HDyT06}, as sketched in figure
\ref{ch3_fig:sketch-experiment}.  The force generally increases with
the length as an indication of the resistance of the biomolecule to
stretch under the applied mechanical load. However, at certain values
of the length, there are almost vertical ``force rips'', marking the
unfolding of one of the units: its abrupt unfolding entails a force
relaxation, similar to the one found when untying a knot in a
rope. Probably, due to its length and relatively stiff nature, one of
the most paradigmatic force-extension curve is that of
homopolyproteins comprising several immunoglobulin domains of titin,
which is the largest known protein in vertebrates \cite{KyL11}.

\subsection{Theoretical developments} \label{ch1_sec:1_th_dev}

Biomolecules are particularly appealing systems from a statistical
mechanics perspective. Consider that $N$ is the number of atoms that
the system comprises.  In biomolecules, we have $1 \ll N \ll N_{A}$,
 $N_A$ being the Avogadro number.  Since relative fluctuations typically
scale with $1/\sqrt{N}$, theorists are interested in biomolecules as a
perfect laboratory for the development of the thermodynamics of small
systems. Herein, we have enough constituent particles to use
statistical mechanics arguments, but the fluctuations are still really
important \cite{Hi02}.

One of the most relevant achievements made by the thermodynamics of
the small systems is the derivation of fluctuation theorems. They link
equilibrium observables of the system with work functionals in
irreversible, nonequilibrium, processes. The first of these theorems
is given by Jarzynski equality \cite{Ja97,Ja97b}, which was later
generalized by Crooks \cite{Cr98,Cr99}. Starting from work
measurements in single-molecule experiments with biomolecules
\cite{CRJSTyB05}, these relations have been used to reconstruct
their free energy landscapes.

\begin{figure}
  \centering
 \includegraphics[width=0.75 \textwidth]{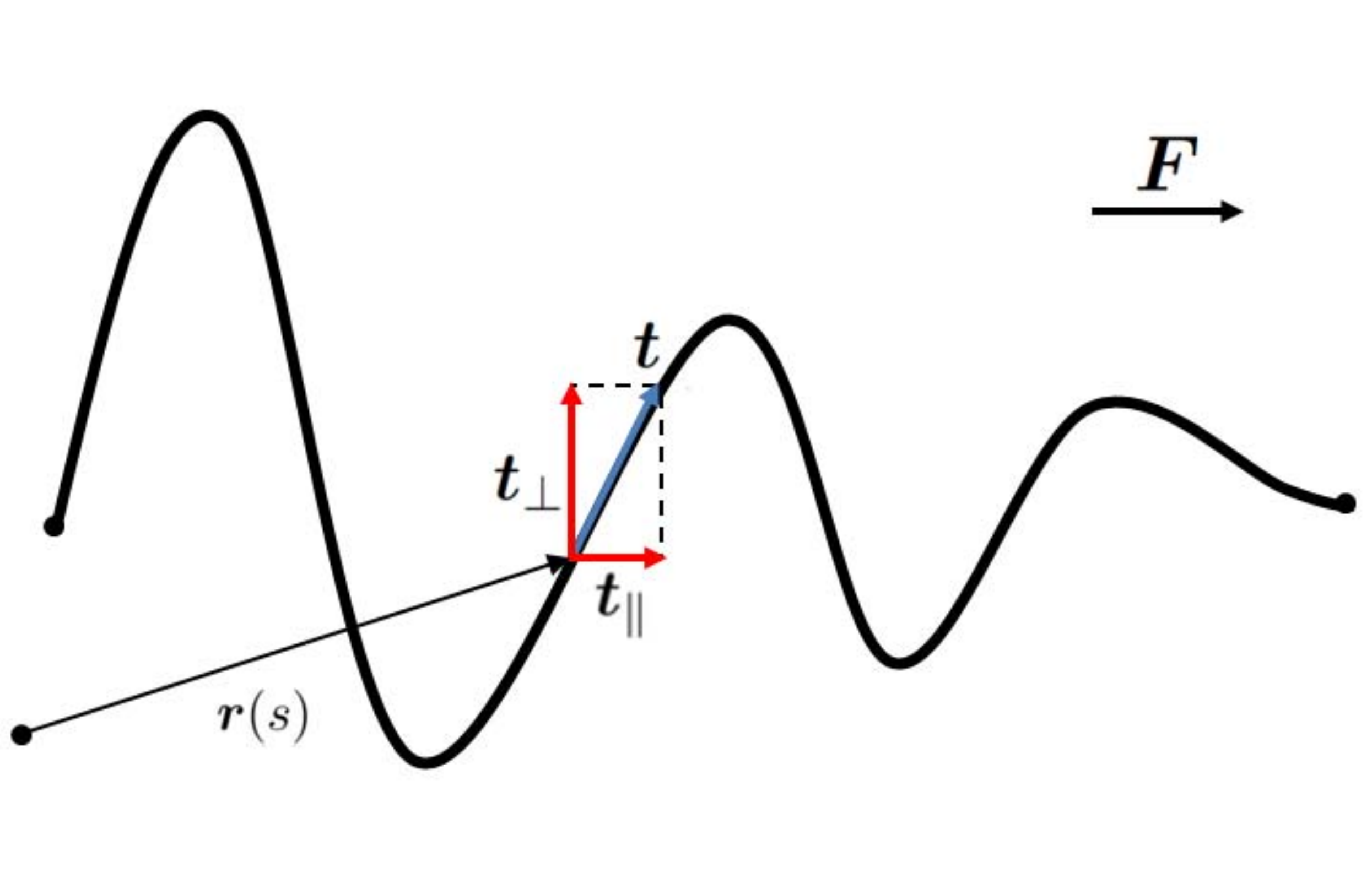}
 \caption{Sketch of the WLC model. The protein chain tends to align
 with the pulling force $\bm{F}$.}
  \label{ch1_fig:WLC_sketch}
\end{figure}

Polymer physics provides the two most paradigmatic elasticity models
of biomolecules: the freely-jointed chain (FJC) and the worm-like
chain (WLC) \cite{Fl89,Ru03}.  The main goal of these models is to
give an equilibrium force-extension curve for the system. The FJC
model considers a concatenation of rigid rods of fixed length with no
internal interaction at all, whereas the WLC emerges after considering
a continuous chain with elastic energy due to its bending,
as sketched in figure \ref{ch1_fig:WLC_sketch}.  Let
$\bm{r}(s)$ be the parametrization of the curve describing the polymer
as a function of its arc length,  the 
unitary tangent vector of the chain is given by
\begin{equation}
\bm{t}=\frac{\partial \bm{r}}{\partial s}.
\end{equation}
This vector can be decomposed into the perpendicular and parallel to
the force directions, with components $\bm{t}_{\perp}$ and 
$\bm{t}_{\parallel}$, respectively. Specifically, we have that
\begin{equation}
 \bm{t}_{\parallel}=(\bm{t}\cdot\bm{u}_{\parallel})\bm{u}_{\parallel},
 \qquad \bm{u}_{\parallel}=\frac{\bm{F}}{|\bm{F}|}, 
\end{equation}
and $\bm{t}_{\perp}=\bm{t}-\bm{t}_{\parallel}$. The curvature $\kappa$
is defined by
\begin{equation}
\kappa\equiv \left| \frac{\partial\bm t}{\partial s} \right|=\left| \frac{\partial^2 \bm{r}}{\partial s^2} \right|.
\end{equation}

The energy of the WLC model is given by
\begin{equation} \label{ch1_eq:WLC-energy}
H=\frac{1}{2} k_BT P \int_{0}^{L_c} \!\!\!\! ds \, \kappa^2(s)-F L.
\end{equation}
The first term stands for the energy due to the bending of the
polymer. Therein, $k_B$ is the Boltzmann constant, $T$ is the
temperature, and $P$ is a parameter called the persistence length that
gives the characteristic length scale for bending. The longer the
persistence length, the larger the bending contribution of the energy. 
The maximum value of $s$ is the contour length $L_c$, which
corresponds to the length of the fully extended polymer. The second
term on the rhs of \eqref{ch1_eq:WLC-energy} stands for the energy
associated with the pulling force, where
\begin{equation}
L= \int_{0}^{L_{c}} ds \, \bm{t}\cdot\bm{u}_{\parallel}
\end{equation}
is the projection of the length of the polymer onto the force
direction. Of course, $|L|$ is upper bounded by the contour length
$L_{c}$, $L=L_{c}$ in the fully extended configuration for which
$\bm{t}_{\perp}=\bm{0}$ ($\kappa=0$). 

Both the equilibrium force-extension curves of the FJC and WLC models
give a harmonic response for small enough stretching, that is
$F \propto L$. On the contrary, in the limit of strong pulling the
extension of the system approaches its contour length $L_c$ and the
force diverges as either $(L-L_c)^{-1}$ for the FJC model, or
$(L-L_c)^{-2}$ for the WLC model.  Both the FJC and WLC models have
been used for fitting real experiments with biomolecules, giving
reasonably good results \cite{MyS95,SFyB92,BMSyS94}. 
Specifically, the following WLC fit \cite{MyS95}
\begin{equation}\label{eq:WLC-force-ext}
\frac{FP}{k_B T}= \frac{L}{L_c}+ \frac{1}{4}\left( \frac{L_c}{L_c-L} \right)^2-
\frac{1}{4}, 
\end{equation}
is asymptotically valid along all the length range, and it is
intensively employed in the literature.

However, the aforementioned paradigmatic models do not take into
account the internal structure of the chain. In fact, the internal
structure of the biomolecule is responsible for its different states,
folded or unfolded for instance.  A coarse-grained modeling usually
involves considering each unit within a macromolecule as a two-state
system, which can be in either a folded or an unfolded state. To
account for this, some models \cite{RFyG98,ByS05,MHyM01} consider a
WLC with several possible values of the contour lengths: each branch
of the force extension curve is fitted by a WLC model with a different
value of contour length.  Transitions between folded and unfolded
states typically follow the development of Kramers theory
\cite{Kr40,HTyB90} carried out by Bell \cite{Be78} and Evans
\cite{EyR97}. The Bell-Evans expression provides the transition rates
between states, given the applied force and the details of the free
energy barrier.

\begin{figure}
  \centering
 \includegraphics[width=0.9 \textwidth]{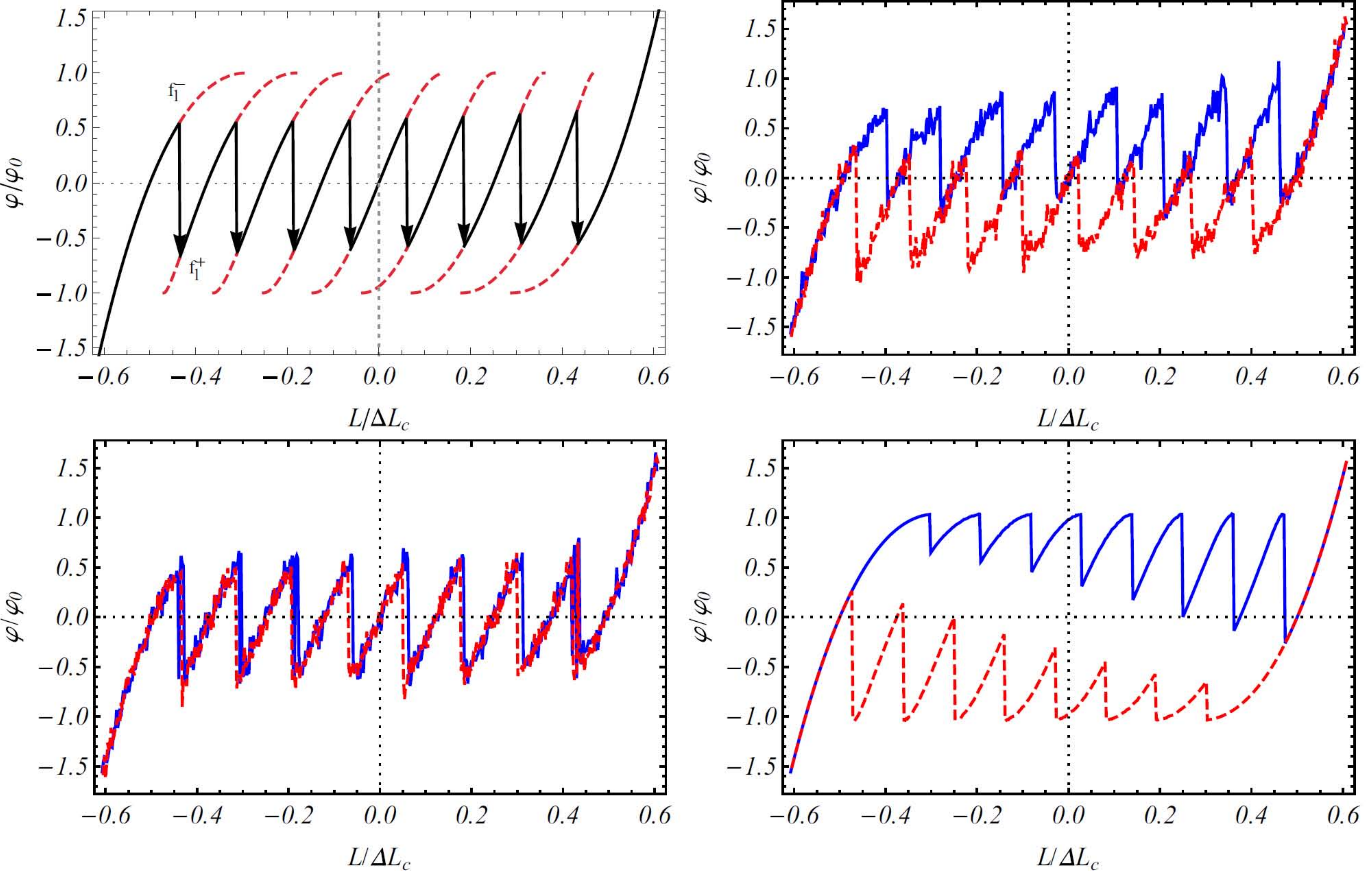}
 \caption{(Top left) Equilibrium force rips in the force-extension
   curve for a system with 8 units (length-control). Different colors
   are used to represent the stability: solid black for stable parts,
   dotted red for metastable parts and black arrows for force rips. In
   a quasistatic pulling process, the system follows the solid black
   curve with a series of first-order transitions in the force.  (Top
   right) Hysteresis cycle for a system composed of 8 units, at a
   relatively high pulling speed and moderate temperature.  Two traces
   are plotted: solid blue for unfolding and dashed red for refolding.
   (Bottom left) Same plot as in the top right panel, but for a
   smaller pulling velocity. Aside from thermal fluctuations, the
   system almost sweeps the equilibrium curve: the pulling is
   basically quasistatic (Bottom right) The same plot as in the
   bottom left panel, but for an almost vanishing temperature. Thermal
   fluctuations are so small that the system approaches the $T=0$
   behavior (adiabatic pulling): therein, the branches are swept up to
   the end of the metastability region. Taken from \cite{BCyP15}.}
  \label{ch1_fig:Prados-sawtooth}
\end{figure}

Quite recently, some more theoretical models
\cite{PCyB13,BCyP14,BCyP15}, closer to the approach that will be
followed in this thesis, have been proposed to analyze the elasticity
of modular biomolecules. These models successfully explain the
sawtooth pattern observed in the experiments: interestingly, an
equilibrium-statistical-mechanics theory is sufficient to understand
their emergence. In a nutshell, the models start from a free energy
where each module gives an additive contribution thereto, the
individual contributions being double well functions of the
corresponding unit's extension. By maximizing the probability
distribution function within the right statistical ensemble
(force-control or length-control), or equivalently minimizing the
corresponding free energy, the equilibrium force-extension curve of the
system is obtained. As a consequence of the different metastable
configurations (folded or unfolded) for each module, the
force-extension curve presents different branches, see top left panel
of figure \ref{ch1_fig:Prados-sawtooth}.

When dynamics is incorporated, hysteresis processes might appear. In figure \ref{ch1_fig:Prados-sawtooth}, some force-extension
curves of a system of 8 units for different velocities are shown. The
hysteretic behavior strongly depends on the pulling rate
employed. More specifically, an interesting interplay between the
pulling velocity and the temperature is observed. At a certain value
of the temperature, the system mainly sweeps the equilibrium
force-extension curve if the pulling speed is slow enough: this is the
quasistatic regime of pulling (bottom left). However, for higher
velocities, hysteresis comes about (top right). This phenomenon is
accentuated as the temperature is lowered, since thermally activated
transitions are not possible in ``cold'' systems.  This allows the
system to sweep the whole metastable branches (bottom right), a regime
that is usually said to correspond to the ``maximum hysteresis path''
\cite{BZDyG16} (also adiabatic pulling \cite{BCyP15}). 

In fact, the maximum hysteresis path regime described above stems from
the interplay between the pulling velocity and the temperature: the
pulling velocity must be slow enough to allow the system to move over
the (metastable) equilibrium branches but fast enough to prevent the
system from going from the folded to the unfolded basin (or vice
versa) by thermal activation. This regime will be of crucial relevance
in the development of the theory presented in part \ref{part:part-bio}
of this thesis.

\subsection{Unfolding pathway and its pulling dependence} \label{ch1_sec:unfold_path}

The unfolding pathway is, roughly, the order and the way in which the
structural blocks of a macromolecule unfold.  The force-extension
curve obtained in single-molecule experiment characterizes the
elastomechanical behavior of the macromolecule and provides basic and
essential information about the unfolding pathway.  Some studies show
that the pulling velocity plays a relevant role in determining the
unfolding pathway, see for example
\cite{HDyT06,LyK09,GMTCyC14,KHLyK13}. 

Different unfolding pathways have been observed depending on (i)
pulling direction, that is, which of the ends (C-terminus or
N-terminus) the molecule is pulled from and (ii) the pulling
speed \cite{HDyT06,LyK09,GMTCyC14,KHLyK13}.
Intuitively, it has been claimed that it is the inhomogeneity in the
distribution of the force across the protein, for high enough pulling
speeds, that causes the unfolding pathway to change
\cite{GMTCyC14}. Nevertheless, to the best of our knowledge, a theory that
explains this crossover is still lacking.

\begin{figure}
  \centering
 \includegraphics[width=0.9 \textwidth]{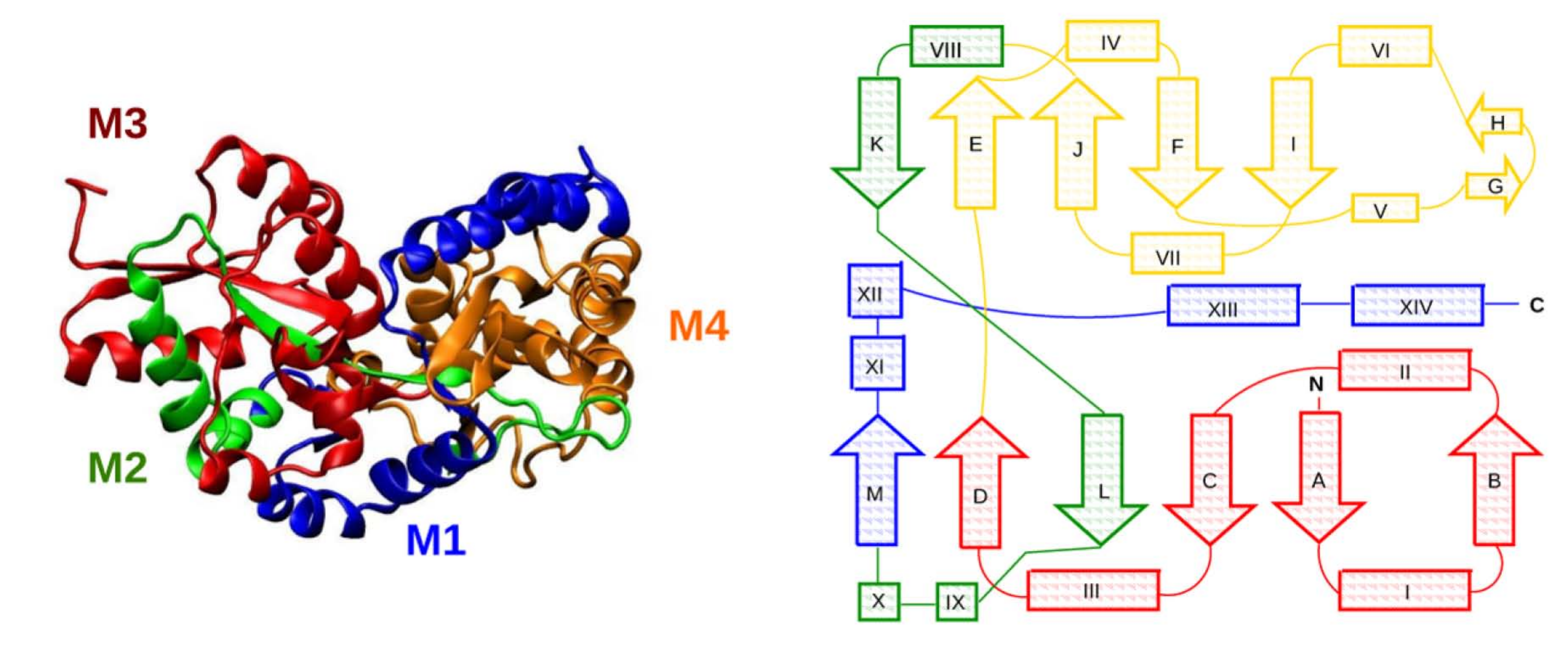}
 \caption{Structure of the Maltose Binding Protein (PDB ID: 4MBP). The
   color code identifies the four unfoldons: M1 blue; M2 green; M3
   red; M4 gold. (Left) Crystallographic structure. (Right)
   Topological diagram. Taken from \cite{GMTCyC14}.}
  \label{ch1_fig:Guardiani_0}
\end{figure}

Our interest in this problem was triggered by the dissenting unfolding
pathway observed in the maltose binding protein (MBP) in experiments
\cite{ByR08} and simulations for high pulling velocity
\cite{GMTCyC14}. This molecule unfolds in four steps, which led Bertz
and Rief to identify four internal substructures they call
``unfoldons'', see figure \ref{ch1_fig:Guardiani_0}.  More
specifically, AFM experiments allowed them to characterize these four
unfoldons in the MBP, labeling them as M1, M2, M3 and M4, with M1
being the closest to the C-terminus and M3 the closest to the
N-terminus \cite{ByR08}.  Pulling the molecule at a typical speed of
$10^{-9}$ nm/ps, Bertz and Rief found a well-defined unfolding
pathway: the weakest unfoldon (M1), that is, the one characterized by
the lowest opening force, unraveled first. Thereafter, the remainder
of the unfoldons opened sequentially, from weakest to strongest (M4).

\begin{figure}
  \centering
 \includegraphics[width=0.8 \textwidth]{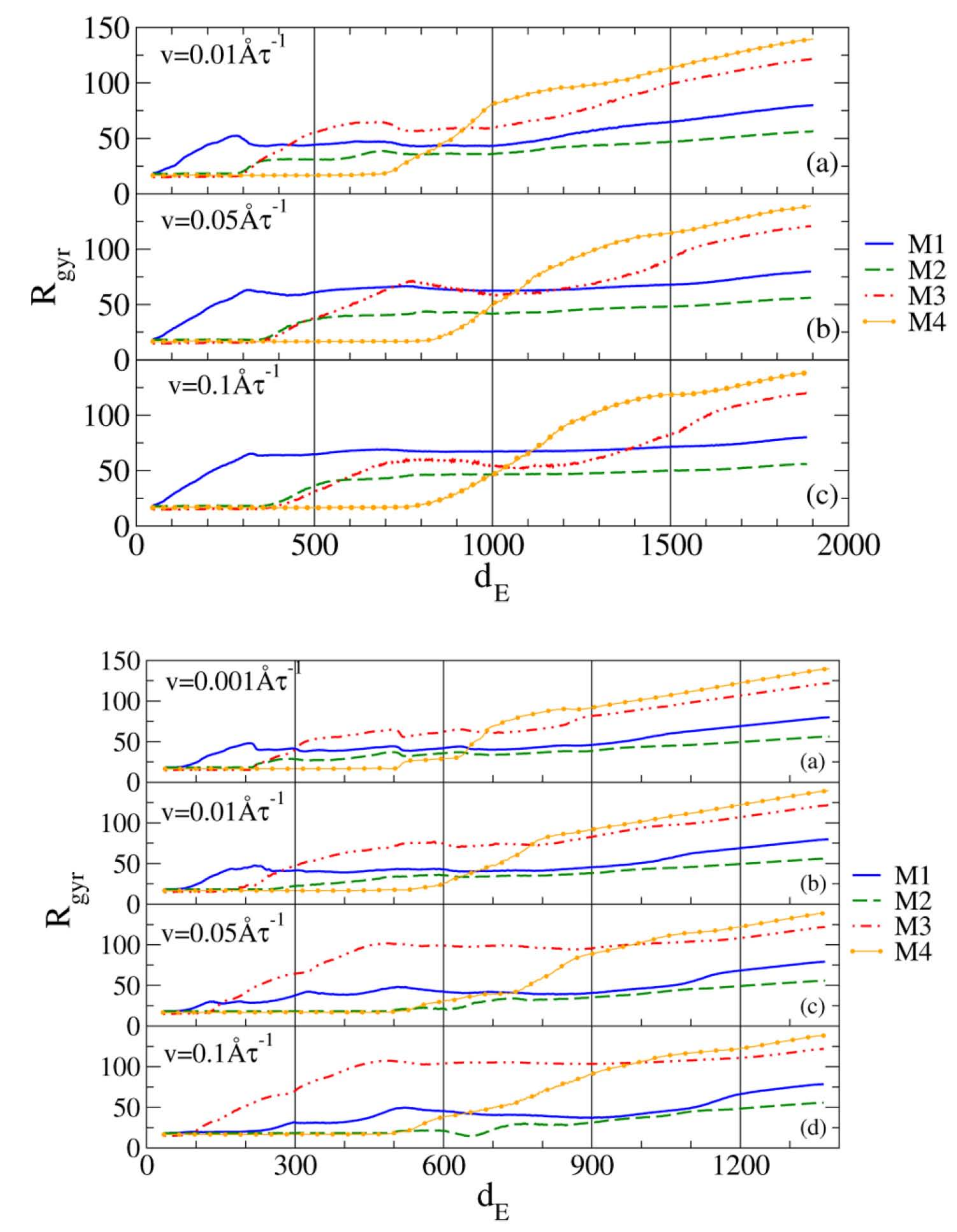}
 \caption{Evolution of the gyration radius for unfoldons M1, M2, M3,
   and M4 as functions of the end-to-end distance (\AA) at different
   pulling speeds.  The curves show averages over 50 runs of
   simulations of the G$\overline{\mbox{o}}$ model, for C-pulling
   (above) and N-pulling (bottom). For C-pulling, it is always M1, the
   weakest unfoldon, that opens first (solid blue
   line). Notwithstanding, for N-pulling, the first unfoldon that
   unravels depends on the pulling velocity: for slow pulling
   ($v=0.001 \text{\AA} \tau^{-1}$) it is still M1, but for fast enough
   pulling ($v=0.1 \text{\AA} \tau^{-1}$) it is M3, the closest to the pulled
   end as depicted in figure \ref{ch1_fig:Guardiani_0} (dot-dashed red
   line). Taken from \cite{GMTCyC14}.}
  \label{ch1_fig:Guardiani}
\end{figure}

Later, Guardiani et al. had a more detailed look into the unfolding of the
MBP \cite{GMTCyC14}. They showed, by means of a combination of
G$\overline{\mbox{o}}$ model simulations and steered molecular
dynamics, that the unfolding pathway is more complex and seems to
depend on both the velocity and direction of pulling. C-pulling
simulations of the G$\overline{\mbox{o}}$ model always showed a
pathway compatible with Bertz and Rief's experiment, see top panel of
figure \ref{ch1_fig:Guardiani}.  However, N-pulling simulations of the
G$\overline{\mbox{o}}$ model displayed a different behavior: for small
velocities, again Bertz and Rief’s pathway was found, but for high
enough pulling speed it was M3, the closest to the N-terminus, that
opened first, as shown in the bottom panel of figure
\ref{ch1_fig:Guardiani}. Steered molecular dynamics simulations
at a pulling speed of $5\cdot 10^{-3}$ nm/ps gave results that were
consistent with those from the G$\overline{\mbox{o}}$ model at high
velocities, showing the different pathways (M1 vs. M3 opening first)
depending on the pulled terminus (C vs. N).

In this context, Guardiani et al. introduced in \cite{GMTCyC14} a toy
model to qualitatively explain the observed pathways, which is the
starting point of part \ref{part:part-bio} of this thesis. This simple
model is akin to those employed in \cite{PCyB13,BCyP14,BCyP15} to
investigate the force-extension curves of modular proteins, their main
difference stemming by the incorporation---in the simplest way---of
the spatial structure of the chain into Guardiani et al.'s
model. Numerical simulations of the latter presented a phenomenology
that was compatible with both the G$\overline{\mbox{o}}$ model and the
steered molecular dynamics results \cite{GMTCyC14}.  One of our aims
is to develop a theoretical framework for this simple model, in order
to get a deeper understanding of the dependence of the unfolding
pathway on the pulling velocity and direction.

\subsection{Summary of part \ref{part:part-bio}}

Despite the large number of models developed to unravel the nature of
biomolecules, not everything is neat. In part \ref{part:part-bio}, we
attempt both to better understand and to predict the unfolding pathway
of modular biomolecules.

Our approach to this problem follows the philosophy presented above,
we put forward a simple model with the minimal ingredients to grasp
the essence of the system. We do so in chapter
\ref{ch:model_pbio}. Therein, we introduce the model, which portrays
the conformation of the protein into a 1d chain with different units.
Each unit contributes to the global free energy with a function that
only depends on its own extension with a double-well shape. The
perturbative solution of the dynamical system leads to our predicting
of the first module that  unfolds.  Numerical results are in excellent
agreement with our theoretical predictions.  Additionally, we carry
out some modifications of the model, in order to get closer to real
experiments.  The results are unchanged at the lowest (leading) order,
thus proving the robustness of our analysis.

Finally, we test the theoretical scheme developed in chapter
\ref{ch:model_pbio} in a simple biomolecule, which comprises two
coiled-coil structures, in chapter \ref{ch:moldyn_pbio}. This analysis
is carried out by means of steered molecular dynamics simulations of
the coiled-coil construct. First, characterizing the molecule is
required: in particular, we need to introduce a criterion for
considering it unfolded or folded. Such a criterion allows us to make a
systematic comparison with the theoretical framework.  A thorough
statistical study of the simulations output provides a significant
test of theory and validates the usefulness of the approach.

Some technical details that are omitted in the main body of this part
of the thesis are given in appendix \ref{ch2:appA}.

\section{Granular gases}

\begin{figure}
  \centering
 \includegraphics[width=0.8 \textwidth]{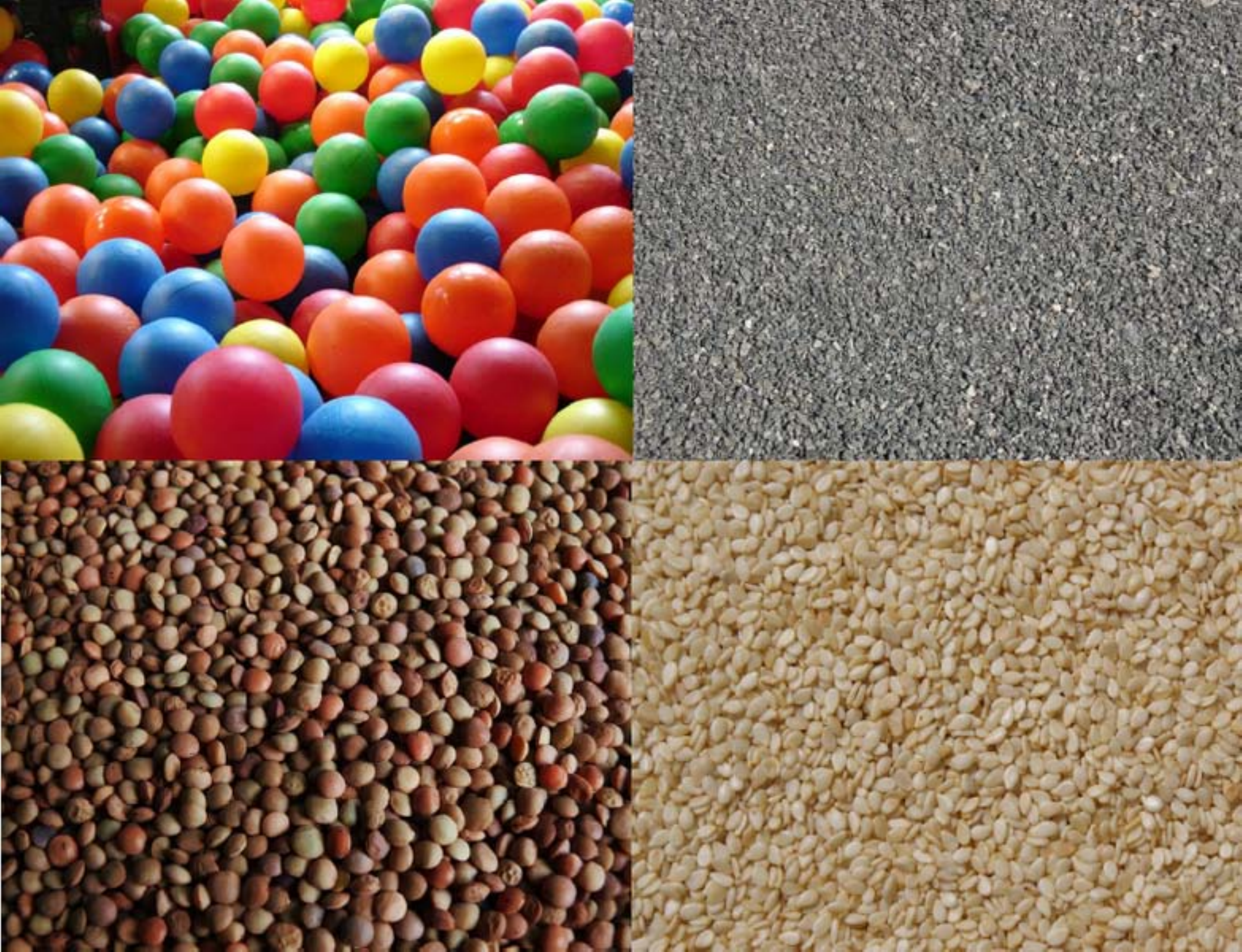}
 \caption{Some examples of granular matter: balls, gravel, lentils and
   sesame seeds.  Taken from \cite{fig-granular-url}.}
  \label{ch1_fig:granular_ex}
\end{figure}

A granular material is made of macroscopic particles that are called
grains \cite{JNyB96,PyL01}. They can be found almost everywhere, for
example dust, sand, seeds, pills, iceberg groups or asteroid
populations are all instances of granular matter. Improving our
knowledge of granular matter has a clear technical, industrial, and
even economic interest. To support this statement, it suffices to turn
our thoughts to transport and storage industry, agriculture or
construction. In figure \ref{ch1_fig:granular_ex}, some typical
examples of granular material are shown.

All granular materials share some typical properties. First, grains
are solid and macroscopic, that is, their dynamics is governed by
classical mechanics laws and they fill a space that is excluded to the
other grains. Second, their interactions are nonconservative in the
following sense: when two grains collide, some of their energy is
``lost'' into internal degrees of freedom, mainly due to deformation,
as heat. Third, the characteristic energy of a grain is much greater
than the thermal energy. Therefore, the temperature of the medium in
which the grains are immersed is largely irrelevant: the typical
energy to lift a grain by its own diameter is several orders of
magnitude larger than the thermal energy \cite{OLLyN02}.  Indeed, in
granular fluids one can define a relevant granular temperature from
kinetic theory, linked to velocity fluctuations. This granular
temperature has nothing to do with the ``conventional'' temperature,
which as already stated plays no role, but is related to the energy
injection mechanism that is needed to keep grains moving.

The result of enclosing a granular material and shaking it rapidly is
a fluidized granular material: the granular fluid that we have just
referred to above.  The amount of available space and the intensity of
the shaking determine the regime of fluidization \cite{Pu14}. When
interactions are dominated by two-particle instantaneous
(hard-core-like) collisions, one usually speaks of a \textit{granular
  gas}. This regime is typically achieved in
experiments when the packing fraction is of the order of $\sim\! 1\%$
or less, and the peak acceleration is many times the gravity
acceleration. 

The gas regime has played a crucial role in the
development of granular kinetic theory: in the dilute limit, one may
retrace the classical molecular kinetic theory after having relaxed
the constraint of energy conservation \cite{ByP04}. Granular
collisions are, in fact, inelastic: this occurs because each grain is
approximated as a rigid body and the collisional internal dynamics is
replaced by an effective energy loss, usually characterized by a
normal restitution coefficient $\alpha \in (0,1]$. This is the smooth
hard-particle model: particles only interact when at contact, then the
component of the relative velocity along the direction joining their centers is
reversed and shrunk by a factor $\alpha$, whereas the other components
remain unchanged. Therefore, $\alpha=1$ corresponds to elastic
collisions, whereas $\alpha \to 0$ describes the completely inelastic
limit.

Most of granular kinetic theory rests upon many variants of the basic
model of smooth hard particles. Important variants include roughness
and rotation \cite{LHMyZ98,SKyS11,RSyK14,GSyK18,Sa18} as well as the
consideration of velocity-dependent inelasticity \cite{BSHyP96}.  
Notwithstanding, the simple model of smooth
inelastic hard spheres suffices to explain
the basic phenomenology of granular gases.  In this context, the
Boltzmann equation for inelastic hard spheres constitutes the
foundation of many investigations in the realm of granular phenomena,
with both numerical and analytical approaches \cite{NyE98}.  

Granular fluids exhibiting separation between fast microscopic scales
and slow macroscopic ones have led to several procedures to build a
granular hydrodynamics \cite{LSJyC84,BDKyS98}.  Note that the scale
separation hypothesis is less clear in the granular case, as compared
to molecular gases.  First, one has the spontaneous tendency of
granular gases to develop strong inhomogeneities even at the scale of
a few mean free paths. Second, a granular system is typically of
``small'' size: it is usually constituted by a few thousands of
grains~\cite{Go99,Ka99}. The latter limitation cannot be easily
relaxed even in theoretical studies: stability analyses have shown
that spatially homogeneous states are unstable for too large sizes in
some typical states \cite{NyE00}.

The intrinsic ``small'' size of granular gases makes it essential to
address another task: an adequate and consistent description of
fluctuations, which are always important in a small
system~\cite{Ei04,OyM53,LyL80}. The number of particles $N$, ranging
from $10^{2}$ to $10^{4}$, is large enough to make it possible to
apply the methods of statistical mechanics, but definitely much
smaller than the Avogadro number. Interestingly, this is also the case
for biomolecules, as stated before in section \ref{ch1_sec:1_th_dev};
the special relevance of fluctuations is a point that links the two
parts of this thesis.

Unfortunately, there is no general theory currently available for
mesoscopic fluctuations out-of-equilibrium.  Notwithstanding,
important steps in the deduction of a consistent fluctuating
hydrodynamics for inelastic hard spheres have been recently taken in
the context of kinetic theory \cite{BMyG09}. In this regard, the
quite broad framework of Macroscopic Fluctuation Theory \cite{BSGJyL01}
cannot be employed. In its current state of development, Macroscopic
Fluctuation Theory does not include macroscopic equations with
advection terms and momentum conservation, such as those in the
``granular'' Navier-Stokes equations.

In the last decades, lattice models have proved to be a flexible tool
to identify the essential steps in a rigorous approach to the
hydrodynamic limit, both at the average \cite{KyL99,KMyP82} and
fluctuating \cite{BSGJyL01} levels of description. Fluctuating
hydrodynamics in linear and nonlinear lattice diffusive models have
been recently investigated, both in the conservative
\cite{HyG09,HyG10,HyG09b,HyK11} and in the nonconservative cases for
the energy field \cite{SyL04,PLyH12a,PLyH11a,PLyH13,PLyH16}.  Later, a
lattice model, which in some simplified way mimics the velocity field
of a granular gas, has been put forward to incorporate
momentum conservation \cite{LMPyP15}.  This model is the central
pillar of our investigations in part \ref{part:granular} of this
thesis.

\subsection{Granular hydrodynamics}
\label{ch1_sec:hcs} 

Evolution equations for the ``slow'' fields, in space $\rr$ and time
$t$, density $n(\rr,t)$, velocity $\bm{u}(\rr,t)$, and granular
temperature $T(\rr,t)$ constitute the full granular
hydrodynamics. These equations can be derived from the Boltzmann
equation for inelastic hard spheres, through a Chapman-Enskog
procedure closed at the Navier-Stokes order \cite{BDKyS98,PyL01}.  For
generic dimension $d$, they are given by
\begin{subequations}
\label{ch1_eq:brey}
\begin{align}
\tder{n}+\nabla \cdot (n \bm{u})&=0,\\
\tder{\bm{u}}+\bm{u}\cdot \nabla \bm{u}+(nm k_B)^{-1}\nabla\cdot\prest&=0,\\
\tder{T}+\bm{u}\cdot\nabla T+\frac{2}{d n k_B}\left[\prest:(\nabla \bm{u})+\nabla \cdot \bm{q} \right]+\zeta T&=0,
\end{align}
\end{subequations}
in which $m$ is the mass of the particles.
The energy dissipation rate  is $\zeta(\rr,t)>0$, being $\zeta
\to 0$ in the elastic limit, while the pressure tensor $\prest(\rr,t)$ and the
heat flow $\bm{q}(\rr,t)$ read
\begin{equation} \label{linflux}
\prest_{ij}=p\delta_{ij}-\eta\left(\nabla_iu_j+\nabla_j u_i-\frac{2}{d}\delta_{ij}\nabla \cdot \bm{u} \right),\quad
\bm{q}=-\kappa \nabla T-\mu \nabla n.
\end{equation}
Additionally, the bulk pressure $p$, the
shear viscosity $\eta$, the heat-temperature conductivity $\kappa$ and
the heat-density conductivity $\mu$ are given by constitutive
relations \cite{PyL01}. We briefly
recall that the transport coefficients depend on the hydrodynamic
fields. Specifically, for hard-spheres one has that  $\eta \sim \sqrt{T}$, $\kappa \sim \sqrt{T}$, $\mu \sim
n^{-1} T^{3/2}$ and $\zeta \sim n \sqrt{T}$ \cite{PyL01}. 

This thesis is not an exhaustive investigation of granular
hydrodynamics and its rich catalog of possible stationary and
nonstationary regimes \cite{ByP04,Pu14}.  Our aim is to study the
model introduced in part \ref{part:granular}, validating its use for
the simplified investigation of some peculiar states in the granular
realm.  With this intention, we highlight three essential aspects that
give some contact points of the aforementioned model with actual
granular fluids. First, the existence of a spatially homogeneous
nonstationary solution, that is, the ``Homogeneous Cooling State''
(HCS). Second, the instability of such a state with respect to
perturbations with long enough wavelength (small enough
wavenumber). And third, the existence of the Uniform Shear Flow (USF)
stationary state, in which the energy loss due to collisions is
balanced---on average---by the heating brought about by the velocity
difference, that is, the shear, imposed between the system boundaries
\cite{GyS13}. All such aspects stem from a key difference with respect
to the hydrodynamics of molecular fluids, which is the presence of the
energy sink term $\zeta T$ in \eqref{ch1_eq:brey}.

When spatial homogeneity is assumed along with
periodic boundary conditions, and initial conditions $n(\rr,t=0)=n$,
$u(\rr,t=0)=0$, and $T(\rr,t=0)=T(0)$,~\eqref{ch1_eq:brey} are
reduced to 
\begin{equation}
\label{ch1_eq:dTdt}
\dot{T}(t)=-\zeta(t) T(t).
\end{equation}  
Since $\zeta(t) \propto T(t)^{1/2}$ for hard-spheres,
\eqref{ch1_eq:dTdt} leads to the well known Haff's law \cite{Ha83}
\begin{equation}
\label{cooling_temperature}
T_{\HCS}(t)=T(0)\left[1+\frac{\zeta(0) t}{2}\right]^{-2}.
\end{equation}
A different collisional model that is often used to simplify the
kinetic theory approach is the so-called gas of pseudo-Maxwell
molecules~\cite{Er81}: its peculiarity is that $\zeta(t)=\zeta(0)$
is constant and therefore Haff's law simplifies to an exponential decay,
 \begin{equation}
\label{cooling_temperature2}
T_{\HCS}(t)=T(0)\exp[-\zeta(0) t].
 \end{equation}
This spatially homogeneous solution, with 
monotonically decreasing temperature, is
generally called ``Homogeneous Cooling State'' (HCS). Of course, it can
be predicted at the more fundamental and general level of
the Boltzmann~\cite{BRyC96} or even Liouville\cite{BPGyM07} equations.

The HCS is not stable if the system size exceeds some critical
value. Spatial perturbations of the velocity and density fields are
amplified when the system is large enough \cite{MN93,NyE00}.  A linear
stability analysis shows that the fastest amplification usually occurs for
shear modes, which correspond to a transverse perturbation of the
velocity field. For instance, a nonzero $y$ component of $\bm{u}$
modulated along the $x$ direction, that is, $u_y(x,t)$. The critical
wavelength $L_c$ separating the stable from the unstable regime
depends upon the restitution coefficient as
\begin{equation} \label{ch1_eq:Lc}
L_c^{2} \propto (1-\alpha^2)^{-1}.
\end{equation}
Velocity perturbations are not really amplified, because the amplitude
of their fluctuations (temperature) always decay: the instability is
observed only when the rescaled velocity field
$u(\rr,t)/\sqrt{T_{\HCS}(t)}$ is considered.  Perturbations in the
other fields (density, longitudinal velocity and temperature), the
evolution of which are coupled with the velocity field, are also
amplified, but with a slower rate and for longer wavelengths.

There is a range of system sizes such that the only linearly unstable
mode is the shear mode~\cite{ByP04}. This entails that the velocity
field is incompressible and density does not evolve from its initial
uniform configuration. Such a regime may be observed for a certain
amount of time, longer and longer as the elastic limit is
approached. In two dimensions, \eqref{ch1_eq:brey} is obeyed with
constant density and, for instance, $u_x=0$ whereas the hydrodynamic
fields $u_{y}$ and $T$ only depend on $x$. In this situation, we have
that \eqref{ch1_eq:brey} reduces to
\begin{subequations}
\label{ch1_eq:brey2}
\begin{align}
\tder{u_y(x,t)}&=(nm)^{-1}\partial_x [\eta\partial_{x} u_y(x,t)],\\
\tder{T}(x,t)&=\frac{1}{n k_B}\eta[\partial_x u_y(x,t)]^2+\frac{1}{n}\partial_x [\kappa \partial_x T(x,t)]-\zeta T.
\end{align}
\end{subequations}
In section~\ref{sec:hydro-eq} we will see that our lattice model is well
described, in the continuum limit, by completely analogous equations.

It is interesting to put in evidence a particular stationary solution
of the system \eqref{ch1_eq:brey2}.  Seeking time-independent
solutions thereof, one finds
\begin{equation}
\label{ch1_eq:Couette-USF}
  \partial_x [\eta\partial_{x} u_y^{(\text{s})}(x)]=0, \quad
  \frac{\eta}{k_B}[\partial_x u_y^{(\text{s})}(x)]^2=-\partial_x [\kappa \partial_x T^{(\text{s})}(x)]+n\zeta T^{(\text{s})}(x).
\end{equation}
The general situation is that both the average velocity and
temperature profiles are inhomogeneous: this is the so-called Couette
flow state, which also exists in molecular fluids.  Nevertheless, in granular
fluids, there appears a new steady state in which the temperature is
homogeneous throughout the system, $T^{(\text{s})}(x)=T$, and the average
velocity has a constant gradient, $\partial_{x}u=a$: this is the
Uniform Shear Flow (USF) state, characterized by the equations
\begin{equation}
\label{ch1_eq:only-USF}
  \partial_{x}^{2} u_y^{(\text{s})}(x)=0,\quad
  \eta[\partial_x u_y^{(\text{s})}(x)]^2=nk_B\zeta T^{(\text{s})}.
\end{equation}
Such a steady state is peculiar of granular gases where the viscous
heating term is locally compensated by the energy sink term.  In
granular gases of hard spheres, it has been proven that the USF state
is linearly stable for perturbations in the direction of the shear
\cite{Ga06}.

\subsection{Irreversibility: $H$-theorem}\label{ch1_sec:H-th}

In thermodynamics and statistical mechanics, proving the global
stability of the equilibrium state---or, in general, of the relevant
stationary state---usually involves the introduction of a suitable
Lyapunov functional \cite{Ly92}. A Lyapunov functional of the
probability distribution function (PDF) has the following three
properties:
\begin{itemize}
\item[(i)] It is bounded from below.
\item[(ii)] It monotonically decreases with time.
\item[(iii)] Its time derivative vanishes only when the PDF is the
  equilibrium one.
\end{itemize} 
Therefore, in the long time limit, the Lyapunov functional must tend
to a finite value and thus its time derivative vanishes. As a
consequence, any PDF, corresponding to an arbitrary initial
preparation, tends to the equilibrium PDF. This rigorously proves that
the equilibrium state is irreversibly approached and said to be
globally stable.

One of the most relevant examples of such a Lyapunov functional is the
renowned Boltzmann $H$-functional. At the Boltzmann level of
description, the nonequilibrium behavior of a dilute gas is completely
encoded in the one-particle velocity distribution function
$f(\rr,\vv,t)$. After introducing the Stosszahlansatz or molecular
chaos hypothesis, Boltzmann derived a closed nonlinear
integro-differential equation for $f(\rr,\vv,t)$ governing its time
evolution \cite{Bo95}. Also, for spatially homogeneous states, he
showed that the functional
$H_{B}[f]=\int d\vv f(\vv,t) \ln\! f(\vv,t)$ has the three properties
of a Lyapunov functional.  This $H$-theorem shows that all solutions
of the Boltzmann equation tend in the long time limit to the Maxwell
velocity distribution. 

Thus, irreversibility naturally stems from a reversible
molecular picture \cite{Le93,Le93b}. Indeed, a key point for deriving
the $H$-theorem is the reversibility of the underlying microscopic
dynamics. This almost paradoxical interplay between reversibility and
irreversibility has not been entirely absent of controversy
\cite{Le99,Pr99,Ru99}. In an inhomogeneous situation, one has to
consider the spatial dependence of the one-particle distribution
function $f(\rr,\vv,t)$, and the above functional must be generalized
to
\begin{equation}\label{H-Boltzmann}
H_{B}[f]=\int d\rr \, d\vv f(\rr,\vv,t) \ln\! f(\rr,\vv,t).
\end{equation}
With an additional assumption about the smoothness of the walls of the
gas container, in order to avoid energy transport through them, it can
also be shown that \eqref{H-Boltzmann} is a nonincreasing Lyapunov
functional in the conservative case \cite{CyC90}.

In the realm of
Markovian stochastic processes,
we find another example of Lyapunov functionals. 
Therein, the stochastic process $X(t)$
is completely determined by its conditional probability density
$P_{1|1}(X,t|X_{0},t_{0})$ of finding the system in state $X$ at time
$t$, given it was in state $X_{0}$ at time $t_{0}$, and the
probability density $P(X,t)$ of finding the system in state $X$ at
time $t$ \cite{vK92}. Both probability densities
satisfy a master equation, but
with different initial conditions: the first verifies
$P_{1|1}(X,t_{0}|X_{0},t_{0})=\delta(X-X_{0})$, whereas for the latter
$P(X,t_{0})=P_{\ini}(X)$, with $P_{\ini}(X)$ corresponding to the
arbitrary initial preparation. 

When the Markovian stochastic process under scrutiny is irreducible or
ergodic, that is, every state can be reached from any other state by a
chain of transitions with nonzero probability, there is only one
stationary solution of the master equation. In physical systems, this
steady solution must correspond to the
equilibrium-statistical-mechanics distribution $P_{\eq}(X)$. Moreover,
a Lyapunov functional can be constructed,
\begin{equation}\label{H-function-master-eq}
  \mathcal{H}[P]=\int dX P_{\text{eq}}(X) \, g\!\left[\frac{P(X,t)}{P_{\eq}(X)}\right],
\end{equation}
where $g(x)$ is any positive-definite convex function
($g''(x)\geq 0$). It must be stressed that the proof of this
$H$-theorem for master equations rely only on the ergodicity of
the underlying microscopic dynamics. It is not necessary to assume 
detailed balance, which is connected with the reversibility  
of the underlying microscopic dynamics \cite{vK92}.

The most usual choice for $g$ is $g(x)=x\ln x-x+1$, which leads to the 
Kullback-Leibler divergence \cite{KyL51}
\begin{equation}\label{H-function-add}
\mathcal{H}[P]=\int dX P(X,t)\, \ln\!\left[\frac{P(X,t)}{P_{\eq}(X)}\right].
\end{equation}
This choice has the physical advantage of $\mathcal{H}[P]$ being ``extensive'': if the system at hand comprises two independent subsystems
$A$ and $B$, so that $dX\equiv dX_{A}dX_{B}$ and
$P(X)=P_{A}(X_{A})P_{B}(X_{B})$, one has that
$\mathcal{H}[P]=\mathcal{H}_{A}[P_{A}]+\mathcal{H}_{B}[P_{B}]$. 
This feature is desirable since usually one considers
$-\mathcal{H}$ to define a nonequilibrium entropy $S$.
In this way, the nonincreasing behavior of $\mathcal{H}$ leads
to a nondecreasing time evolution of $S$. Moreover, in this way $\mathcal{H}[P]$ remains
invariant upon a change of variables $Y=f(X)$ 
\cite{MPyV13,GMMMRyT15}.

Although the Boltzmann equation is not a master equation, we may
wonder why the expressions for $H_{B}$ in \eqref{H-Boltzmann}
and $\mathcal{H}[P]$ in  \eqref{H-function-add} are
different. Specifically, we may wonder why not writing
\begin{equation}\label{H-one-particle}
H[f]=\int d\rr \, d\vv f(\rr,\vv,t) \ln\! \left[\frac{f(\rr,\vv,t)}{f_{\eq}(\vv)}\right]
\end{equation}
for the Boltzmann equation, instead of $H_{B}[f]$. Up to now, we have
been implicitly considering the ``classic'' problem with elastic
collisions between particles, in which the system eventually reaches
thermodynamic equilibrium. Therein, the answer is trivial: since
$\ln f_{\eq}(\vv)$ is a sum of constants of motion, $H[f]-H_{B}[f]$ is
constant and both are utterly equivalent.\footnote{When nonconservative external forces are present and thus there is not an equilibrium state, the ``good'' Lyapunov functional has been shown to be of the form given by \eqref{H-one-particle}, but $f_{\eq}(\vv)$ has to be substituted with a time-dependent reference state $f_\text{R}(\vv,t)$ \cite{GSyB90}.}

The problem about the existence of an extensive $H$-functional is
quite relevant in nonequilibrium statistical mechanics.  If there is
one, it makes it possible to define a monotonically increasing
nonequilibrium entropy $-H$ that extends the Clausius inequality to
nonequilibrium states, as already stated above.  In this context, it
is important to stress that the final state is, in general, not an
equilibrium one but a nonequilibrium steady state. Thus, stationarity
holds but nonvanishing currents are allowed in the system.  In
addition, the equilibrium distribution $f_{\eq}$ in $H$ has to be
substituted with the stationary one $f_{\st}$. Due to its
intrinsically nonequilibrium nature, granular fluids is a benchmark
for these investigations.

In granular fluids, functionals $H[f]$ and $H_{B}[f]$ are no longer
equivalent: with the stationary PDF, $\ln f_{\st}$ is no longer a sum
of constants of motion. Indeed, for granular gases described by the
inelastic Boltzmann equation \cite{PyL01,Vi06}, there are some results
that hint at $H_{B}$ not being a ``good'' Lyapunov functional. Within
the first Sonine approximation, it has been proven that the time
derivative of $H_{B}$ does not have a definite sign in the linear
approximation around the steady state \cite{BCDVTyW06}.

\begin{figure}
  \centering
 \includegraphics[width=0.8\textwidth]{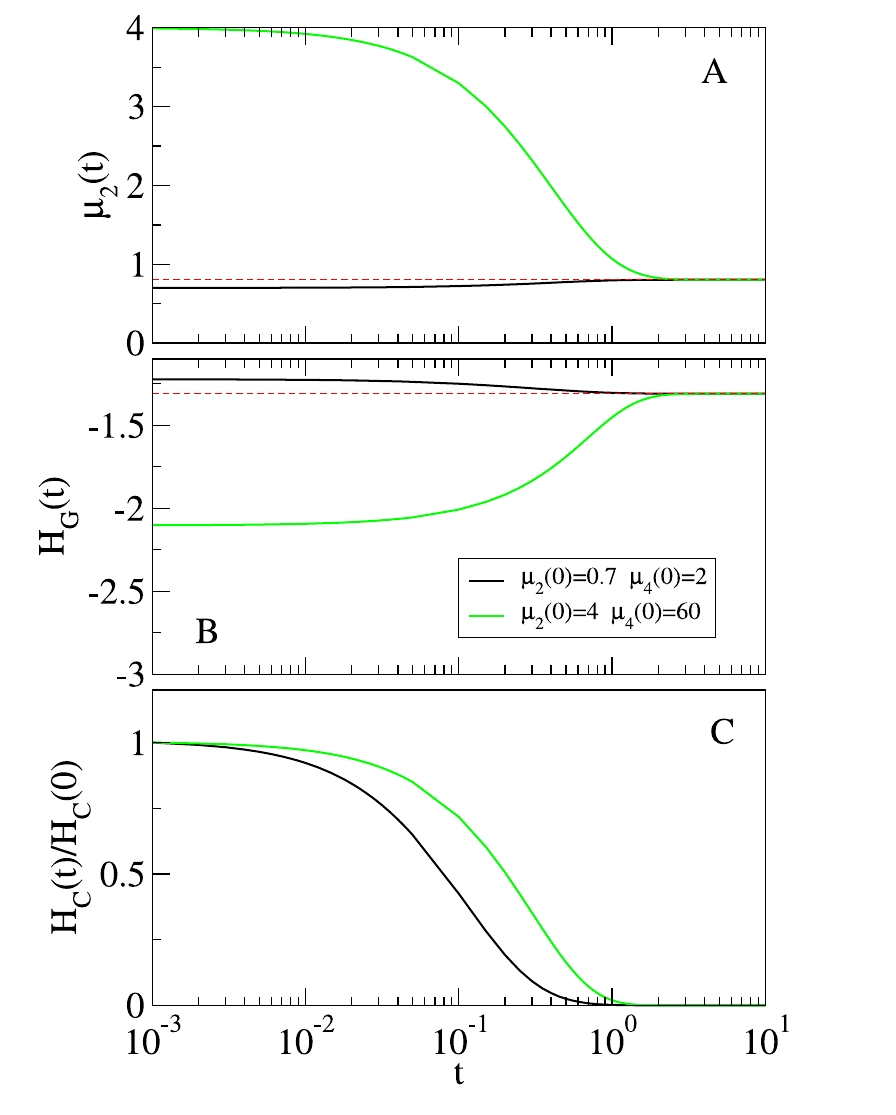}
 \caption{Time evolution of different candidates for a Lyapunov
   functional in a uniformly heated inelastic Maxwell
   model. Boltzmann's $H_{B}$ ($H_{G}$ in the notation of
   Ref.~\cite{MPyV13}) is shown in panel B , $H$ ($H_{C}$) is shown in
   panel C, whereas the evolution of the granular temperature to its
   steady value is displayed in panel A. These numerical results show
   that $H_{B}$ does not decrease for some situations, in contrast a
   monotonic decrease of $H$ is always observed. Taken from
   \cite{MPyV13}.}
  \label{ch1_fig:MPyV13}
\end{figure}

Moreover, Marconi et al.~have numerically shown that $H_{B}$ is
nonmonotonic and even steadily increases from certain initial
conditions \cite{MPyV13}. They have also put forward some numerical
evidence, further reinforced by Garc\'ia de Soria et al.'s
work~\cite{GMMMRyT15}, in favor of $H$ being a ``good'' Lyapunov
functional. Numerical results for the specific case of a uniformly
heated inelastic Maxwell model, taken from \cite{MPyV13}, are
reproduced in figure \ref{ch1_fig:MPyV13}.  In addition, it should be
stressed that $H$ is found to be a nonincreasing function in
\cite{GMMMRyT15} by a combination of three different simulation methods:
spectral method, direct simulation Monte Carlo \cite{Bi63,Bi13},
 and molecular dynamics.

These numerical evidences in favor of $H$ being a ``good'' Lyapunov
functional \cite{MPyV13,GMMMRyT15} call for further theoretical
work. Some attempts at proving such a result have been carried out by
Garc\'ia de Soria et al.~\cite{GMMMRyT15}. Specifically, they have shown
that an $H$-theorem holds at the level of the $N$-particle PDF
(Kac-like description), with an $H$-functional similar to that in
\eqref{H-function-add}. Undoubtedly, this proof at the $N$-particle
level is a neat step forward in the right direction. However, to the
best of our knowledge, there is no rigorous mathematical proof for the
$H$-theorem at the level of the one-particle description.  In
addition, only spatially homogeneous situations, in which the
$\rr$-dependence of $f$ and thus the integration over $\rr$ may be
dropped, have been analyzed \cite{MPyV13,GMMMRyT15}.

Wrapping things up, an analytical proof of either global stability or
the $H$-theorem is currently unavailable at the level of the kinetic
one-particle description for granular gases. This is true even for
simple collision terms, such as those corresponding to hard-spheres or
the cruder Maxwell molecules model, which are considered in
\cite{MPyV13,GMMMRyT15}.  Therefore, it seems worth investigating this
subject in simplified models, like the one introduced in part
\ref{part:granular} of this thesis, for which analytical calculations
are expected to be more feasible.

\subsection{Memory effects: Kovacs experiment}\label{ch1_sec:kovacs}

The equilibrium state of physical systems is characterized by the
value of a few macroscopic variables, for instance pressure, volume
and temperature in molecular fluids. These macroscopic variables
provide a full characterization of the system: different samples
sharing the same values respond identically to an external
perturbation.  On the contrary, a system in a nonequilibrium state,
even if it is stationary, is not fully characterized by the value of
the macroscopic variables: the response to an external perturbation
may depend also on additional variables or, equivalently, on its
entire thermal history. This behavior unavoidably leads to the
emergence of memory effects.

Kovacs carried out a pioneering work in the field of memory effects in
nonequilibrium systems \cite{KAHyR79}. The Kovacs experiment showed
that pressure, volume and temperature did not fully characterize the
state of a sample of polyvinyl acetate that had been aged for a long
time at a certain temperature $T_{1}$. The pressure was fixed during
the whole experiment, and the time evolution of the volume was
recorded. After a waiting time $t_{w}$, the temperature was suddenly
changed to $T$, for which the equilibrium value of the volume equaled
its instantaneous value at precisely $t_{w}$. Counterintuitively---from
an equilibrium perspective---the volume did not remain
constant. Instead, it displayed a hump, passing through a maximum
before tending back to its initial equilibrium value.

\begin{figure}
\centering
\includegraphics[width=0.8 \textwidth]{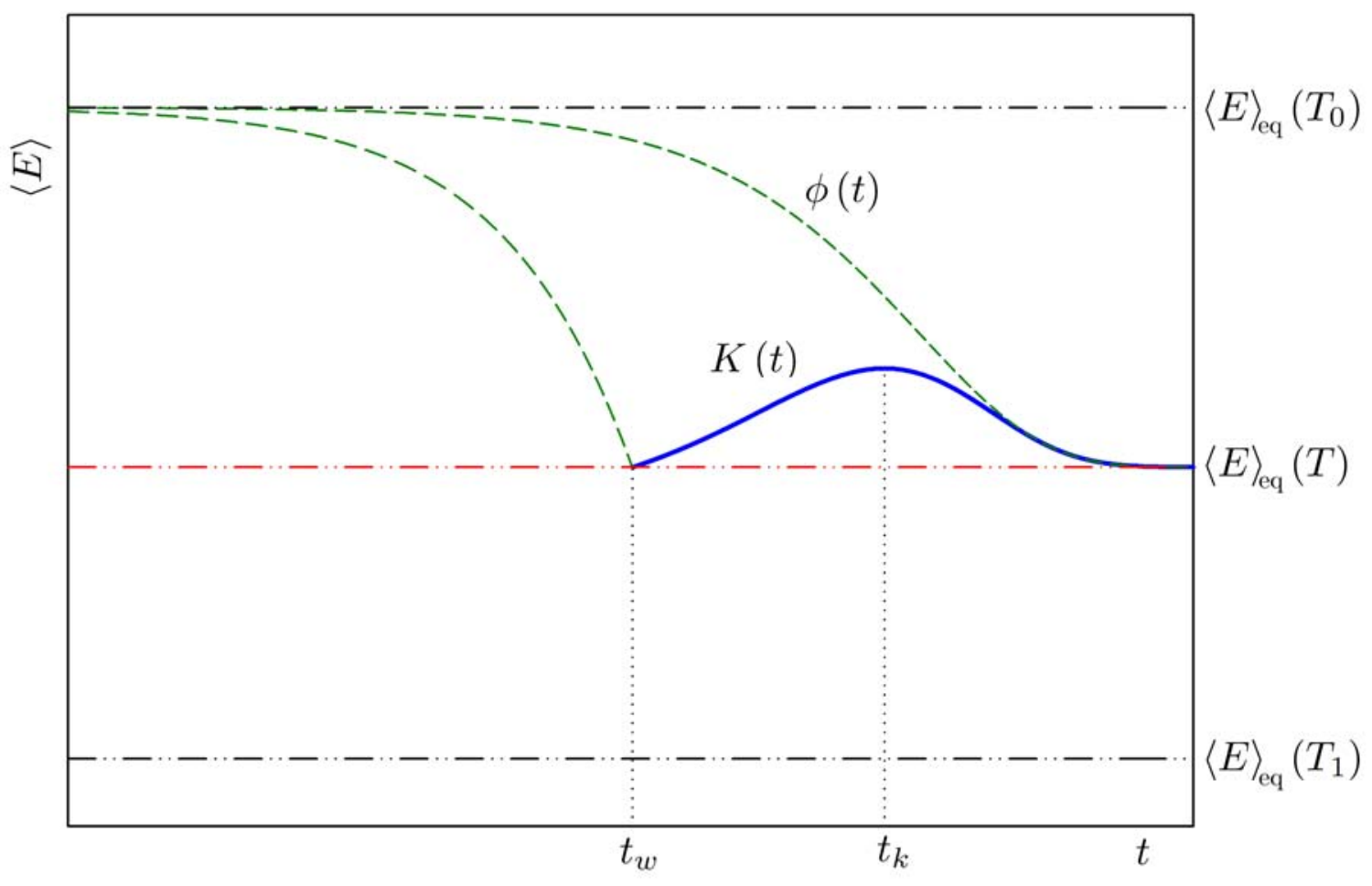}
\caption{Scheme of the Kovacs experiment. The
 dashed curve on the right, labeled by $\phi(t)$, represents the
 direct relaxation from $T_0$ to $T$. The dashed curve on the left
 stands for the part of the relaxation from $T_0$ to $T_1$ that is
 interrupted at $t=t_{w}$ by the second temperature jump, changing
 abruptly the temperature from $T_1$ to $T$. After this second jump,
 the system follows the nonmonotonic response $K(t)$, given by the solid line,
 which reaches a maximum at $t=t_k$ and, afterwards, approaches
 $\phi(t)$ for very long times. 
 \label{ch8_fig1}}
\end{figure}

This effect has extensively been  studied  in glassy systems
 \cite{BBDyG03,Bu03,MyS04,ALyN06,PyB10,DyH11,RyP14}. Therein, the
relevant physical variable is the energy instead of the volume. 
First, the system is equilibrated at a ``high'' temperature $T_0$. Then, at $t=0$,
the temperature is suddenly quenched to a lower temperature $T$, after
which the relaxation function $\phi(t)$ of the energy $E$ is
recorded. Specifically,
$\phi(t)=\langle E(t)\rangle-\langle E\rangle_{\eq}$, where
$\langle E\rangle_{\eq}$ is the average equilibrium energy at
temperature $T$. Alternatively, a similar procedure is followed, equilibrating
the system again at $T_{0}$, but at $t=0$, the temperature is changed
to an even lower value $T_{1}$, $T_{1}<T<T_{0}$. The system relaxes
isothermally at $T_{1}$ for a certain waiting time $t_{w}$, such that
$\langle E\rangle(t=t_{w})$ equals $\langle E\rangle_{\eq}$. At this
time $t_{w}$, the temperature is increased to its corresponding equilibrium value
$T$. However, the energy does
not remain constant, but displays a hump behavior represented by a function $K(t)$.
 At first, $K(t)$ increases from zero until a maximum
is attained for $t=t_{k}$, and only afterwards, it goes back to zero.
Similarly to the relaxation function, we have defined
$K(t)=\langle E(t)\rangle-\langle E\rangle_{\eq}$, for $t\geq t_{w}$.
Note that $K(t)\leq \phi(t)$ for all times, with the equality being
only asymptotically approached in the long time limit.
A qualitative plot of the Kovacs effect is depicted in figure
\ref{ch8_fig1}.

For molecular (thermal) systems, the equilibrium distribution is the
canonical one, and it has been shown that, in linear response theory
\cite{PyB10},
\begin{equation}\label{ch8_eq:kovacs-thermal-linear}
K(t)=\frac{T_{0}-T_{1}}{T_{0}-T} \phi(t)-\frac{T-T_{1}}{T_{0}-T}
\phi(t-t_{w}),
\end{equation}
where the final temperature $T$ and the waiting time $t_{w}$ are
related by
\begin{equation}\label{ch8_eq:T-tw-relation} \frac{T-T_{1}}{T_{0}-T_{1}}=\frac{\phi(t_{w})}{\phi(0)}.
\end{equation}
In linear response, the relaxation function $\phi(t)$ decays
monotonically in time because it is proportional to the equilibrium
time correlation function, as predicted by the fluctuation-dissipation theorem.
Therefrom,
\begin{equation}
\phi(t)\propto \langle E(0) E(t)\rangle_{\eq}-\langle E\rangle_{\eq}^{2}=\sum_{i}
c_{i} \exp(\lambda_{i}t),
\end{equation}
with $c_{i}>0$ and $\lambda_{i}<0$ for all $i$
\cite{vK92}.

The linear response results above make it possible to understand the
phenomenology observed in the Kovacs experiments
\cite{PyB10}: (i) the inequality
$0\leq K(t)\leq \phi(t)$, which assures that the hump always has a
positive sign (from now on, ``normal'' behavior), (ii) the existence
of only one maximum in the hump and (iii) the increase of the maximum
height and the shift of its position to smaller times as $t_{w}$ is
decreased. Nevertheless, it must be noted that the experiments, both
real \cite{KAHyR79} and numerical
\cite{BBDyG03,Bu03,MyS04,ALyN06,PyB10,DyH11},
are mostly done out of the linear response regime. Therefore, it seems that
the validity of these results extends beyond expectations. In fact, it
has been checked that the linear approximation still
gives a fair description of the hump for not-so-small temperature
jumps in simple models \cite{RyP14}.

More recently, the Kovacs memory effect has been investigated in
granular gases.  The simplest case is that of granular gases
considered in \cite{PyT14,TyP14}, uniformly heated by the
\textit{stochastic thermostat} introduced in
\cite{Wi96,WyM96,SBCyM98,NyE98}. The value of the kinetic energy, or granular
temperature, at the nonequilibrium steady state (NESS) $T_{s}$ is
controlled by the driving intensity $\xi$ of the 
thermostat, $T_{s}=T_{s}(\xi)$. Therefore, a Kovacs-like protocol can
be implemented in a completely analogous way to the one described
above, with the the granular temperature $T$ and driving intensity
$\xi$ playing the role of energy and conventional temperature,
respectively.

\begin{figure}
\centering
\includegraphics[width=0.8 \textwidth]{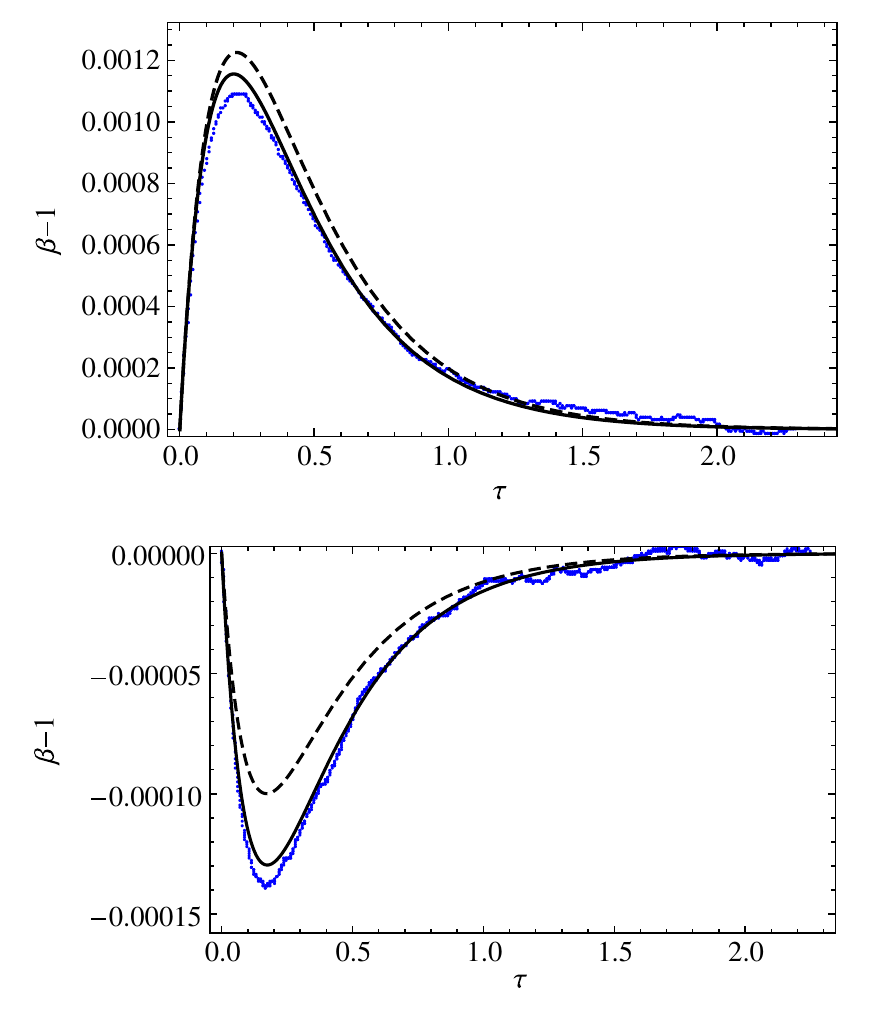}
\caption{Emergence of anomalous Kovacs response in a uniformly
  heated granular gas. Specifically, $\beta=\sqrt{T_{s}/T}$, where
  $T_{s}$ is the steady value of the temperature, and $\tau$ measures
  time in the number of collisions per particle. Therefore, a maximum
  in the temperature corresponds to a minimum in $\beta$ and vice
  versa. (Top) Highly inelastic case (restitution coefficient
  $\alpha=0.3$), for which the anomalous Kovacs response is
  clearly observed. (Bottom) Inelasticity is decreased ($\alpha=0.8$,
  closer to the elastic case $\alpha=1$) and the Kovacs response becomes
  normal. The crossover from normal to anomalous response takes place
  at $\alpha=1/\sqrt{2}$; a detailed discussion can be found in
  \cite{PyT14}, from which the figure is taken. 
  \label{fig:kovacs-granular}}
\end{figure}

In the simplest protocol, the driving intensity is first decreased
from $\xi_{0}$ to $\xi_{1}=0$, and after the waiting time $t_{w}$
increased to $\xi$, with the instantaneous value of the granular
temperature verifying $T(t_{w})=T_{s}(\xi)$. Then, the granular gas is
freely cooling in the waiting time window. One of the main results
found in \cite{PyT14,TyP14} is the emergence of ``anomalous'' Kovacs
behavior for large enough inelasticity, when $K(t)$ becomes negative
and displays a minimum instead of a maximum, see figure
\ref{fig:kovacs-granular} for details. For smaller inelasticities,
however, the response becomes normal and $K(t)$ is positive as in
molecular systems.

It must be stressed that these results have been obtained in the
nonlinear regime, that is, for driving jumps $\xi_{0}-\xi$,
$\xi_{0}-\xi_{1}$ that are not small. The main implication of the
Kovacs-like behavior is, once more, the necessity of incorporating
additional variables to have a complete characterization of
nonequilibrium states. In the granular gas, this additional
information are the non-Gaussianities of the velocity distribution
function, basically encoded in the so-called excess kurtosis
\cite{PyT14,TyP14}. Finally, we note that similar anomalous Kovacs
humps have been found for other energy injection mechanisms
\cite{BGMyB14}, which undoubtedly show that
their emergence is not an artifact introduced by the use of the
\textit{stochastic thermostat}.

Kovacs-like behavior has also been reported in other, more complex,
athermal systems. This is the case of disordered mechanical systems
\cite{LGAyR17} and active matter \cite{KSyI17}. In the latter, a
``giant'' Kovacs hump has been observed, with the numerically observed
maximum being much larger than the one predicted by the extrapolation
of the linear response expression \eqref{ch8_eq:kovacs-thermal-linear}
to the considered nonlinear protocol.  Moreover, an alternative
derivation of \eqref{ch8_eq:kovacs-thermal-linear} has been provided
in the supplemental material of \cite{KSyI17}. This derivation holds
for athermal systems, since it does not make use of either the
explicit form of the probability distribution or the relationship
between response functions and time correlations at the steady
state. Nevertheless, it is restricted to discrete-time dynamics at the
macroscopic (average) level of description. In chapter
\ref{ch:Kovacs_pgran}, we proceed to generalize these results for
continuous time dynamics and also for the mesoscopic level of
description.  Therein, the dynamics is governed by a master equation
for the probability distribution function, from which the macroscopic
description can be obtained in the appropriate limit.

\subsection{Summary of part \ref{part:granular}}

As advanced above, part \ref{part:granular} is devoted to the thorough
analysis of a lattice model that attempts to catch the essential
phenomenology of the shear modes of granular fluids.

In each chapter, we explore different aspects of the model. We start
by introducing the model in chapter \ref{ch:finsize_pgran}. Therein,
we focus on the continuous hydrodynamic-like limit that, to the lowest
order, leads to equations completely analogous to those in
\eqref{ch1_eq:brey2}. In that limit, we study different relevant
physical states. Specifically, we analyze the Homogeneous Cooling
State and the Uniform Shear Flow state through its one-particle
velocity distribution.  We also go beyond the aforementioned lowest
order analysis in two ways: (i) looking into the behavior of the
fluctuating fields and (ii) solving exactly the Homogeneous Cooling
State on the lattice.

In chapter \ref{ch:Hth_pgran}, we turn our attention to the stability
of the NESS of this model system.  Not only have we proven the global
stability of a quite general family of states, but also clarified the
inadequacy of $H_B$, given by \eqref{H-Boltzmann}, as a Lyapunov
functional. We finish the chapter with a rigorous proof of an
$H$-theorem. To the best of our knowledge, our result constitutes the
first proof, even in simple models, of an $H$-theorem within the
context of systems with nonconservative interactions.

Finally, chapter \ref{ch:Kovacs_pgran} is devoted to the analysis of
Kovacs-like memory effects.  We develop a general theoretical
framework for the linear response analysis in athermal systems,
starting from either the master equation for the probability
distribution function (mesoscopic description) or the evolution
equations for the macroscopic moments (macroscopic description). Our
results are particularized for a variant of our lattice model of
granular gas, and they show an excellent agreement with
simulations. Although we test the theory in our specific model, it is
worth noting the quite broad range of physical systems that our
developed theoretical framework can be directly applied to.

Appendices \ref{ch5-6_app-a}-\ref{app_final_glob_stab} deal
with some technicalities that are skipped within the main text of this
part of the thesis.
{\clearpage \thispagestyle{empty}}
\part{Predicting the unfolding pathway of modular systems with toy models}
\label{part:part-bio}
\chapter{The basics of modelling modular systems}
\label{ch:model_pbio}
\newcommand{\sgn}{\text{sgn}}
\newcommand{\deltaf}{\delta\! f}

Over the course of this first part of the thesis, we study in depth a simple model 
of elasticity for biomolecules. This model, introduced by 
Guardiani et al. \cite{GMTCyC14} for simulation purposes, 
lacked a thorough theoretical analysis.

Herein, we  specifically look  for a theory
capable of predicting the 
unfolding pathway in modular biomolecule submitted to mechanical pulling. We expect
that this theory could also explain the unfolding pathway observed in
experiments \cite{ByR08} and simulations \cite{GMTCyC14} of the maltose binding
protein. As described in the introduction chapter, the unfolding pathway depends
on the pulling velocity: at very low pulling rates, it is the weakest
 unit that unfolds first, while at higher rates the first 
 unit to unravel is the pulled one. 

This chapter is dedicated to the development of the aforementioned
theory and its plan is detailed below.
We start by introducing
the basics of the model in section \ref{ch2_sec:model}. The system
dynamical response to mechanical pulling is obtained by means of a perturbative
approach in section \ref{sec:real_chain}. Section \ref{sec:critical_speed} 
is devoted to the obtention of the solution of the dynamical equations, which
allows us to study the emergence of a set of critical velocities at which
the unfolding pathway changes. Section \ref{ch2_sec:comp} is dedicated
to the comparison of the theoretical predictions with numerical results of the
model equations. We seek more realistic variants of the model
in section \ref{sec:stiff}. Finally, we propose a possible experiment for testing
our theory in section \ref{sec:experiments}.

\section{The model fundamentals} \label{ch2_sec:model}

\begin{figure}
\centering
  \includegraphics[width=0.85 \textwidth]{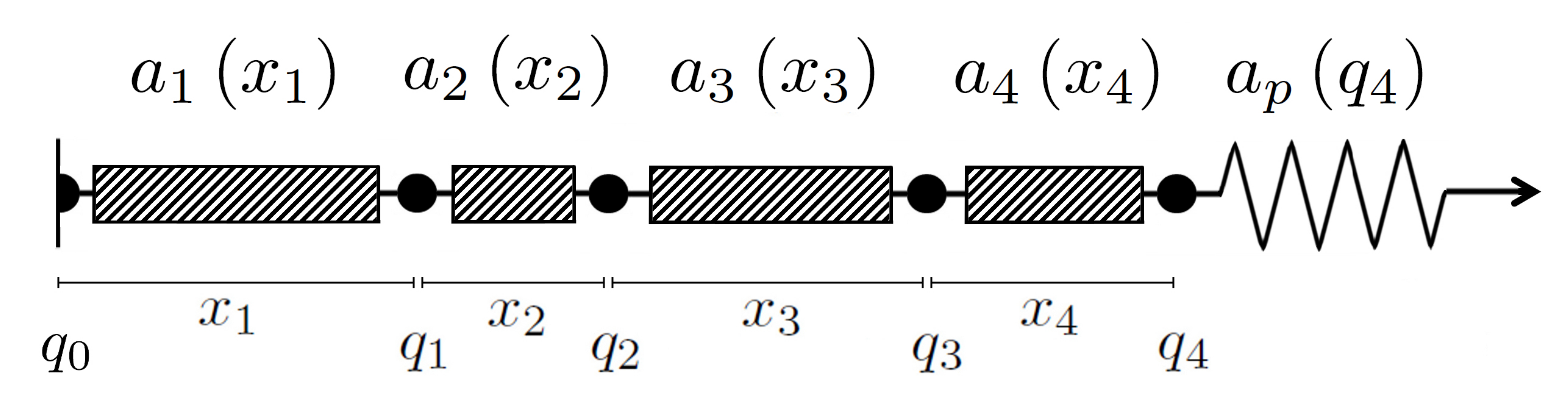}
\caption{\label{fig:1ch2}
   Sketch of the basic model for a protein with four units. Each unit
     is depicted by a rectangle with nonharmonic free energy $a_i(x_i)$.
     The beads mark the
     coordinates $q_{i}$ of their endpoints, so that
	 the $i$-th unit's extension     
     $x_{i}=q_{i}-q_{i-1}$ (by definition, $q_{0}=0$). Finally, the
     spring represents the device attached to the pulled end $q_4$,
     which controls either the applied force 
     (force control) or the end-to-end distance $q_4$ (length control). The
     device contribution to the free energy is
     $a_p(q_4)$, see \eqref{ch2_eq:4} and \eqref{ch2_eq:6b}.
}
\end{figure}

Let us consider a certain modular biomolecule comprising $N$ modules. The paradigmatic example is a polyprotein composed of $N$, possibly different, modules. Notwithstanding, we may also be considering a protein domain with $N$ unfoldons. From now
on, we will refer to these modules or unfoldons, indistinctly, as modules or units.  When the
system is submitted to an external force $F$, the simplest
description is to portray it as a one-dimensional chain in the direction of the
force. We denote the
end-to-end extension of the $i$-th unit
by $x_{i}$.  In a real AFM experiment, the molecule is
attached as a whole to the AFM device and stretched.  Following
Guardiani et al. \cite{GMTCyC14}, we model this system with a sequence
of nonlinear bonds, as in figure \ref{fig:1ch2}. Therein, the endpoints of
  the $i$-th unit are denoted by $q_{i-1}$ and $q_{i}$, so that its
  extension $x_{i}$ is
\begin{equation}
  \label{ch2_eq:3}
  x_{i}=q_{i}-q_{i-1}, \quad i=1,\ldots,N.
\end{equation}
In this basic model, we consider that the left end of the first unit is fixed, that 
is,  $q_{0}=0$ for all times. 

We assume that the inertia terms can be neglected and the evolution of the system follows the coupled overdamped
Langevin equations
\begin{equation}
  \label{ch2_eq:5}
  \gamma \dot{q}_{i}=-\frac{\partial}{\partial q_{i}} A(q_{0},\ldots,q_{N})+\eta_{i},
\end{equation}
in which $\eta_{i}$ are Gaussian white noise terms. They verify
\begin{equation}
\langle \eta_{i}(t)\rangle=0, \quad
\langle \eta_{i}(t) \eta_{j}(t')\rangle=2 \gamma k_{B} T \delta_{ij}
\delta(t-t'),
\end{equation}
with $\gamma$ being the friction coefficient of each unit (the
same for all),  $T$ the temperature of the fluid in which the protein is
immersed, and $k_B$ the Boltzmann constant.  The global
free energy function of the system is
\begin{equation}
  \label{ch2_eq:4}
  A(q_{0},\ldots,q_{N})=\sum_{i=1}^{N} a_{i}(q_{i}-q_{i-1})+a_{p}(q_{N}) \, .
\end{equation}
In \eqref{ch2_eq:4}, $a_{p}(q_{N})$ is the contribution to the free
energy introduced by the force control or length control device, see
below, while $a_i(x_i)$ is the contribution to $A$ stemming from the $i$-th unit, which is only function of its own extension $x_i$.

The total length of the system is given by
\begin{equation}
  \label{ch2_eq:6}
  \sum_{i=1}^{N} x_{i}=q_{N}.
\end{equation}
In force control experiments, the applied force $F$ is a given
function of time, whereas in length control experiments the device 
(portrayed by the spring in figure \ref{fig:1ch2})
tries to keep the total length $q_{N}$ equal to the desired value $L$,
also a certain function of time. 
The corresponding contributions to the free energy are
\begin{subequations}\label{ch2_eq:6b}
\begin{equation}
\label{ch2_eq:6b1}
  a_{p}(q_{N})=-F q_{N}, \qquad \text{force control},
\end{equation}
\begin{equation}
\label{ch2_eq:6b2}
  a_{p}(q_{N})=\frac{1}{2}k_{c}(q_{N}-L)^{2}, \qquad \text{length control},
\end{equation}
\end{subequations}
in which $k_{c}$ stands for the stiffness of the length control
device. The length is perfectly controlled in the limit $k_{c}\to\infty$,
when $q_{N}=L$ for all times. For the sake of a common notation, we have not used 
different letters for Helmholtz or Gibbs free energies. It has to be understood that, 
on the one hand, introducing \eqref{ch2_eq:6b1} in \eqref{ch2_eq:4}, we obtain a 
Gibbs free energy. On the other hand, if we use \eqref{ch2_eq:6b2} in 
\eqref{ch2_eq:4}, and the limit $k_c \to \infty$ (taking into account that $a_p 
\to 0$ since the force $k_c(L-q_n)$ goes to a constant Lagrange multiplier), 
the result is a Helmholtz free energy whose 
minimum, restrained to the total length constraint, gives
the equilibrium configuration of the system.

An apparently similar system, briefly discussed in section \ref{ch1_sec:1_th_dev},
in which each module of the chain
follows the Langevin equation
$\gamma \dot{x}_{i}=-\partial A/\partial x_{i}+\eta_{i}$,  has been
analyzed in the literature \cite{BCyP14,BCyP15}. In this approach, the modules
are completely independent in force control experiments, the global free energy is the sum of individual ones and,  because of
this, Langevin equations completely neglect the spatial structure of
the chain. While this simplifying assumption poses no problem for the
characterization of the force-extension curves in \cite{BCyP15}, it is
not suited for the investigation of the unfolding pathway. In this context,
the spatial structure plays an essential role. The spatial
  structure of biomolecules can be described in quite a realistic way
  by using the framework proposed by Hummer and Szabo several years ago
  \cite{HyS03}, but our simplified
  picture in figure \ref{fig:1ch2} makes an analytical
  approach feasible.

Now, we look into the unfolding pathway of this system. As the
evolution equations are stochastic, this pathway may vary from one
trajectory of the dynamics to another.  Nevertheless, in many
experiments \cite{GMTCyC14,LyK09,ByR08} a
quite well-defined pathway is observed, which suggests that thermal fluctuations
do not play an important part in its determination. Physically, this
  means that the free energy barrier separating the unfolded and
  folded conformations at coexistence---that is, at the critical force,
  see below---is expected to be much larger than the typical energy
  $k_{B}T$ for thermal fluctuations. Therefore, we expect the thermal
noise terms in our Langevin equations to be negligible and,
consequently, they will be dropped in the remainder of our
  theoretical approach. Of course, if the
  unfolding barrier for a given biomolecule were only a few $k_{B}T$s,
  the thermal noise terms in the Langevin equations could not be
  neglected and our theoretical approach would have to be changed.

In order to undertake a theoretical analysis of the stretching
dynamics, we introduce one further simplification of the problem.
 We consider that the device controlling the length is
perfectly stiff, thus the total length $q_{N} = L$ does not fluctuate.
  We expect this assumption to have little impact on the
unfolding pathway: otherwise, the latter would be more a property of
the length control device than of the chain.  In fact, we  show in
section \ref{sec:numerics} that the unfolding order is not affected by
this simplification.  For perfect length control, the mathematical
problem is identical to that of the force control situation, but now
the force $F$ is an unknown (Lagrange multiplier) that must be
calculated at the end by imposing the constraint
$q_{N}=\sum_{i}x_{i}=L$.  Therefore, the extensions $x_i$'s obey the
deterministic equations
\begin{subequations}
\label{ch2_eq:7}
\begin{align}
  \gamma\dot{x}_{1}  =&  -a'_{1}(x_{1})+a'_{2}(x_{2}), \\
  \gamma\dot{x}_{i}  =&  -2a'_{i}(x_{i})+a'_{i+1}(x_{i+1})+a'_{i-1}(x_{i-1}),
     \quad 1<i<N, \\
  \gamma\dot{x}_{N}  =&  -2a'_{N}(x_{N})+a'_{N-1}(x_{N-1})+F, \\  
  F =& \, \, \gamma v_{p}+a'_{N}(x_{N}).
\end{align}
\end{subequations}
We have introduced the pulling speed 
\begin{equation}
  \label{ch2_eq:8}
  v_{p}\equiv \dot{L},
\end{equation}
which is usually time independent.

We assume that $a_i(x_i)$ allows for bistability in a certain range of
the external force $F$, in the sense that $a_{i}(x_{i})-F x_i$ is a double-well potential with two minima, see figure
\ref{fig:2ch2}. Therefore, in that force range, each unit may be
either folded, if $x_{i}$ is in the well corresponding to the minimum
with the smallest extension, or unfolded, when $x_{i}$ belongs to the
well with the largest extension. 
If the length is kept constant ($v_{p}=0$), there is an equilibrium 
solution of \eqref{ch2_eq:7},
\begin{equation}
  \label{ch2_eq:17}
  a'_{1}(x_{1}^{\text{st}})=a'_{2}(x_{2}^{\text{st}})=\cdots=a'_{N}(x_{N}^{\text{st}})=F^{\text{st}},
\end{equation}
and $F^{\text{st}}$ is calculated with the constraint $\sum_{i}x_{i}^{\text{st}}=L$. 
This solution is stable as long as $a''_{i}(x_{i}^{\text{st}})>0$ for all $i$. 

If all the units are identical, $a_{i}(x) = a(x)$, 
the metastability regions of each module---the range of forces for 
which the equation $a'_{i}(x)=F$ has several solutions---coincide.
 
Therefore, as briefly introduced in section \ref{ch1_sec:1_th_dev},
we obtain stationary branches corresponding
to $J$ unfolded units and $N-J$ folded units that have been analyzed
in detail in \cite{BCyP15,PCyB13}. If all the modules are not
identical, the metastability regions do not perfectly
overlap since the units are not equally strong: the weakest one is
that for which the equation $a'_{i}(x)=F$ ceases to have multiple
solutions at a smaller force value.

It is important to note that if we change all the forces
$a'_{i}(x_{i})$ to $\tilde{a}'_{i}(x_{i})=a'_{i} (x_{i})-F_{0}$ and $F$ to
$\varphi=F-F_{0}$, we have the same system \eqref{ch2_eq:7} but with
$\tilde{a}'_{i}$ and $\varphi$ instead of $a'_{i}$ and $F$, respectively.
Then, we may use the free energies for any common value of the force
$F_{0}$ and interpret the Lagrange multiplier as the excess force from
this value to be applied to the system. A similar result is
  also found if the length is controlled by using a device with a
  finite value of the stiffness $k_{c}$. A constant force only
  shifts the equilibrium point of a harmonic oscillator:
  $(q_{N}-L)$ must be substituted by $(q_{N}-L-F_{0}/k_{c})$.

\section{Pulling the system: perturbative solution}
\label{sec:real_chain}

We write the $i$-th unit's free energy as
\begin{equation}\label{ch2_eq:9}
 a_{i}(x) = a(x) +\xi\,\delta a_{i}(x),
\end{equation}
in which $a(x)$ is the ``main'' part, common to all the units, and
$\xi \delta a_{i}(x)$ represents the separation from this main
contribution. If all the units are perfectly identical,
$a_{i}(x)=a(x)$ for all $i$ or, equivalently, $\delta a_{i}(x)=0$.  In
principle, in an actual experiment, the splitting of the free
  energy in \eqref{ch2_eq:9} can be done if the free energy $a_{i}$ of
  each unit is known: we may define the common part as the ``average''
  free energy over all the units,
  $a(x)\equiv \overline{a}(x)\equiv N^{-1}\sum_{i=1}^{N} a_{i}(x)$,
  and $\xi \delta a_{i}(x)\equiv a_{i}(x)-\overline{a}(x)$. From a
  physical point of view, the dimensionless parameter $\xi>0$
  measures the importance of the heterogeneity in the free energies.
Our theory could be applied to a situation in which the free
  energy deviations $\delta a_{i}$ were stochastic and followed a
  certain probability distribution, for instance to represent the
  slight differences among very similar units, as done in
  \cite{BCyP15} to analyze the force-extension curves. Also
the forces $a'_{i}(x)$ in the evolution equations are split as
\begin{equation}
\label{ch2_eq:10}
 a'_{i}(x)=a'(x)+\xi \, \deltaf_{i}(x), \quad  
\deltaf_{i}(x)\equiv\delta a'_{i}(x).
\end{equation}

\begin{figure}
  \centering
  \includegraphics[width=0.8 \textwidth]{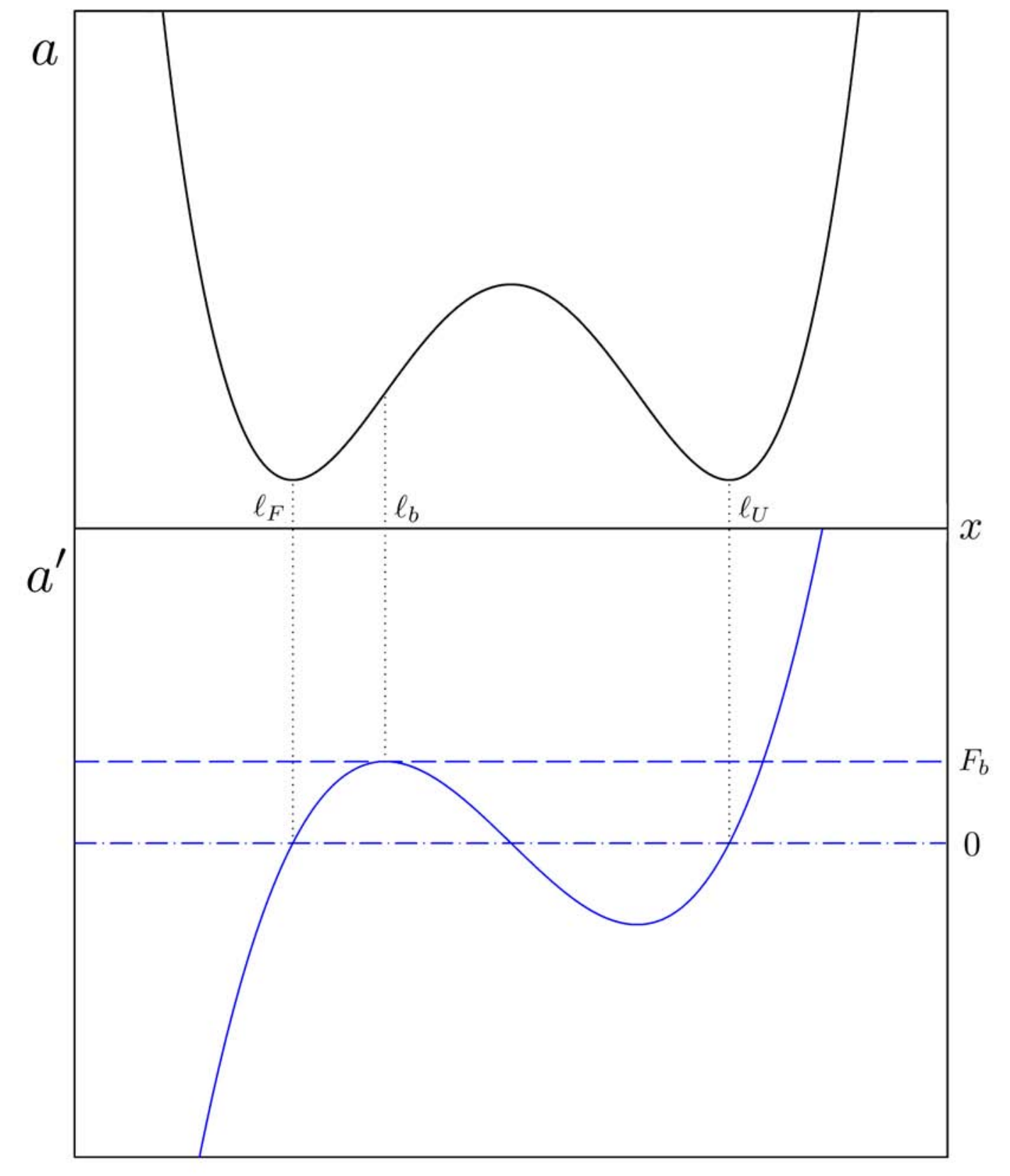}
  \caption{Qualitative behavior of the main contribution to the free
    energy $a(x)$ (top) and its associated 
    force $a'(x)$ (bottom) as functions of the extension. 
    Specifically, the plots correspond to the critical force, for which the two 
    minima of the free energy are equally deep.  The values of the
    lengths at the folded and unfolded minima are $\ell_{F}$ and
    $\ell_{U}$, respectively. The threshold length $\ell_{b}$
    stands for the length at the limit of stability, with
    $F_{b}$ being the corresponding force.}
  \label{fig:2ch2}
\end{figure}

As already noted above, we can use the free energies for any
common value of the force $F_{0}$, and interpret $F$ as the extra
applied force from this value.  In what follows, 
we consider the main part $a(x)$ with two, equally deep, 
minima corresponding to the folded (F)
and unfolded (U) configurations. 
Therefore, our ``origin of force'' $F_0$ corresponds to the 
critical force for the main, common, contribution $a$ to the units' free energies.
 Figure \ref{fig:2ch2} presents a
qualitative picture of the free energy and its derivative. The two
minima correspond to lengths $\ell_{F}$ and $\ell_{U}$, with
$\ell_{F}<\ell_{U}$. Also the point $\ell_{b}$ 
at which $a''(\ell_{b})=0$ is marked.

It is the condition $a''(\ell_{b}) = 0$ that essentially determines
the stability threshold, as it provides the limit force
$F_{b}=a'(\ell_{b})>0$ at which the folded basin ceases to exist for
the ``main'' potential.  
In the deterministic approximation considered here, thermal
  fluctuations are  neglected and, for $F<F_{b}$, the folded unit
  cannot jump over the free energy barrier hindering its unfolding: it
  has to wait until, at $F=F_{b}$,  the only possible extension is that
  of the unfolded basin. Of course, neglecting thermal noise restricts
  in some way the range of applicability of our results, see
  section~\ref{sec:critical_speed} for a more detailed discussion and
  also the numerical section \ref{sec:numerics}.

Keeping the above discussion in mind, now we analyze the limit of
stability of the different units. The asymmetry correction
  $\delta f_{i}$ shifts the threshold force for the different units and
the extension $x_{i,b}$ at which the $i$-th unit loses its stability
is obtained by solving the equation
$a''_i(x_{i,b}) = a''(x_{i,b}) + \xi \delta f'_{i}(x_{i,b}) = 0$. 
Linearizing in both the displacement $x_{i,b}-\ell_{b}$ and $\xi$
one gets
\begin{equation}
a''(\ell_{b}) + a'''(\ell_{b})(x_{i,b} - \ell_{b}) + \xi \delta f'_{i}(\ell_b) = 0.
\end{equation}
Noting that $a''(\ell_{b}) = 0$, we obtain
\begin{equation}
  \label{ch2_eq:10b}
  x_{i,b}=\ell_{b}-\xi \,\frac{\delta f'_{i}(\ell_{b})}{a'''(\ell_{b})}.
\end{equation}
The corresponding force  is
\begin{equation}
  \label{ch2_eq:10c}
  F_{i,b}  \equiv a'_i (x_{i,b}) =F_{b}+\xi\,\deltaf_{i}(\ell_{b}),
\end{equation}
in which we have consistently dropped terms of the order of $\xi^{2}$.  Then,
units with $\deltaf_{i}(\ell_{b})<0$ ($\deltaf_{i}(\ell_{b})>0$) are
weaker (stronger) than average. See appendix \ref{ch2:appA} for more details.

When the system is pulled, the total length 
of the system $L$ has been shown to be a good reaction
coordinate \cite{ACDNKMNLyR10} and, on physical grounds, it
is reasonable to use $L$ to measure time. Therefore, we write the evolution equations
\eqref{ch2_eq:7} as
\begin{subequations}\label{ch2_eq:11}
  \begin{align}
  \gamma v_{p} \frac{d x_{1}}{dL}  =&  -a'(x_{1})+a'(x_{2})+\xi[-\delta
  f_{1}(x_{1})+\deltaf_{2}(x_{2})],
\\
 \gamma v_{p} \frac{d x_{i}}{dL}  =&  -2a'(x_{i})+a'(x_{i+1}) +a'(x_{i-1})
     +\xi[-2\deltaf_{i}(x_{i})+\deltaf_{i+1}(x_{i+1})+ \delta
   f_{i-1}(x_{i-1})], \nonumber \\
& \qquad \qquad \qquad \qquad \qquad \qquad \qquad \qquad \qquad \qquad \qquad \qquad \qquad \qquad 1<i<N,
\\
 \gamma v_{p} \frac{d x_{N}}{dL}  =&  -2a'(x_{N})+a'(x_{N-1})+F 
+\xi[-2\delta
  f_{N}(x_{N}) 
  +\deltaf_{N-1}(x_{N-1})],
\\ 
 F =& \, \, \gamma v_{p}+a'(x_{N})+\xi\,\deltaf_{N}(x_{N}). 
  \end{align}   
\end{subequations}
Moreover, this change of variable makes the pulling speed $v_{p}$
  appear explicitly in the equations, allowing us to consider $v_{p}$
  as a perturbation parameter for slow enough pulling processes.

Now, we consider a system such that (i) the asymmetry in the free energies
is small and (ii) it is slowly pulled. Equations \eqref{ch2_eq:11} 
are solved by means of a perturbative expansion in powers
of the pulling velocity $v_{p}$ and the disorder parameter $\xi$, that
is,
\begin{subequations}\label{ch2_eq:12}
  \begin{align}
    \label{ch2_eq:12a}
    x_{i}(L)=& \,\, x_{i}^{(0)}(L)+\xi \delta x_{i}(L)+v_{p} \Delta x_{i}(L), \\
    \label{ch2_eq:12b}
    F(L)=& \,\, F^{(0)}(L)+ \xi \delta F (L)+v_{p} \Delta F (L),
  \end{align}
\end{subequations}
up to the linear order in both $v_{p}$ and $\xi$.

The zero-th (lowest) order corresponds to the chain of identical units,
$\xi=0$, with a given constant length $L$, $v_{p}=0$. Namely,
$x_{i}^{(0)}$ and $F^{(0)}$ obey the equations
\begin{subequations}\label{ch2_eq:13}
  \begin{align}
   0  =&  -a'(x_{1}^{(0)})+a'(x_{2}^{(0)}),  \\
   0  =& -2a'(x_{i}^{(0)})+a'(x_{i+1}^{(0)})+a'(x_{i-1}^{(0)}), \quad 1<i<N, \\
   0  =&  -2a'(x_{N}^{(0)})+a'(x_{N-1}^{(0)})+F^{(0)}, \\  
   F^{(0)} =& \,\, a'(x_{N}^{(0)}),
\end{align}
\end{subequations}
which have the straightforward solution
\begin{equation}
  \label{ch2_eq:16}
  a'(x_{i}^{(0)})=F^{(0)}.
\end{equation}
The force is equally distributed among all the units of the chain in
equilibrium, as expected. 

If we start the pulling process from a
configuration in which all the units are folded and the force is
outside the metastability region, that is, the usual situation, the units
extensions and the applied force are
\begin{equation}
  \label{ch2_eq:14}
  x_{i}^{(0)}=\ell\equiv \frac{L}{N},  \quad \forall i, \qquad F^{(0)}=a'(\ell),
\end{equation}
to the lowest order. To calculate the linear corrections in $\xi$ and
$v_{p}$, we have to substitute \eqref{ch2_eq:12} and \eqref{ch2_eq:14}
into \eqref{ch2_eq:11}, and equate terms
proportional to $\xi$ and $v_{p}$, respectively. 
This is done below in
two separate sections: first, for the asymmetry contribution $\delta
x_{i}$, and second, for the ``kinetic'' contribution $\Delta x_{i}$.

\subsection{Asymmetry term}
\label{sec:asymmetry}
All the modules are not characterized by the same free energy. Here, 
we calculate the first order correction introduced by this
``asymmetry'' in the modules.
The asymmetry corrections $\delta x_{i}$ obey the system of equations
\begin{subequations}\label{ch2_eq:18}
\begin{align}
\delta x_{2}-\delta x_{1}
= & \,\, \frac{\deltaf_{1}(\ell)-\deltaf_{2}(\ell)}{a''(\ell)},\\
\delta x_{i+1}+\delta x_{i-1}-2\delta x_{i}  = & \, \, 
\frac{2\deltaf_{i}(\ell)-\deltaf_{i+1}(\ell)-\deltaf_{i-1}(\ell)}{a''(\ell)},
\quad  1<i<N, \label{ch2_eq:18b}\\
\delta x_{N-1}-2 \delta x_{N}  = & \, \,
\frac{2\deltaf_{N}(\ell)-\deltaf_{N-1}(\ell) -\delta F}{a''(\ell)}, \\
\delta F  = & \,\, a''(\ell) \delta x_{N}+\deltaf_{N}(\ell),
\end{align}
\end{subequations}
which is linear in the $\delta x_{i}$'s, and thus can be analytically
solved. Note that our expansion breaks down when
$a''(\ell)=0$. This was expected, since the
stationary branch with all the modules folded is unstable when
$a''_{i}$ becomes negative for some unit $i$, and to the lowest order
this takes place when $a''(\ell)=0$.

The solution of \eqref{ch2_eq:18} is obtained
by standard methods for solving difference equations \cite{ByO99},  with the
result
\begin{equation}
  \label{ch2_eq:19}
  \delta x_{i}=\frac{\overline{\deltaf}(\ell)-\deltaf_{i}(\ell)}{a''(\ell)}, \;
  \forall i,
  \quad \delta F=\overline{\deltaf}(\ell)=\frac{1}{N} \sum_{i=1}^{N} \delta f_i(\ell).
\end{equation}
Interestingly, the force is homogeneous across the chain, since
  to first order in $\xi$ we have that
\begin{equation}
  a'_{i}(x_{i}) =a'(x_{i}^{(0)})+\xi [a''(x_{i}^{(0)})\delta x_{i}+
  \deltaf_{i}(x_{i}^{(0)})]  =  a'(\ell)+\xi  \overline{\deltaf}(\ell)=F^{(0)}+\xi \delta F,
\label{ch2_eq:20}
\end{equation}
where we have made use of \eqref{ch2_eq:10}, \eqref{ch2_eq:14} and 
\eqref{ch2_eq:19}. Equation \eqref{ch2_eq:20}
is nothing but the stationary solution \eqref{ch2_eq:17}, up
  to first order in the disorder. If the zero-th order free
    energy were the average of the $a_{i}$'s, no correction for the
    Lagrange multiplier (applied force) would appear to the first
    order. This is logical, up to the first order, the force expression
    coincides with the spatial derivative of the average potential,
    that is,
    $F^{(0)}+\xi \delta F=a'(\ell)+\xi
    \overline{\deltaf}(\ell)=\overline{a'}(\ell)$.
  Moreover, \eqref{ch2_eq:19} implies that there are units with
  $\delta x_{i}>0$ and others with $\delta x_{i}<0$, depending on the
  sign of $\overline{\deltaf}(\ell)-\deltaf_{i}(\ell)$. This is a
  consequence of the length constraint
  $\sum_{i}x_{i}^{(0)}=L$ for all times, as given by
  \eqref{ch2_eq:14}, from which $\sum_{i}\delta x_{i}=0$.

Let us remember that we denote by $\ell_{b}$ the value of the
extension at which the common main free energy reaches its limit of
stability, see figure \ref{fig:2ch2}. Taking into account only the
asymmetry correction, it is the weakest unit that unfolds first: since
the most negative $\deltaf_{i}(\ell)$ leads to the largest positive
$\delta x_{i}$ which is the one that verifies the
condition $x_{i}=\ell+\xi\delta x_{i}=\ell_{b}$ for the shortest time. 
For a more detailed
discussion, see appendix \ref{ch2:appA}. An alternative way of looking at this
is to recall that the force corresponding to the limit of stability
is smallest for the weakest unit: since the force is homogeneously
distributed along the chain, it is the weakest module that first
reaches its stability threshold.

\subsection{Kinetic term}
\label{sec:pulling_speed} 
Now we look into the ``kinetic'' correction that stems from the finite pulling
speed $v_{p}$. The zero-th order solution is given by
\eqref{ch2_eq:14}, so that $dx_{i}^{(0)}/dL=N^{-1}$ for all $i$, and
we have 
\begin{subequations}
  \label{ch2_eq:21}
\begin{align}
\Delta x_{2}-\Delta x_{1}=& \,\, \frac{\gamma}{N a''(\ell)},\\
\Delta x_{i+1}+\Delta x_{i-1}-2\Delta x_{i}  =& \, \, 
\frac{\gamma}{N a''(\ell)}, \quad 1<i<N, \label{ch2_eq:21b}\\
\Delta x_{N-1}-2 \Delta x_{N}  =& \, \, 
\frac{1}{a''(\ell)}\left[\frac{\gamma}{N}-\Delta F \right], \\
\Delta F  =& \, \, \gamma + a''(\ell)\Delta x_{N}. 
\end{align}
\end{subequations}
The solution to this system of linear difference equations is  again obtained by 
employing standard methods \cite{ByO99}, with the result 
\begin{subequations}\label{ch2_eq:22}
\begin{align}
  \label{ch2_eq:22a}
\Delta x_{i} =& \, \, \frac{\gamma}{2Na''(\ell)} \left[i(i-1)-\frac{(N+1)(N-1)}{3}
\right],\\
\label{ch2_eq:22b}
\Delta F =& \, \,\frac{(N+1)(2N+1)\gamma}{6N}.  
\end{align}
\end{subequations}
Also, $\sum_{i}\Delta x_{i}=0$ because the zero-th order solution
  \eqref{ch2_eq:14} gives the total length, $\sum_{i}x_{i}^{(0)}=L$ for
  all times. Note that
\eqref{ch2_eq:22a} is reasonable on intuitive
grounds: the kinetic correction $\Delta x_{i}$ increases with $i$
because the last module is the one closer to the pulled end. Therefore,
on the basis of only the kinetic correction, it is the last module
that would unfold first, since $\Delta x_{N}$ is the largest  and
the condition $x_{i}=\ell+v_{p}\Delta x_{i}=\ell_{b}$ is first
verified for $i=N$.

It is interesting to highlight that the force was equally distributed for
the asymmetry correction, as expressed by \eqref{ch2_eq:20}, but this
is no longer true if we incorporate the kinetic correction.  Up to the 
the first order,
\begin{equation}
a'_{i}(x_{i})=a'(x_{i}^{(0)}+\xi\delta x_{i}+v_{p}\Delta
x_{i})+\xi \deltaf_{i}(x_{i})\simeq a'(\ell)+\xi
\overline{\deltaf}(\ell)+v_{p}a''(\ell) \Delta x_{i}.
\end{equation} 
Therefore, the force
$a'_{i}(x_{i})$ depends on the unit $i$: for all times, it is smaller
the further from the pulled unit we are. Again, there is an
alternative way of understanding the reason why the last unit would unfold first
if we were considering perfectly identical units ($\xi=0$): for any
time, it would be the one suffering the largest force, and 
thus the first to reached their common limit of stability $F_{b}$.

\section{The critical velocities}
\label{sec:critical_speed}
If the last unit is not the weakest, there is a competition between
the asymmetry and the kinetic corrections. For very low pulling
speeds, in the sense that $v_{p}/\xi \to 0$, the term
  proportional to $v_{p}$ can be neglected and it is the weakest unit
  (the one with the largest $\delta x_{i}$) that unfolds first, as
  discussed in section \ref{sec:asymmetry}. On the other hand, for
  very small disorder, in the sense that $\xi/v_{p} \to 0$, the
  term proportional to $\xi$ is the one to be neglected and it is the
  last unit (the one with the largest $\Delta x_{i}$) that unfolds
  first, as also discussed in
  section \ref{sec:pulling_speed}. Therefore, different unfolding
    pathways are expected as the pulling speed is changed.

Collecting all the contributions to the extensions, we have that
\begin{equation}
x_{i}=\ell+ \frac{\xi\overline{\deltaf}(\ell)-v_{p}\gamma\dfrac{N^2-1}{6N}}{a''(\ell)}
   + \frac{v_{p}\gamma \dfrac{ i(i-1)}{2N}-\xi \deltaf_{i}(\ell)}{a''(\ell)} .
\label{ch2_eq:23}
\end{equation}
We have rearranged the terms in $x_{i}$ in such a way that the first
two terms on the rhs are independent of the unit $i$, all the
dependence of the length of the module on its position across the
chain has been included in the last term. We are expanding
the solution in powers of $v_{p}$ around the ``static'' solution,
  which is obtained by putting $v_{p}=0$ in \eqref{ch2_eq:23}. Thus, the
  ``static'' solution corresponds to the stationary one the
  system would reach if we kept the total length constant and equal to
  its instantaneous value at the considered time.  It is essential to
  realize that \eqref{ch2_eq:23} is only valid for very slow pulling, as
  long as the corrections to the ``static'' solution are small, and 
this is the reason why the limit of stability is basically unchanged
as compared to the static case.  In order to be more precise, we refer
to this kind of very slow pulling as \textit{adiabatic} pulling. One of our main results is that, even for the case of adiabatic pulling,
there appear different unfolding pathways depending on the value of
the pulling speed.

In the adiabatic limit we are considering here, the pulling
  process has to be slow enough to make the system move very close to
  the stationary force-length branches, but not so slow to give the
  system enough time to escape from the folded basin
  by thermal activation. As discussed
  in \cite{BCyP15}, there is an interplay between the pulling velocity
  and thermal fluctuations. For very slow pulling velocities, the
  system has enough time to surpass the energy barrier separating the
  two minima, which leads to the typical logarithmic dependence of the
  ``unfolding force'' $F_{U}$ on the pulling speed, specifically
  $F_{U}\propto (\ln v_{p})^{c}$ \cite{RGPCyS13,DHyS06}. The
    parameter $c$ is of the order of unity, and its particular value
    depends on the specific shape of the potential (linear-cubic,
    cuspid-like, ...)  considered \cite{DHyS06}.  On the other
  hand, as already argued at the beginning of section
  \ref{sec:real_chain}, for \textit{adiabatic pulling}, the units
  unfold not because they are able to surpass the free energy barrier
  but because the folded state ceases to exist at the force $F_{b}$
  corresponding to the upper limit of the metastability region.  Therefore, our predictions are expected to be valid in an intermediate range of velocities: high enough to avoid thermal activation, but low enough to allow for a perturbative analysis of the dynamical equations.

The unit that unfolds first is the one for
which $x_{i}=\ell_{b}$ for the shortest time. In light of the above,
it is natural to investigate whether it is possible to determine
which module is the first to unfold for a given pulling speed. To
put it another way, we would like to calculate the ``critical''
velocities separating velocity intervals inside which a specific
module unfolds first. Let us assume that, for a given pulling speed
$v_{p}$, it is the $i$-th module that unfolds first. All the modules
$j$ to its left, that is, with $j<i$, will not open first if the
pulling velocity is further increased because the difference between
the kinetic corrections $\Delta x_{i}-\Delta x_{j}$ increases with
$v_{p}$. Therefore, the first module $j$ that unfolds when the velocity
surpasses some critical value is always to its right.  More specifically, velocity
$v^{i}(j)$ for which each couple of modules $(i,j)$, $j>i$, reach
  simultaneously the stability threshold verifies
\begin{equation}
  \label{ch2_eq:stab}
  x_{i}(\ell_{c})=x_{j}(\ell_{c}) = \ell_{b},
\end{equation}
which yields both the value of $\ell_{c}$
(or time $t_{c}$) at which the stability threshold is reached and the
relationship between $v_{p}$ and $\xi$.  Equations 
\eqref{ch2_eq:23} and \eqref{ch2_eq:stab}  imply that
\begin{equation}
  \label{ch2_eq:25a}
  -\xi \delta f_{i}(\ell_{c})+\gamma v^{i}(j)
  \frac{i(i-i)}{2N} =-\xi \deltaf_{j}(\ell_{c})+ 
  \gamma v^{i}(j) \frac{j(j-1)}{2N}.
\end{equation}
We already know that the length corresponding to the limit of
stability is very close to the threshold length $\ell_{b}$, its
distance thereto being of the order of $\sqrt{\xi}$, as shown in
 appendix \ref{ch2:appA}. Therefore, to the lowest order, $\ell_{c}$ can be
approximated by $\ell_{b}$ and we get
\begin{equation}
  \label{ch2_eq:vcij}
  \frac{\gamma v^{i}(j)}{\xi}=\frac{2N[\deltaf_{j}(\ell_{b})-
    \deltaf_{i}(\ell_{b})]}{j(j-1)-i(i-1)}, \quad j>i.
\end{equation}

Clearly, the minimum of these velocities $v^{i}(j)$ is the one that matters.
  Let us denote by $j^{(i)}_{\text{min}}$ the position of the module for which
  $v^{i}(j)$ reaches its minimum value $v_{\text{min}}^{i}$,
\begin{equation}
  \label{ch2_eq:vci}
  v_{\text{min}}^{i}=v^{i}(j^{(i)}_{\text{min}})=\min_{j} v^{i}(j)
\end{equation}
for $v_{p}$ just below $v_{\text{min}}^{i}$, it is the $i$-th
  module that unfolds first, but for $v_{p}$ just above
  $v_{\text{min}}^{i}$, it is the $j^{(i)}_{\text{min}}$-th module that unfolds first.
 Let us denote the weakest module by $\alpha_{1}$, that is,
  $\deltaf_{i}(\ell_{b})$ is smallest for $i=\alpha_{1}$. If $v_p$ is
  smaller than $v_{\text{min}}^{\alpha_{1}}$, the first unit to reach the
  stability limit is the weakest one. Then, we rename the latter
  velocity $v_c^{(1)}$, that is,
\begin{equation}
v_c^{(1)} = v_{\text{min}}^{\alpha_{1}} , \quad
\deltaf_{\alpha_{1}}(\ell_{b})=\min_{i}\deltaf_{i}(\ell_{b}),
\label{ch2_eq:vcw}
\end{equation}
because it is the first one of a (possible) series of critical
  velocities separating different unfolding pathways, as detailed below.

Let us denote by $\alpha_{2}$ the module which unfolds first in the
  ``second'' velocity region, $v_{p}$ just above $v_{c}^{(1)}$, that
  is, $\alpha_{2}=j_{\text{min}}^{(\alpha_{1})}$. This unit ceases to be the
first to unfold for the velocity
\begin{equation}
v_c^{(2)} = v_{\text{min}}^{\alpha_{2}}.
\label{ch2_eq:vc1}
\end{equation}
The successive changes on the unfolding pathway take place at the critical 
velocities
\begin{equation}
v_c^{(k)} = v_{\min}^{\alpha_{k}},
\label{ch2_eq:vck}
\end{equation}
at which $\alpha_{k+1}=j_{\text{min}}^{(\alpha_{k})}$.  This
succession ends when $\alpha_{k+1}=N$: in that case, for
$v_{p}>v^{(k)}_c$, the first unit to unfold is always the pulled one.
This upper critical velocity $v_c^{\text{end}}$ can be computed in a
more direct way,\footnote{In our original publication \cite{PCCyP15},
  this equation had a typo, specifically an extra ``min'' in the
  subindex.}
\begin{equation}
v_c^{\text{end}} = \max_{j} v^{j}(N).
\label{ch2_eq:vce}
\end{equation}
Consistency of the theory requires that $v_c^{(k+1)}>v_c^{(k)}$. This can be proved right away. The consistency condition implies that
\begin{equation}
  \label{ch2_eq:29}
  \deltaf_{\alpha_{k+2}}(\ell_{b})>\frac{\deltaf_{\alpha_{k+1}}(\ell_{b})
    (\nu_{k+2}-\nu_{k}) -\deltaf_{\alpha_{k}}(\ell_{b})
    (\nu_{k+2}-\nu_{k+1})}
    {\nu_{k+1}-\nu_{k}},
\end{equation}
in which $\nu_k=\alpha_{k}(\alpha_{k}-1)$. Due to \eqref{ch2_eq:vck}, $\alpha_{k+1}$ minimizes
$v^{\alpha_{k}}(j)$. Therefore, in particular,
$v^{\alpha_{k}}(\alpha_{k+1})<v^{\alpha_{k}}(\alpha_{k+2})$, which is readily
shown to be equivalent to \eqref{ch2_eq:29} and proves the inequality.

We have a trivial case for $\alpha_{1}=N$, when the pulled unit is
precisely the weakest and it is always the first to unfold for any
pulling speed. The simplest nontrivial case appears when all the
modules have the same free energy with the exception of the weakest,
and $\alpha_{1} \neq N$, \eqref{ch2_eq:vcij},
\eqref{ch2_eq:vcw} and \eqref{ch2_eq:vce} reduce to
\begin{equation}
  \label{ch2_eq:26a}
  \frac{\gamma v^{(1)}_{c}}{\xi} = 
  \frac{\gamma v^{\text{end}}_{c}}{\xi}=\frac{2N[\deltaf_{N}(\ell_{b})- \deltaf_{\alpha_{1}}(\ell_{b})]}{N(N-1)-\alpha_{1}(\alpha_{1}-1)}.
\end{equation}
Note that the situation is quite simple, since there exist a single
critical velocity $v_c = v_c^{(1)} = v_c^{\text{end}}$.  For
$v_p < v_c$ the weakest module unfolds first whereas for $v_p > v_c$
the last one unfolds first.  If more units have different
free energies, the situation may be more complex, as shown in the
previous paragraph. There appear intermediate critical velocities,
which define pulling speed windows where neither the weakest
unit nor the last one is the first to unfold.  In order to obtain
these regions, we need to recursively evaluate \eqref{ch2_eq:vck}.

\section{Comparing our theory with simulations}
\label{ch2_sec:comp}
\subsection{Free energies of the units: shape and physical parameters}
\label{ch2_sec:shape}

Different shapes for the double-well potentials have been
  considered in the literature. They can be classified in, mainly, two different 
  classes: simple Landau-like quartic
  potentials which are employed to understand the basic mechanisms underlying the
  observed behaviors \cite{GMTCyC14,BCyP15,PCyB13}, and more complex
  realistic potentials, when the aim is 
  obtaining a more detailed, closer to quantitative, description of the
  experiments \cite{BCyP14,BCyP15,BGUKyF10,BHPSGByF12}. 
  
For the simplest modeling of a double well, we use a quartic potential $a_q(x)$
\begin{equation}
\label{ch2_eq:quartic-pot}
a_q(x)=\frac{\varepsilon}{4} \left[ \left(x - \sigma \right)^2 - \alpha^2 \right]^2.
\end{equation}  
The physical meaning of the parameters are straightforward: $\varepsilon$ scales the shape of the potential, $x=\sigma$ gives the position of the maximum of the barrier, whereas the minima are at $x=\sigma \pm \alpha$. This simple dependence allows us to compute analytically the borders of the metastability region, specifically one can obtain the stability threshold 
\begin{equation}
\ell_b=\sigma -\frac{\alpha}{\sqrt{3}},
\end{equation}
 and the forces within the metastability region fulfill 
 \begin{equation} 
|F|<F_b=a'(\ell_b)=\frac{2\sqrt{3}}{9} \varepsilon \alpha^3.
\end{equation}

Below, we use a nondimensional version of this potential. It is done by defining a length scale $[x]$ and a energy scale $\varepsilon [x]^4$,
\begin{equation}
\label{ch2_eq:quartic-pot-dimless}
a_q(x)=\frac{1}{4} \left[ \left(x - \sigma \right)^2 - \alpha^2 \right]^2
\end{equation}
where we have not used any special notation for the nondimensional variables in order not to clutter our formulae.

The more realistic proposal we use in this work has been put forward by Berkovich 
et al. some years ago \cite{BGUKyF10,BHPSGByF12}. Therein, the free energy of a
module is represented by the sum of a Morse
potential, which mimics the enthalpic minimum of the folded
state, and a WLC term \cite{MyS95}, which accounts for the entropic
contribution to the elasticity of the unfolded
state. Specifically, Berkovich's free energy $a_B(x)$ is written as
 \begin{equation}
\label{ch2_eq:berk-pot}
 a_B(x)=U_0 \left[ \left( 1 - e^{-2b\frac{x-R_c}{R_c}} \right)^2 -1 \right] + \frac{k_BT}{4P}L_c
 \left( \frac{1}{1-\frac{x}{L_c}} -1 -\frac{x}{L_c}+\frac{2x^2}{L_c^2} \right).
 \end{equation} 
 This shape has shown to be useful for some pulling experiments with
 actual proteins as titin I27 or ubiquitin
 \cite{BGUKyF10,BHPSGByF12}. Therein, each parameter has a neat
 physical interpretation. First, in the WLC part, we recall (see
 section \ref{ch1_sec:1_th_dev}) that we have: (i) the contour length
 $L_c$, which is the maximum length for the totally extended protein,
 and (ii) the persistence length $P$, which measures the
 characteristic length over which the chain is flexible. Both of them,
 $L_c$ and $P$, can be thought in terms of the number of amino acids
 in the chain.  Note that the derivative of the WLC term leads
 precisely to the force-extension curve given by
 \eqref{eq:WLC-force-ext}, with the identification
 $L\leftrightarrow x$.  Second, for the Morse contribution, we have:
 (iii) $R_c$, which gives the location of the enthalpic minimum and
 (iv) $U_0$ and $b$, which measure the depth and the width (in a
 nontrivial form) of the folded basin. The explicit function giving
 the stability threshold $\ell_b$ in terms of the parameters in
 Berkovich's potential cannot be obtained. However, we can always
 estimate $\ell_b$ numerically, solving $a_B''(\ell_b)=0$ for a
 specific set of parameters.

Once more, we use below nondimensional variables. In order to do so, we define a force scale $[F]$ and take $L_c$ as the length unit. Accordingly, dimensionless variables are introduced with
the definitions $\mu=U_0/(L_c[F])$, $\beta=2bL_c/R_c$, $\rho=R_c/L_c$,
$A=k_B TL_c/(4PU_0) $. 
Thus, the corresponding dimensionless potential reads
\begin{equation}
a_B(x)= \mu\!\left\{\left[1-e^{-\beta(x-\rho)}\right]^2-1  
 +A\!\left(\frac{1}{1-x}-1-x+2x^2\right) \right\}.    \label{ch2_eq:berk-dimless}
\end{equation}

\subsection{Numerical results}
\label{sec:numerics}

Here, we check the agreement between our
theory and the numerical integration of the evolution
equations. First, we discuss the validity of the simplifications
introduced in the development of the theory, namely (i) negligible
thermal noise and (ii) perfect length control. Second, we look into
the critical pulling speed, showing that there appears such a critical
speed in the simulations and comparing this numerical value 
with our theory.

We consider a system composed of $N=4$ units, 
such as the maltose binding protein \cite{GMTCyC14}, see section 
\ref{ch1_sec:unfold_path}. First,
each unit is assumed to be  characterized by a quartic bistable free energy. 
In reduced variables, the free energies have the 
form $a_{i} (x)= \epsilon_{i} a(x)$, 
with $\epsilon_{i}=1$ for $i\neq 1$, $\epsilon_{1}< 1$, where $a(x)=a_q(x)$ 
as given by \eqref{ch2_eq:quartic-pot-dimless}, with 
$\sigma=0$ and $\alpha=3$. Here, the value of $\sigma$ is
  different from the one in \cite{GMTCyC14} ($\sigma=8$). Its
  only effect is a shift of the origin of the extensions, our choice
  implies that a positive (negative) sign of the extension corresponds
  to an unfolded (folded) configuration.
The value of the friction coefficient is, also in reduced variables, 
$\gamma=1$. We use these dimensionless reduced variables to 
make it easier to compare our results with those in \cite{GMTCyC14}.
 
As stated above, the function \eqref{ch2_eq:quartic-pot-dimless} is one of the simplest, but
reasonable, choices to describe the free energy of different unit
of the same modular biomolecule.  
Using the notation introduced in \eqref{ch2_eq:10}, we have
\begin{equation}
\deltaf_{i}(x)=0, \; i\neq 1, \qquad \deltaf_{1}(x)=-\xi a_q'(x),
\end{equation}
with $\xi=1-\epsilon_1$. Equations
\eqref{ch2_eq:10b} and \eqref{ch2_eq:10c} give us the limits of
stability up to first order in the asymmetry $\xi$,
\begin{equation}
  \label{ch2_eq:34}
  x_{i,b}=\ell_{b},\;\forall i, \quad F_{i,b}=F_{b}, \; i\neq 1, \qquad F_{1,b}=(1-\xi_{1})F_{b}.
\end{equation}
For this simple example, \eqref{ch2_eq:34} is exact. 
The weakest unit is the first one, because $F_{1,b}$ is the minimum
value of the force at the limit of stability. For the values of
  the parameters we are using, $\ell_{b}=-\alpha/ \sqrt{3}=-1.73$ and 
$F_{b}= a'(\ell_b) = 2\sqrt{3} \alpha^3/9 = 10.4$. Since we are writing the
free energies for a common given value of the force, 
all the units have their two minima equally deep at the same
force. This assumption is made to keep things simple: the main
ingredient for having an unfolding pathway that depends on the pulling
speed is to have different values of the forces $F_{i,b}$ at the
stability threshold for the different units.

In the case we are considering, the weakest unit is the first one,
while the others share the same free energy. This means that we have
the simplest scenario for the critical velocity in our theoretical
approach: either the weakest (for $v_{p}<v_{c}$) or the last
(for $v_{p}>v_{c}$) unit that opens first, as discussed at the end of
the section \ref{sec:real_chain}. Here, (\ref{ch2_eq:26a}) for $\alpha_{1}=1$ and
$N=4$ reduces to
\begin{equation}\label{ch2_eq:simple}
\frac{\gamma v_{c}}{\xi}=\frac{2}{3} F_{b}.
\end{equation}

\begin{figure}
\centering
\includegraphics[width=0.8\textwidth]{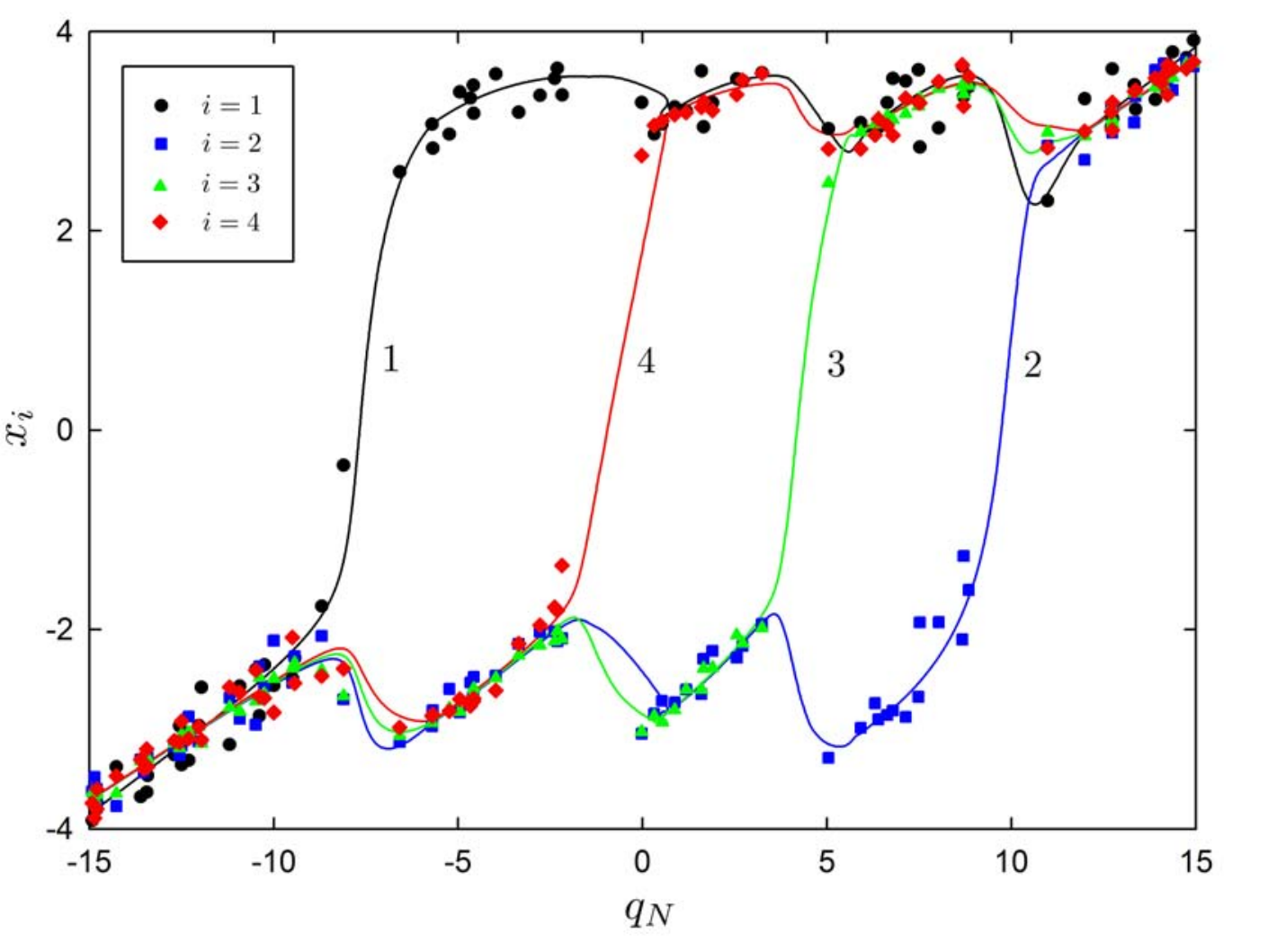}
\caption{\label{fig:3ch2} Evolution of the units' extensions 
  as a function of the system length
  $q_{N}$. The pulling speed is $v_{p}=0.38$ and the
    length control device has a stiffness $k_{c}=5$.
  The symbols correspond to a typical realization of the
    Langevin process \eqref{ch2_eq:5} with $T=1$, whereas the lines
  correspond to the deterministic (zero noise) approximation.  }
\end{figure}

To start with, we consider the relevance of the noise terms in
\eqref{ch2_eq:5}. In figure \ref{fig:3ch2}, we plot the integration of the
Langevin equations together with the deterministic approximation
\cite{vK92} for a particular case: the first unit's free energy 
 corresponds to $\epsilon_{1}=0.8$ ($\xi=0.2$), the
  stiffness of the length control device  is $k_{c}=5$, the
  temperature is $T=1$, and the pulling speed is $v_{p}=0.38$. For
these values of the parameters, taken from \cite{GMTCyC14}, the
critical velocity in \eqref{ch2_eq:simple} is $v_{c}=1.4$; thus we are
considering a subcritical velocity, $v_{p}<v_{c}$. Thermal
fluctuations are small, which entail that the same unfolding pathway is
observed in the deterministic and the majority of the stochastic
  trajectories.

\begin{figure}
  \centering
\includegraphics[width=0.8\textwidth]{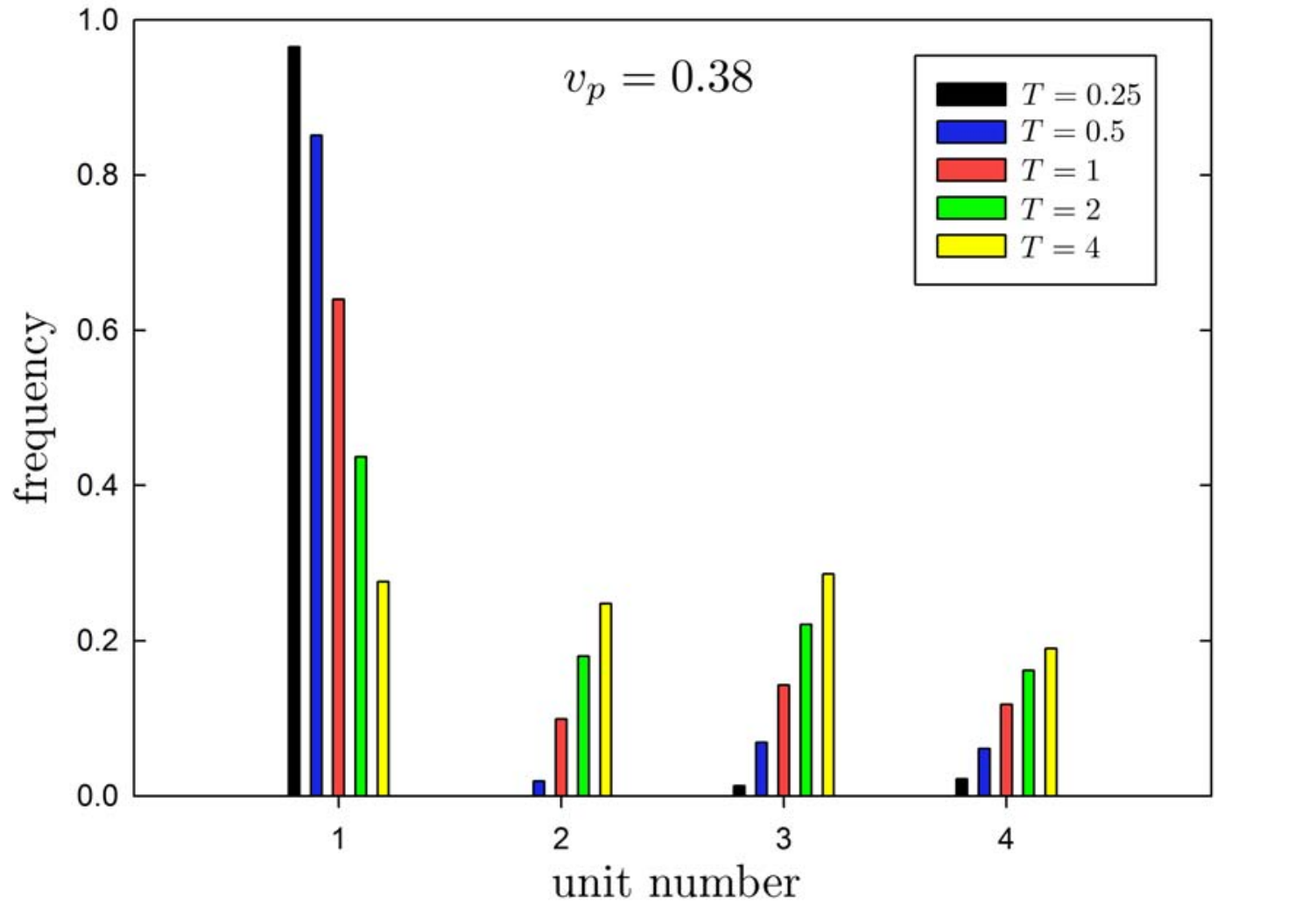}
\includegraphics[width=0.8\textwidth]{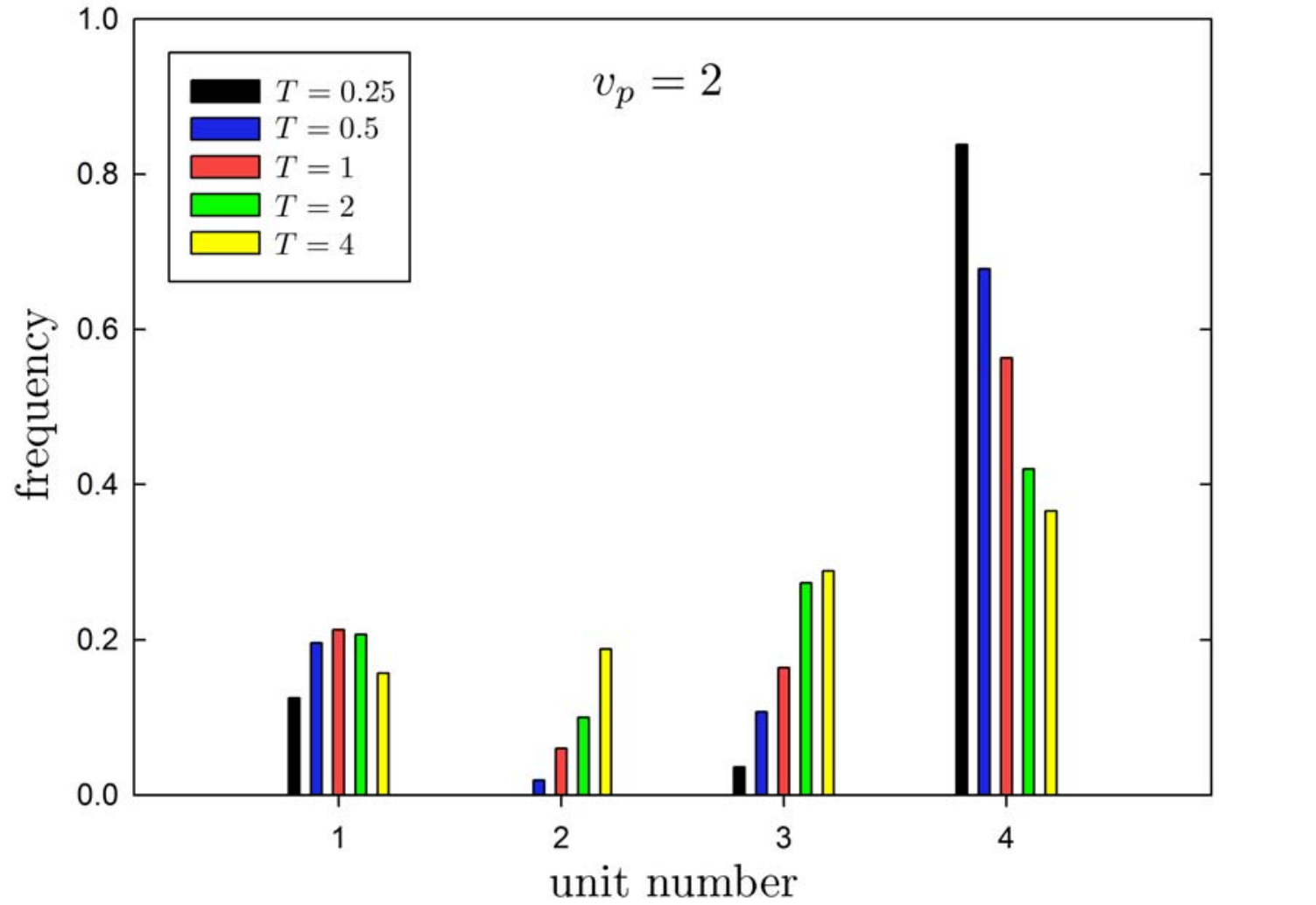}
\caption{First-unfolding frequency for each of the units.
  These frequencies have been obtained by integrating  the
  Langevin equations with perfect length control, for
  different values of the temperature.  (Top) Numerical frequencies
  obtained in $1000$ trajectories, for a subcritical pulling speed
  $v_{p}=0.38<v_{c}$.  (Bottom) The same as in the top panel, but for
  a supercritical pulling speed $v_{p}=2>v_{c}$. As the temperature
  decreases, the frequency of the deterministic unfolding pathway
  approaches unity in both cases.}
  \label{fig:newch2}
\end{figure}

Let us consider in more detail the relevance of thermal noise:
  from a physical point of view, it may be inferred by looking at the
  height of the free energy barrier at the critical force in terms of
  the thermal energy $k_{B}T$. For the values of the parameters we are
  using, this barrier is around $20k_BT$, which  explains
  why thermal noise is basically negligible in figure \ref{fig:3ch2}. If the temperature is decreased from $T=1$ to $T=0.25$, the
  barrier is so high, around $80$ times the thermal energy, that
  essentially all the stochastic trajectories coincide with the
  deterministic one. On the other hand, if the temperature is
  increased to $T=4$, the barrier in only a few $k_B T$,
   and we expect that the deterministic approximation
  ceases to be valid. 
  
  In order to further clarify the role played by the temperature, we
  present figure~\ref{fig:newch2}. Both panels display bar graphs with
  the frequencies with which each unit unfolds first in the stochastic
  trajectories. Specifically, the statistics shown has been
  obtained with $1000$ trajectories of the Langevin
  equations \eqref{ch2_eq:5} with perfect length control, and 
  several different
  values of the temperature. In the top panel, a subcritical velocity
  $v_{p}=0.38<v_{c}$ is considered, so that the weakest (first) unit
  is expected to unfold first. In the bottom panel, the numerical data
  for a supercritical velocity $v_{p}=2>v_{c}$ are shown, for which the
  pulled (fourth) unit would unfold first. The effect of thermal noise
  is quite similar in both cases. For the lowest temperature $T=0.25$,
  the frequency of the deterministic pathway is close to unity and,
  for the temperature in figure~\ref{fig:3ch2}, $T=1$, its frequency is
  still very large, clearly larger than any of the others. On the
  other hand, for the highest temperature, $T=4$, thermal
  noise is no longer negligible.

\begin{figure}
  \centering
  \includegraphics[width=0.8  \textwidth]{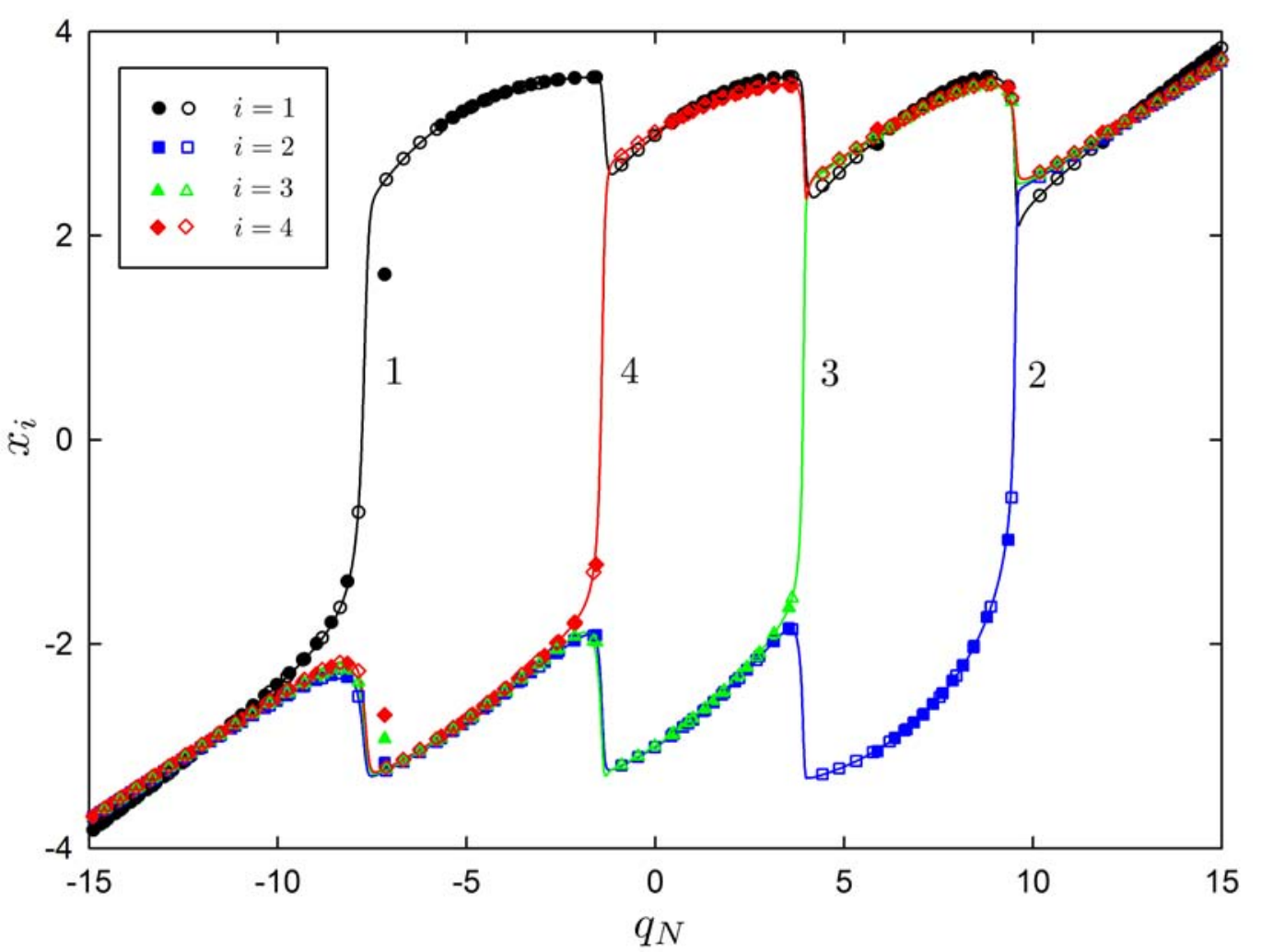}
  \includegraphics[width=0.8 \textwidth]{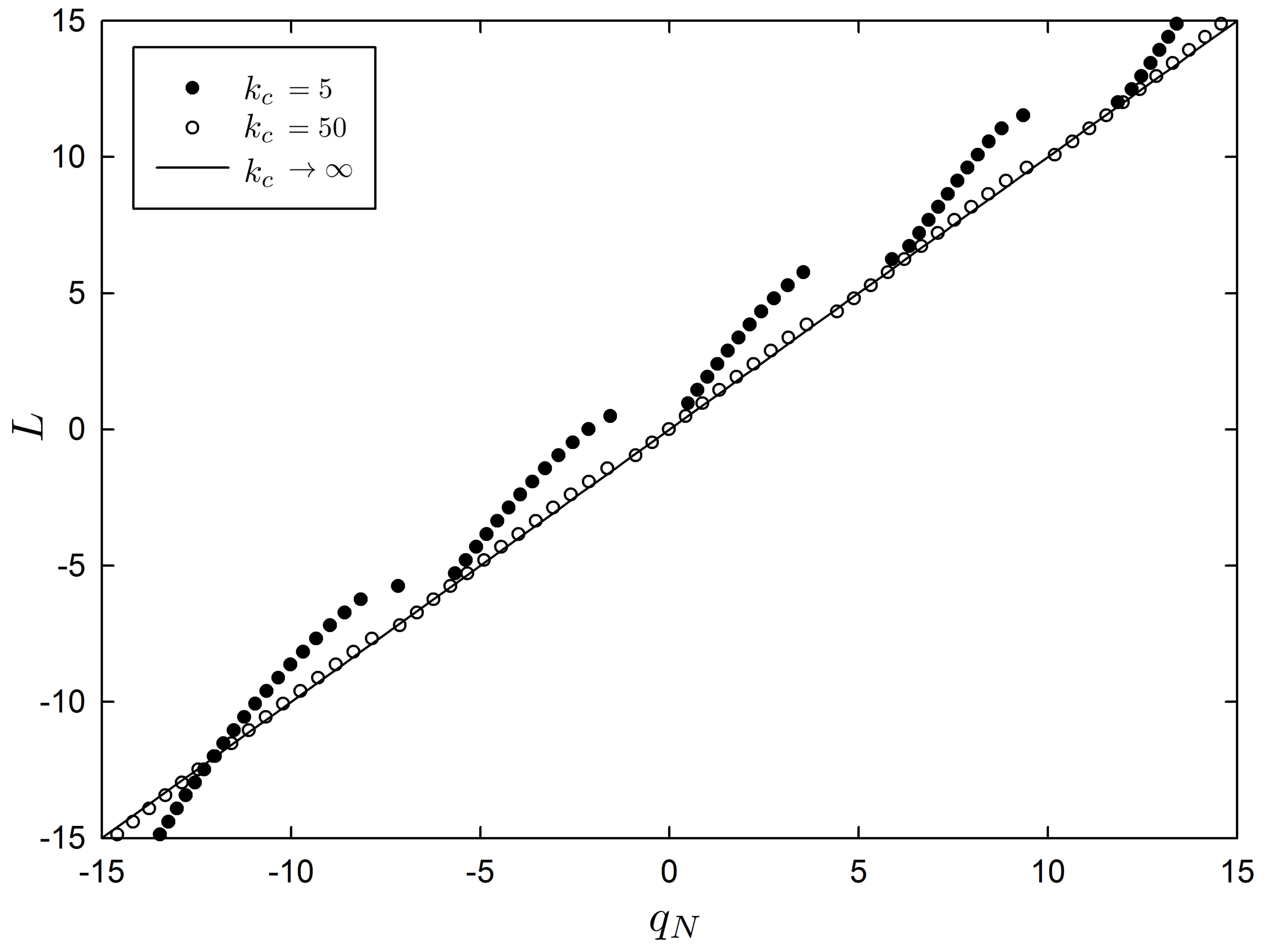}
  \caption{(Top) Evolution of the units' extensions 
   as a function of the system length 
    $q_{N}$. The symbols correspond to the integration of the
    deterministic equations, for $k_{c}=5$ (filled) and
    $k_{c}=50$ (empty), whereas the line corresponds to the
    limit of perfect length control,
    $k_{c}\to\infty$. The pulling speed is the same as in
    figure~\ref{fig:3ch2}, that is, $v_{p}=0.38$. (Bottom) Comparison
    between the desired and actual lengths, $L$ and $q_{N}$.
    Clearly, the length control improves as $k_{c}$ increases.
  }
  \label{fig:4ch2}
\end{figure}

In the following, we restrict the analysis to the physically
  relevant case in which the deterministic approximation gives a good
  description of the first unfolding event. In figure \ref{fig:4ch2}
(top panel), we look into the same pulling experiment as before, but
now we compare the deterministic evolution of the extensions for two
finite values of the stiffness to the $k_{c}\to\infty$
limit. Consistently with our expectations, the unfolding pathway is
not affected by this simplification. 
For the smaller values of $k_c$, the length is not perfectly controlled, but the 
length control improves as $k_c$ increases, as seen in the bottom panel. 
Notwithstanding, 
 the curves in the top panel, which correspond to
different values of $k_{c}$, are almost perfectly superimposed when
plotted as a function of the real length of the system $q_{N}$ (but
not of the desired length $L$). This means that the real length
$q_{N}$ is a good reaction coordinate, as already stated in section
\ref{sec:real_chain}.

\begin{figure}
  \centering
  \includegraphics[width=0.8  \textwidth]{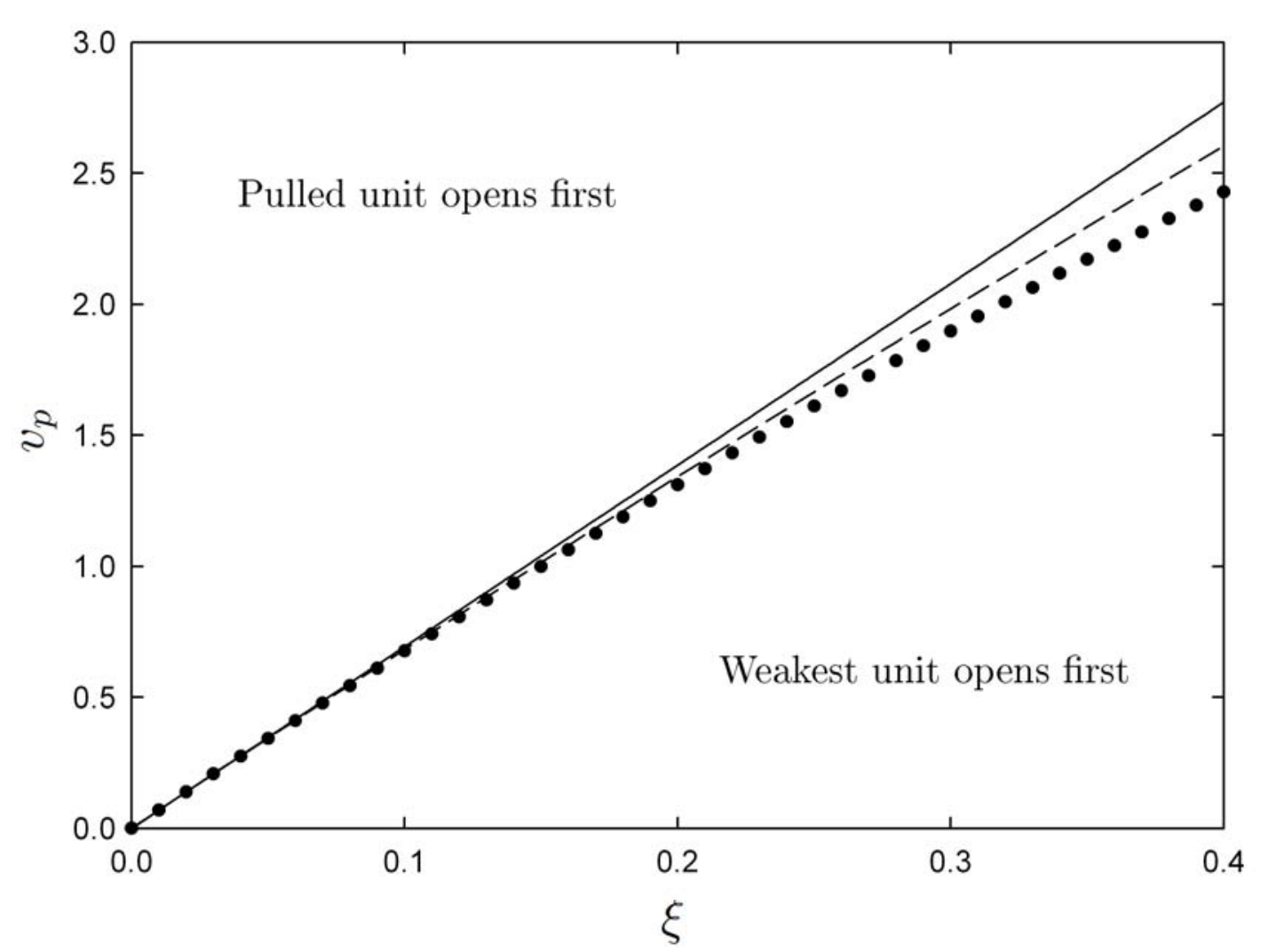}
  \caption{Phase diagram for the unfolding pathway in the pulling
    velocity-asymmetry plane for the quartic potential.
    Two well-defined regions are observed. These regions are separated by
    the curve critical velocity $v_{c}$ vs.
    asymmetry $\xi$  of the first unit. The
    numerical values for $v_{c}$ (circles) are compared to the
    theoretical expression \eqref{ch2_eq:simple} (solid line). The
    dashed line corresponds to the alternative approach discussed in
    the text, which improves the agreement with the numerical
    results for $\xi>0.1$. Error bars have been omitted
    because they are smaller than point size.  }
  \label{fig:5ch2}
\end{figure}

We have integrated the deterministic approximation  \eqref{ch2_eq:7} (zero noise) of the
Langevin equations  for different values of the
pulling speed, and extracted from them the numerical value of the
critical velocity as a function of the asymmetry
$\xi=1-\epsilon_{1}$. In order to obtain this numerical prediction, we initially set $v_{p}$ equal to the theoretical critical velocity given by \eqref{ch2_eq:simple}. Then, we recursively shift it by a small amount $\delta v_{p}$, such that $\delta v_{p}/v_c=0.0001$, until the pathway changes. We compare the values so obtained to the
theoretical expression \eqref{ch2_eq:simple}, in figure \ref{fig:5ch2}. We
find an excellent agreement for $\xi\lesssim 0.1$, for $\xi>0.1$ there
appear some quantitative discrepancies. They stem
from two points: (i) the perturbative expansion used for
obtaining \eqref{ch2_eq:26a} from \eqref{ch2_eq:stab} and (ii) the
intrinsically approximate character of \eqref{ch2_eq:stab}, since
$\ell_{b}$ gives rigorously the limit of stability only for the static
case $v_{p}=0$.  Therefore, we have looked for the solution of
\eqref{ch2_eq:stab} in the numerical integration of the deterministic
equations. This is the dashed line in figure \ref{fig:5ch2}, which
substantially improves the agreement between theory and numerics
because we have eliminated the deviations arising from point (i)
above. In fact, for the case we have studied in the previous
figures, which corresponds to a not so small asymmetry $\xi=0.2$, the
improved theory gives an almost perfect prediction for the critical
velocity. Note that the new numerical estimate of $v_c$ by solving 
\eqref{ch2_eq:stab} is always below that given by \eqref{ch2_eq:26a}. 
This can be easily understood: since the value $\ell_c$ at which $x_1$ and $x_4$ 
intersect is lower than $\ell_b$, the pulling velocity needed for the crossing 
is lower than \eqref{ch2_eq:26a}. 

Now we consider the  more realistic 
Berkovich's potential, given by \eqref{ch2_eq:berk-pot} and
 \eqref{ch2_eq:berk-dimless}. We take the values of the
parameters from \cite{BCyP15,ber10}, namely
\begin{equation}\label{ch3_eq:parameter}
 P=0.4\text{ nm}, \quad L_c=30\text{ nm},\quad R_c=4\text{ nm},\quad b=2,\quad U_0=100\text{ pN}\,\text{ nm},  
\end{equation} 
and $T=300$ K. Defining the force scale $[F]=100$ pN, 
the values of the nondimensional parameters in \eqref{ch2_eq:berk-dimless} are $\mu=0.0333$, $\beta=30$,
$\rho=0.133$ and $A=0.776$. In dimensionless variables, $F_{b}=0.527$
($52.7$ pN) and $\ell_{b}=0.157$ ($4.70$ nm). The relevant time scale
is set by the friction coefficient $\gamma$, $[t]= \gamma L_c/[F]$. In
turn, $\gamma$ is given by the Einstein relation $D=k_B T/\gamma$,
where $D$ is the diffusion coefficient for tethered proteins in
solution. We consider a typical value $D=1500$ nm$^2$/s, 
also taken from \cite{BGUKyF10}, so that 
$\gamma=0.0028$ pN nm$^{-1}$ s. 

\begin{figure}
  \centering
  \includegraphics[width=0.8  \textwidth]{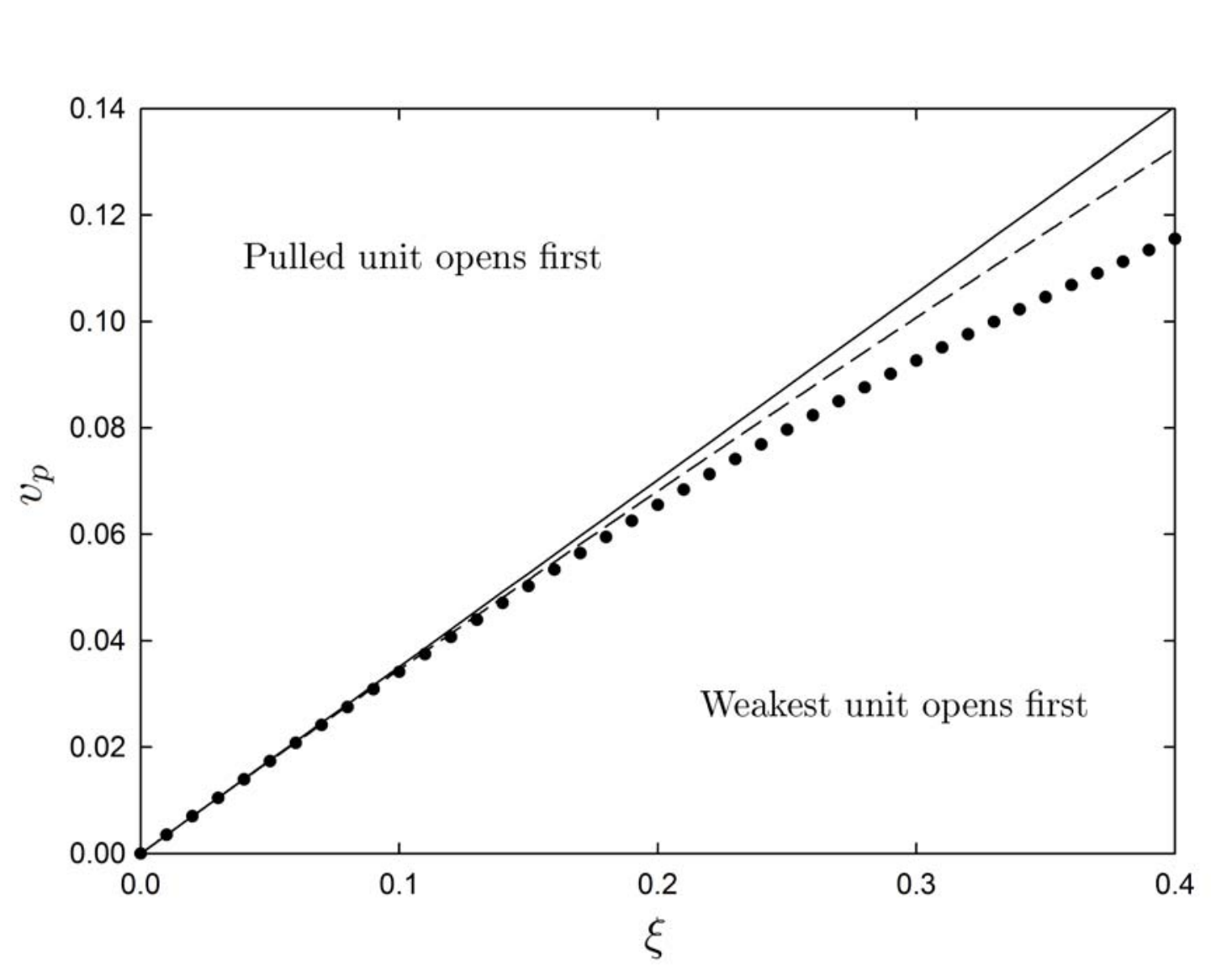}
  \caption{Phase diagram for the unfolding pathway in the pulling
    velocity-asymmetry plane for the Berkovich potential
    \eqref{ch2_eq:berk-dimless}. As in figure \ref{fig:5ch2}, there
    appear two well-defined regions, separated by the curve
    critical velocity $v_{c}$ vs.
    asymmetry $\xi$. Numerical values for $v_{c}$ (circles)
    compare very well with our theoretical prediction 
    \eqref{ch2_eq:26a} (solid line).
    Again, the dashed line corresponds to the
    alternative approach discussed in the text,
    which once more improves the agreement
    theory-simulation as $\xi$ increases.  }
\label{fig:6ch2}
\end{figure}

We consider a system of $4$ units, again with all the units but the
first being identical. Then, $a_{i}(x)=a_B(x)$, $i\neq 1$, and the first
unit is the weakest because $a_{1}(x)=(1-\xi)a_B(x)$. The situation is
then similar to the one we have already analyzed with the quartic
potential \eqref{ch2_eq:quartic-pot-dimless}, but there is a difference that should be
noted: here, $a_B(x)$ is the free energy at zero force, whereas for the
quartic potential $a_q(x)$ was the free energy at the critical force $F_{0}$ for
which the folded and unfolded minima were equally deep. Then, the force
here must not be interpreted as the extra force from $F_{0}$, but as
the whole force applied to the polyprotein. On the basis of
our theory, we expect the simplest situation with only one critical
velocity $v_{c}$, below (above) which the weakest unit (the pulled
unit) unfolds first. This is also indeed the case in the numerical
simulations, and we compare the theoretical and numerical critical
velocities in figure \ref{fig:6ch2}. A very good agreement is found
again, up to values of the asymmetry $\xi$ of the order $0.1-0.2$.

The above discussion shows that the validity of the theory presented
here is not restricted to simple potentials like the quartic one; on
the contrary, it can be confidently applied to situations in which
the units are described by realistic potentials. For the
typical parameters we are using, the theoretical critical velocity
$v_{c}$ for the Berkovich potential equals $1270$nm/s for an asymmetry
$\xi=0.1$. The latter can be regarded as a conservative estimate of
the largest asymmetries for which our theory gives an almost perfect
account of the unfolding pathway. Interestingly, this pulling
speed corresponds to the upper range of velocities employed in
AFM experiments, for instance see table I of \cite{HyD12}. Therefore,
testing our theory in real AFM experiments with modular proteins
should be achievable, see also section \ref{sec:experiments}.

Finally, we consider a more complex situation, in which more than one
unit is different from the rest. Therefore, there may exist more than one
critical velocity as discussed in section \ref{sec:critical_speed}. To be concrete, we have considered a system with
$4$ units in which $a_{2}(x)=a_{3}(x)=a_q(x)$, $a_{1}(x)=(1-\xi)a_q(x)$, as
before, but the pulled unit free energy is changed to 
$a_{4}(x)=(1+3\xi/2)a_q(x)$. In this situation, we have two different
critical velocities: for very low pulling speeds, the weakest unit is
the first to unfold, but there appears a velocity window inside which
neither the weakest nor the pulled unit is the first to unfold. This
stems from the fact that the first and the third unit reach
simultaneously the limit of stability for a velocity
$v^{1}(3)=4\xi \gamma^{-1} F_{b}/3$ that is smaller than the velocity
$v^{1}(4)=5\xi \gamma^{-1}F_{b}/3$ for which the first and the last
would do so. The physical reason behind this is the pulled unit's 
threshold force being larger enough than that of the third one.  We
recall that $v^{i}(j)$ is the velocity for which the $i$-th and the
$j$-th unit reach simultaneously their limits of
stability. Afterwards, the third unit and the fourth attain the limit
of stability in unison for a velocity $v^{3}(4)=2\xi\gamma^{-1}F_{b}$,
and the following picture emerges from our theory. Using the notation
introduced in section \ref{sec:critical_speed}, we define two critical
velocities,
  \begin{equation}
    \label{ch2_eq:1}
    \frac{\gamma v_{c}^{(1)}}{\xi}=\frac{4 F_{b}}{3}, \qquad
    \frac{\gamma v_{c}^{(2)}}{\xi}=2 F_{b},
  \end{equation}
  such that: (i) for $v_{p}<v_{c}^{(1)}$, it is the weakest unit that
    unfolds first, (ii) for $v_{c}^{(1)}<v_{p}<v_{c}^{(2)}$, it is the
    third unit that unfolds first, and (iii) for
    $v_{p}>v_{c}^{(2)}$, the first unit to unfold is the pulled one. 

\begin{figure}
  \centering
  \includegraphics[width=0.8\textwidth]{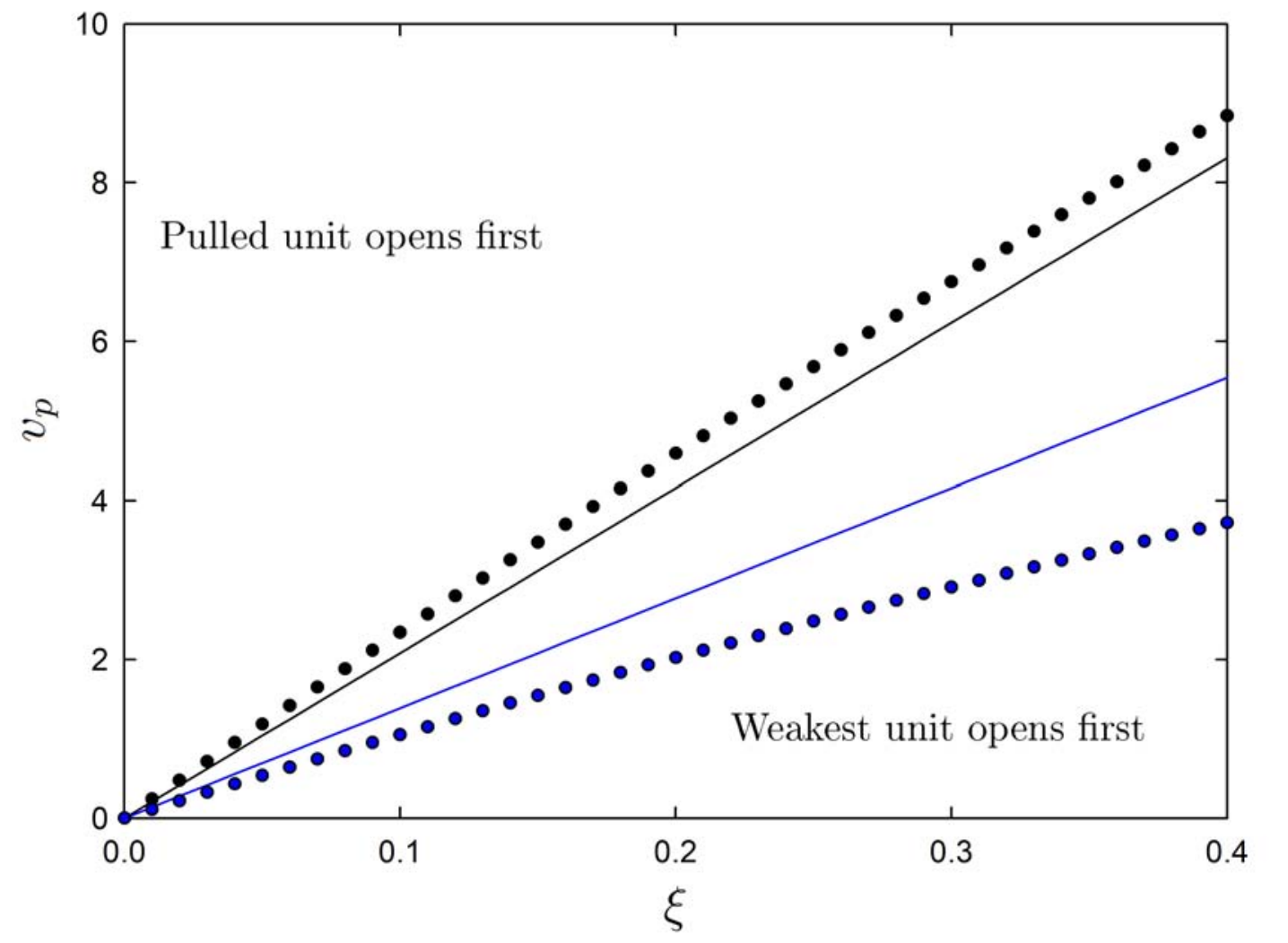}
  \caption{Phase diagram for the unfolding pathway in the pulling
    velocity-asymmetry plane for the more complex situation 
    with two critical velocities.
    Now, we have three well-defined regions,
    separated by two curves, respectively, critical velocities
    $v_{c}^{(1)}$ (blue) and $v_{c}^{(2)}$ (black) vs. asymmetry
    $\xi$. Note that our theoretical approach is able to reproduce
    the existence of the three different pulling regimes. Nevertheless, the
    discrepancies between theoretical and numerical values for
    the critical velocities are larger than those in figures \ref{fig:5ch2} and  
    \ref{fig:6ch2}.}
\label{fig:7ch2}
\end{figure}

    We check the more complex scenario described in the previous
    paragraph in figure \ref{fig:7ch2}. On the one hand, our
    theory correctly predicts the existence of the three pulling
    regimes described above. On the other hand, even for very small asymmetries,
    there appear some noticeable discrepancies between theory and
    simulation. The validity of the perturbative expansion for
    obtaining the critical velocities, expressed by condition
    \eqref{ch2_eq:stab}, is strongly supported by the accurateness of the
    theoretical prediction for the simplest case with 
    only one critical velocity, see figures
    \ref{fig:5ch2} and \ref{fig:6ch2}. Then, we believe that 
    this discrepancy stems from the
    intrinsically approximate character of the condition $a''_{i}=0$
    for determining the stability threshold  for finite 
    pulling velocity $v_{p}\neq 0$. Thus,
    improving the present theory should involve the
    derivation of a more accurate condition for the
    stability threshold in this case. This refinement to our theoretical framework,
    which probably makes a multiple scale analysis necessary
    for lengths close to the condition $a''=0$, is an open question that 
    deserves further investigation.

\section{Moving closer to the experiment}
\label{sec:stiff}

Now, we try to get closer to the experiment. We do so by introducing some variants 
of the basic model we have analyzed in the previous sections. The main idea is to 
sophisticate the model to make it more realistic, and test the robustness of our 
theoretical results.

In a real AFM experiment, the stiffness is finite and, as a result,
the control over the length is not perfect. Furthermore, the position
that is externally controlled is, usually, that of the platform and
the main elastic force stems from the bending of the tip of the
cantilever, as depicted in figure \ref{ch3_fig:sketch-experiment}.

Some authors \cite{HyS03} have used other elastic reactions that
reflect the attachment by means of flexible linkers among the platform
and the pulled end, and between consecutive modules. Here, we
will consider a perfect absorption, in order to keep the model as
simple as possible. 

Also, the procedure followed in the previous sections for making the free energy of 
one unit different from the rest may be considered a little bit artificial. 
Therefore, we also introduce here a more physical way of perturbing the free 
energies. Specifically, we do so by changing the contour length of the 
corresponding module.

This section is structured as follows. 
We study the effect on the unfolding pathway of the finite value of the stiffness 
and the location of the spring, in sections \ref{ch2_sec:finit} and 
\ref{ch2_sec:loc}, respectively. Finally, 
section \ref{ch2_sec:depe} is devoted to analyze the perturbation 
of the free energy of one unit 
brought to bear by the change of its contour length.

\subsection{Finite stiffness}\label{ch2_sec:finit}

Here, we still consider the basic model depicted in figure
\ref{fig:1ch2}, with the spring located at the pulled end.
Notwithstanding, we assume unperfect length control, that is, the
stiffness $k_c$ of the spring is finite.  Still, we consider the
macroscopic equations (zero noise), which are
\begin{subequations}
\label{ch3_eq:7v2}
\begin{align}
  \gamma\dot{x}_{1}  =&  -a'_{1}(x_{1})+a'_{2}(x_{2}), \\
  \gamma\dot{x}_{i}  =&  -2a'_{i}(x_{i})+a'_{i+1}(x_{i+1})+a'_{i-1}(x_{i-1}),
     \quad 1<i<N, \\
  \gamma\dot{x}_{N}  =&  -2a'_{N}(x_{N})+a'_{N-1}(x_{N-1})+ k_c \left(L- \sum_{k=1}^{N} x_k \right).
\end{align}
\end{subequations}
This system differs from that in \eqref{ch2_eq:7}
because, in the last equation, the Lagrange multiplier
$F$ is substituted by the harmonic force $k_c(L- \sum_{k} x_k)$. As in the previous case,
this system is analytically solvable by means of a perturbative
expansion in $v_p$ and $\xi$. The approximate solution for the
extension $x_i$ is
\begin{align}
x_{i}=\ell +& \frac{\xi N k_c \overline{\delta f}(\ell)-v_{p}\gamma k_c \dfrac{[3U''(\ell)+k_c(N-1)]N(N+1)}{6[Nk_c+U''(\ell)]}}{U''(\ell)[Nk_c+a''(\ell)]}  \nonumber
\\
+& \frac{v_{p}\gamma k_c i(i-1)-2\xi [Nk_c+a''(\ell)] \delta f_{i}(\ell)}{2a''(\ell)[Nk_c+a''(\ell)]}.
\label{ch3_eq:24}
\end{align}
Here $\ell \neq L/N$, it stems from the relation
\begin{equation}
\label{zeroth}
a'(\ell)=k_c(L-N\ell).
\end{equation}
We can see easily how we reobtain \eqref{ch2_eq:23} taking the limit
$k_c \to \infty$ in \eqref{ch3_eq:24}, as it should be. Although the
solution is slightly different, it still breaks down when $a''(\ell)$
vanishes, that is, when $\ell\to \ell_{b}$. Therefore, to the lowest
order, again we have to seek a solution of \eqref{ch2_eq:stab}, with the
extensions given by \eqref{ch3_eq:24}, and substitute
$\ell_c \simeq \ell_b$ therein. This leads to the same critical
velocities found for the infinite stiffness limit.

\begin{figure}
\centering
  \includegraphics[width=0.85 \textwidth]{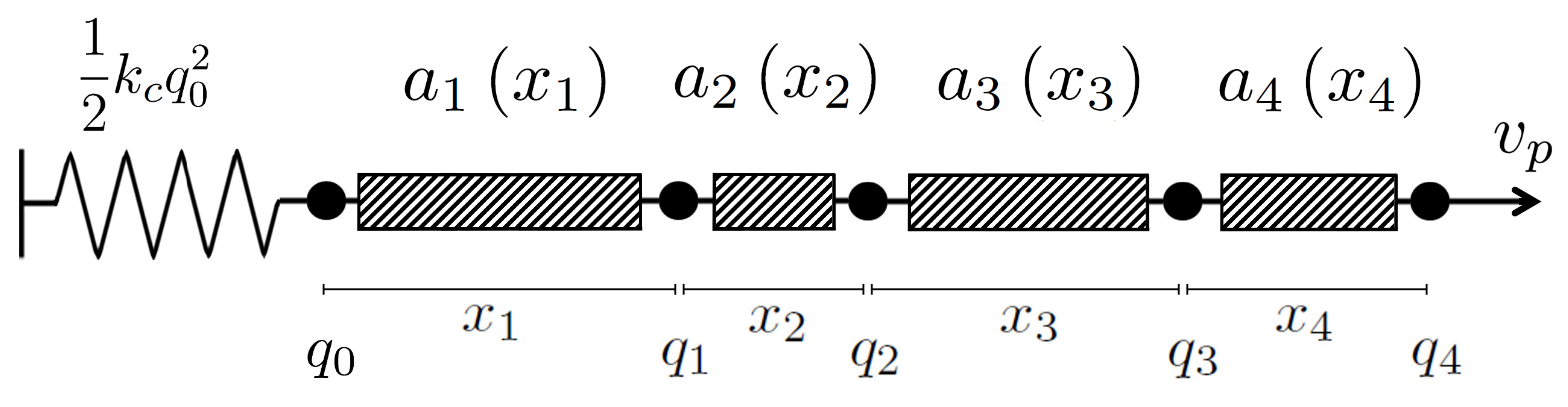}
  \caption{\label{ch3_fig:2} Sketch of the model for a protein with four
    units. It is identical to figure \ref{fig:1ch2}, except for the position
    of the length control device, which is now located at the
    fixed end.  }
\end{figure}

\subsection{Location of the elastic reaction}\label{ch2_sec:loc}

As depicted in figure \ref{ch3_fig:sketch-experiment}, in an AFM experiment
the distance between the moving platform and the fixed cantilever is
the controlled quantity.  Then, the model sketched in figure \ref{ch3_fig:2}
is closer to the experimental setup: the left end corresponds to the
fixed cantilever, with $q_0$ standing for $\Delta X$, and the
right end represents the moving platform. Thus, the free energy of
this setup is given by
\begin{equation}
\label{freeG2}
  A(q_{0},\ldots,q_{N})=\sum_{i=1}^{N} a_{i}(q_{i}-q_{i-1})+\frac{1}{2}k_c q_0^2 \, .
\end{equation}
From the free energy \eqref{freeG2}, we derive the Langevin
equations by making use of \eqref{ch2_eq:5}. The macroscopic
equations (zero noise) read
\begin{subequations}\label{ch3_eq:7v3}
\begin{align}
  \gamma\dot{x}_{1}  =&  -2 a'_{1}(x_{1})+a'_{2}(x_{2})+ k_c \left(L- \sum_{k=1}^{N} x_k \right), \label{ch3_eq:7v3-a} \\
  \gamma\dot{x}_{i}  =&  -2a'_{i}(x_{i})+a'_{i+1}(x_{i+1})+a'_{i-1}(x_{i-1}),
     \quad 1<i<N, \\
  \gamma\dot{x}_{N}  =&  -a'_{N}(x_{N})+a'_{N-1}(x_{N-1})+ \gamma v_p.
\end{align}
\end{subequations}

In the infinite stiffness limit, $k_c \to \infty$, the
harmonic contribution in \eqref{ch3_eq:7v3-a} tends 
to a new Lagrange multiplier $F$
such that $\sum_i x_{i}=L$. By summing up all the equations,
 it is obtained that
$F=a_1'(x_1)$ and the resulting system is exactly equal to
  that in \eqref{ch2_eq:7}. This is logical: if the spring
is totally stiff and then the control over the length is perfect, the
two models are identical. It is worth emphasizing that
  the two variants of the model, with the spring at either the
    fixed or moving end, have the same number of degrees of
  freedom. In the original model, the left end is fixed,
   $q_0=0$ and our degrees of freedom are
  $q_i$, $i=1, \ldots, N$, whereas in figure \ref{ch3_fig:2} we have the
  dynamical constraint $q_N=L$ and the degrees of freedom are $q_i$,
  $i=0,\ldots, N-1$. In the limit as $k_c \to \infty$, we
  have the constraints $q_{0}=0$ and $q_{N}=L$ in both
  models, making it obvious that they are identical.

The system \eqref{ch3_eq:7v3} with finite stiffness 
can be solved in an analogous way,
  by means of a perturbative expansion in the asymmetry
  $\xi$ and the pulling velocity $v_{p}$. The result is
\begin{align}
x_{i}=\ell +& \frac{\xi N k_c \overline{\delta f}(\ell)-v_{p}\gamma k_c\dfrac{[3a''(\ell)+k_c(N-1)]N(N+1)}{6[Nk_c+a''(\ell)]}}{a''(\ell)[Nk_c+a''(\ell)]}  \nonumber
\\
+& \frac{v_{p}\gamma k_c i\left(i-1+\dfrac{2a''(\ell)}{k_c}\right)-2\xi [Nk_c+a''(\ell)] \delta f_{i}(\ell)}{2a''(\ell)[Nk_c+a''(\ell)]},
\label{ch3_eq:25}
\end{align}
where $\ell$ is again given by \eqref{zeroth}. Of course, we
can reobtain \eqref{ch2_eq:23} by taking the infinite stiffness
limit in \eqref{ch3_eq:25}. Although the final solution for the extension
is different from the previous one, when we look for the critical
velocities and make the approximation $\ell_c \simeq \ell_b$ we get
the same analytical results for them.

The main conclusion of the last two sections, 
\ref{ch2_sec:finit} and \ref{ch2_sec:loc},
is that the existence of a set of critical
velocities, setting apart regions where the first unit to unfold is
different, is robust. In particular, it is not an artificial effect of either the 
limit $k_c \to \infty$
or the location of the spring.
Indeed, at the lowest order, all the variants of the model
give the same critical velocities. This robustness is an
appealing feature of our theory, which makes it reasonable to seek the
predicted phenomenology in real experiments.

\subsection{Units with different contour lengths}\label{ch2_sec:depe}

In the experiments, the observation of the unfolding pathway is not
trivial at all. The typical outcome of AFM experiments is a
force-extension curve  in which the identification of the
unfolding events is, in principle, not possible when the modules are
identical. Thus, in order to test our theory, molecular engineering
techniques that manipulate proteins adding some extra structures, such
as coiled-coil \cite{LSyM14} or Glycine \cite{SKSNyR03} probes, 
come in handy. For instance, a polyprotein in which all the modules
except one have the same contour length may be constructed in this
way. A reasonable model for this situation is a chain with modules
described by Berkovich's potentials \eqref{ch2_eq:berk-pot}
 with the same parameters for
all the modules, with the exception of the contour length of one of
them. According to our discussion in section \ref{sec:critical_speed}, 
this configuration is one of the simplest in which a critical velocity,
as given by \eqref{ch2_eq:26a}, emerges. In addition, this peculiar behavior
may be observed in real experiments, because the unfolding of the unit
that is different can be easily identified in the force-extension curve, 
see next section.

Then, we consider that the free energy of the different unit 
is perturbed in the more physical way described above. 
Accordingly, we change the module's free energy by considering that 
the contour length of the module is slightly increased, from $L_c$ to $L_c +\Delta
$. Consistently, we use $a(x)=a_B(x;L_c)$ to represent the free
energy of each of the identical modules, and $a_{1}(x)=a_B(x;L_c+\Delta)$ 
for that of
the first one. We have explicitly introduced in the notation that the only 
difference between the first unit and the rest is  the
slightly different contour length. Thus, we can
linearize $a_{1}(x)$ around $a(x)$, using the natural, dimensionless, asymmetry
parameter $\xi = \Delta/ L_c\ll 1$. Therefore,
\begin{equation} \label{ch3_eq:alin}
a_{1}'(x) \simeq a_B'(x;L_{c}) + \xi \delta f_{1}(x;L_{c}),
\end{equation}
where 
\begin{equation}
\label{ch3_eq:deltaf}
\delta f_{1}(x;L_{c}) \equiv L_c \frac{\partial a_B'(x;L_{c})}{\partial L_{c}}=-\frac{k_B T}{2P} \left[ \frac{\frac{x}{L_c}}{\left( 1-\frac{x}{L_c} \right)^3} +\frac{2x}{L_c} \right].
\end{equation}

The linearization in \eqref{ch3_eq:alin} is 
useful for the direct application of our theory
to  some  engineered systems, see next section. We would like to emphasize that $
\delta f_1(x;L_{c})<0$, since the function between brackets is 
always positive for $0<x<L_c$. That means that, for two units described by 
Berkovich's free energy
\eqref{ch2_eq:berk-pot} with the same values for all parameters except the contour 
length, the weakest unit is the longest one.

\section{Experimental prospect}
\label{sec:experiments}

Let us consider an example of a possible real experiment for a polyprotein with $N=10$ modules. We characterize the
modules by Berkovich's potentials with the parameters introduced in \eqref{ch3_eq:parameter}, with $k_BT=4.2\text{ pN}\,\text{ nm}$
and friction coefficient $\gamma=0.0028$pN$\,$nm$^{-1}$s
\cite{BGUKyF10}. We call this system M$_{10}$: since all the modules
are equal in M$_{10}$, it is not a very interesting system from the
point of view of our theory.  Nevertheless, 
 now we can resort to the ideas we have just put forward. 
 Thus, we consider a mutant
species M$'_{10}$ that is identical to M$_{10}$, except for the module
located in the first position (the fixed end), which has an insertion
adding $\Delta$ to its contour length. Our theory gives an estimate
for the critical velocity $v_c$ by inserting \eqref{ch3_eq:deltaf}
into \eqref{ch2_eq:26a} with $\alpha_1=1$.

\begin{figure}
\centering
\includegraphics[width=0.8 \textwidth]{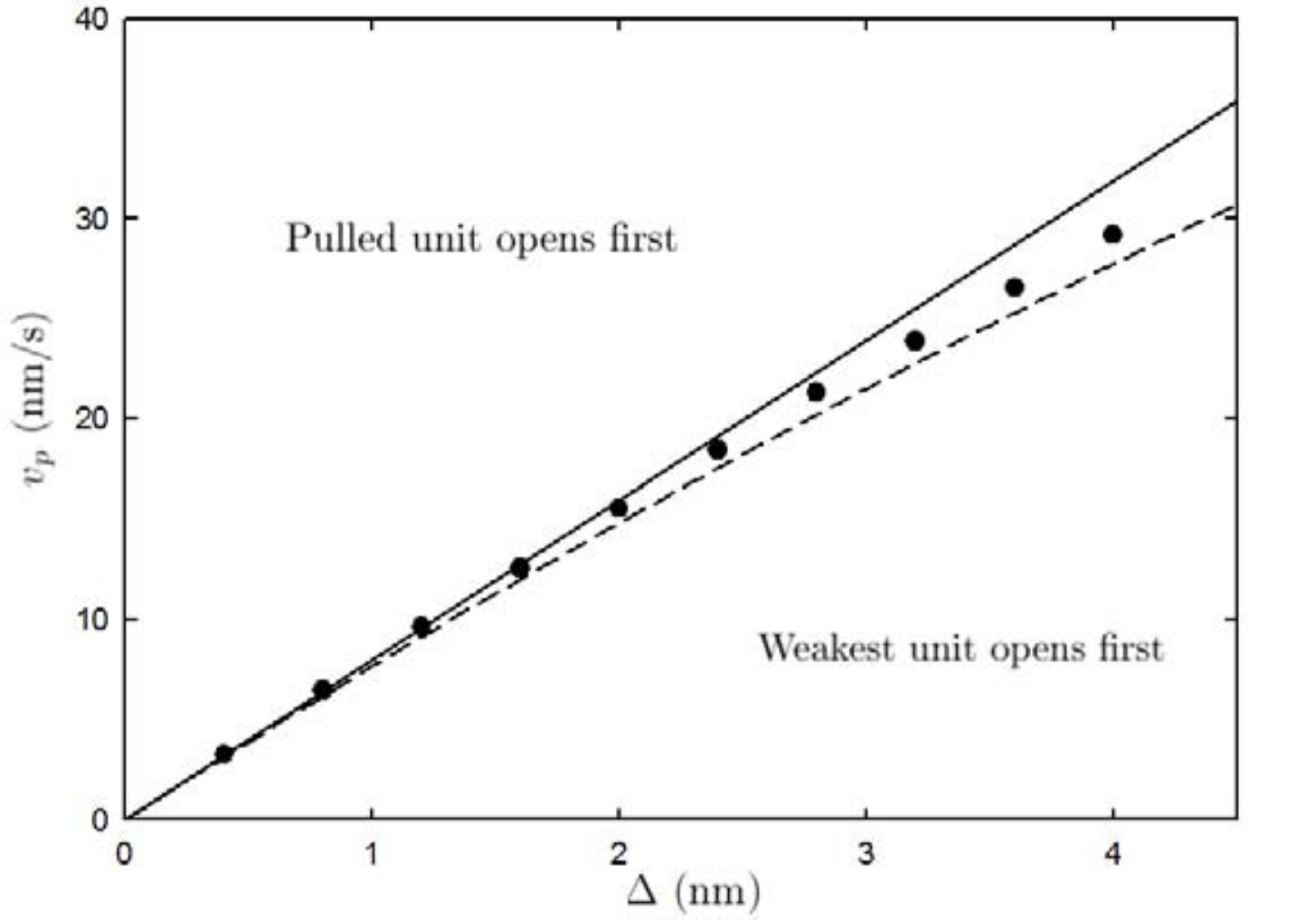}
\caption{\label{ch3_fig:phase} Critical velocity in the M$'_{10}$
  system. The parameter $\Delta$ is the first module's additional contour
  length. Numerical values (circles) are compared
  with two theoretical results:
  ``complete'' (dashed line) and linear (solid line).}
\end{figure}

In figure \ref{ch3_fig:phase}, we compare the theoretical estimate for the
critical velocity with the actual critical velocity obtained by
integration of the dynamical system \eqref{ch3_eq:7v3}. Specifically, we
have considered a system with spring constant $k_c=100$pN/nm. The
numerical strategy to determine $v_c$ has been quite similar to that in section \ref{sec:numerics}: starting
from a completely folded state we let the system evolve obeying
\eqref{ch3_eq:7v3}, with a ``high'' value of $v_p$---well above the
critical velocity---, up to the first unfolding. We tune $v_p$ down
until it is observed that the first module that unfolds is the weakest
one: this marks the actual critical velocity. There are two
theoretical lines: the solid line stems from the rigorous application
of \eqref{ch2_eq:26a}, with $\delta f_{1}$ given by \eqref{ch3_eq:deltaf}, and
$v_{c}$ is a linear function of $\xi$, whereas the dashed line
corresponds to the substitution in \eqref{ch2_eq:26a} of
$\xi \delta f_{1}(x)$ by $a_B'(x;L_{c}+\Delta)-a_B'(x;L_{c})$, without
linearizing in the asymmetry $\xi$.  Note the good agreement between
theory and numerics, especially in the ``complete'' theory where, for
the range of plotted values, the relative error never exceeds
5$\%$. Interestingly, most of the computed values of the critical velocity lie
in the range of typical AFM pulling speeds, from $10$ nm/s to $10^4$
nm/s \cite{AyA06}.

Below the critical velocity $v_{c}$, it is always the weakest 
unit that unfolds first. Above
$v_c$, the unit that unfolds first is the pulled one. For the sake of
concreteness, from now we consider a specific molecule M$'_{10}$
fixing $\Delta=2$ nm. Inserting the linear estimation \eqref{ch3_eq:deltaf} into
\eqref{ch2_eq:26a}, we get a critical velocity $v_c \simeq 16$ nm/s, which
is in the range of typical pulling speeds in AFM
experiments.

\begin{figure}
\centering
\includegraphics[width=0.8 \textwidth]{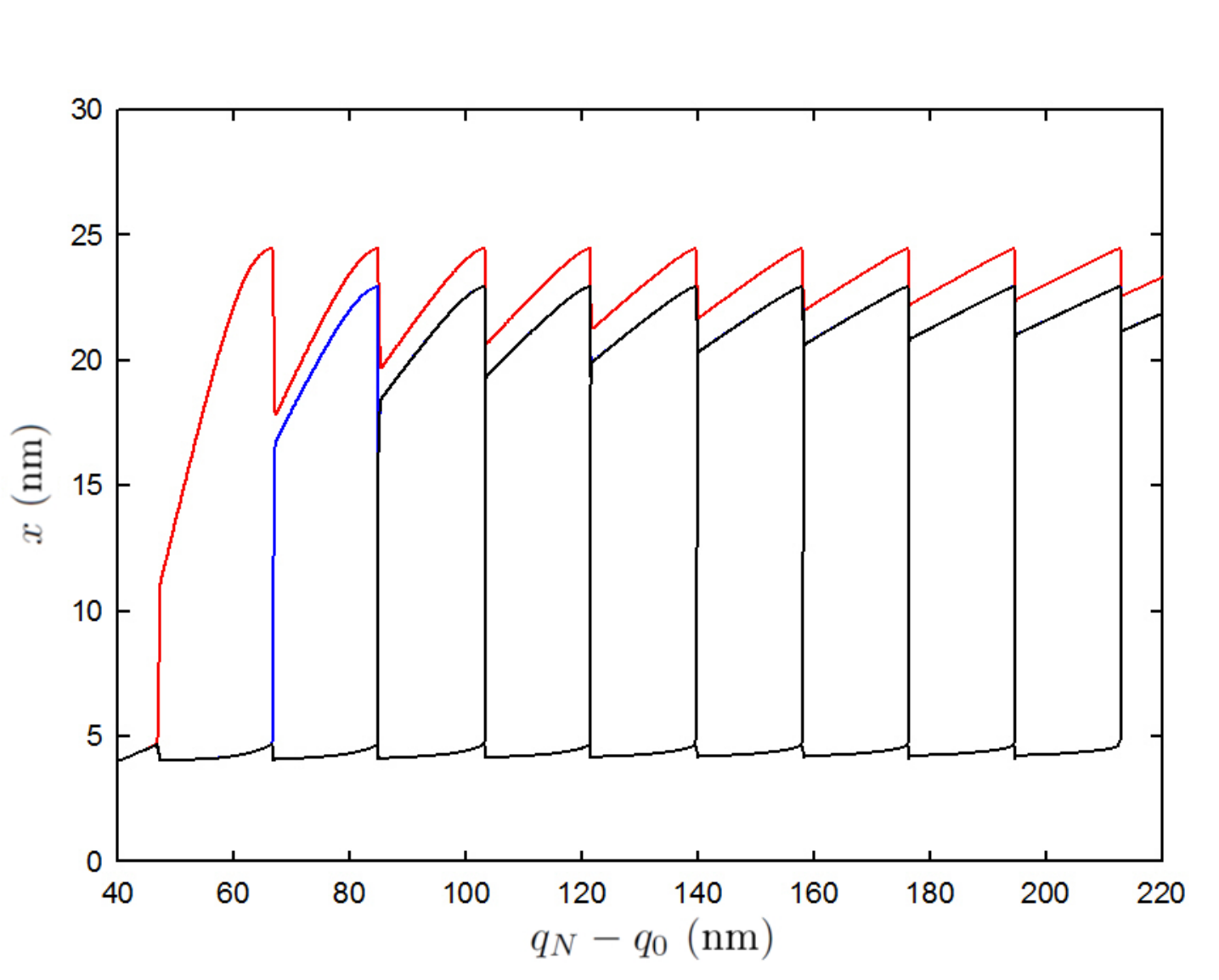}
\includegraphics[width=0.8 \textwidth]{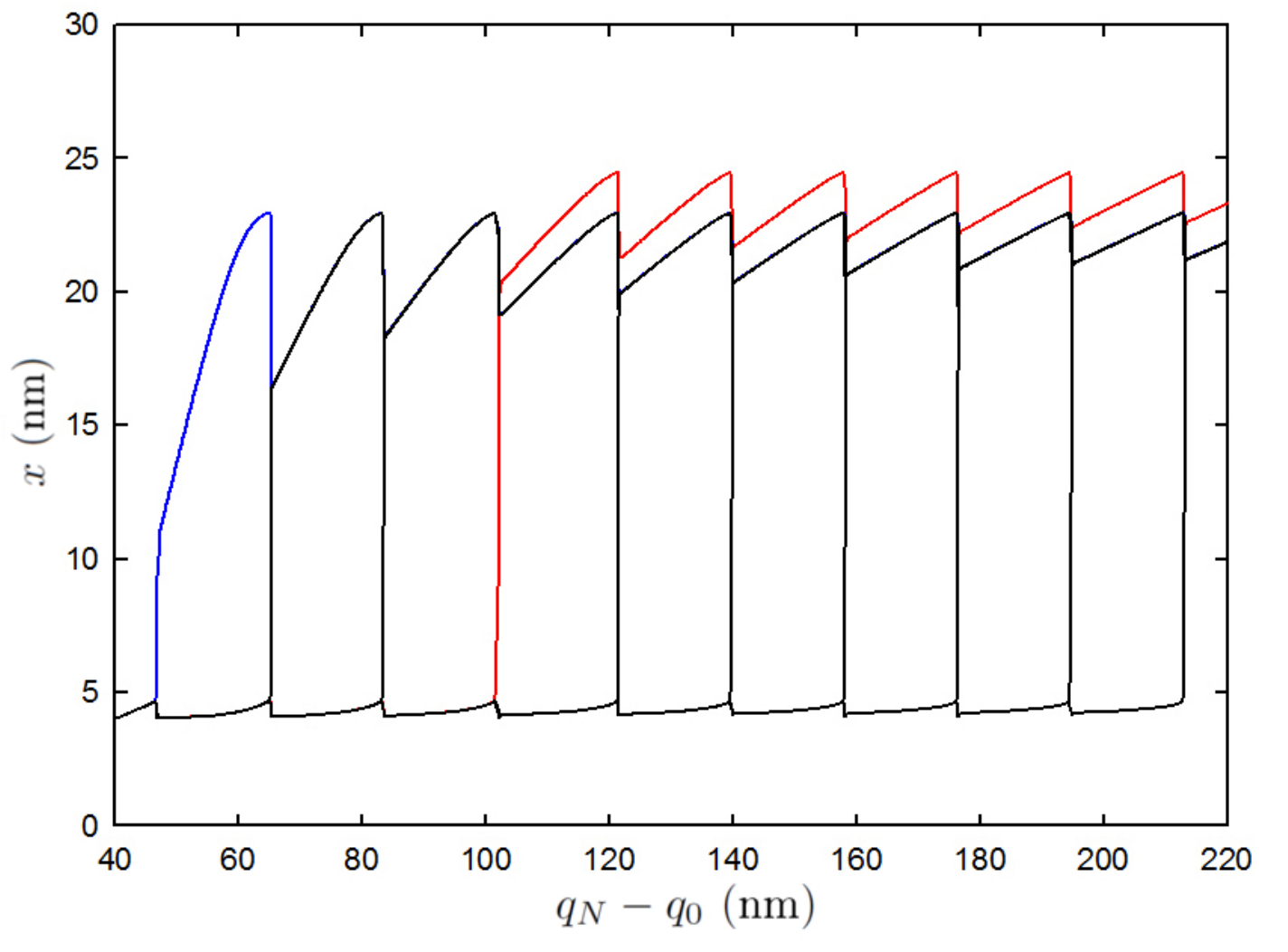}
\caption{\label{ch3_fig:3} Evolution of the units' extensions 
  as a function of the system length $q_{N}-q_{0}$. 
  The potential parameters are given in
  \eqref{ch3_eq:parameter}, and the pulling speeds are $v_{p}=10$ nm/s
  $<v_c$ (top) and $v_{p}=22$ nm/s $>v_c$ (bottom). The stiffness is
  $k_c= 100$ pN/nm, which lies in the range of typical AFM values.
  The red line corresponds to the weakest unit and the blue line to
  the pulled one.  }
\end{figure}

In figure \ref{ch3_fig:3}, we plot the extension of each unit vs the total
extension $q_N-q_0$ in our notation ($L$ in
figure \ref{ch3_fig:sketch-experiment}).
We have numerically integrated
 \eqref{ch3_eq:7v3} for two values of $v_{p}$: one below and one above
$v_c$, namely $v_p=10$ nm/s and $v_p=22$ nm/s. The red trace stands
for the weakest unit extension whereas the blue one corresponds to the
pulled module. We can see that, for $v_p=10\text{ nm/s}<v_c$, the
first unit that unfolds is the weakest one, whereas for
$v_p=22\text{ nm/s}>v_c$ that is no longer the case. Specifically, the
first unit that unfolds is the pulled one, and the weakest unfolds in
the fourth place.

\begin{figure}
\centering
\includegraphics[width=0.8 \textwidth]{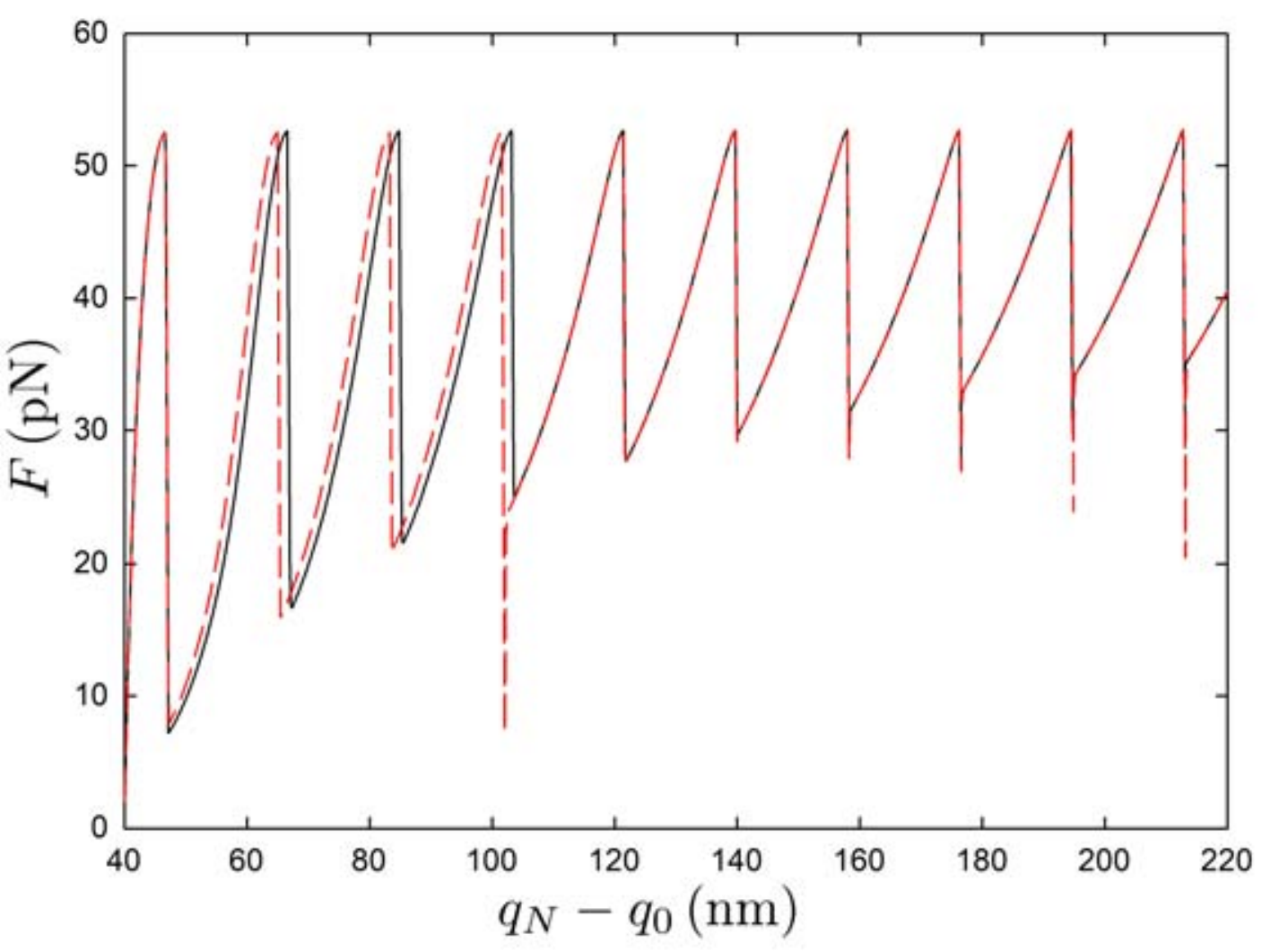}
\includegraphics[width=0.8 \textwidth]{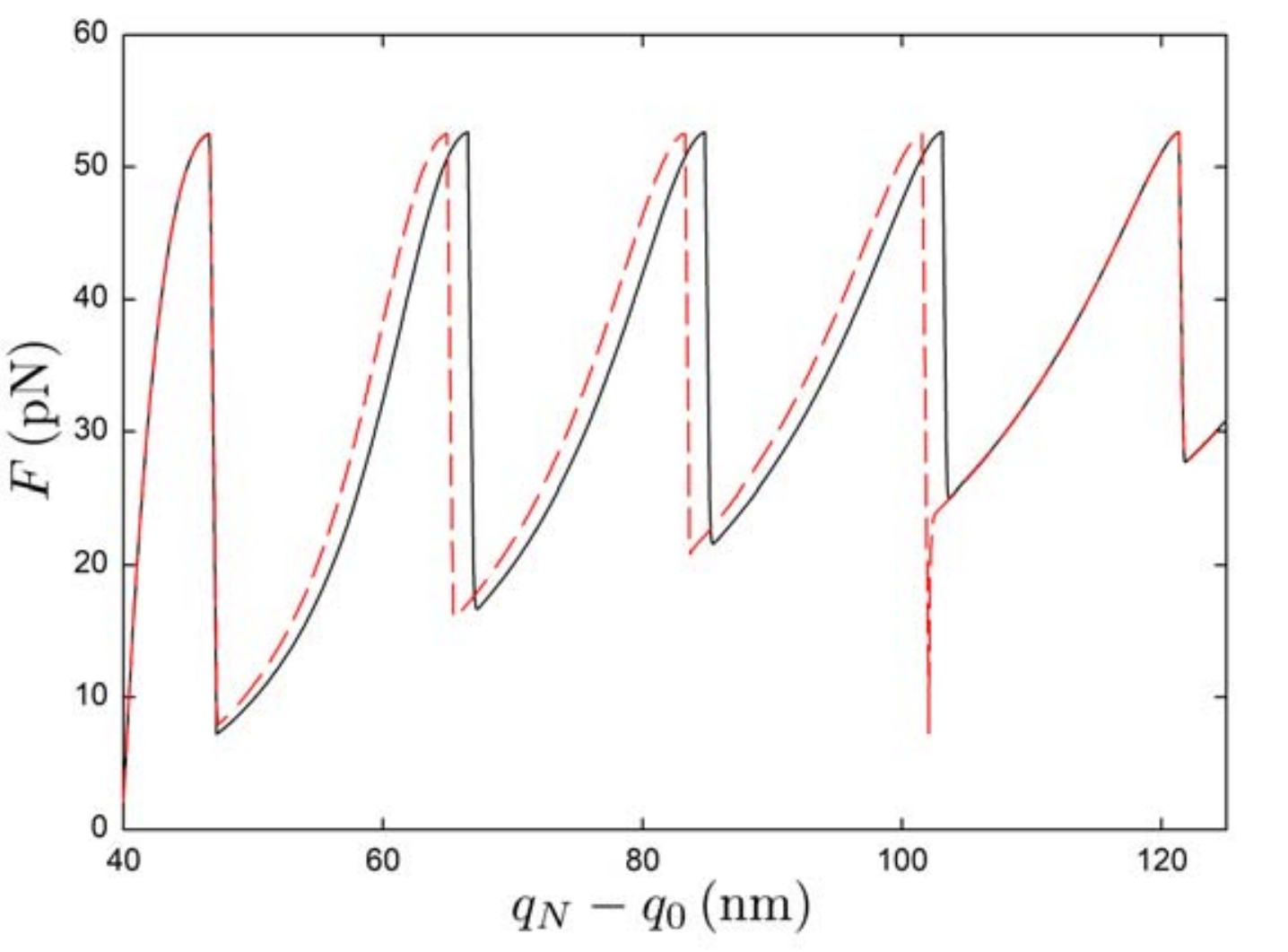}
\caption{\label{ch3_fig:4} (Top) Force-extension curve 
  corresponding to the pulling experiment in
  figure \ref{ch3_fig:3}. Specifically, we consider two pulling velocities
  $v_{p}=10$ nm/s (subcritical, solid) and $v_{p}=22$ nm/s
  (supercritical, dashed). (Bottom) Zoom of the region of interest,
  which clearly shows the shift between the peaks stemming from the increased
  contour length of the mutant unit.  }
\end{figure}

The plots in figure \ref{ch3_fig:3} are the most useful in order to detect
the unfolding pathway of the polyprotein. Unfortunately, they
are not accessible in real experiments, for which the
typical output is the force-extension curve, as already stated above.
As a consequence, we have
also plotted the force-extension curve in order to bring to
light the expected outcome of a real experiment. In figure \ref{ch3_fig:4}, we show the force-extension curve for the two
considered velocities in the same graph (solid line for the lower speed
and dashed line for the higher one).

The force-extension curves in figure \ref{ch3_fig:4} are superimposed until the first force
rip, which corresponds to the first unfolding event: that of the
mutant module for the slower velocity and that of the pulled unit for
the faster one. As the mutant unit has a longer contour length than
the rest, a shift between the curves in the next three spikes is found.
This shift stems from the effective contour length of the polyprotein 
having an extra
contribution of $2$ nm. Reasonably, for the higher velocity, this
shift disappears when the mutant module unfolds. Thenceforth, 
the force-extension curves are
once again superimposed. This plot clearly shows how the emergence of
a critical velocity  could be sought in a real experiment.

%

\chapter{Testing the model with molecular dynamics}
\label{ch:moldyn_pbio}
The theoretical approach developed in the previous chapter is clearly a drastic 
simplification of reality. In fact, the final result of our theory is 
deterministic, in the sense that the unfolding pathway is a definite one, the 
randomness coming from thermal fluctuations being effectively ``suppressed'' by the 
fast enough pulling velocity. In reality, as also discussed 
in chapter \ref{ch:model_pbio}, the 
unfolding pathway does have some degree of stochasticity, stemming from the 
interactions between  the molecule under study and the 
fluid where it is immersed, which are encoded in the Gaussian white noises of our 
Langevin description.

Testing the theory is a mandatory step of scientific method. The task of 
performing real pulling experiments with a modular protein that matches all the 
requirements of our theoretical framework is not simple at all. 
Therefore, in this chapter, we focus on steered molecular dynamic
(SMD) of engineered systems. Since the first investigations made by
Grubm{\"u}ller \cite{GHyT96} and Schulten \cite{LIKVyS98}, 
these computational techniques have shown to be
of crucial relevance in the current development of biophysics 
\cite{Is11,Zu10,FyS01}.
To test our theoretical predictions, we
consider a particularly simple structure composed of coiled coils. This kind
of structures is common in nature, which makes it extremely
useful as a model system \cite{PByS08,RMAVWyG10,TyL16}.

The organization of the chapter is as follows. Section \ref{sec:cand_to_test} is 
devoted to stress the relevance of the hypotheses assumed in the theory developed 
in the previous chapter. Also, we discuss its applicability to specific molecules
in specific ranges of velocities. Therein, we introduce the construct we work
with throughout this chapter, which comprises two 
consecutive coiled-coil structures.
The simulation procedure along with the method employed for the data analysis are 
presented in section \ref{ch4_sec:SMD}. 
Specifically, we explain how we acquire the initial configurations from which the 
construct is pulled. 
In section \ref{ch4_sec:pull-CC}, the 
results of our simulations are put forward, and they are compared 
with our theoretical prediction. 
A key point in our numerical analysis is our assumption about the independence of 
the initial conditions chosen for the pulling stage.
Then, in section \ref{ch4:appA}, we prove this data acquisition to be 
actually uncorrelated, within the accuracy of our statistics.

\section{The candidate protein}
\label{sec:cand_to_test}

\begin{figure}
\centering
 \includegraphics[width=0.95 \textwidth]{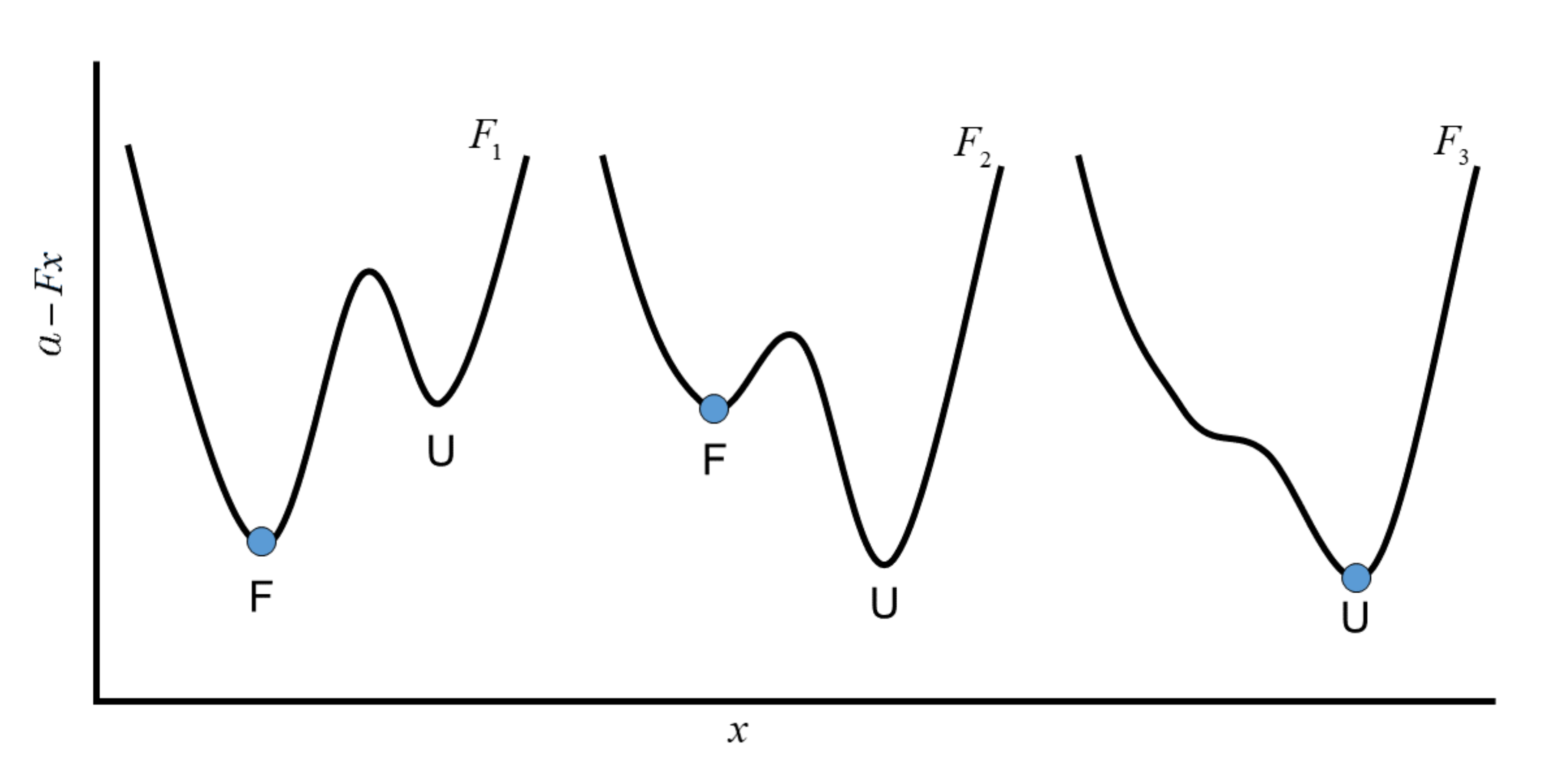}
 \includegraphics[width=0.7 \textwidth]{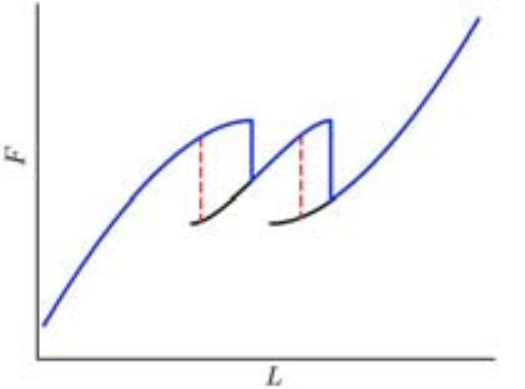}
  \caption{(Top) Schematic free energy landscape of a single repeat for three 
  different forces $F_1 < F_2 < F_3$. The system starts in the folded (F) state, 
  which is the
  absolute minimum for $F_1$. If thermal noise is negligible, the repeat remains in 
  the folded state for $F_2$ even when the unfolded state is more stable. For
  $F_3$, the F state disappears, and the repeat finally unfolds.
  (Bottom) Qualitative picture of the stability branches in a modular system with 
  two units. The blue
  line shows the unfolding pathway followed in the limit of the so-called ``maximum  
  hysteresis path'', when the pulling speed is high enough to make the
  system sweep the whole branches, including their metastable parts. In this limit, the jumps between consecutive
  branches take place by the mechanism shown in the top panel
  because the system does not have enough time to jump over the barrier separating
  the folded and unfolded states. In other words, the ``fast enough'' pulling speed
  effectively suppresses thermal fluctuations. Conversely, in the quasistatic limit
  the transition from folded to unfolded takes place at the
  lengths (dashed red lines) at which the branch with one more unfolded unit
  becomes more stable, that is, when its free energy becomes smaller.
  For the quasistatic case,  the system has always  time to find its way 
  through the barrier.}
\label{ch4_fig:max_hyst}
\end{figure}

We expect our theoretical approach to hold for some molecules within a specific range of pulling velocities. One of the obvious requirements our protein candidate must meet is a negligible interaction between repeats, since we have assumed no nearest-neighbor interaction terms in the global free energy. Regarding the range of velocities, if pulling is very slow and quasistatic, the
first unfolding event occurs at the length value for which the
free energy minima over the branch with all the units folded
and over the branch with only one unit unfolded are equally
deep \cite{PCyB13}. Therein, the jump between branches occurs by
thermal activation over the free energy barrier separating
them. As a consequence, the completely folded branch is only partially
swept, as marked by the dashed vertical (red) lines in figure \ref{ch4_fig:max_hyst}. In contrast, there is a range of fast pulling
velocities that do not give the system enough time to be
thermally activated over the barrier, but are slow enough to
allow it to sweep completely the part of the branches that
corresponds to metastable equilibrium states. In this case, the
jump between branches comes about at the limit of
metastability, only when the folded minimum disappears \cite{BCyP15},
as depicted in the top panel of figure \ref{ch4_fig:max_hyst}. This is
marked by the solid vertical (blue) lines in the bottom panel of
figure \ref{ch4_fig:max_hyst}. This range of adiabatic velocities we 
are interested in, as defined in chapter \ref{ch:model_pbio}, are also
said to lead to the ``maximum hysteresis path'' \cite{BZDyG16}, regime briefly 
introduced at the end of section \ref{ch1_sec:1_th_dev}.

\begin{figure}
\centering
 \includegraphics[width=0.55 \textwidth]{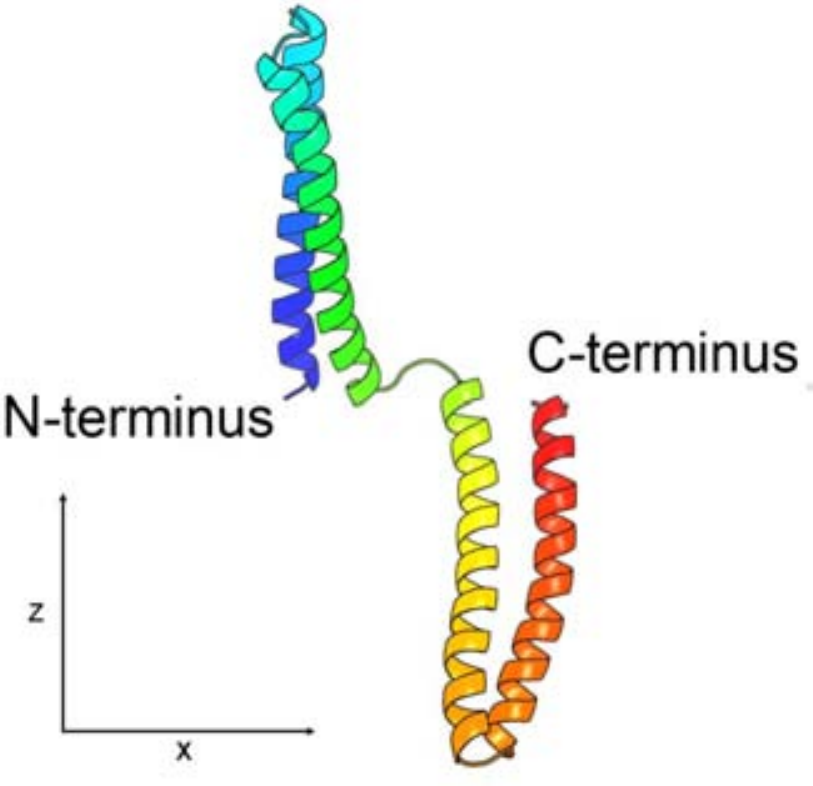}
  \caption{Initial conformation of the homopolyprotein comprising two CCs in the SMD simulations. The pulling direction is aligned with the $x$-axis, whereas the axial directions of the CCs are aligned with the $z$-axis.}
\label{ch4_fig:2CCs}
\end{figure}

We have designed a simple homopolyprotein, which we employ below to test whether it 
fits our theoretical description. We have extracted the structure of an 
antiparallel coiled-coil motif (CC) from the archeal box C/D sRNP core protein 
(Protein Data Bank entry 1NT2), which comprises 67 residues and whose N-terminus 
and C-terminus are, respectively, arginine (ARG) and isoleucine (ILE) 
\cite{ARCPDyL03}. This structure has been proven to be useful as a mechanical 
folding probe \cite{LSyM14}. We use this CC as the building blocks of the molecule: 
our system is simply a concatenation of two CC motives connected by a linker, which 
is composed of two consecutive pairs of alternated residues of glycine and serine. 
We expect this linker not to introduce any significant interaction between the two 
domains. The initial conformation of the constructed model structure and 
orientation of the two CC repeats is shown in figure \ref{ch4_fig:2CCs}. The end-
to-end vector points from the N-terminus to the C-terminus, aligned with the 
$x$-axis, whereas both axial directions of the two CC structures are located as 
parallel as possible to the $z$-axis.

According to the theoretical framework we have developed along this work, since we 
have two identical units, we expect that if we pull from one end of the designed 
molecule, the first repeat to unfold will be precisely the closest to the moving 
end.  We perform SMD simulations to analyze the degree of agreement of the obtained 
numerical results with the theory.

\section{All-atom molecular dynamics simulation} \label{ch4_sec:SMD}
Our molecular dynamics simulations start from the initial conformation shown in 
figure \ref{ch4_fig:2CCs}. First, we add hydrogen atoms using VMD Automatic PSF 
Builder \cite{HDyS96}. Then, we create a water box of size 
$300 \textrm{\AA} \times 70 \textrm{\AA} \times  120 \textrm{\AA}$, 
in the $x$-, $y$-, and $z$-axes, respectively, which is long enough in the 
direction of pulling (the $x$-axis) to contain the unfolded protein. Also, NaCl is 
introduced in the system replacing water molecules until the concentration reaches 
$150\,$mM/L and the charge is neutralized. Finally, simulations are performed using 
NAMD2 2.10 \cite{PBWGTVCSKyS05}, with two different stages: (i) the ``equilibration 
stage'' at $310\,$K and (ii) the ``pulling stage'' with velocity 
$v_0=1.4 \cdot10^{-2} \,$nm/ps and stiffness $4860 \,$pN/nm. We have also 
considered faster pulling velocities, namely $2v_0$ and  $5v_0$, as detailed below. 
The molecule's behavior for these faster pulling velocities has been investigated 
in order to elucidate whether or not the unfolding pathway becomes more 
deterministic as the pulling speed is increased.

These typical pulling speeds in steered molecular dynamics simulations are higher 
by several orders of magnitude than the experimental ones. However, they are 
necessary to investigate this kind of system with the available computer power. In 
addition, this high velocity range is especially relevant for our present purposes, 
since we are interested in exploring the maximum hysteresis path limit. Note that 
the considered value for the stiffness of the elastic reaction is also two orders 
of magnitude higher than the typical ones in AFM experiments, and thus closer to 
the perfect length control situation assumed in the theory developed in chapter 
\ref{ch:model_pbio}.

\subsection{Pulling trajectories}
\label{ch4_sect-pill-traj}

A notable number of pulling trajectories $N_T$ are needed in order to obtain a 
meaningful statistical analysis of the unfolding pathway. The final configuration 
of the molecule in the equilibration stage is taken as the initial condition for 
the pulling stage.  In order to generate different initial conditions for pulling, 
we have considered one ``long'' trajectory in the equilibration stage and collected 
the molecule configurations at several different times $t_k$, with 
$t_{k+1}-t_k>0.1 \,$ns, as the initial conditions for the different pulling 
trajectories $k=1,\ldots,N_T$. We have checked the ``independence'' of the 
trajectories obtained from these initial conditions, in the sense that the 
unfolding pathway from two consecutive initial conditions, corresponding to $k$ and 
$k+1$, are not correlated. Details are given in section \ref{ch4:appA}. 

The duration of the pulling stage $\Delta t_p$ is chosen to allow the molecule to 
unravel. The size of a single CC motif in its axial direction is around 5 nm, 
specifically 4.82 nm between the two C$\alpha$s most separated in the axial 
direction. Therefore, a motif can be considered as completely unfolded when its 
end-to-end distance, measured between their C$\alpha$s in the terminal residues ARG 
and ILE, exceeds 10 nm.  For the ``base'' velocity $v_0=1.4 \cdot 10^{-2} \,$nm/ps, 
we have chosen $\Delta t_p=1.6 \,$ns, so that the total length increment is 
$v_0 \Delta t_p=22.4\,$nm. 
For the faster pulling velocities, $2v_0$ and $5v_0$, we have decreased 
$\Delta t_p$ accordingly.

In SMD simulations, the length of each repeat can be measured as a function of time 
and thus we may introduce the basic (and the simplest) classification of 
trajectories by labeling them as ``good'' (G) if the CC motif closest to the pulled 
end unfolds sooner than the furthest, and ``bad'' (B) otherwise. Clearly, it is 
G-trajectories that agree with the prediction of the theory
developed in chapter \ref{ch:model_pbio}, when particularized
for identical units. The above basic classification of SMD trajectories as G or B 
can be refined, so as to have a more accurate description of the unfolding pathway,
as done below in section \ref{ch4_sec:unf-crit}. In our theoretical framework, the 
unfolding is sequential: when the first unit unfolds, the second one has not 
reached its limit of stability and thus remains in the folded state. However, in 
the simulations, this perfectly sequential unfolding is not always found: when the 
first unit unfolds, the second one can be partially unfolded, see also section 
 \ref{ch4_sec:pull-CC}.

\subsection{Unfolding criterion}\label{ch4_sec:unf-crit}

\begin{figure}
\centering
 \includegraphics[width=0.7 \textwidth]{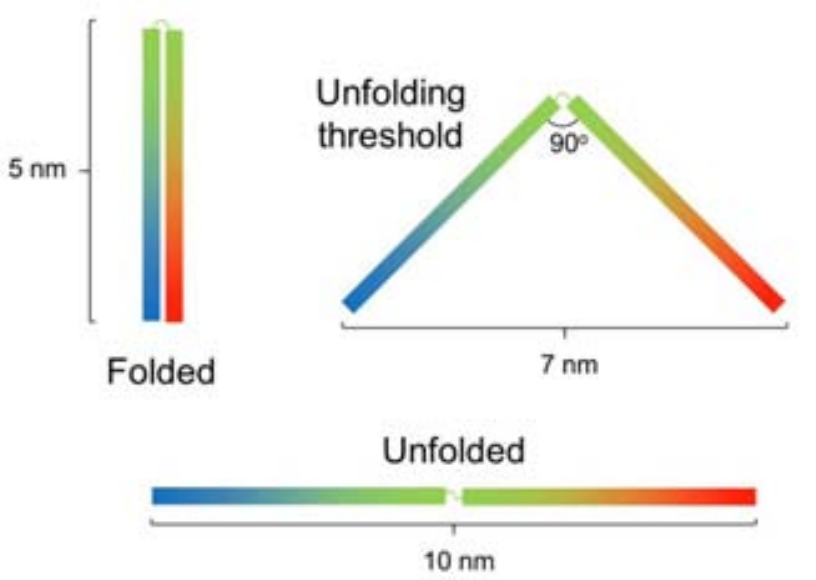}
  \caption{Simple unfolding criterion for the CC under study. 
  This criterion is based on a rigid-rod picture. Depending on the value of its end-to-end distance, we consider the molecule as (i) folded if it is shorter than 7 nm, (ii) partially unfolded when it is between 7 and 10 nm, and (iii) completely unfolded if it is longer than 10 nm.}
\label{ch4_fig:criterion}
\end{figure}

Following the discussion in the previous paragraph and for the sake of accuracy, we 
define a criterion for distinguishing between different subtypes of trajectories. 
Therefore, we incorporate a quantitative measurement of the degree of unfolding for 
each repeat; first in a geometric way, and second taking into account the fraction 
of native contacts \cite{BHyE13}. As already said above, the size of the CC motif 
in its axial direction is approximately $5\,$nm and thus we consider a motif to be 
completely unfolded when its end-to-end distance is greater than $10\,$nm. Also, we 
define when a motif is partially unfolded in our simulations by introducing an 
``unfolding threshold'', that is, a length below which we consider the unit to be 
still folded. Specifically, we take this unfolding threshold to be $7\,$nm, which 
corresponds to an opening angle of $90^{\circ}$ in a rigid rods picture, as 
depicted in figure \ref{ch4_fig:criterion}.

\begin{figure}
\centering
 \includegraphics[width=0.6 \textwidth]{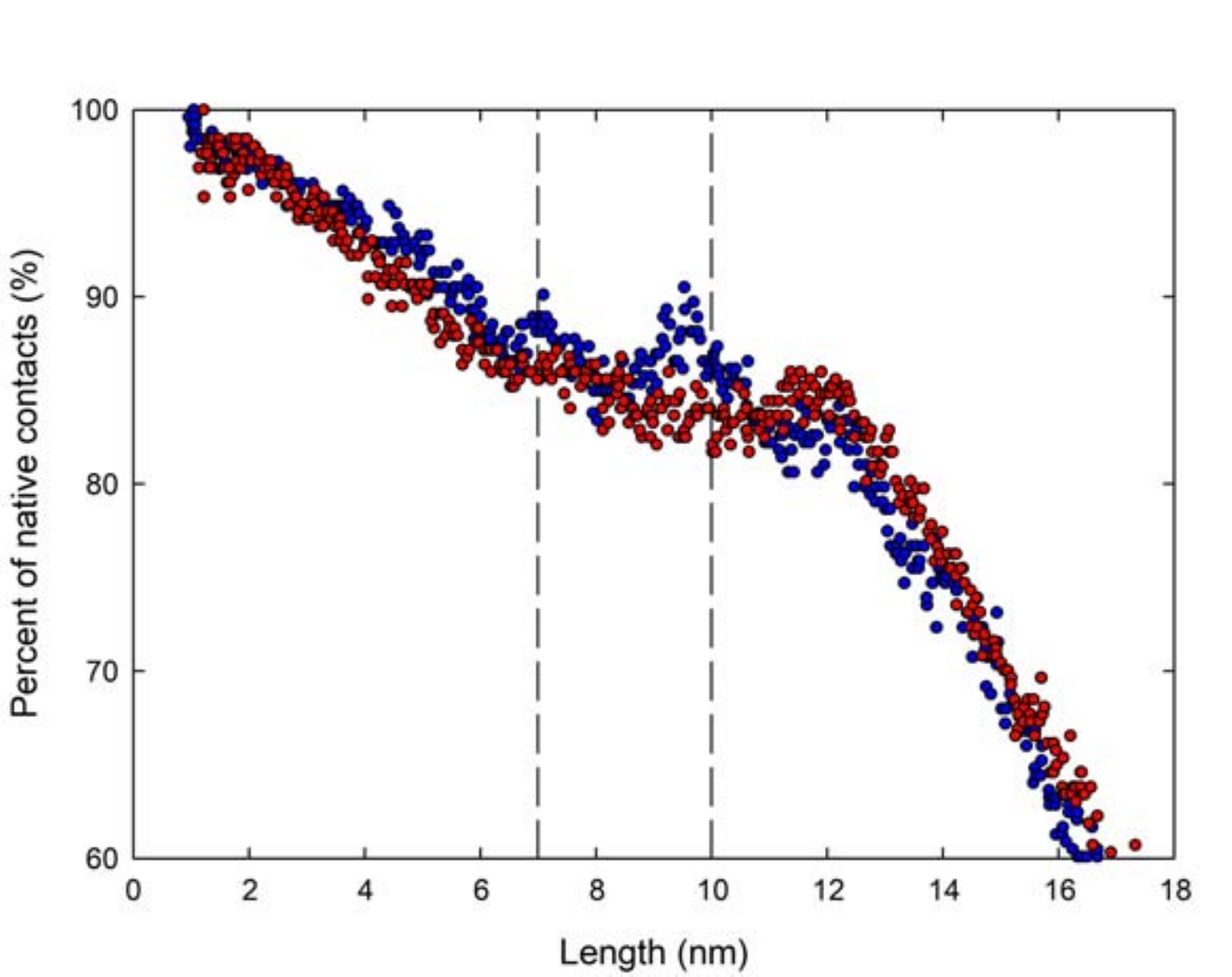}
  \caption{Percentage of native contacts as a function of the total length. Both the data from C-pulling (blue) and N-pulling (red), at a speed $3.75 \cdot 10^{-3} \,$nm/ps, are plotted. The borders of the plateau agree quite well with the thresholds up to (from where) we consider the molecule to be folded (unfolded) in the geometrical picture, marked with vertical dashed lines.}
\label{ch4_fig:native}
\end{figure}

The above ``geometric'' choice for the unfolding threshold has some degree of 
arbitrariness, especially in relation to the length (or angle between rigid-rods) 
chosen for the unfolding threshold. In order to give a physical basis for this 
choice, we look into the fraction of native contacts \cite{BHyE13} as a function of 
the total length for a single CC motif in figure \ref{ch4_fig:native}. For this 
purpose, we have performed a SMD simulation in which a single CC motif, which is 
initially folded, is pulled at a speed of $3.75\cdot10^{-3} \,$nm/ps from either 
its C-terminus (blue) or its N-terminus (red). It can be observed how the number of 
native contacts decreases along the trajectory, with a well-defined plateau arising 
between (approximately) 7 and 10 nm. Clearly, the borders of this plateau demarcate 
the region where the main bonds that keep the double-stranded CC folded are broken. 
Therefore, the above-defined geometric thresholds for considering the molecule 
folded/partially unfolded/completely unfolded agree with the corresponding limits 
in the fraction-of-native-contacts picture.

Consistently with the above-described criteria, let us index the different subtypes 
of trajectories by (I, II, III, IV), attending to their degree of agreement with 
the theoretical prediction.  G-trajectories are split into I and II subtypes: when 
the pulled motif is the first that unfolds, the other unit can be either still 
folded (type I) or partially unfolded (type II). Similarly, B-trajectories are 
divided into III and IV subtypes: when the non-pulled motif unfolds first, the 
pulled unit can be either partially unfolded (type III) or still folded (type IV). 
The explicit distinction between the different cases, for a simulation in which the 
molecules is pulled from its C-terminus, is shown in table \ref{ch4_table:1}. 
Obviously, if the molecule is pulled from its N-terminus, the same classification 
of trajectories applies but with the role of the repeats C and N reversed.

\begin{table}[]
\centering
\caption{Definition of the different trajectory types in a
SMD C-pulling simulation. Types I and IV are the closest trajectories to a deterministic pathway,
agreeing and disagreeing, respectively, with the prediction of our
model.}
\label{ch4_table:1}
\begin{tabular}{ccccc}
\hline
Type     & Subtype & \begin{tabular}[c]{@{}c@{}}First repeat that\\ unfolds\end{tabular} & \begin{tabular}[c]{@{}c@{}}State of the\\ other repeat\end{tabular} \\ \hline
Good (G) & I       & C-terminus                                                          & folded                                                                                                                          \\
Good (G) & II      & C-terminus                                                          & partially unfolded                                                  \\
Bad (B)  & III     & N-terminus                                                          & partially unfolded                                                   \\
Bad (B)  & IV      & N-terminus                                                          & folded                                                               \\ \hline
\end{tabular}
\end{table}

\section{Pulling the CC construct} \label{ch4_sec:pull-CC}

In this section, we present the SMD results corresponding to the pulling of the CC 
construction described above.
We consider the two possible experiments, pulling from the C-terminus (C-pulling) 
and from the N-terminus (N-pulling), separately.

\subsection{C-pulling}

\begin{figure}
\centering
 \includegraphics[width=0.99 \textwidth]{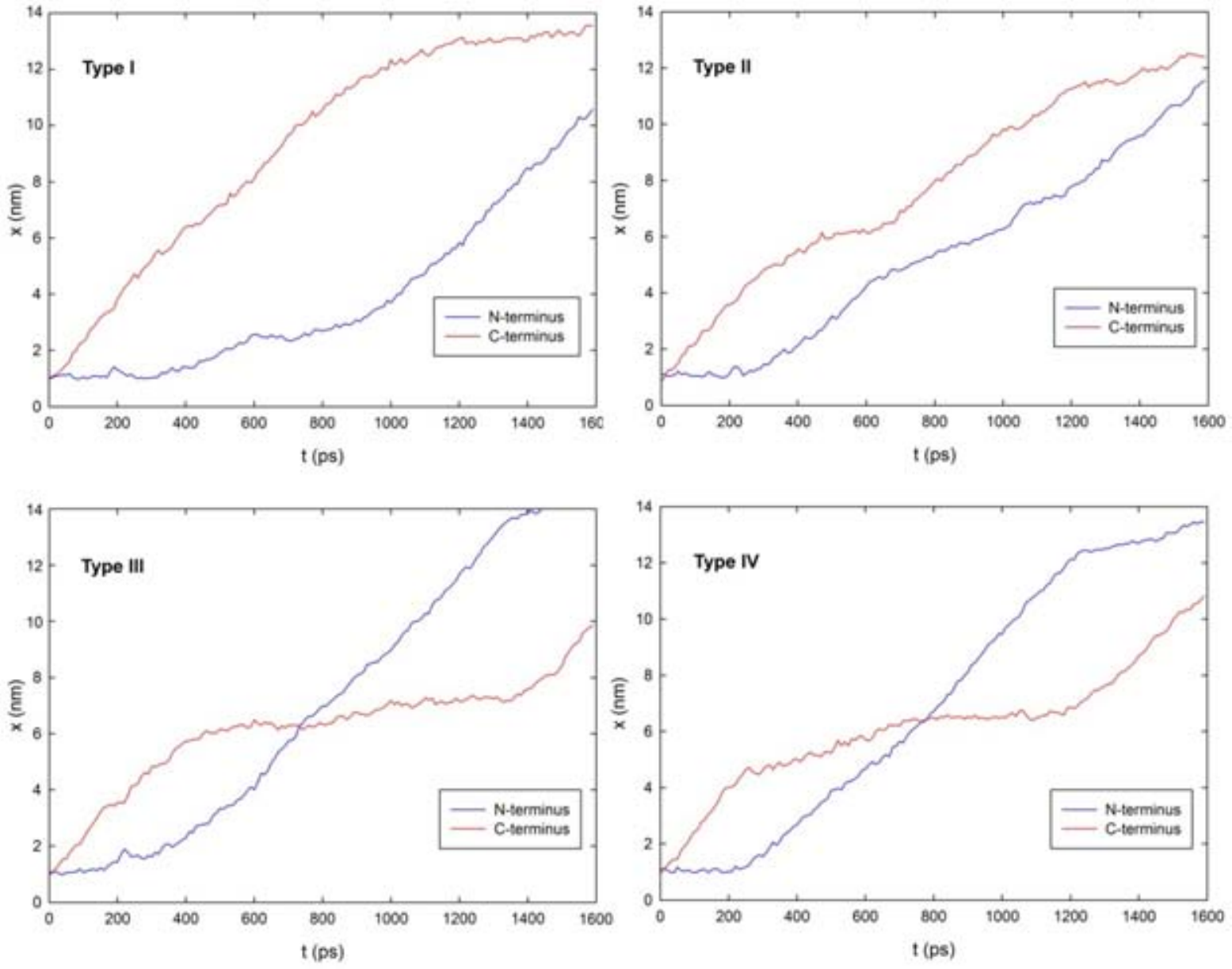}
  \caption{Representative plots of the different types of trajectories for 
  SMD C-pulling simulations. Each panel corresponds to a given type, as labeled.  
  Both repeats' extensions are plotted: N-repeat (blue) and C-repeat (red). Our
   model predicts that the pulled repeat (C-terminus) is the first that unfolds.
    Videos corresponding to each of the trajectory types can be found in the
     supporting information to \cite{PSMyP18}.}
  \label{ch4_fig:types}
\end{figure}

First, we pull the molecule from its C-terminus at the base velocity 
$v_0=1.4\cdot 10^{-2} \,$nm/ps. We plot the evolution of the distance between the 
end terminals of each repeat in this C-pulling experiment in figure 
\ref{ch4_fig:types}. The red line stands for the pulled repeat (C-terminus) whereas 
we plot in blue the length of the other repeat (N-terminus). It can be seen how the 
pulled repeat clearly unfolds first in type I. Although from different categories, 
types II, III and IV seem to share a common feature. In the initial part of the 
trajectory, it is the pulled repeat the fastest to lengthen but its unfolding comes 
to a standstill before being completed, and the second repeat takes advantage of 
this impasse to increase its extension.

Due to thermal fluctuations, we do not expect to obtain a perfect agreement with 
our theory, but a preponderance of the deterministic (type I) trajectories. As 
discussed before, a ``fast'' pulling velocity is necessary to be in the ``maximum 
hysteresis path'' limit, in which our theory is expected to hold. Therefore, in 
addition to the base velocity $v_0=1.4\cdot 10^{-2} \,$nm/ps, we have carried out 
simulations at velocity $2v_0$. Specifically, we have done 31 trajectories for each 
velocity and collected their statistical information in table \ref{ch4_table:2}, 
which is completely compatible with our theoretical expectation. There is already a 
preponderance of type I trajectories at the base velocity $v_0$, at which they 
represent almost half of the total number of trajectories, against a reduced 
fraction, only 16$\%$, of type IV trajectories. Moreover, the prevalence of type I 
trajectories increases with the pulling velocity, as expected in our theoretical 
framework: at velocity $2v_0$, more than two thirds of the SMD runs are of type I 
and there are no type IV trajectories. 

We can get additional insight by calculating the frequencies of G (sum of I+II) and 
B (sum of III+IV) trajectories. This simplification of the types of events shows, 
even more clearly, the preponderance of the deterministic path as the velocity is 
increased: the frequency of G-trajectories shows a clear increase from 61$\%$ at 
velocity $v_0$ to 87$\%$ at velocity $2v_0$ (the frequency of B-trajectories 
decreases accordingly from 39$\%$ to 13$\%$). This is the reason why we have not 
considered even higher velocities for C-pulling.

The above results show that increasing the pulling speed effectively diminishes the 
relevance of thermal activation effects and makes the unfolding more 
``deterministic'', in the sense of increasing the prevalence of type I 
trajectories. It should also be stressed that detailed analysis of the rest of type 
trajectories highlights a branching from type I due to an impasse of the length of 
pulled repeat, as already stated at the beginning of this section. Wrapping things 
up, our theory seems to predict the unfolding mechanism displayed in the SMD of 
this CC homopolyprotein---at least, when it is pulled from the C-end.

\begin{table}[]
\centering
\caption{Statistical analysis of the output of 31 runs of
SMD C-pulling simulations of the two CCs construct. At the base velocity $v_0$, there is already a significant preponderance of type I
trajectories (45$\%$) as compared to type IV ones (16$\%$). Type I
trajectories clearly prevail as the pulling velocity is increased to $2v_0$, at
which its frequency is boosted up to 67$\%$. In addition, no type IV
trajectories are found at this higher velocity. This behavior clearly agrees with our
theoretical prediction.}
\label{ch4_table:2}
\begin{tabular}{ccc}
\hline
Type           & Occurrence at $v_0$ ($\%$) & Occurrence at $2 v_0$ ($\%$) \\ \hline
I              & 45                   & 68                     \\
II             & 16                   & 19                     \\
III            & 23                   & 13                     \\
IV             & 16                   & 0                      \\ \hline
Good = I + II  & 61                   & 87                     \\
Bad = III + IV & 39                   & 13                     \\ \hline
\end{tabular}
\end{table}

\subsection{N-pulling}

Of course, our one-dimensional theory is completely left-right symmetric, since the 
free energy only depends on the extensions. Therefore, if we perform the same kind 
of SMD simulations but pulling from the N-terminus, we expect similar results 
(within statistical errors for the limited number of trajectories). Thus, the 
unfolding should start, preponderantly, from the unit closer to the N-terminus. 
Nevertheless, we show below that the situation is more complex, which we understand 
as a signature of anisotropy in the considered molecule.

Table \ref{ch4_table:3} presents the statistics of the different types of 
trajectories for SMD simulations in which we pull from the N-terminus. In terms of 
the same base velocity $v_0=1.4\cdot 10^{-2} \,$nm/ps, we have conducted numerical 
experiments at velocities $v_0$, $2v_0$ and $5v_0$. Specifically, the statistics 
have been obtained again from 31 runs for each pulling velocity. The situation is 
much more complex than for C-pulling, since for both $v_0$ and $2v_0$ there is no 
clear preponderance of the deterministic pathway. On the one hand, the fraction of 
G-trajectories increases from less than half to more than two thirds as the pulling 
velocity is increased from $v_0$ to $2v_0$, which seems to indicate that the 
pathway is becoming more deterministic. On the other hand, the frequency of type I 
trajectories decreases in favor of those of type II. This complex behavior makes it 
necessary to consider a higher velocity, so as to ascertain the tendency of the 
pathway as the velocity is increased. Specifically, we have incorporated 
simulations with pulling velocity $5v_0$. For this velocity, almost every 
trajectory (97$\%$) is type I, again in agreement with our theoretical framework.

\begin{table}[]
\centering
\caption{Statistical analysis of the output of 31 runs of
SMD N-pulling simulations of the two CCs construct. In addition to the pulling velocities $v_0$ and $2v_0$, already analyzed for C-pulling,
we have considered a faster velocity $5v_0$.}
\label{ch4_table:3}
\centering
\begin{tabular}{cccc}
\hline
Type & \begin{tabular}[c]{@{}c@{}}Occurrence\\ at $v_0$ ($\%$)\end{tabular} & \begin{tabular}[c]{@{}c@{}}Occurrence\\ at $2v_0$ ($\%$)\end{tabular} & \begin{tabular}[c]{@{}c@{}}Occurrence\\ at $5v_0$ ($\%$)\end{tabular} \\ \hline
I              & 29                   & 13                     & 97                     \\
II             & 13                   & 55                     & 3                      \\
III            & 23                   & 29                     & 0                      \\
IV             & 35                   & 3                      & 0                      \\ \hline
Good = I + II  & 42                   & 68                     & 100                    \\
Bad = III + IV & 58                   & 32                     & 0                      \\ \hline
\end{tabular}
\end{table}

Anisotropy of biological systems has been extensively studied \cite{DBByR06}. 
Indeed, in the work of Gao et al. \cite{GSyZ11}, a coiled-coil system very similar 
to ours presents different unfolding kinetics depending on the direction of the 
pulling: N- or C-pulling. Therein, the observed N-pulling transition rates between 
folded and unfolded states were much higher than the C-pulling rates. This property 
is compatible with our observations: if transition rates are higher in N-pulling, 
thermally activated jumps from the folded to the unfolded state could be relevant 
for the slower pulling velocities, although they were not for C-pulling. In this 
situation, we expect the range of velocities in which the ``deterministic'' or 
``maximum hysteresis'' prevails to depend on the end from which the molecule is 
pulled. Specifically, we expect it to be higher for N-pulling,
in agreement with our observations. Anyhow, the 
deterministic pathway becomes largely preponderant when the pulling velocity is 
high enough, as confirmed by our N-pulling simulations with the fastest velocity 
$5v_0$.

\section{Independence of the initial conformations for the pulling stage}\label{ch4:appA}

A possible issue with our simulations stems from our choice of initial 
conformations for the pulling stage. Instead of equilibrating the system many 
times, we have chosen to run a long trajectory and pick different 
conformations along it as the initial conditions for the subsequent pulling. One 
may wonder, quite reasonably, whether these initial conditions for the pulling 
stage are really uncorrelated, as we implicitly are assuming, or not. In this 
section, we thoroughly discuss this issue.

Let us denote by $\Gamma_k$ the different $k=1,\ldots,N_T$ initial conformations 
used as initial conditions for the pulling stage. To check the independence of 
these initial conformations $\Gamma_k$ invoked in section \ref{ch4_sect-pill-traj}, 
we have analyzed possible correlations between the observed pathways in consecutive 
runs $k$ and $k+1$. In particular, we have focused our attention on the four 
possible pairs of ``good'' (G) and ``bad'' (B) trajectories (GG, GB, BG, BB) for 
each case we have considered, i.e. for given pulled terminus and pulling speed. 

The $k$-th pulling trajectory, for given pulled terminus and pulling speed, starts 
from conformation $\Gamma_k$. Let us introduce a stochastic variable $\zeta_k$ to 
identify the type to which the $k$-th trajectory belongs. Thus, $\zeta_k$ can have 
two values that are denoted by $\alpha$, that is, $\alpha$ is equal to either G or 
B. In consecutive runs, $k$ and $k+1$, the possible values of the pair 
($\zeta_k$,$\zeta_{k+1}$) are denoted by $\alpha \beta$, that is, the four 
possibilities (GG, GB, BG, BB). We denote the number of pairs of consecutive 
trajectories with a certain outcome $\alpha \beta$ by $n_{\alpha \beta}$, which is 
\begin{equation}
n_{\alpha \beta}= \sum_{k=1}^{N_p}\delta_{\zeta_k,\alpha} \delta_{\zeta_{k+1},\beta}
\end{equation}
where $N_p=N_T-1$ is the number of consecutive pairs and $\delta_{ij}$ is Kronecker 
delta.

The probability that any trajectory corresponds to a given type (G or B) can be 
formally written as
\begin{equation}
\label{ch4_app_eq:2}
\left\langle \delta_{\zeta_k,\alpha} \right \rangle = p_{\alpha}.
\end{equation}
We are denoting the corresponding probabilities by $p_{\text{G}}$ and 
$p_{\text{B}}$, respectively, and $p_G+p_B=1$ because these events are mutually 
exclusive, $\delta_{\zeta_k,\text{G}}=1-\delta_{\zeta_k,\text{B}}$. Now, assuming 
the outcomes of consecutive trajectories to be independent, we can ask ourselves 
the following questions in an ensemble of simulations comprising $N_T$ trajectories 
(corresponding to a given pulled terminus and pulling speed). 
\begin{enumerate}
\item 	What is the expected number of each pair type $ \left\langle n_{\alpha \beta} \right\rangle$?
\item 	What are their corresponding standard deviations $\sigma_{\alpha \beta}$?
\end{enumerate}

The mean value is directly obtained by taking into account \eqref{ch4_app_eq:2}, 
for all $k$, and the assumed statistical independence of the variables $\zeta_k$ 
and $\zeta_{k+1}$. Therefore, one gets straightforwardly that
\begin{equation}
\left\langle n_{\alpha \beta} \right\rangle =N_p p_\alpha p_\beta.
\end{equation}
The derivation of the expression for the fluctuations is lengthier. We start by 
writing
\begin{equation}
n_{\alpha \beta}^2=\sum_{k=1}^{N_p} \sum_{l=1}^{N_p} \delta_{\zeta_k,\alpha} \delta_{\zeta_{k+1},\beta}  \delta_{\zeta_l,\alpha} \delta_{\zeta_{l+1},\beta},
\end{equation}
and we split the above expression into four contributions, corresponding to the 
cases (i) $l=k$, (ii)-(iii) $l=k \pm 1$, and (iv) all the other values of $l$. 
Thus, we can write
\begin{subequations}
\label{ch4_app_eq:B5}
\begin{align}
n_{\alpha \beta }^2 =& \sum_{k=1}^{N_p} \delta_{\zeta_k,\alpha} \delta_{\zeta_{k+1},\beta}+ \delta_{\alpha \beta} \sum_{k=1}^{N_p-1} \delta_{\zeta_k,\alpha} \delta_{\zeta_{k+1},\alpha} \delta_{\zeta_{k+2,\beta}}  \nonumber
\\ &+ \delta_{\alpha \beta} \sum_{k=2}^{N_p}\delta_{\zeta_{k-1},\alpha} \delta_{\zeta_k,\alpha} \delta_{\zeta_{k+1,\beta}}+ \sum_{k=1}^{N_p} \sum_{l \neq k, k\pm1 }^{N_p} \delta_{\zeta_k,\alpha}  \delta_{\zeta_{k+1},\beta}  \delta_{\zeta_l,\alpha} \delta_{\zeta_{l+1},\beta} \tag{\ref{ch4_app_eq:B5}}.
\end{align}
\end{subequations}
\begin{table}[]
\centering
\caption{Frequencies (empirical and theoretical) and expected statistical errors for individual events and pairs of consecutive events. The values for the frequencies of G- and B-trajectory types are taken from tables \ref{ch4_table:2} and \ref{ch4_table:3}.}
\label{ch4_ap_table:1}
\begin{tabular}{cccccc}
\hline
Frequency & C-$v_0$ & C-$2v_0$ & N-$v_0$ & N-$2v_0$ & N-$5v_0$ \\ \hline
G         & 0.61   & 0.87    & 0.42   & 0.68    & 1.00    \\
B         & 0.39   & 0.13    & 0.58   & 0.32    & 0.00    \\
GG        & 0.37   & 0.73    & 0.13   & 0.50    & 1.00    \\
GG-th     & 0.38   & 0.76    & 0.18   & 0.46    & 1.00    \\
GG-err    & 0.12   & 0.10    & 0.09   & 0.12    & 0.00    \\
GB        & 0.23   & 0.13    & 0.27   & 0.17    & 0.00    \\
GB-th     & 0.24   & 0.11    & 0.24   & 0.22    & 0.00    \\
GB-err    & 0.05   & 0.05    & 0.05   & 0.05    & 0.00    \\
BG        & 0.23   & 0.13    & 0.27   & 0.17    & 0.00    \\
BG-th     & 0.24   & 0.11    & 0.24   & 0.22    & 0.00    \\
BG-err    & 0.05   & 0.05    & 0.05   & 0.05    & 0.00    \\
BB        & 0.17   & 0.00    & 0.33   & 0.17    & 0.00    \\
BB-th     & 0.15   & 0.02    & 0.34   & 0.10    & 0.00    \\
BB-err    & 0.08   & 0.02    & 0.11   & 0.07    & 0.00    \\ \hline
\end{tabular}
\end{table}
Now we compute the average. We have four contributions on the rhs of the equation 
above: in the first sum, there are $N_p$ terms, each of them with average 
$p_\alpha p_\beta$; in both the second and third sums, there are $N_p-1$ terms, 
each of them with average $p_\alpha^2 p_\beta$; and in the fourth sum we have the 
remaining $N_p^2-(3N_p-2)$ terms, each of them with average $p_\alpha^2 p_\beta^2$. 
Once more, we have assumed the independence of the variables $\zeta_k,\zeta_l$ for 
$l\neq k$. Thus, we get 
\begin{equation}
 \left\langle n_{\alpha\beta}^2 \right\rangle = N_p p_\alpha p_\beta+2 \delta_{\alpha \beta} \left(N_p-1  \right) p_\alpha^2 p_\beta+\left[N_p^2- \left(3N_p-2 \right)\right] p_\alpha^2 p_\beta^2.
\end{equation}
The variance is thus given by
\begin{equation}
\sigma_{\alpha \beta}^2 \equiv \left\langle n_{\alpha \beta}^2 \right\rangle - \left\langle n_{\alpha\beta} \right\rangle^2=N_p p_\alpha p_\beta (1-3p_\alpha p_\beta )+2p_\alpha^2 p_\beta^2+2 \delta_{\alpha\beta}  \left(N_p-1 \right) p_\alpha^2 p_\beta.
\end{equation}
For the sake of clarity, we list below the average values and the variances for the 
three pairs of outcomes leading to different values; note that 
$\left\langle n_{\text{GB}} \right\rangle = \left\langle n_{\text{BG}} \right\rangle$ and also $\sigma_{\text{GB}}=\sigma_{\text{BG}}$. 
Specifically,
\begin{subequations}
\label{ch4_app_eq:8}
\begin{align}
&\left\langle n_{\text{GG}} \right\rangle = N_p p_{\text{G}}^2,& \quad &\sigma_{\text{GG}}^2=p_\text{G}^2 \left(1-p_{\text{G}} \right)\left[N_p+p_{\text{G}} \left(3N_p-2\right)\right],&\\
&\left\langle n_{\text{BB}} \right\rangle = N_p p_{\text{B}}^2,& \quad &\sigma_{\text{BB}}^2=p_{\text{B}}^2 \left(1-p_{\text{B}} \right)\left[N_p+p_{\text{B}} \left(3N_p-2\right)\right],&\\
&\left\langle n_{\text{GB}} \right\rangle=\left\langle n_{\text{BG}} \right\rangle=N_p p_{\text{G}} p_{\text{B}}, & \quad &\sigma^2_{\text{GB}}=\sigma^2_{\text{BG}}=N_p p_{\text{G}} p_{\text{B}} \left(1-3p_{\text{G}} p_{\text{B}} \right)+2p_{\text{G}}^2 p_{\text{B}}^2.&
\end{align}
\end{subequations}
As expected, the variances would vanish if the process were purely deterministic 
and $p_{\text{G}}=1$ or, equivalently, $p_{\text{B}}=0$, since all the pairs would 
correspond to the GG case.

From an empirical point of view, we can identify $p_{\text{G}}$ and $p_{\text{B}}$ 
with the frequencies of G- and B-trajectories for the considered experiment (given 
pulled terminus/pulling speed). After doing that, we can count the actual number of 
pairs $n_{\alpha \beta}$ for each pair type in the ensemble of $N_T$ trajectories, 
and check if it lies within the theoretical expectation, for these empirical values 
of $p_\alpha$. What we show below is that for all our simulations we have that 
$\left|n_{\alpha \beta}- \left\langle n_{\alpha \beta} \right\rangle \right| \leq \sigma_{\alpha \beta}$,
 that is, the assumption of considering the initial conformations as independent is 
 really good.

In order to give all the results in table \ref{ch4_ap_table:1}, we list the 
empirical frequencies for each ensemble of trajectories, corresponding to a given 
pulled terminus and pulling speed. For example, C-$v_0$ means that the data in the 
corresponding column is for C-pulling at velocity $v_0$. For each column, the rows 
correspond to the frequencies for different events: (i) in the G and B rows, we 
give the empirical frequencies for good and bad trajectories, taken from tables 
\ref{ch4_table:2} and \ref{ch4_table:3}; (ii) in the subsequent 
$\alpha \beta$ rows, the empirical frequencies for two consecutive trajectories of 
type $\alpha \beta$; (iii) in the $\alpha \beta$-th row, the corresponding 
theoretical prediction for that frequency, calculated as 
$\left\langle n_{\alpha \beta} \right\rangle/N_p$ from \eqref{ch4_app_eq:8}; 
(iv) in the $\alpha \beta$-err, we provide the theoretical prediction for the 
standard deviation of that frequency, calculated as $\sigma_{\alpha \beta}/N_p$ 
from \eqref{ch4_app_eq:8}. For the theoretical values of 
$\left\langle n_{\alpha \beta} \right\rangle /N_p$ and $\sigma_{\alpha \beta}/N_p$,
 we take $p_{\text{G}}$ and $p_{\text{B}}$ equal to the empirical frequencies of 
 G- and B-trajectories, respectively, as already stated above.

{\clearpage \thispagestyle{empty}}
\part{Understanding granular matter with simple lattice models}
\label{part:granular}

\chapter{Continuum limit, fluctuating hydrodynamics and finite size effects}
\label{ch:finsize_pgran}
\newcommand{\tauder}[1]{\partial_{\tau}{#1}}
\newcommand{\xder}[1]{\partial_{x}{#1}}
\newcommand{\sder}[1]{\partial_{s}{#1}}
\newcommand{\sigder}[1]{\partial_{\sigma}{#1}}
\newcommand{\sigdertot}[1]{\frac{d}{d\sigma}{#1}}
\newcommand{\xxder}[1]{\partial_{xx}{#1}}
\newcommand{\eps}{\varepsilon}
\newcommand{\tmax}{t_{\text{max}}}
\newcommand{\calL}{\mathcal{L}}
\newcommand{\calS}{\mathcal{S}}
\newcommand{\llangle}{\left\langle}
\newcommand{\rrangle}{\right\rangle}

Throughout this part of the thesis, we mainly investigate a lattice model for granular media inspired 
by two previous models on the lattice, specifically those
in \cite{BByP02,PLyH11a}. The model differs from
these two previous proposals in a few crucial
aspects. First, in \cite{BByP02}, the velocity field evolved under the
enforcement of the so-called kinematic constraint, which
is disregarded here. Second, in \cite{PLyH11a}, only the energy field was
considered, and therefore momentum conservation was absent.

Lattice models have been of paramount importance for understanding rigorously the 
conditions needed to have a hydrodynamic description, both at the average 
\cite{KyL99,KMyP82} and fluctuating \cite{BSGJyL01} levels. 
Recently, fluctuating hydrodynamics has been employed 
to derive the large deviation function in the context of energy-conserving 
\cite{HyG09,HyG10,HyG09b,HyK11}
 and even in energy-dissipating \cite{PLyH11a,PLyH12a,PLyH13,PLyH16}
 models. 
 Both in the conservative and 
nonconservative case, momentum conservation has not been taken into account. This 
shortcoming may be relevant, since it is known that momentum conservation is linked 
to the appearance of long-ranged correlations in out-of-equilibrium systems 
\cite{GLyS90,GLMyS90}.
 More specifically, spatial long-range velocity correlations in the homogeneous 
cooling state can be partly explained by fluctuating hydrodynamics, but require a 
more refined treatment to be fully investigated. 

This chapter is dedicated to introduce our model, from the basics to more involved
issues. Our approach intends to contain the essential ingredients to investigate 
granular fluids, but reducing the complexity of the mathematical framework. In this 
way, we expect to get a transparent interpretation of the physical results. In 
particular, we aim to elucidate the ``perturbative'' nature of the continuum limit 
and calculate the corrections thereto \cite{Sp80}. 
Such corrections give interesting 
information about the structure---in space and time---of the correlated granular 
fluctuations and reveal new phenomena, which are peculiar of inelastic collisions.

The organization of this chapter is the following.
In section \ref{sec:model}, we briefly revise the 
main aspects of the model, focusing on its  continuum,
hydrodynamic-like, limit. A more detailed description of the model can
be found in \cite{LMPyP15,MPLPyP16,AM_thesis}. The evolution of the one-particle
 distribution function and some  physically relevant stationary states 
 are thoroughly analyzed, respectively, in sections \ref{ch5_sec:1PDF} and \ref{ch5_sec:have}. Section \ref{sec:meso-fluc-th} is devoted to the study of the fluctuations 
 of the system, including a discussion of their relation with some instabilities 
 that appear in the model. Finally, we 
 present  an exact solution for the two-particle correlation
  function of the system in section \ref{sec:HCS-exact}, which is valid
  for arbitrary size and provides further insight into the instabilities.

\section{Model and previous results}
\label{sec:model}

\subsection{Evolution equations}
\label{ch5_ev_eq}

Let us consider a 1d lattice with $N$ sites. First, we define the
dynamics in discrete time. After the $p$-th step of the dynamics, the
particle at the $l$-th site has a velocity $v_{l,p} \in$
$\mathbb{R}$, and the configuration for the system at time $p$ is denoted
as $\vv_p \equiv \{v_{1,p},...,v_{N,p}\}$. One individual trajectory
of the stochastic process is built in the following way:
the configuration of the system changes from time $p$ to time $p+1$
because a pair of nearest neighbors $\left( l,l+1 \right)$ is chosen
at random and collides inelastically, that is, $\vv_{p+1} = \hat{b}_l
\vv_p$ where the operator $\hat{b}_{l}$ transforms the pre-collisional
velocities $(v_{l,p},v_{l+1,p})$ into the post-collisional ones
$(v_{l,p+1},v_{l+1,p+1})$ and leaves all other sites unaltered. The
post-collisional velocities are given by
\begin{subequations}\label{coll_rule}
\begin{eqnarray}
v_{l,p+1} &=& v_{l,p}-\frac{1+\alpha}{2}\Delta_{l,p}, \\
v_{l+1,p+1} &=& v_{l+1,p}+\frac{1+\alpha}{2}\Delta_{l,p},
\end{eqnarray}
\end{subequations}
where $\Delta_{l,p}=v_{l,p}-v_{l+1,p}$ and the normal restitution coefficient is $\alpha \in (0,1]$. Note that, by its own definition, this stochastic process is Markovian. In the
following, we use a notation such that the evolution operator
$\hat{b}_{l}$ acts naturally on {\em observables}, for example, $\hat{b}_l
v_{l,p} = v_{l,p+1}$. Momentum is always conserved,
\begin{equation}
(\hat{b}_{l}-1)(v_{l,p}+v_{l+1,p})=0,
\end{equation}
whereas energy, if $\alpha
\neq 1$, is not,
\begin{equation}
(\hat{b}_{l}-1)(v^{2}_{l,p}+v^{2}_{l+1,p})=(\alpha^2-1)\Delta_{l,p}^{2}/2<0.
\end{equation}
The collision rule \eqref{coll_rule}, which corresponds to the simplest
one used in granular fluids \cite{PyL01}, is valid for bulk sites. It
must be complemented with suitable evolution equations for the sites next to the system boundaries, which depend on the physical situation at hand. See below for details.

The evolution equation for the velocities can be cast in the form
\begin{equation}\label{eq:mom}
v_{l,p+1}-v_{l,p}=-j_{l,p}+j_{l-1,p}, \quad j_{l,p}=\frac{1+\alpha}{2}\Delta_{l,p}\delta_{y_p,l},
\end{equation}
which is nothing but a discrete continuity equation. Therein, $j_{l,p}$
is the momentum current from site $l$ to site $l+1$ at time $p$, $\delta_{y_{p},l}$ is Kronecker's delta, and $y_{p}$
is a homogeneously distributed random integer in $[1,L]$, where $L$ is
the number of possible colliding pairs. For periodic
boundary conditions, $L=N$, whereas for thermostatted boundaries
$L=N+1$. 

We have only kinetic energy, ${\cal K}_{p}=\sum_{l=1}^{N} e_{l,p}$ at time $p$, where
$e_{l,p}=v_{l,p}^{2}$. By squaring (\ref{eq:mom}), the evolution
equation for the energy at site $l$ reads
\begin{eqnarray}\label{eq:en}
 e_{l,p+1}-e_{l,p}=-J_{l,p}+J_{l-1,p}+d_{l,p}.
\end{eqnarray}
There are two contributions to the evolution of the energy: (i)  the ``flux'' term $-J_{l,p}+J_{l-1,p}$, and (ii) a sink term
$d_{l,p}$ stemming from the inelasticity of collisions. The energy
current $J_{l,p}$ from site $l$ to site $l+1$ and energy dissipation $d_{l,p}$
at site $l$ are 
\begin{equation}\label{micro-ener-dis}
J_{l,p}=(v_{l,p}+v_{l+1,p})j_{l,p}, \quad
d_{l,p}= \frac{\alpha^2-1}{4}\left[\delta_{y_p,l}\Delta_{l,p}^2+\delta_{y_p,l-1}\Delta_{l-1,p}^2 \right]<0,
\end{equation}
respectively.

The above stochastic dynamics generates the trajectories
corresponding to the Markov process described by the following master equation in
continuous time \cite{MPLPyP16,AM_thesis},  
\begin{equation} \label{eq:cma2}
\partial_\tau P_N(\vv,\tau|\vv_0,\tau_0)=\omega \sum_{l=1}^L \left| \Delta_{l} \right|^{\beta} \left[ \frac{P_{N}(\hat{b}_l^{-1} \vv,\tau|\vv_0,\tau_0)}{\alpha^{\beta+1}}  -  P_N(\vv,\tau|\vv_0,\tau_0) \right],
\end{equation}
where 
  $P_{N}(\vv,\tau|\vv_0,\tau_0)$ is the conditional probability density  of finding the system in state $\vv$
  at time $\tau$ provided it was in state $\vv_{0}$ at time
  $\tau_{0}$.  Above, $\omega$ is a constant with dimensions of frequency that
determines the time scale. Moreover, the operator $\hat{b}_l^{-1}$ is the
inverse of $\hat{b}_{l}$, that is, $\hat{b}_l^{-1}$ changes the
post-collisional velocities into the pre-collisional ones for the
colliding pair $(l,l+1)$.  At the $p$-th step of each dynamical trajectory, the continuous time $\tau$ is incremented by
\begin{equation}
\delta\tau_{p}=-\Omega_p(L)^{-1}\ln z, \quad \Omega_p(L)=\omega \sum_{l=1}^{L}\left| v_{l,p} - v_{l+1,p} \right|^{\beta},
\end{equation} 
in which $z$ is a stochastic variable homogeneously distributed in the
interval $(0,1)$ and $\beta \geq 0$ is a parameter that affects the collision rate. On the one hand, the collision rate becomes
independent of the relative velocity for $\beta=0$, similarly to the case of
pseudo-Maxwell molecules~\cite{BByP02,ByK03}. On the other hand, $\beta=1$ and $\beta=2$ are analogous to the hard
core \cite{BRyC96} and ``very hard-core''~\cite{Er81,ETyB06}
collisions, respectively. From now on, we focus on the Maxwell case $\beta=0$, although the case with $\beta\neq 0$ is considered in chapter  \ref{ch:Kovacs_pgran}.

The initial condition for \eqref{eq:cma2} is clearly $P_{N}(\vv,\tau_0|\vv_0,\tau_0)=\delta(\vv-\vv_0)$. Moreover, the one-time probability distribution $P_N(\vv,\tau)$ verifies the same equation but with an arbitrary (normalized) initial condition $P_N(\vv,0)$.

\subsection{Physical interpretation}
\label{ch5_sec:interp}

Literally taken, there is no mass transport in the model,
 particles are at fixed positions on the lattice and they only exchange
momentum and kinetic energy. As discussed in section~\ref{ch1_sec:hcs},
this can be a valid assumption in an {\em incompressible} regime which
is expected when the velocity field is divergence free, for instance
during the first stage of the development of the shear instability, or
in the so-called Uniform Shear Flow. We are also disregarding the
so-called kinematic constraint, at difference with the approach 
in~\cite{BByP02,BPyP18}: a colliding pair is chosen regardless of
the sign of its relative velocity. 

\begin{figure}
\centering
\includegraphics[angle=0,width=0.8\textwidth]{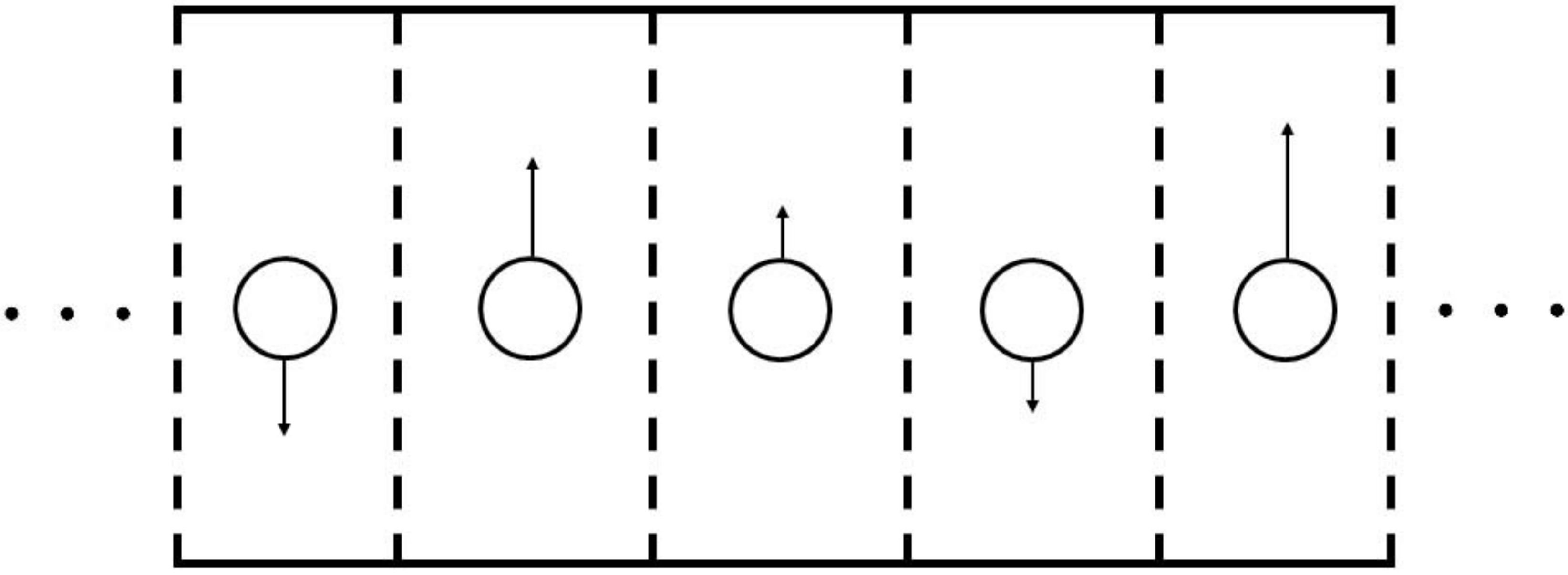}
\caption{Sketch of the lattice model of granular media. Sites are characterized by a  transverse-to-the-lattice velocity  and exchange momentum and energy through nearest-neighbor collisions.}
\label{fig:sketch_granular}
\end{figure}

Our model has a nice
physical interpretation: the dynamics occurs inside an elongated 2d or
3d channel, the lattice sites represent positions on the long axis,
while the transverse (shorter) directions are ignored; the velocity of
the particles do not represent their motion along the lattice axis but
rather along a perpendicular one, as depicted in figure \ref{fig:sketch_granular}. Let us easily imagine that the
(hidden) component along the lattice axis is of the order of the
perpendicular component, but in random direction. On the one hand, this justifies 
our disregarding of the kinematic constraint, while on the other, the collision
rate may still be considered proportional to some power $\beta$ of the
relative velocity.  A fair confirmation of this interpretation comes from the
average hydrodynamics equations derived below.
As anticipated in section~\ref{ch1_sec:hcs}, they replicate the transport
equations~\eqref{ch1_eq:brey2} for granular gases in $d>1$, but restricted to
the shear (transverse) velocity field.

\subsection{Average fields in the continuum limit}\label{sec:hydro-eq}

In the large system size limit as $L\to\infty$, a continuum limit may
be introduced by considering that the average velocity
$u_{l,p}=\langle v_{l,p}\rangle$ and energy
$E_{l,p}=\langle v_{l,p}^{2}\rangle$ are smooth functions of space and
time. Of course, the local temperature $T_{l,p}=E_{l,p}-u_{l,p}^{2}$
is also assumed to be smooth.  Specifically, we introduce hydrodynamic
continuous space and time variables, respectively,\footnote{Note that
  the definition of the continuous space variable below is slightly
  different from that given in our original publications
  \cite{MPLPyP16,PMLPyP16}.  This choice has no consequences for the
  equations derived in the continuum limit for the bulk sites, but
  allows us to simplify the derivations of some specific boundary
  conditions.}
\begin{equation}
\label{eq:hydro-scales-x-t}
x=\frac{l-1}{L}, \quad t=\frac{\omega\tau}{L^{2}}. 
\end{equation} 

For $\beta=0$, the balance equations for the average
velocity $u(x,t)$ and energy 
\begin{equation} \label{eq:definitio_energy}
E(x,t)=u^{2}(x,t)+T(x,t)
\end{equation} 
read
\begin{subequations}\label{eq:av-u-E}
\begin{align}
\partial_t u(x,t) &= -\partial_x j_{\av}(x,t) , \\
\partial_t E(x,t) &=-\partial_x J_{\av}(x,t) + d_{\av}(x,t).
\end{align}
\end{subequations}
Therein, the average momentum and energy currents,  $j_{\av}(x,t)$ and 
$J_{\av}(x,t)$, respectively,   are given by
\begin{equation}\label{eq:av-j-J}
  j_{\av}(x,t)=-\partial_{x} u(x,t), \quad
  J_{\av}(x,t)=-\partial_{x} E(x,t),
\end{equation}
and the dissipation field $d_{\av}(x,t)$ is
\begin{equation}\label{eq:av-d-nu}
  d_{\av}(x,t)=-\nu T, 
  \end{equation}
  with
  \begin{equation} \label{eq:nu}
      \nu=(1-\alpha^{2})L^{2}.
\end{equation}
In order to write the average dissipation we have made use of the
\textit{molecular chaos} assumption or \textit{Stosszahlansatz}. Specifically,
we have assumed that
\begin{equation}
\langle v_{l,p} v_{l\pm 1,p}\rangle = u_{l,p} u_{l\pm 1,p} + \mathcal{O}(L^{-1}).
\end{equation} 
In \eqref{eq:nu}, we have introduced the macroscopic dissipation coefficient $\nu$,
which is the relevant parameter in the hydrodynamic space and time
scales \cite{MPLPyP16,AM_thesis}. Note that if $\nu$ is of order unity, the collisions have to be quasi-elastic $1-\alpha^2 \ll 1$.
It is straightforward to combine
\eqref{eq:av-u-E}, \eqref{eq:av-j-J} and \eqref{eq:av-d-nu} to write
closed equations for the hydrodynamic fields: average velocity and
temperature, 
\begin{subequations}\label{eq:hydroMM}
\begin{align}
\partial_{t}u(x,t)&=\partial_{xx} u(x,t), \label{eq:hydroMMu} \\
\partial_{t}T(x,t)&=-\nu T(x,t)+ \partial_{xx}T(x,t)+2\left[\partial_{x}u(x,t)\right]^2 . \label{eq:hydroMMT}
\end{align}
\end{subequations}
These equations must be solved submitted to suitable boundary
conditions, which depend on the physical state under scrutiny. Note the resemblance between our hydrodynamic equations and those in \eqref{ch1_eq:brey2}.

\subsection{Fluctuating hydrodynamics}\label{sec:fluc-hydro}

The balance equations~\eqref{eq:av-u-E} may also be written
at the fluctuating level of description, by considering that $v(x,t)$
and $e(x,t)$ are fluctuating quantities, whose averages are $u(x,t)$
and $E(x,t)$. In this way, fluctuating balance equations are
written for both $v(x,t)$ and $e(x,t)$, which are the continuum limit
versions of the microscopic balance equations \eqref{eq:mom} and
\eqref{eq:en}, namely
\begin{subequations}\label{eq:fluct-hydro-v-e}
\begin{align}
\label{eq:fluct-hydro-v-e-a}
&\tder v(x,t)=-\partial_{x} j(x,t),& 
&j(x,t)=-\partial_{x}v(x,t)+\xi^{(j)}(x,t),& \\
\label{eq:fluct-hydro-v-e-b}
&\tder e(x,t)=-\partial_{x}J(x,t)+d(x,t),&  
&J(x,t)=-\partial_{x}e(x,t)+\xi^{(J)}(x,t).&
\end{align}
\end{subequations}
In the equations above, $(j,J)$ are the fluctuating currents for
momentum and energy, and $(\xi^{(j)},\xi^{(J)})$ are their
corresponding noises.  These noises have been shown to be Gaussian and white~\cite{MPLPyP16,AM_thesis}. The
amplitudes of their correlations $\langle
\xi^{(\gamma)}\xi^{(\gamma')}\rangle$ can be cast in matrix form, 
\begin{equation}
  \label{eq:noise-corr}
  \langle
  \xi^{(\gamma)}(x,t)\xi^{(\gamma')}(x',t')\rangle=L^{-1}
  \bm{\Xi}^{(\gamma\gamma')}\delta(x-x')\delta(t-t'),
\end{equation}
 where $(1,2)$ for $\gamma$ or $\gamma'$ correspond to
$(j,J)$. These amplitudes have been computed
  within the local equilibrium approximation in~\cite{MPLPyP16}, with the result
\begin{equation}\label{eq:corr-matrix}
   \bm{\Xi}=2T(x,t)
\begin{pmatrix}
    1 & & 2u(x,t) \\
   2u(x,t) & & 2[T(x,t)+2 u^2(x,t)]
 \end{pmatrix}.
\end{equation}
The average velocity $u(x,t)$ and the temperature $T(x,t)$ must be
calculated in the state corresponding to the physical situation of
interest.


Finally, the dissipation field $d(x,t)$ is given by
\begin{equation}
  \label{eq:fluct-d}
  d(x,t)=-\nu \theta(x,t)=-\nu \left[e(x,t)-v_{R}^{2}(x,t)\right],
\end{equation}
where $v_{R}^{2}$ is the regular part of $v^{2}$, defined as
\begin{equation}\label{vR-main-text}
  v_{R}^{2}(x,t)\equiv v^{2}(x,t)-L^{-1}\theta(x,t)
  \lim_{\Delta x\to 0}\delta(\Delta x).
\end{equation}
This regular part of the velocity field has the property
$\langle v_{R} ^{2}(x,t)\rangle =u^{2}(x,t)$, as shown in
appendix~\ref{ch5-6_app-a}. Equation~\eqref{eq:fluct-d} tells us that the
fluctuations of the dissipation field are enslaved to those of the
fluctuating temperature field $\theta(x,t)$.  This is so because the
dissipation noise $\xi^{(d)}$ is subdominant as compared to the
current noises, since it scales as $L^{-3}$ instead of as $L^{-1}$,
as proven in \cite{MPLPyP16}.

\section{Dynamics of the one-particle distribution function}\label{ch5_sec:1PDF}

Here, we apply the usual procedure of kinetic theory and map the
master equation into a BBGKY hierarchy. In particular, we focus on the
evolution equation for the one-particle distribution function at site
$l$ and at time $\tau$, which we denote by $P_{1}(v;l,\tau)$. By definition,
\begin{equation}
  \label{eq:f(v,t)}
 P_{1}(v;l,\tau)=\int d\vv P_{N}(\vv,\tau) \delta(v_{l}-v).
\end{equation}
It is easy to show that none of the terms in the sum \eqref{eq:cma2} contribute to the
time evolution of $P_1$ except those corresponding to $l-1$ and
$l$, because the collisions involving the pairs $(l-1,l)$ and $(l,l+1)$
are the only ones which change the velocity at site $l$. Therefore,
\begin{eqnarray} \label{eq:pseudoBoltzmann}
&& \partial_\tau P_1(v;l,\tau) = \omega \nonumber\\
&&  \times  \left\{ \int_{-\infty}^{+\infty}\!\!
  dv_{l-1}|\Delta_{l-1}|^\beta \left[
   \frac{P_{2}(\hat{b}_{l-1}^{-1}\{v_{l-1},v\};l-1,l,\tau)}{\alpha^{\beta+1}}
    -  P_2(v_{l-1},v;l-1,l,\tau) \right]\right. \nonumber\\
&&\left.
\;+\int_{-\infty}^{+\infty}\!\! dv_{l+1}|\Delta_{l}|^\beta \left[ \frac{P_{2}(\hat{b}_{l}^{-1}\{v,v_{l+1}\};l,l+1,\tau)}{\alpha^{\beta+1}}  -  P_2(v,v_{l+1};l,l+1,\tau) \right]\right\},\nonumber\\
\end{eqnarray}
where $P_{2}(v,v';l,l+1,\tau)$ is  the two-particle
probability distribution  for finding
the particles at the $l$-th and $(l+1)$-th sites with velocities $v$
and $v'$, respectively.  For the special case $\beta=0$, the evolution
equation for $P_{1}$ can be further simplified, because the terms on
the rhs of \eqref{eq:pseudoBoltzmann} coming from the loss (negative)
terms of the master equation can be integrated. We get
\begin{multline} \label{eq:pseudoBoltzmann-MM}
\partial_\tau P_1(v;l,\tau) =
\omega \left[ -2 P_1(v;l,\tau) 
+ \frac{1}{\alpha}\int_{-\infty}^{+\infty}
  dv_{l-1}
   P_{2}(\hat{b}_{l-1}^{-1}\{v_{l-1},v\};l-1,l,\tau)\right.
    \\
+\left.\frac{1}{\alpha}\int_{-\infty}^{+\infty} dv_{l+1} P_{2}(\hat{b}_{l}^{-1}\{v,v_{l+1}\};l,l+1,\tau)\right].
\end{multline}

The equation for $P_{1}$, either \eqref{eq:pseudoBoltzmann} for a
generic $\beta$ or \eqref{eq:pseudoBoltzmann-MM} for $\beta=0$, could
be converted to a closed equation for $P_{1}$ by introducing the
\textit{molecular chaos} assumption, which in our present context
means that
\begin{equation}
  \label{eq:mol-chaos}
P_{2}(v,v';l,l+1,\tau)=P_{1}(v;l,\tau)P_{1}(v';l+1,\tau)+\mathcal{O}(L^{-1}).
\end{equation}
By neglecting the $\mathcal{O}(L^{-1})$ terms in \eqref{eq:mol-chaos}, we obtain
a pseudo-Boltzmann or kinetic equation for $P_{1}$, which determines
the evolution of the one-time and one-particle averages under the
assumption of $\mathcal{O}(L^{-1})$ correlations.  Note that
  this ``smallness'' of two-particle correlations do not prevent them
  from being long-ranged.

In the continuum (hydrodynamic-like) limit defined in \eqref{eq:hydro-scales-x-t}, the
evolution equation for the one-particle distribution function
\eqref{eq:pseudoBoltzmann} can be written in a simpler form. The main
idea is the quasi-elasticity of the microscopic dynamics, stemming
from \eqref{eq:nu}, i.e. $\alpha=1-L^{-2}\nu/2+\mathcal{O}(L^{-4})$. Then,
\begin{align}
  \label{eq:3}
P_{2}\left(\hat{b}^{-1}_{l-1}\{v_{l-1},v\};l-1,l,\tau\right)
=P_{2}\left(v-\frac{\nu}{4L^{2}}\Delta_{l-1},v_{l-1}+\frac{\nu}{4L^{2}}\Delta_{l-1};l-1,l,\tau \right)
  \nonumber\\
=\left[1+\frac{\nu}{4L^{2}}\Delta_{l-1}(\partial_{v_{l-1}}-\partial_{v})\right]P_{2}(v,v_{l-1};l-1,l,\tau)+\mathcal{O}(L^{-4}).
\end{align}
Moreover, we make use of the molecular chaos assumption
\eqref{eq:mol-chaos} and identify
\begin{equation}
  P_{1}(v;l,\tau)=P_{1}(v;x=(l-1)/L,t=\omega\tau/L^{2}),
\end{equation}
that is, we consider $P_{1}$ to be a smooth function of the
hydrodynamic space and time variables $x$ and $t$. 

Under the hypotheses outlined above, a lengthy but straightforward
calculation gives for arbitrary $\beta$ that
\begin{align}
  \label{eq:P1-hydrobeta}
  \partial_{t}P_{1}(v;x,t)=& \,\, \partial_{x} \int_{-\infty}^{+\infty} dv' 
|v'-v|^{\beta}
  \left[P_{1}(v';x,t)\partial_{x}P_{1}(v;x,t)\right. \nonumber \\                          &\qquad\qquad\qquad\qquad\qquad- \left.P_{1}(v;x,t)\partial_{x}P_{1}(v';x,t)\right] \nonumber
  \\
& -\frac{\nu}{2} \partial_{v} \int_{-\infty}^{+\infty} dv' (v'-v)|v'-v|^{\beta}P_{1}(v';x,t)P_{1}(v;x,t).
\end{align}
The most important corrections to this equation emanate from finite
size effects that give rise to nonzero correlations, i.e. the
corrections to molecular chaos hypothesis that are expected to be
of the order of $L^{-1}$, see section \ref{sec:HCS-exact} for details. This equation must
be supplemented with suitable boundary conditions depending on the
physical state under scrutiny. Note that the divergence
  structure of the rhs of \eqref{eq:P1-hydrobeta} stems from the fact
  that $P_{1}(v;x,t)$ is a locally conserved quantity.

If we took moments in \eqref{eq:P1-hydrobeta}, we would obtain the
hydrodynamic equations for a generic value of $\beta$. It is clear,
from the structure of this equation, that these hydrodynamic equations
would be not closed and constitutive relations for the momentum and
energy currents and the dissipation fields would be needed. We also recall that in our model
there is no mass transport and therefore the ``field density''
$n(x,t)$ is uniform and constant in time,
\begin{equation}
  \label{eq:density}
  n(x,t)=\int_{-\infty}^{+\infty}dv\, P_{1}(v;x,t)=1,
\end{equation}
which is consistent with the
result $\partial_t n(x,t)=0$ obtained by marginalizing $v$ in~\eqref{eq:P1-hydrobeta}.

Again, for the case $\beta=0$, the equation for $P_{1}$ in the
continuum limit can be simplified,
\begin{equation}
  \label{eq:P1-hydroMM}
\partial_{t}P_{1}(v;x,t)=\partial_{xx}P_{1}(v;x,t)+\frac{\nu}{2}\partial_{v}\left[(v-u(x,t))
   P_{1}(v;x,t)\right]
\end{equation}
The above equation is not linear for $P_{1}$, since the average
momentum is a functional thereof,
$u(x,t)=\int_{-\infty}^{+\infty} dv \, v P_{1}(v;x,t)$. Note that, consistently,
by taking moments in \eqref{eq:P1-hydroMM}, the average hydrodynamic
equations \eqref{eq:hydroMM} are reobtained.  
Moreover,  the evolution equations for all the moments are closed (at any order) under  the molecular chaos assumption, without further knowledge of the one-particle distribution function $P_1$.

Note that  an appealing physical picture for
$P_{1}(v;x,t)$ arises in the continuum limit. In
fact,
\begin{multline}
  \label{eq:P1-interpretation}
  P_{1}(v;x,t)\,dv\,dx=\sum_{l=1}^{N}P_{1}(v;l,t)\,dv\,\Delta x\,
                          \Theta(L^{-1}l-x) \Theta(x+dx-L^{-1}l)  \\ = L^{-1}\sum_{l=1}^{N}P_{1}(v;l,t)\,dv \,\Theta(L^{-1}l-x) \Theta(x+dx-L^{-1}l),
\end{multline}
in which $\Theta(x)$ is Heaviside step function. The product of
Heaviside functions selects the range of $l$'s corresponding to the
interval $(x,x+dx)$. Thus, $P_{1}(v;x,t)dv\,dx$ can be interpreted as
the fraction of the total number of particles with velocities in the
interval $(v,v+dv)$ and positions in the interval $(x,x+dx)$, which
makes it neater the connection with the usual kinetic approach.

\section{The one-particle distribution function for some physical states}
\label{ch5_sec:have}

In this section, we restrict ourselves to the pseudo-Maxwell case 
$\beta=0$. Therein, we analyze two physically relevant states that are
typical of dissipative systems such as granular fluids. Specifically,
we investigate the Homogeneous Cooling State (HCS) and the Uniform Shear
Flow (USF) state. Although the model also admits stationary solutions of Couette Flows for appropiate boundary conditions \cite{MPLPyP16,AM_thesis}, they are not treated here. 
The theoretical results
below are compared to numerical results, which have been obtained by means of Monte Carlo simulations described in appendix \ref{ch5-6_app-b}. Specifically, we always 
plot the scaled one-particle distribution  defined as
\begin{equation}\label{P1-scaled-num}
\varphi(c;x,t)=\sqrt{T(x,t)} P_1(v;x,t), \quad c=\frac{v-u(x,t)}{\sqrt{T(x,t)}},
\end{equation}
in order to avoid visualizing the much sharper distributions $P_1$ that
arise in some situations as a consequence of the cooling.

\subsection{The Homogeneous Cooling State}
\label{ch5_sec:haveHCS}

Let us consider a system with periodic boundary conditions. Moreover, the initial condition is ``thermal'', that is, the random variables $v_{l,0}$ are Gaussian with zero average and unit variance. This means that we are choosing the uniform initial temperature to be the temperature unit. From this initial condition, the system evolves into  the state known as the Homogeneous Cooling State (HCS), in which the system remains homogeneous and the temperature decays in time. At the (average) hydrodynamic level, one has \cite{MPLPyP16,AM_thesis}
\begin{equation}
u(x,t) = 0, \qquad
T_{\HCS}(x,t) = T(t=0) e^{-\nu t} .
\label{HCS}
\end{equation}
Since the collision frequency of our model is velocity-independent, we are dealing with pseudo-Maxwell molecules and thus the temperature decays exponentially in time. This is the expected behavior, which replaces the typical algebraic decay for hard particles, known as Haff's law~\cite{Ha83}.

As stated in \ref{ch1_sec:hcs}, 
the HCS is known to be unstable: it breaks down in too large or too
inelastic systems \cite{MN93,NyE00}. In our model and in the
continuum limit, this condition (studied in~\cite{LMPyP15}) reads $\nu>\nu_c$, or, equivalently, $L>L_c$, where
\begin{equation}
  \label{ch5_eq:2}
\nu_{c}=8\pi^{2}, \quad   L_{c}=2\pi\sqrt{2}\left(1-\alpha^{2}\right)^{-1/2}.
\end{equation}
Note that the scaling of the critical length $L_c$ is the same as that 
introduced in \eqref{ch1_eq:Lc}.
When $\nu<\nu_c$ (or $L<L_c$) there is no unstable {mode}.
This instability mechanisms is completely analogous to the one found in
granular gases for the shear mode~\cite{NEByO97}.  Note that the amplification appears in the rescaled velocity
$\tilde{u}(x,t)=u(x,t)/v_{\text{th}}(t)$, being $v_{\text{th}}(t)=\sqrt{T_{\HCS}(t)}$, and not in the velocity $u(x,t)$.

Interestingly, the one-particle distribution function can be exactly
calculated in the HCS. Taking into account the homogeneity of the
state and the vanishing of the average velocity $u(x,t)$,
\eqref{eq:P1-hydroMM} for $P_1(v;t)$  simplifies to
\begin{equation}
  \label{ch5_eq:P1-MM-HCS}
\partial_{t}P_{1}(v;t)=\frac{\nu}{2}\partial_{v}\left[v
   P_{1}(v;t)\right].
\end{equation}
This equation can be integrated right away to give
\begin{equation}
\label{ch5_eq:P1-sol-HCS}
  P_{1}(v;t)=e^{\nu t/2}P_{1}(v e^{\nu t/2};t=0).
\end{equation}
Now, we employ the definition of the scaled 
distribution function \eqref{P1-scaled-num}, that here reduces to
\begin{equation}
  \label{ch5_eq:phi-def}
  \varphi(c;t)=v_{\thr}(t)P_{1}(v;t), \quad c=v/v_{\thr}(t).
\end{equation}
By
combining \eqref{ch5_eq:P1-sol-HCS} and \eqref{ch5_eq:phi-def}, we find that
$\varphi(c;t)$ does not evolve, that is,
\begin{equation}\label{ch5_eq:P1-sclng-HCS}
  \varphi(c;t)=\sqrt{T(0)}P_{1}(c\sqrt{T(0)};t=0)=\varphi(c;0).
\end{equation}
Therefore, the one-particle distribution function would remain
Gaussian for all times if it were so initially, as is usually the
case. In general, the shape of the initial distribution of velocities
is not altered, and it only ``shrinks'' with the thermal velocity.  A
similar behavior was found for elastic Maxwell molecules with
annihilation starting from the Boltzmann equation
\cite{GMSByT08}. This is a peculiarity of Maxwell molecules, in
  which the probability that a given pair collides is independent of
  its relative velocity.

\begin{figure}
\centering
\includegraphics[width=0.8\textwidth]{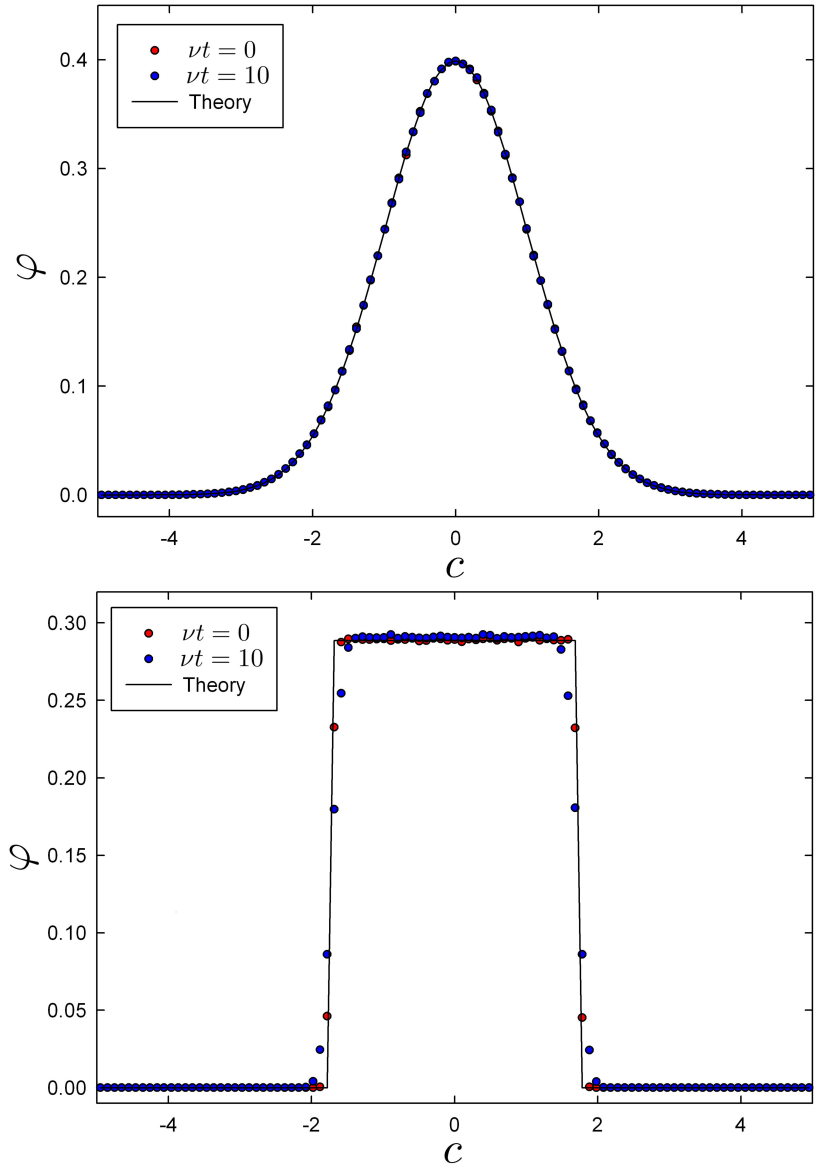}
\caption{Time evolution of the scaled
  one-particle distribution function $\varphi (c)$ in the HCS.
  The shape of the distribution
  remains unaltered when plotted as a function of the
  scaled velocity $c=v/v_{\thr}(t)$. The
  two panels correspond to different initial conditions
  $P_{1}(v;0)$: Gaussian (top) and square (bottom).  Note that deviations from the initial shape are
  barely observable in the Gaussian case, and remain very small for the square 
  shape. We have averaged over $10^{4}$
  realizations, in a system with $N=500$ and $\nu=20$.}
\label{fig:P1-HCS-scaling}
\end{figure}

We have simulated the
homogeneous cooling state of
a system made of $10^3$ particles, with periodic boundaries and
starting from a flat velocity profile $u(x,0) \equiv 0$ with unit
variance $T(x,0) \equiv T_0 = 1$. It can be observed how the one-particle distribution function $P_{1}(v;t)$
conserves its initial shape, as given by~\eqref{ch5_eq:P1-sclng-HCS}. We
check numerically this result for $\nu=20$ by
considering two initial velocity distributions, Gaussian and square,
that is,
\begin{equation}
  \label{ch5_eq:square-dist}
  P_{1}(v;t=0)=(2v_{0})^{-1}\Theta(v_{0}- |v|),
\end{equation}
 $\Theta(v)$ being the Heaviside step function. The parameter $v_{0}$
is adjusted in order to have unit variance ($v_{0}=\sqrt{3}$).  In
figure \ref{fig:P1-HCS-scaling}, we compare the theoretical prediction
with numerical results, finding excellent agreement except for very
small finite-size corrections.

\subsection{The Uniform Shear Flow steady state}
\label{ch5_sec:haveUSF}

Here, a velocity difference $a$ (shear rate) is imposed between the velocities
at the left and right boundaries of the system. Specifically, the boundary conditions for the hydrodynamic equations are
\begin{subequations} \label{ch5_eq:lees-edwards}
\begin{align}
  u(1,t)-u(0,t)&=a, \; &\left. \partial_x u(x,t)\right|_{x=0}&=\left. \partial_x u(x,t) \right|_{x=1}, \\
  T(0,t)&=T(1,t), \; &\left. \partial_x T(x,t)\right|_{x=0}&=\left. \partial_x T(x,t) \right|_{x=1},
\end{align}
\end{subequations}
that is, of Lees-Edwards type \cite{LyE72}. The corresponding stationary solution of the hydrodynamic equations \eqref{eq:hydroMM} is known as the Uniform Shear Flow (USF) state,
\begin{equation}\label{USF-profiles}
u_{s}(x)=a(x-1/2), \quad T_{s}=2a^{2}/\nu,
\end{equation}
that is, the velocity profile is linear whereas the temperature remains homogeneous. This steady state is peculiar of
dissipative systems, the continuous energy loss in collisions
compensates the viscous heating. The rheological effects described by
Santos et al.~\cite{SyG07,SGyD04} are not present in our system because
the microscopic dynamics is quasi-elastic.

In chapter \ref{ch:Hth_pgran}, we show that the USF is globally stable, that is, the system monotonically approaches the USF state from any initial condition. This is done by proving an $H$-theorem \cite{vK92,Re77} at the level of the one-particle distribution function. Physically, this is reasonable because the energy injection allows the system to fully explore its phase space, which hints at the validity of the $H$-theorem for the master equation at the $N$-particle level \cite{MPyV13,GMMMRyT15}. 
This is consistent with the (linear) stability
of the USF state of a dilute granular gas of hard spheres \cite{Ga06}
with respect to perturbations in the velocity gradient
introduced in section \ref{ch1_sec:hcs}. Note that in our 
1d model mimicking the shear modes, these perturbations are the only possible
ones to consider.

For the USF state, the stationary solution of the one-particle
distribution function can be solved: we seek a time-independent
solution of \eqref{eq:P1-hydroMM} with the ``scaling'' form
\begin{equation}
  \label{ch5_eq:P1-sclng-USF}
  P_{1}^{(\st)}(v;x)=T_{s}^{-1/2} \varphi(c), \qquad c=\frac{v-u_{s}(x)}{T_{s}^{1/2}}.
\end{equation}
By doing so, the probability distribution verifies the
appropriate boundary conditions for the USF state, that is,
\begin{equation}
  \label{ch5_eq:P1-bc-USF}
  P_{1}(v;x=1,t)=P_{1}(v-a;x=0,t), \quad \left. \partial_{x} P_{1}(v;x,t)\right|_{x=1}=\left. \partial_{x}P_{1}(v-a;x,t)\right|_{x=0}.
\end{equation}
Therefrom, the Lees-Edwards conditions directly
follow \eqref{ch5_eq:lees-edwards}. The resulting equation for
$\varphi(c)$ is 
\begin{equation}
\varphi''(c)+[c\varphi(c)]'=0,
\end{equation}
in which the prime stands
for the derivative with respect to $c$. Thus, the physical solution is
$\varphi(c)\propto \exp(-c^{2}/2)$ and 
\begin{equation}\label{USF-P1}
P_{1}^{(\st)}(v;x)=(2\pi T_{s})^{-1/2} \exp\left\{-\frac{[v-u_{s}(x)]^{2}}{2T_{s}}\right\},
\end{equation}
that is, the steady one-particle velocity distribution for the USF state is a 
Gaussian with average local velocity $u_{s}(x)$ and
temperature $T_{s}$. In \cite{MPLPyP16,AM_thesis}, the equation for higher order 
central moments of the velocity have been derived. Of course, the steady values for 
those moments are in agreement with the Gaussian shape obtained here.

\begin{figure}
 \centering
    \includegraphics[width=0.8\textwidth]{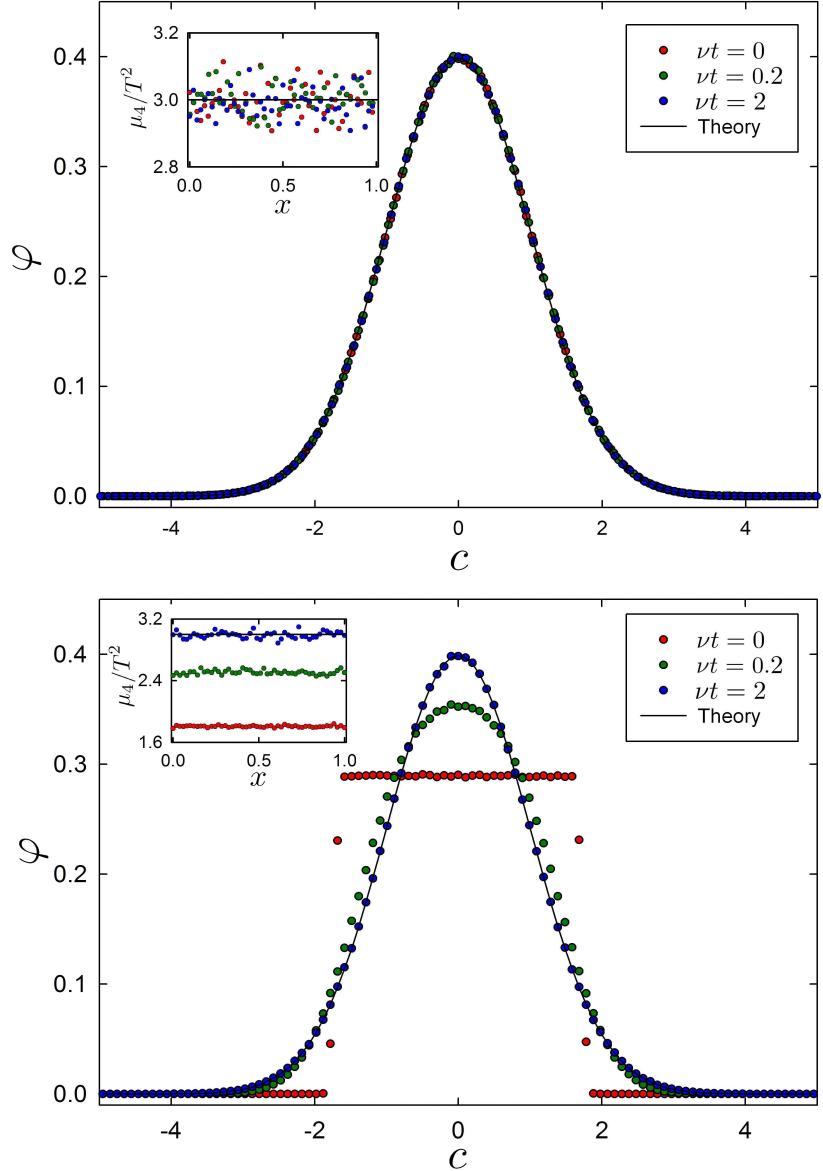}
    \caption{Evolution of the scaled one-particle
      velocity distribution  in the
      USF. Specifically, we show how the
      Gaussian steady distribution \eqref{USF-P1} is approached. In both panels,
      the initial velocity profile is already the steady one $u_{s}(x)$ but
      the initial homogeneous temperature $T_{0}=1\neq T_{s}$. We have considered, 
      a shear rate $a=5$ in a system with $N=500$ and $\nu=20$. Different initial
       shapes of the distribution have been chosen for each panel:
       Gaussian (top) and square (bottom). In
      the insets, we show the time evolution of the fourth central
      moment $\mu_{4}$ over $T^{2}$, which equals 3 for a Gaussian distribution,
       In all cases, averages over
      $M=10^4$ simulation trajectories  have been performed.}
\label{f:P1_USF}
\end{figure}

The USF described above can
be simulated by introducing appropriate collision rules for the boundary pairs.
When the pair $(1,N)$ is chosen to collide at time $p$,
there are two separate collisions: particle $1$ ($N$) undergoes a
collision with a particle with velocity $v_{N,p}-a$
($v_{1,p}+a$). These boundary collision rules introduce a shear rate
$a$ between the left and right ends of the system, and at the
hydrodynamic level are represented by the Lees-Edwards conditions
\eqref{ch5_eq:lees-edwards}. This can be readily shown by considering the
special evolution equations for $v_{1,p}$ and $v_{N,p}$ with the above
boundary collision rules in the continuum limit, see appendix \ref{app-lees}. 

In figure~\ref{f:P1_USF}, we check numerically the tendency
of the system to approach the steady Gaussian one-particle velocity
distribution of the USF state, given by \eqref{USF-P1}. We do so in
two cases: in the top panel, we start from a Gaussian distribution
with the steady velocity profile but with an initial value of the
temperature $T_{0}=1\neq T_{s}$. No time evolution is apparent in the
scaled variables, since the distribution remains a Gaussian of unit
variance for all times.  In the bottom panel, our simulation starts
from a square distribution. It is clearly observed that the Gaussian
shape is approached as time increases. In the inset of both panels, we
show the fourth central moment $\mu_{4}$ over $T^{2}$ at the same times, which also 
tends to its Gaussian value.

For the longest time in figure~\ref{f:P1_USF}, $\nu t=2$, the
one-particle velocity distribution has already reached the predicted
Gaussian shape. It is worth pointing out that this is so although the
temperature is still $10$\% below its steady value (not shown, see \cite{MPLPyP16,AM_thesis}). This fact suggests the
existence of a two-step approach to the steady state. In a first
stage, the one-particle distribution function forgets its initial
conditions and tends to a ``normal'' solution of the kinetic
equation. Afterwards, it is moving over this ``normal'' solution that
the system reaches the steady state. This resembles the so-called
hydrodynamic $\beta$-state reported by Garc\'ia de Soria et al.~in a uniformly
heated granular gas \cite{GMyT12,GMyT13}.

\section{Study of the fluctuations}
\label{sec:meso-fluc-th}

In this section, we focus on fluctuations around the HCS, which have
already been analyzed in the literature for a hard-sphere system
described by the Boltzmann equation close to the shear instability
\cite{BDGyM06}. To do so, it is useful to go to Fourier
  space, by considering that all the fields are written as
\begin{equation}
  \label{eq:Fourier-def}
  y(x,t)=\sum_{n} y_{n}(t) e^{i k_{n}x}, \quad
  y_{n}(t)=\int_{0}^{1} dx\, y(x,t)   e^{-i k_{n}x}, \; k_{n}=2n\pi.
\end{equation}


\subsection{Velocity fluctuations}\label{ch5_sec:fluc}

The equation for the fluctuating velocity \eqref{eq:fluct-hydro-v-e-a} is closed,
\begin{equation}
  \label{eq:fluc-veloc}
 \tder{v}=\partial_{xx}v-\partial_{x}\xi^{(j)},
\end{equation}
and going to Fourier space,
\begin{equation}
  \label{eq:fluc-veloc-Fourier}
  \tder{v}_{n}=-k_{n}^{2}v_{n}-i k\xi_{n}^{(j)}.
\end{equation}
The long time behavior of the solution to
\eqref{eq:fluc-veloc-Fourier} is readily obtained by taking the
initial time $t_{0}$ to $-\infty$, and then
\begin{equation}
  \label{eq:fluc-veloc-longtime}
  v_{n}(t)=-ik_{n} \int_{-\infty}^{t} ds\, e^{-k_{n}^{2}(t-s)} \xi_{n}^{(j)}(s).
\end{equation}
Now, we compute the equal-time velocity correlation in Fourier space,
\begin{equation}
  \label{eq:fluc-veloc-corr-Fourier}
  \langle v_{n}(t) v_{n'}(t)\rangle_{\HCS}=-k^{2}\int_{-\infty}^{t}
  \!\!\! ds \,
  e^{-k^{2}(t-s)}\int_{-\infty}^{t}\!\!\! ds' e^{-k^{2}(t-s')} \langle \xi_{n}^{(j)}(s)\xi_{n'}^{(j)}(s')\rangle_{\HCS}.
\end{equation}
Making use of the time dependence of the temperature in the HCS, that is, Haff's law \eqref{HCS}, we get to the lowest order
\begin{equation}
  \label{eq:veloc-corr-final}
  \langle v_{n}(t) v_{n'}(t)\rangle_{\HCS}=\frac{T_{\HCS}(t)}{L}
  \frac{2k_{n}^{2}}{2k_{n}^{2}-\nu}\delta_{n,-n'}=\frac{T_{\HCS}(t)}{L} \left(1+\frac{\nu}{2k_{n}^{2}-\nu}\right)\delta_{n,-n'},
\end{equation}
provided that $2k_{n}^{2}-\nu>0$. Thus, these correlations are valid
for all $n$ when $\nu<\nu_{c}=8\pi^{2}$ since at
$\nu=\nu_{c}$ we have that $\langle v_{1}(t)v_{-1}(t)\rangle$
diverges. 

The above correlations allow us to calculate the spatial integral of
$v^{2}(x,t)$. At the fluctuating level, we have that
\begin{equation}
  \label{eq:parseval}
  \int_{0}^{1} dx \, v^{2}(x,t)=\sum_{n=-\infty}^{+\infty} v_{n}(t) v_{-n}(t),
\end{equation}
which is Parseval's theorem for the Fourier transform. By taking
averages, we readily see that $v^{2}$ has a singular contribution,
because the sum of the correlations $\langle v_{n}(t)v_{-n}(t)\rangle$~diverges. This
stems from the $\delta(0)$ contribution in \eqref{vR-main-text}, the
average value of which in the HCS is
\begin{equation}
  \label{eq:sing-part-HCS}
  \langle L^{-1}\theta(x,t)\lim_{\Delta x\to 0}\delta(\Delta x)\rangle=L^{-1}T_{\HCS}(t)\sum_{n}1,
\end{equation}
since $\delta(x-x')=\sum_{n }\exp[i k_{n}(x-x')]$. Therefore,
\begin{subequations}\label{eq:reg-psi-HCS}
\begin{align}
  \label{eq:reg-HCS} 
\hspace{12mm}  \int_{0}^{1}dx\, \langle
v_R^{2}(x,t)\rangle &=
\frac{T_{\HCS}(t)}{L}\, \psi_{\HCS}, \\
\label{eq:psi-HCS} 
 \psi_{\HCS}(\nu)\equiv\sum_{n}
\frac{\nu}{2k_{n}^{2}-\nu}&=-\frac{\sqrt{\nu}}{2\sqrt{2}}
\cot\left({\frac{\sqrt{\nu}}{2\sqrt{2}}}\right).
\end{align}
\end{subequations}
Of course, the spatial integral of the regular part has a finite
value. The shear instability of the HCS is clearly observed within the
framework of the fluctuating hydrodynamic description: at
$\nu=\nu_{c}=8\pi^{2}$, we have that
\begin{equation}
  \label{eq:instability}
  \lim_{\nu\to\nu_{c}}\psi_{\HCS}(\nu)=\infty,
\end{equation}
and the spatial integral of $v_{R}^{2}$ diverge. In particular, it is
$\langle v_{1}(t)v_{-1}(t) \rangle$ that diverges, as readily seen from
\eqref{eq:veloc-corr-final} and said above.

\subsection{Effect of velocity fluctuations on the total energy}

Here, we consider the fluctuations of the total energy per particle, defined by
\begin{equation}
  \label{eq:total-energy}
  e(t)=\int_{0}^{1} dx\, e(x,t).
\end{equation}
At the mesoscopic fluctuating level, we have that
\begin{equation}
  \label{eq:total-energy-evolution}
  \frac{d}{dt}e(t)=\int_{0}^{1}dx \, d(x,t)=-\nu\, e(t) +\nu\!\int_{0}^{1}dx\,v_{R}^{2}(x,t),
\end{equation}
consistently with \eqref{eq:fluct-hydro-v-e} and
\eqref{eq:fluct-d}.

We introduce a rescaled dimensionless total energy by 
\begin{equation}\label{meso-sc-e}
 \tilde{e}(t)=\frac{e(t)}{T_{\HCS}(t)},
\end{equation}
which verifies the evolution equation
\begin{equation}
\label{eq:total-energy-sc-evol}
\frac{d}{dt}\tilde{e}(t)=\nu\!\int_{0}^{1}dx\,\tilde{v}_{R}^{2}(x,t),
\end{equation}
in which $\tilde{v}_{R}^{2}(x,t)=v_{R}^{2}(x,t)/T_{\HCS}(t)$. Now, we take averages and make use of \eqref{eq:reg-psi-HCS} to write 
\begin{equation}
  \label{eq:total-energy-sc-evol-average}
  \frac{d}{dt} \tilde{E}(t)=\psi_{\HCS}\frac{\nu}{L},
\end{equation}
which has to be integrated with the initial condition
  $\tilde{E}(0)=1$. We have omitted the $\nu$-dependence of
$\psi_{\HCS}$ in order not to clutter our formulae. Therefore, up to
order of $L^{-1}$, we have
\begin{equation}
  \label{eq:total-energy-sc-sol}
  \tilde{E}(t)=1+\delta\tilde{E}(t), \quad
  \delta{\tilde{E}}(t)=\psi_{\HCS} \frac{\nu\,t}{L}.
\end{equation}
which is expected to be valid as long as $\nu\psi_{\HCS} t/L\ll 1$. Equation \eqref{eq:total-energy-sc-sol} can be also obtained by solving perturbatively the evolution equations for the temperature and two-particle correlation function \cite{PMLPyP16,AM_thesis}.

There is a critical dissipation value $\nu_{\psi}$ such that
$\psi_{\HCS}$ vanishes, that is,
\begin{equation}\label{eq:nu-psi}
\nu_{\psi}=\nu_{c}/4=2\pi^{2}, \qquad \psi_{\HCS}(\nu_{\psi})=0,
\end{equation}
and the finite-size correction in~\eqref{eq:total-energy-sc-sol}
changes sign. Therefore, at this point we find a change in the
time-derivative of $\delta\tilde{E}(t)$.  For large system sizes, the
energy decays faster (slower) than the Haff's law for $\nu<\nu_{\psi}$
($\nu>\nu_{\psi}$) because $\psi_{\HCS}<0$ ($\psi_{\HCS}>0$). Comparisons between these theoretical results and Monte Carlo simulations give an excellent agreement \cite{PMLPyP16,AM_thesis}.

\section{Finite size effects: exact solution of the HCS}\label{sec:HCS-exact}

In this section, we further analyze the velocity correlations. The average equation
for the granular temperature (or the energy) in the HCS is closed only
when the correlation $\langle v_{l}v_{l+1}\rangle$ is neglected, since
it is expected to be of the order of $L^{-1}$. In other words, the
evolution equation for the temperature is closed in the molecular
chaos approximation. Interestingly, for the case of Maxwell
  molecules we are considering in this work, we can account for the
effect of the correlations in an exact way, thus going beyond
molecular chaos.

We assume that the system is in a spatial-translation-invariant state,
such as the HCS. We define the set of spatial
correlations of the velocity at time $\tau$ as
\begin{equation}\label{eq:C-k}
C_{k}(\tau)=\langle v_{j}(\tau)v_{j+k}(\tau)\rangle.
\end{equation} 
Here, $k$ represents the distance between the involved sites in the
correlation. Note that the average temperature at any site $j$ is given by $C_{0}$,
\begin{equation}
  \label{eq:C0-energy}
  T(\tau)\equiv C_{0}(\tau)=\langle v_{j}^{2}(\tau)\rangle.
\end{equation}
As a consequence of momentum conservation, in the center of mass frame
we have the ``sum rule''
\begin{equation}\label{eq:sum-rule}
C_{0}(\tau)+2 \sum_{k=1}^{\frac{L-1}{2}} C_{k}(\tau)=0, \quad \forall\tau,
\end{equation}
where we have considered an odd $L$.
 
The evolution equation of the correlations is readily
obtained from the master equation \eqref{eq:cma2},\footnote{
Equation \eqref{hier2} had a typo in our original publication
\cite{PMLPyP16}. Specifically, the prefactor of $(C_{0}-C_{1})$ was
$(1-\alpha^{2})$ instead of $(1-\alpha^{2})/2$, as pointed out by the
authors of \cite{PSDyN17}.}
\begin{subequations}\label{hier}
\begin{align}
\omega^{-1}\tauder{C_{0}}  = & \,\, (\alpha^2-1)(C_{0}-C_{1}), \label{hier1} \\
\omega^{-1}\tauder{C_{1}}  = & \,\, \frac{1-\alpha^2}{2}(C_{0}-C_{1})+(1+\alpha)(C_{2}-C_{1}),\label{hier2}  \\
\omega^{-1}\tauder{C_{k}}  = & \,\, (1+\alpha)(C_{k+1}+C_{k-1}-2C_{k}), \quad 2\leq k\leq (L-1)/2, \label{hier3} \\
\hspace{7mm} C_{\frac{L+1}{2}}= & \,\, C_{\frac{L-1}{2}}, \quad \forall\tau. \label{hier4}
\end{align}
\end{subequations}
Above, we have omitted the
$\tau$-dependence of the correlations to keep our notation simple, 
and written the evolution  equations for odd $L$, because the ``upper'' boundary
condition (for the maximum value of $k$) is simpler to write. For even $L$, the boundary condition would read $C_{\frac{L}{2}+1}=  C_{\frac{L}{2}-1}$.  Of course, our choice of $L$ as odd is irrelevant in the large system size limit $L\gg 1$. 

The hierarchy~\eqref{hier} can be exactly solved by reducing it to the
eigenvalue problem of a certain matrix. We carry out
this approach to the problem also for odd $L$.  The problem for
an even number particles may be solved by following an utterly similar
strategy, but the boundary conditions are a little more involved to
write. We do not present here these calculations because they do not
provide any additional physical insight.

First, it is useful to introduce a change of variables in order to make
the matrix symmetric. Specifically, we define
\begin{equation}
c_0=C_0 , \qquad c_k=\sqrt{2}\,C_k , \quad 1 \leq k \leq (L-1)/2.
\end{equation}
Second, we rewrite the  hierarchy~\eqref{hier} as
\begin{subequations}\label{hierv2}
\begin{align}
 \omega^{-1}(1+\alpha)^{-1}\partial_{\tau} c_{0} & =  -(1-\alpha)c_{0}+ \frac{1-\alpha}{\sqrt{2}}c_{1}, \label{hierv21} \\
 \omega^{-1}(1+\alpha)^{-1}\partial_{\tau}c_{1} & = \frac{1-\alpha}{\sqrt{2}} c_{0}-\frac{3-\alpha}{2}c_1+c_2, \label{hierv22} \\
 \omega^{-1}(1+\alpha)^{-1}\partial_{\tau} c_{k} & =  c_{k-1}-2c_{k}+c_{k+1}, \quad 2\leq k\leq (L-3)/2, \label{hierv23} \\
\omega^{-1}(1+\alpha)^{-1}\partial_{\tau} c_{\frac{L-1}{2}} & = c_{\frac{L-3}{2}} - c_{\frac{L-1}{2}}, \label{hierv24}
\end{align}
\end{subequations}
in which we have extracted the common factor $(1+\alpha)$ on the rhs
of \eqref{hier} and made use of \eqref{hier4} to write \eqref{hierv24}
for $c_{\frac{L-1}{2}}$. 

Now, we can solve the system above by a standard eigenvector method,
that is, we seek solutions of the form
\begin{equation}
c_k=e^{\lambda (1+\alpha)\omega\tau} \phi_k.
\end{equation} 
We denote the eigenvalues by $\lambda$ and its corresponding eigenvector by
$\phi$, $\phi_{k}$ is thus the $k$-th component thereof. In this way, we reach the system
\begin{subequations}\label{eigen}
\begin{align}
\hspace{3mm} \lambda \phi_{0}  &=  -(1-\alpha) \phi_{0}+ \frac{(1-\alpha)}{\sqrt{2}} \phi_{1}, \label{eigen1} \\
\hspace{3mm} \lambda \phi_{1}  &= \frac{(1-\alpha)}{\sqrt{2}} \phi_{0}-\frac{(3-\alpha)}{2} \phi_1+ \phi_2, \label{eigen2} \\
\hspace{3mm} \lambda \phi_{k}  &=  \phi_{k-1}-2 \phi_{k}+ \phi_{k+1}, \quad 2\leq k\leq (L-3)/2, \label{eigen3} \\
\lambda \phi_{\frac{L-1}{2}}  &=  \phi_{\frac{L-3}{2}} - \phi_{\frac{L-1}{2}}. \label{eigen4}
\end{align}
\end{subequations}

Equations \eqref{eigen} are a system of second-order difference
equations for $\phi_k$ with constant coefficients, in
which~\eqref{eigen3} is the general equation and \eqref{eigen2} and
\eqref{eigen4} are their boundary conditions. On top of that,
\eqref{eigen1} acts as an extra condition that ensures momentum
conservation, as shown below (see also \cite{PMLPyP16,AM_thesis}). The general solution of~\eqref{eigen3} is
of the form $\phi_{k>0}=r^k$ \cite{ByO99}, which substituted into
\eqref{eigen3} has two solutions $(r_1,r_2)$ that verify
\begin{subequations}\label{eigenr}
\begin{align}
 r_1 r_2  &=  1, \label{eigenr1}\\
r_1 + r_2  &=  2 + \lambda. \label{eigenr2}
\end{align}
\end{subequations}
We introduce a new variable $q\in[0,\pi]$ such that $r_1=e^{iq}$ and
$r_2=e^{-iq}$, as suggested by \eqref{eigenr1}. Note that
$|r_{1}|=|r_{2}|=1$, if one of the roots were larger than one it would
lead to correlations increasing with $k$, which is physically
absurd. Moreover, from a purely mathematical point of view,
restricting ourselves to $|r_{1}|=|r_{2}|=1$ leads to a complete set
of eigenvectors.  From \eqref{eigenr2}, we obtain
\begin{equation}
\lambda(q)=2(\cos q - 1),
\end{equation} 
and the corresponding eigenvector is given by
\begin{subequations}\label{eigenphi}
\begin{align}
\phi_{k>0}(q) & =  A \,e^{ikq} + B \, e^{-ikq}, \\
 \phi_{0}(q)  &=  \frac{1-\alpha}{\sqrt{2} \left(2\cos q-1-\alpha\right)}  (A\, e^{iq} + B\, e^{-iq}).
\end{align}
\end{subequations}

\begin{figure}
  \centering
  \includegraphics[width=0.8\textwidth]{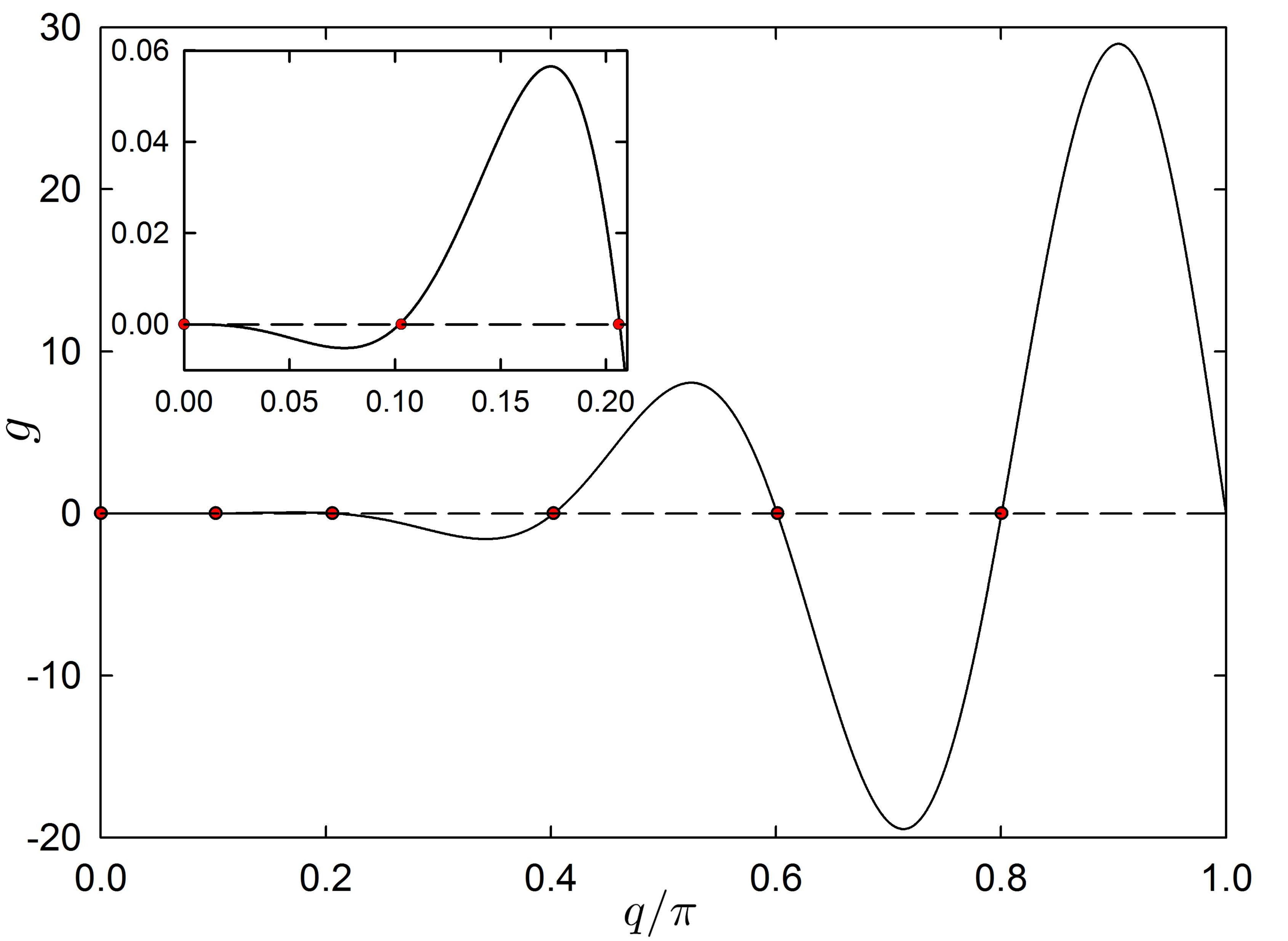}
 \caption{Plot of the function $g(q)$ defined in~\eqref{eigenfunction}. 
 The system size is 
    $L=11$. Its zeros $q_{n}$  determine the
   eigenvalues, as given by \eqref{eq:lambda-qn}. The first
   zero is always $q_{0}=0$, and there are $(L-1)/2$ additional 
   zeros $q_{i} \neq 0$, $i=1,\ldots,(L-1)/2$.  The inset shows a zoom of 
   the small $q$ region, to facilitate the identification of $q_1$ and $q_2$.}
   \label{fig:g(q)}
\end{figure}

The boundary conditions~\eqref{eigen2} and \eqref{eigen4} determine
the constants $A$ and $B$, and also the allowed values of the ``index'' $q$. The determinant of the linear system for
$A$ and $B$ must be zero, which is equivalent to impose that $q$ must
be a zero of the function
\begin{align}
\label{eigenfunction}
g(q)=& \, \, 2 \sin \left( \frac{L+3}{2} q \right) -(5+3\alpha) \sin \left( \frac{L+1}{2} q \right) +(5+7\alpha) \sin \left( \frac{L-1}{2} q \right) \nonumber\\
 &-(3+5\alpha) \sin \left( \frac{L-3}{2} q \right) +(1+\alpha) \sin \left( \frac{L-5}{2} q \right).
\end{align} 
This function has $(L+1)/2$ different zeros in the half-open interval
$[0,\pi)$, which we denote by $q_{n}$: $q_{0}=0$, $q_n$ is the $n$-th
nonvanishing zero of $g(q)$, $n=1,\ldots, (L-1)/2$. In addition $q=\pi$ also makes $g(q)$ vanish, but it does not correspond to an eigenvalue because the associated eigenvector would be identically zero, as shown below. In this way, we
find $(L+1)/2$ eigenvalues
\begin{equation}
  \label{eq:lambda-qn}
  \lambda_{n}=2(\cos q_{n}-1),
\end{equation}
the corresponding eigenvectors of which give a complete set for our
problem. In figure~\ref{fig:g(q)}, we plot the function $g(q)$
for $L=11$, which has six zeros in the interval $[0,\pi)$.

The proportionality relation obtained between $A$ and $B$ makes it
possible to write the eigenvector corresponding to the eigenvalue
$\lambda_n$ up to a normalization constant $\mathcal{N}_{n}$,
\begin{subequations}\label{eigenvector}
\begin{align}
\phi^{(n)}_{k>0} & =  \mathcal{N}_n \cos\left[ \left( \frac{L}{2} - k \right) q_n \right], \\
\phi^{(n)}_{0} & =  \frac{(1-\alpha) \mathcal{N}_n}{\sqrt{2} \left(2 \cos q_{n} - 1-\alpha\right)}  \cos\left[ \left( \frac{L}{2} - 1 \right)q_n \right].
\end{align}
\end{subequations}
The above expressions clearly show that there is no eigenvector for
$q=\pi$, since all its components are zero (recall that $L$ is
odd). The constant $\mathcal{N}_{n}$ is chosen to obtain
a orthonormal set of eigenvectors, in the sense that
\begin{equation}
\sum_{k=0}^{\frac{L-1}{2}} \phi^{(n)}_k \phi^{(n')}_k=\delta_{nn'}. 
\end{equation} 
We do not give the explicit expression for $\mathcal{N}_{n}$ because
it is quite involved and unnecessary for our purposes. The eigenvector corresponding to $q_{0}=0$ is particularly simple,
\eqref{eigenvector} implies that
\begin{equation}
  \label{eq:0-eigenvector}
  \phi_{0}^{(0)}=\frac{\mathcal{N}_{0}}{\sqrt{2}}, \quad
  \phi_{k>0}^{(0)}=\mathcal{N}_{0}, \qquad \mathcal{N}_{0}=\sqrt{\frac{2}{L}}.
\end{equation}
Then, the orthogonality relation of $\phi^{(0)}$ and $\phi^{(n)}$
($n\neq 0$) makes it possible to write a ``sum rule'' for the
components of the latter eigenvectors,
\begin{equation}
  \label{eq:sum-rule-eigenv}
  \phi_{0}^{(n)}+\sqrt{2}\sum_{k=1}^{\frac{L-1}{2}} \phi_{k}^{(n)}=0,
  \quad n>0.
\end{equation}
This sum rule is connected with ~\eqref{eq:sum-rule}, which stemmed
from momentum conservation. It also allows us to write
$\phi_{0}^{(n)}$ in a more convenient form for some calculations,
\begin{equation}
  \label{eq:phi0-conv}
\phi_{0}^{(n)}=-\frac{\mathcal{N}_{n}}{\sqrt{2}}\csc\left(\frac{q_{n}}{2}\right)\sin\left(\frac{L-1}{2}q_{n}\right),
\end{equation}
which does not depend explicitly on $\alpha$.

Finally, we have all the ingredients to build the general solution of \eqref{hierv2} as the sum
\begin{equation}
\label{eigensum}
c_k=\sum_{n=1}^{\frac{L-1}{2}} a_n \,e^{\lambda_n(1+\alpha)\omega\tau} \phi_k^{(n)},
\end{equation}
where $a_n$ is given in terms of the initial conditions by
\begin{equation}
a_n = \sum_{k=0}^{\frac{L-1}{2}} \phi_k^{(n)} c_k(0).
\end{equation}
The sum in~\eqref{eigensum} starts from $n=1$ because $a_{0}=0$, since
\begin{equation}
a_0= \mathcal{N}_0 \left[ \frac{c_0(0)}{\sqrt{2}} + \sum_{k=1}^{\frac{L-1}{2}} c_k(0) \right]=\frac{\mathcal{N}_0}{\sqrt{2}} \left[ C_0(0) + 2 \sum_{k=1}^{\frac{L-1}{2}} C_k(0) \right]=0.
\end{equation}
We have made use of momentum conservation, as expressed by the sum rule \eqref{eq:sum-rule}, to obtain the last equality.

\subsection{Eigenvalues for large systems}

Here, we would like to derive an approximate expression for the
eigenvalue spectrum in the large system size limit $L\gg 1$.
Therefore, we consider that the microscopic dynamics is quasi-elastic
by introducing the macroscopic dissipation coefficient $\nu$,
$(1-\alpha^2)L^2=\nu$, as in \eqref{eq:av-d-nu}. The eigenvalues are
given by the zeros of the function $g(q)$ in \eqref{eigenfunction}, and we expand
this function for $q \ll 1$ by introducing the scaling $Q=qL$, with
the result
\begin{equation}
\tan \left( \frac{Q}{2} \right) \left( \frac{\nu}{2} Q^2 L^{-2} - Q^4 L^{-4} \right) + \frac{1}{2} Q^5 L^{-5} =0.
\end{equation}
We are assuming that $Q$ is of the order of unity and have neglected
terms of the order of $L^{-6}$.

In order to obtain an analytical approximation for the eigenvalues, we
propose an expansion of $Q_{n}=q_{n}L$ in powers of $L^{-1}$,
$Q_n = Q^{(0)}_n+Q^{(1)}_n L^{-1}+\mathcal{O}(L^{-2})$. To the lowest order, we
obtain
\begin{subequations}\label{Qzeroth}
\begin{align}
Q^{(0)}_{1} & =  \sqrt{\frac{\nu}{2}}, \\
Q^{(0)}_{n} & =  2(n-1) \pi, \qquad n=2,\ldots,(L-1)/2.
\end{align}
\end{subequations}
Moreover, the  finite size corrections are
\begin{subequations}\label{Qfirst}
\begin{align}
Q^{(1)}_{1} & =  \frac{\nu}{8 \tan \left( \frac{1}{2}\sqrt{\frac{\nu}{2}} \right)}, \\
Q^{(1)}_{n} & =  \frac{16 (n-1)^3 \pi^3}{8 (n-1)^2 \pi^2 - \nu}, \qquad n=2,\ldots,(L-1)/2.
\end{align}
\end{subequations}
Note that $Q_{1}^{(1)}$ vanishes at $\nu=\nu_{\psi}=2\pi^{2}$ whereas
it diverges at $\nu=\nu_{c}=8\pi^{2}$. The former property is
connected to the change of sign in the finite-size correction to the
cooling rate of the HCS, whereas the latter gives rise to the
instability of the HCS, as discussed in
sections~\ref{sec:meso-fluc-th}.

\begin{figure}
  \centering
\includegraphics[width=0.8\textwidth]{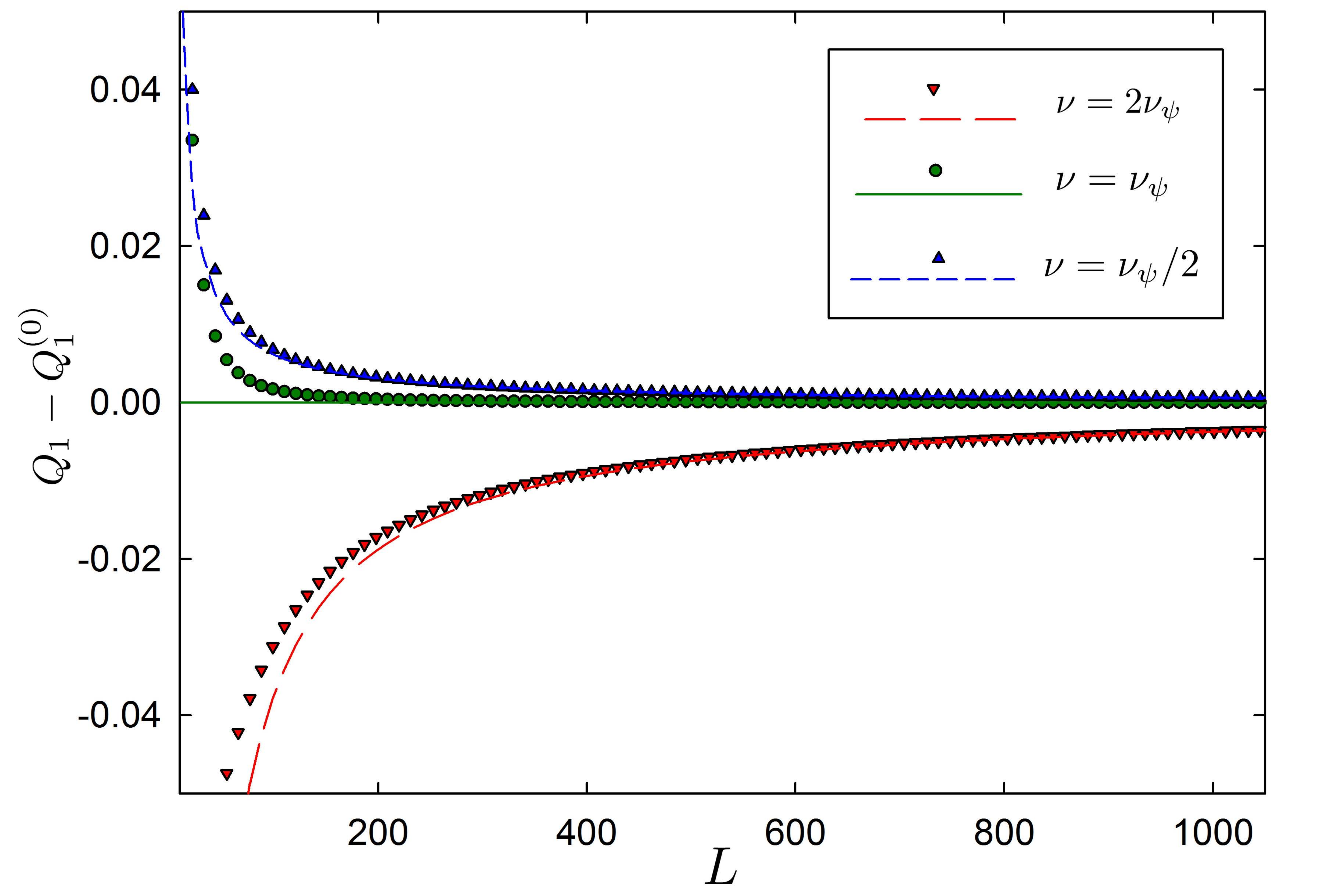}  
  \caption{Plot of the difference $Q_{1}-Q_{1}^{(0)}$ as a function of
  the system size $L$. Specifically, we present the plots for $\nu=\pi^{2}$,
  $\nu=\nu_{\psi}=2\pi^{2}$ and $\nu=4\pi^{2}$. For each of these values, 
  there are two curves: the theoretical
  curve $Q_{1}^{(1)}/L$ (lines) and the numerical estimate of
  $Q_{1}$ (symbols). See the legend for details. 
  The finite-size correction is especially small for $\nu=\nu_{\psi}$.}
 \label{fig:eigenvalues-expansion} 
\end{figure}

In figure \ref{fig:eigenvalues-expansion}, we check the above
expansion for the zeros of the function $g(q)$. Specifically, we do so
for the first zero $q_{1}$: the numerical estimation of $q_{1}$ is
compared with the expansion in \eqref{Qzeroth} and \eqref{Qfirst} by
plotting $Q_{1}-Q_{1}^{(0)}$ as a function of the system size $L$.  It
is observed that this difference tends to zero as the system size
increases, for all the considered values of $\nu$. The finite size
correction is especially small for $\nu=\nu_{\psi}=2\pi^{2}$, for
which the theoretical correction $Q_{1}^{(1)}$ vanishes. Therefore,
finite size corrections are as small as possible for this case, which
makes it particularly adequate to investigate the hydrodynamic
description, as done in section \ref{ch5_sec:have} and  also in \cite{LMPyP15,MPLPyP16,AM_thesis}.

We want to emphasize that the instability of the HCS is reobtained
here as a crossing between the first two nonzero eigenvalues: for
$\nu=\nu_c=8\pi^{2}$, we have that $Q_1^{(0)}=Q_2^{(0)}$. On the one
hand, for $\nu<\nu_{c}$, the largest nonvanishing eigenvalue is $\lambda_{1}$
($\lambda_{1}<0$) and dominates the long-time dynamics of the system:
the energy $C_{0}$ and all the correlations $C_{k}$ decay with
$\exp[\lambda_{1}\omega(1+\alpha)\tau]=\exp(\nu_{\HCS}^{r}t)$, see
below. On the other hand, for $\nu>\nu_c$, the dominant term is the
one corresponding to $Q_{2}\simeq 2\pi$ and the long time behavior of
the system becomes independent of $\nu$.

The large system size limit of the eigenvalues is then
\begin{equation}
  \label{eq:lambda-large-N}  \lambda_{n}=-\frac{{Q_{n}^{(0)}}^{2}}{L^{2}}\left[1+L^{-1}\frac{2Q_{n}^{(1)}}{Q_{n}^{(0)}}+\mathcal{O}(L^{-2})\right].
\end{equation}
Moreover, the exponent in~\eqref{eigensum} controlling the time
dependence of the contribution for each mode is
\begin{equation}
  \label{eq:eigensum-exponent}
  \lambda_{n}(1+\alpha)\omega \tau \sim                                      -2{Q_{n}^{(0)}}^{2}\left[1+L^{-1}\frac{2Q_{n}^{(1)}}{Q_{n}^{(0)}}+\mathcal{O}(L^{-2})\right]t,
\end{equation}
which shows the relevance of the hydrodynamic scale $t$, defined 
in \eqref{eq:hydro-scales-x-t}, in the large system size limit.

\subsection{Long time limit}

Equation \eqref{eigensum} gives  the general time evolution for the
velocity correlations. Here, we show that these correlations tend to
their HCS values in the long time limit, provided that $\nu<\nu_{c}$,
that is, we are below the instability. 

Let us consider the scaled correlations $\tilde{C}_{k}$
\begin{equation}\label{eq:scaled-corr-tau}
\tilde{C}_k(\tau)=\frac{C_k(\tau)}{C_0(\tau)}=\frac{c_k(\tau)}{\sqrt{2} c_0(\tau)}, 
\end{equation}
that is, we scale the correlations with the energy $C_{0}\neq 0$. For
long enough times, the only relevant contribution to~\eqref{eigensum}
stems from the maximum (minimum in absolute value) eigenvalue
$\lambda_1$. Thus, the time dependence for all the correlations
$C_{k}$ (or $c_{k}$) are the same and, consequently, the quotient in
\eqref{eq:scaled-corr-tau} becomes time-independent for long enough
times. Making use of  \eqref{eq:phi0-conv} and \eqref{eigensum},
\begin{equation}\label{eq:scaled-corr-discrete}
\tilde{C}_k= \frac{\phi^{(1)}_k}{\sqrt{2}\phi^{(1)}_0}=-
\sin \left( \frac{q_1}{2} \right) \csc \left(
\frac{L-1}{2} q_1 \right) \cos \left[ \left( \frac{L}{2} -k \right) q_1 \right] ,
\end{equation}
which is nothing but the discrete version of the continuum solution obtained in \cite{PMLPyP16,AM_thesis}.

We can also derive the rate at which the energy and all the
correlations are decaying in the long time limit. Particularizing \eqref{eq:eigensum-exponent} for $n=1$, we have that
\begin{equation}\label{min-eigenvalue}
\lambda_1 (1+\alpha)\omega\tau \sim -\nu t\left[1-L^{-1} \psi_{\HCS}+\mathcal{O}(L^{-2})\right] =-\nu_{\HCS}^{r}t,
\end{equation}
where $\nu_{\HCS}^{r}$ is the ``renormalized'' by fluctuations cooling
rate introduced in \cite{PMLPyP16,AM_thesis}, by applying a multiple scale
analysis to find finite size corrections to the hydrodynamic
description. Thus, the energy is given by
\begin{equation}
C_{0}(t)=T(t=0)\exp(-\nu_{\HCS}^{r}t)
\end{equation}
 and the correlations
$C_{k}$, with $k>0$, follow from~\eqref{eq:scaled-corr-discrete}.

{\clearpage \thispagestyle{empty}}
\chapter{Global stability, derivation of the $H$-theorem}
\label{ch:Hth_pgran}
\newcommand{\vder}[1]{\partial_{v}{#1}}
\newcommand{\vvder}[1]{\partial_{v}^{2}{#1}}
\newcommand{\ra}{\rangle}
\newcommand{\la}{\langle}
\newcommand{\be}{\begin{equation}}
\newcommand{\ee}{\end{equation}}
\newcommand{\diff}{\text{diff}}
\newcommand{\noise}{\text{noise}}
\newcommand{\inel}{\text{inel}}
\newcommand{\gauss}{\text{Gauss}}
\newcommand{\lin}{\text{lin}}
\newcommand{\calH}{\mathcal{H}}
\newcommand{\PR}{\textit{Phys. Rev. }}
\newcommand{\PRL}{\textit{Phys. Rev. Lett.}}
\newcommand{\JMP}{\textit{J. Math. Phys.}}
\newcommand{\JSTAT}{\textit{J. Stat. Mech. (Theor. Exp.)}}

Here, we aim at investigating the global stability and the possibly associated 
$H$-theorem for a generalization of the model introduced in the previous chapter. 
At difference with the approach in 
\cite{MPyV13,GMMMRyT15}, our analysis is not restricted to 
spatially homogeneous situations: we consider the whole space and velocity 
dependence of the one-particle probability distribution function. More 
specifically, a general energy injection mechanism is introduced, in which the 
system may be driven through both the boundaries and the bulk. 

We show that, under quite general conditions, the steady state is globally stable. 
Independently of the initial preparation, the system always ends up in the steady 
state. Interestingly, it is not necessary to have an $H$-theorem to prove this: it 
suffices to show that $H$ is decreasing in the long-time limit, 
not for all times. In 
this regard, we find a situation that is similar to the  proof of the tendency 
towards the equilibrium curve in systems whose dynamics is governed by master 
equations with time-dependent transition rates
 \cite{ByP93b,ByP94,BPyR94,VyR97,VMyR98,PByS00,EyK10}.

Our proof of global stability also makes it possible to show the inadequacy of 
Boltzmann's $H_B$, defined in \eqref{H-Boltzmann}, as a candidate 
for a Lyapunov functional in 
inelastic systems. Not only is this done for the simplified models considered 
throughout this thesis, but for a general collision term that does not conserve 
energy in collisions. Therefore, this result also applies to the inelastic 
Boltzmann or Enskog equations, employed for granular fluids. Specifically, the main 
idea is the possibility of reversing the 
sign of $d H_B /d t$ by a suitable choice of the 
initial PDF. Thus, $d H_B /d t$ cannot have a definite sign. 
In this sense, our result 
can be understood as a generalization of that in \cite{BCDVTyW06}, 
within the first Sonine 
approximation of the inelastic Boltzmann equation, to an arbitrary collision kernel 
with nonconservative interactions.

Having proved global stability by showing that $H$ is a 
nonincreasing functional for 
long times, a natural question arises. Is $H$ a Lyapunov function, that is, a 
nonincreasing functional for all times? There does not seem to be a unique proof, 
valid for any driving mechanism, even within the framework of our simplified model. 
Notwithstanding, we have been able to derive a specific proof for a quite general 
driving mechanism, which includes as limiting cases both the sheared system and the 
uniformly heated system 
\cite{NyE98,MyS00,GMyT09,GMyT12,PTNyE01,MGyT09,PyT14,TyP14,PSyD13}. 
Our proof is based on a suitable expansion of the 
one-particle PDF in Hermite polynomials, which is a generalization of the customary 
Sonine expansion of kinetic theory.

The plan of this chapter is detailed below.
We introduce the aforementioned generalized model and study its stationary states 
in section \ref{ch7_sec:model}. The main difference is the consideration of a
 thermostat, allowing an energy input not only through boundaries, but also
through the bulk. Section \ref{global-stability} is devoted to the proof of the global stability
of the nonequilibrium steady states for this general energy
injection mechanism. The inadequacy of Boltzmann's $H_B$
as a Lyapunov functional for inelastic systems is discussed
in section \ref{sec:inadequacy}. Finally, in section \ref{USF}, 
we consider some concrete physical situations in our model, 
which include the sheared
and the uniformly heated systems. Therein, we prove that the 
functional $H$, which is used in the proof of global stability,  is a monotonically
 decreasing Lyapunov functional. In this way, we prove an $H$-theorem
 for our system with nonconservative interactions.

\section{Basics of the model}\label{ch7_sec:model}

\subsection{The stochastic forcing}
As introduced above, in this chapter we use a generalization of the
model presented in the previous chapter. In addition to collisions
\eqref{coll_rule}, the system is heated by a stochastic force that is
modeled by a white noise that affects all sites, the so-called
\textit{stochastic thermostat}
\cite{NyE98,MyS00,GMyT09,GMyT12,PTNyE01,MGyT09,PyT14,TyP14,PSyD13,Wi96,WyM96,SBCyM98}. Specifically,
for a short time interval, the change of the velocity due to the
heating is given by
\begin{align}
  \left.\Delta v_{i}(\tau)\right|_{\text{noise}}&\equiv \left. v_{i}(\tau+\Delta
                                                       \tau)-v_{i}(\tau)\right|_{\text{noise}} \nonumber \\ &=\left(\xi_{i}(\tau)-\frac{1}{N}\sum_{j=1}^{N}
                                                                                                          \xi_{j}(\tau)\right)\Delta \tau, \label{jump-moments-1}
\end{align}
where $\xi_{i}(t)$ are Gaussian white noises, verifying
\begin{equation}\label{jump-moments-2}
\la \xi_{i}(\tau)\ra_{\text{noise}}=0, \quad \la \xi_{i}(\tau)\xi_{j}(\tau')\ra_{\text{noise}}=\chi
\delta_{ij}\delta(\tau-\tau'),
\end{equation}
for $i,j=1,\ldots,N$. Above, $\chi$ is the amplitude of the noise, and
$\la\cdots\ra_{\text{noise}}$ denotes the average over the different
realizations of the noise. Note that this version of the stochastic
thermostat conserves total momentum, a necessary condition to have a
steady state \cite{MGyT09,PSyD13,PSDyN17}.

We turn our attention to the probability density of finding the
system in state $\vv$ at time $\tau$, $P_{N}(\vv,\tau)$ . The stochastic process
$\vv(\tau)$ is Markovian and the equation governing the time evolution
of $P_{N}(\vv,\tau)$ has two
contributions. First, we have the master equation contribution stemming
from collisions, as given by \eqref{eq:cma2} with $\beta=0$,
\begin{equation} \label{ch7_eq:master-equation}
\left.\partial_{\tau} P_N(\vv,\tau)\right|_{\text{coll}}=\omega \sum_{l=1}^N \left[ \frac{P_{N}(\hat{b}_l^{-1} \vv,\tau)}{\alpha}  -  P_N(\vv,\tau) \right].
\end{equation}
Second, there is a Fokker-Planck
contribution stemming from the stochastic forcing \cite{MPyV13,GMMMRyT15}
\begin{equation} \label{ch7_eq:Fokker-Planck}
\left.\partial_\tau
  P_N(\vv,\tau)\right|_{\text{noise}}=\frac{\chi}{2}
\sum_{i,j=1}^{N}\left(\delta_{ij}-\frac{1}{N}\right)\frac{\partial^{2}}{\partial
v_{i}\partial v_{j}}P_{N}(\vv,\tau).
\end{equation}
The time evolution of $P_{N}(\vv,\tau)$ is obtained by combining
\eqref{ch7_eq:master-equation} and \eqref{ch7_eq:Fokker-Planck}, that is,
\begin{equation}\label{time-evol}
\partial_{\tau}P_{N}(\vv,\tau)=\left.\partial_{\tau}P_{N}(\vv,\tau)\right|_{\text{coll}}+\left.\partial_{\tau}P_{N}(\vv,\tau)\right|_{\text{noise}}.
\end{equation}

We can derive the evolution equation of the one-particle distribution function in the ``hydrodynamic'' continuous space
and time variables $x=(l-1)/L$ and $t=\omega\tau/L^{2}$. As in the previous chapter, we assume molecular chaos to get a closed evolution equation for $P_1(v;x,t)$
\begin{eqnarray}
\partial_{t}P_1(v;x,t)=\partial_{x}^{2}P_1(v;x,t)+\frac{\nu}{2}\partial_{v}\left\{[v-u(x,t)]
   P_1(v;x,t)\right\}+\frac{\xi}{2}\partial_{v}^{2}P_1(v;x,t),
\label{ch7_eq:P1-hydroMM}
\end{eqnarray}
where we recall that $u(x,t)$ is the local average velocity and
$\nu=(1-\alpha^2)L^2$ is the macroscopic dissipation coefficient. The 
macroscopic noise strength $\xi$ is given by
\begin{equation}\label{ch7_eq:nu}
  \xi=\frac{\chi L^{2}}{\omega}.
\end{equation}
This shows that the microscopic noise strength $\chi$ must scale as
$L^{-2}$ in order to have a finite contribution in the continuum
limit. Of course, for $\xi=0$, we recover the kinetic equation for the
case in which there is no stochastic forcing, see \eqref{eq:P1-hydroMM}.

From the kinetic
equation for $P_1(v;x,t)$, one can derive the evolution equations for the profiles $u(x,t)$
and $T(x,t)$,
\begin{subequations}\label{ch7_eq:hydroMM}
\begin{align}
\partial_{t}u&=\partial_{xx} u, \label{ch7_eq:hydroMMu} \\
\partial_{t}T&=-\nu T+ \partial_{x}^{2} T+2\left(\partial_{x}u\right)^2+\xi . \label{ch7_eq:hydroMMT}
\end{align}
\end{subequations}
Note that the only difference with \eqref{eq:hydroMM} is an extra term $\xi$ in the temperature equation, corresponding to the uniform heating. 

\subsection{Nonequilibrium steady states}\label{NESS}

We are interested in the driven cases.
 Therein, energy loss in collisions  is eventually balanced (in average) by the energy input, and in the long time limit
  the system reaches a steady state. These
nonequilibrium steady states (NESS) are described by the corresponding
stationary solutions $P^{(\st)}_{1}(v;x)$ of the kinetic equation, which
verify
\begin{equation}
0=\partial_{x}^{2}P^{(\st)}_{1}(v;x)+\frac{\nu}{2}\partial_{v}\left\{[v-u_{\st}(x)]P^{(\st)}_1 (v;x) \right\} +\frac{\xi}{2}\partial_{v}^{2}P^{(\st)}_1 (v;x),
\label{ch7_eq:P1-steady}
\end{equation}
where $u_{\st}(x)$ is the stationary
average velocity profile. In this chapter, to be concrete, we consider two cases: a
system that is (a) sheared and (b) uniformly heated.

First, let us consider the sheared system  that we analyzed in section \ref{ch5_sec:haveUSF}, for which there is no
stochastic forcing, $\xi=0$. The stationary values for the average velocity and temperature in the USF are those in \eqref{USF-profiles}, whereas the stationary distribution is given by \eqref{USF-P1}, and thus we do not repeat them here.

Second, we address the uniformly heated system, $\xi\neq 0$, but without any shear, $a=0$ in \ref{ch5_eq:lees-edwards} and \ref{ch5_eq:P1-bc-USF}. In other words, we have the usual periodic boundary conditions. 
In the steady state, the system is homogeneous: there is no average
velocity and the temperature is uniform,
\begin{equation}\label{ch7_eq:ave-Unif-heated}
 u_{\st}(x)=0, \quad T_{\st}=\frac{\xi}{\nu}.
\end{equation}
The corresponding stationary PDF is also Gaussian,
\begin{equation}\label{ch7_eq:P1-Unif-heated}
P^{(\st)}_1(v;\cancel{x})=\left(2\pi T_{\st}\right)^{-1/2} \exp\left[-\frac{v^{2}}{2T_{\st}}\right].
\end{equation}
With this ``stochastic thermostat'' forcing, the system remains
homogeneous for all times if it is initially so, as is also the case
of a inelastic gas of hard particles described by the inelastic
Boltzmann equation \cite{NyE98}.

\section{Proof of global stability}
\label{global-stability}

In this section, we analyze the global stability of the nonequilibrium
stationary solutions of the kinetic equation \eqref{ch7_eq:P1-hydroMM}. 
We do so for quite a general class of boundary conditions.  Following
the discussion in section \ref{ch1_sec:H-th}, we define the $H$-functional as
\begin{equation}\label{H-functional-st}
H[P_1]=\int\!\! dx\,dv P_1(v;x,t) \ln\!\left[\frac{P_1(v;x,t)}{P^{(\st)}_1(v;x)}\right],
\end{equation}
note the analogy with \eqref{H-one-particle}.

Let us consider the time evolution of $H[P_1]$. It is directly obtained
that 
\begin{align}
\frac{dH}{dt}=\int\!\! dx\, dv\, \tder{\left[P_1
\ln\left(\frac{P_1}{P^{(\st)}_1}\right)\right]}=\int\!\! dx\, dv\,\left(\mathcal{L}P_1 \right) \,\ln\left(\frac{P_1}{P^{(\st)}_1}\right),
\label{ch7_eq:H-time-ev-1}
\end{align}
where $\mathcal{L}$ stands for the nonlinear evolution operator on the
rhs of the kinetic equation \eqref{ch7_eq:P1-hydroMM}, that is,
$\tder{P_1}=\mathcal{L}P_1$. Now we make use of the following property: if we
define $\Delta P_1=P_1-P^{(\st)}_1$ to be the deviation of the PDF from the
steady state, the linear terms in the deviation vanish, since both
factors in the integrand of \eqref{ch7_eq:H-time-ev-1} are equal to zero
for $P_1=P^{(\st)}_1$. This is a desirable property: were it not true, the
sign of $dH/dt$ could be reversed for initial conditions close enough
to the steady state by simply reversing the initial value of
$\Delta P_1$. Thus, the existence of an $H$-theorem would be utterly
impossible, see also next section.

Then, we can write
\begin{align}
\frac{dH}{dt}=\int\!\! dx\, dv\,\left(\mathcal{L}P_1\right)
  \,\ln\left(\frac{P_1}{P^{(\st)}_1}\right)-\int\!\! dx\,dv\, \left( \mathcal{L}P^{(\st)}_1\right)\, \frac{P_1-P^{(\st)}_1}{P^{(\st)}_1}.
\label{ch7_eq:H-time-ev-2}
\end{align}
Now, the idea is to split the operator $\mathcal{L}$ into the three
contributions on the rhs of \eqref{ch7_eq:P1-hydroMM}: first, the
diffusive one; second, the one proportional to $\nu$, which is intrinsically dissipative; and third,
the one proportional to the noise strength $\xi$:
$\mathcal{L}_{\diff}$, $\mathcal{L}_{\inel}$ and
$\mathcal{L}_{\noise}$, respectively.  Accordingly, we have that the
time derivative of $H$  has
three contributions, 
\begin{equation}\label{ch7_eq:H-time-ev-total}
\frac{dH}{dt}=\left.\frac{dH}{dt}\right|_{\diff}+\left.\frac{dH}{dt}\right|_{\inel}+\left.\frac{dH}{dt}\right|_{\noise},
\end{equation}
obtained by inserting into \eqref{ch7_eq:H-time-ev-2} the relevant
part of the evolution operator $\mathcal{L}$. Note that, although
$\mathcal{L}P^{(\st)}_1=0$, in general $\mathcal{L}_{\diff}P^{(\st)}_1\neq 0$,
$\mathcal{L}_{\inel}P^{(\st)}_1\neq 0$ and
$\mathcal{L}_{\noise}P^{(\st)}_1\neq 0$.

After some tedious but easy algebra, collected in
Appendix \ref{app_gran-dHdt-general}, the following expressions are
derived. Firstly, for the diffusive term,
\begin{equation}\label{ch7_eq:H-time-ev-diff}
\left.\frac{dH}{dt}\right|_{\diff}=-\int\!\! dx\,dv\,P_1
\left(\xder{\ln P_1}-\xder{\ln P^{(\st)}_1}\right)^{2}\leq 0.
\end{equation}
Secondly, for the inelastic term, proportional to $\nu$, 
\begin{equation}\label{ch7_eq:H-time-ev-inel}
\left.\frac{dH}{dt}\right|_{\inel}=-\frac{\nu}{2}\int\!\! dx
\left(u-u_{\st}\right) \int\!\! dv
P_1 \, \vder{\ln P^{(\st)}_1}.
\end{equation}
Finally, the noise term, proportional to $\xi$, reads
\begin{equation}\label{ch7_eq:H-time-ev-noise}
\left.\frac{dH}{dt}\right|_{\noise}=-\frac{\xi}{2}\int\!\! dx\,dv\,P_1
\left(\vder{\ln P_1}-\vder{\ln P^{(\st)}_1}\right)^{2}\leq 0.
\end{equation}
These results, and the following throughout this section, are valid
for a quite general set of boundary conditions, leading to the
cancellation of all the boundary terms arising after integrating by
parts, as detailed in Appendix~\ref{app_gran-dHdt-general}. This set
includes but is not limited to the Lees-Edwards and periodic boundary
conditions corresponding to the sheared and uniformly heated
situations, respectively. For instance, they also apply to the Couette
state, in which the system is driven by keeping its two edges at two
(in general, different) fixed temperatures $T_{L}$ and $T_{R}$.

The inelastic term $dH/dt|_{\inel}$ in \eqref{ch7_eq:H-time-ev-inel}
does not have a definite sign in general. Therefore, it is the
inelastic term that prevents us from proving $H$ to be a
nonincreasing function of time. It must be stressed that the
diffusive, inelastic and noise contributions to $dH/dt$ in
\eqref{ch7_eq:H-time-ev-diff}-\eqref{ch7_eq:H-time-ev-inel} come
exclusively from the diffusive, noise and inelastic contributions in
the kinetic equation, respectively, only once the linear terms have
been subtracted as is done in  \eqref{ch7_eq:H-time-ev-2}, see
Appendix~\ref{app_gran-dHdt-general} for details.

Despite the above discussion, global stability of the steady state can
be established without proving an $H$-theorem. The key point is the
following: the long time limit of $dH/dt$ is nonpositive and thus $H$
has a finite limit, since it is bounded from below. Therefore, $dH/dt$
tends to zero in the long time limit and it can be shown that this is
only the case if
$P_1(v;x,\infty)\equiv\lim_{t\to\infty}P_1(v;x,t)=P^{(\st)}_1(v;x)$. 

The average velocity $u(x,t)$ satisfies a diffusive equation
\eqref{ch7_eq:hydroMMu}, and thus it irreversibly tends to the steady
profile corresponding to the given boundary conditions in the long
time limit. Therefore,
$u(x,\infty)\equiv\lim_{t\to\infty}u(x,t)=u_{\st}(x)$ and taking into
account  \eqref{ch7_eq:H-time-ev-inel},
\begin{equation}\label{ch7_eq:lim-long-time-dH/dt-2}
\lim_{t\to\infty}\left.\frac{dH}{dt}\right|_{\text{inel}}=0
\Rightarrow \lim_{t\to\infty}\frac{dH}{dt}\leq 0.
\end{equation}
Since $H[P_1]$ is bounded from below, the only possibility is
\begin{equation}\label{ch7_eq:lim-long-time-dH/dt}
\lim_{t\to\infty}\frac{dH}{dt}=0,
\end{equation}
and all the contributions to $dH/dt$ in
\eqref{ch7_eq:H-time-ev-diff}-\eqref{ch7_eq:H-time-ev-noise} vanish in
the long time limit. The vanishing of  \eqref{ch7_eq:H-time-ev-diff}
imposes that 
\begin{equation} \label{ch7_eq:phi_stab}
P_1(v;x,\infty)=P^{(\st)}_1(v;x)\phi(v;t),
\end{equation}
where $\phi(v;t)$ is
an arbitrary function of $v$ and $t$. 
For $\xi\neq 0$,
 \eqref{ch7_eq:H-time-ev-noise} implies that $\phi(v;t)$ must be a function depending only of time, independent of $v$, and normalization yields
$\phi(v;t)=1$. For $\xi=0$,  \eqref{ch7_eq:H-time-ev-noise} identically
vanishes but it can be also shown that $\phi(v;t)=1$, see appendix 
\ref{app_final_glob_stab} for details. Thus, for
arbitrary $\xi$, including $\xi=0$, we have that
\begin{equation}\label{ch7_eq:global-stable}
P_1(v;x,\infty)=P^{(\st)}_1(v;x).
\end{equation}
The steady distribution $P^{(\st)}_1(v;x)$ is
globally stable. Each time evolution $P_1(v;x,t)$, corresponding to
a given initial condition, tends to $P^{(\st)}_1(v;x)$ in the long time limit.

\section{Why cannot Boltzmann's $H_B$ be the ``good'' Lyapunov functional?}\label{sec:inadequacy}

Here we prove that Boltzmann's $H_{B}[P_1]$,  given by \eqref{H-Boltzmann},
cannot be used to build a
Lyapunov functional for intrinsically dissipative systems, in
agreement with the numerical results by Marconi et
al.~\cite{MPyV13}. Not only does our proof hold for the
simplified models considered here, but for a general kinetic equation
in which energy is not conserved in collisions, such as the inelastic
Boltzmann or Enskog equations. To keep the notation simple, we still
write $\tder{P_1}=\mathcal{L}P_1$, but now $\mathcal{L}$ stands for the
evolution operator in the considered kinetic description, which is
nonlinear in general.

First, we restrict ourselves to homogeneous
situations and thus drop the integral over $x$,
\begin{subequations}
\begin{align}
H_{B}[P_1]=&\int\!\! dv P_1 \ln P_1, \\
\frac{dH_{B}}{dt}=&\int\!\! dv\, \tder{P_1}
\ln P_1=\int\!\! dv\, \left( \mathcal{L}P_1\right) \,\ln P_1,
\label{ch7_eq:HB-time-ev-1}
\end{align} 
\end{subequations}
 Also, we consider a
system that is initially close to the steady state, such that we can
expand everything in powers of $\Delta P_1=P_1-P^{(\st)}_1$. Then,
\begin{equation}
\mathcal{L} P_1\equiv \mathcal{L}(P^{(\st)}_1+\Delta
P_1)=\cancelto{0}{\mathcal{L}P^{(\st)}_1}+\mathcal{L}_{\text{lin}}\Delta
P_1+O(\Delta P_1)^{2},
\end{equation}
in which $\mathcal{L}_{\text{lin}}$ is the linearized evolution
operator. Neglecting $O(\Delta P_1)^{2}$ terms, the linear approximation
arises,
\begin{align}
\left.\frac{dH_{B}}{dt}\right|_{\text{lin}}=\int\!\! dv\,
  (\mathcal{L}_{\text{lin}}\Delta P_1) \,\ln
  P^{(\st)}_1=\left.\frac{d}{dt}\la\ln P^{(\st)}_1\ra\right|_{\text{lin}}.
\label{ch7_eq:HB-time-ev-linear}
\end{align} 
On the one hand, the linear contribution vanishes in the elastic case:
$\ln P^{(\st)}_1$ is a sum of constants of motion, which are unchanged by
the linearized kinetic operator. Then, $H_{B}$ can be a candidate for
a Lyapunov functional. On the other hand, only mass and linear
momentum are conserved for nonconservative interactions.  Thus, no
longer is $\ln P^{(\st)}_1$ a sum of conserved quantities, and
\begin{align}
\left.\frac{dH_{B}}{dt}\right|_{\text{lin}}\neq 0.
\label{ch7_eq:HB-time-ev-linear-2}
\end{align} 
Therefore, by changing the initial sign of $\Delta P_1=P_1-P^{(\st)}_1$, which
can always be done, the initial sign of $dH_{B}/dt$ is reversed and
$H_{B}$ cannot be a Lyapunov functional. 

\begin{figure}
\centering
  \includegraphics[width=0.8 \textwidth]{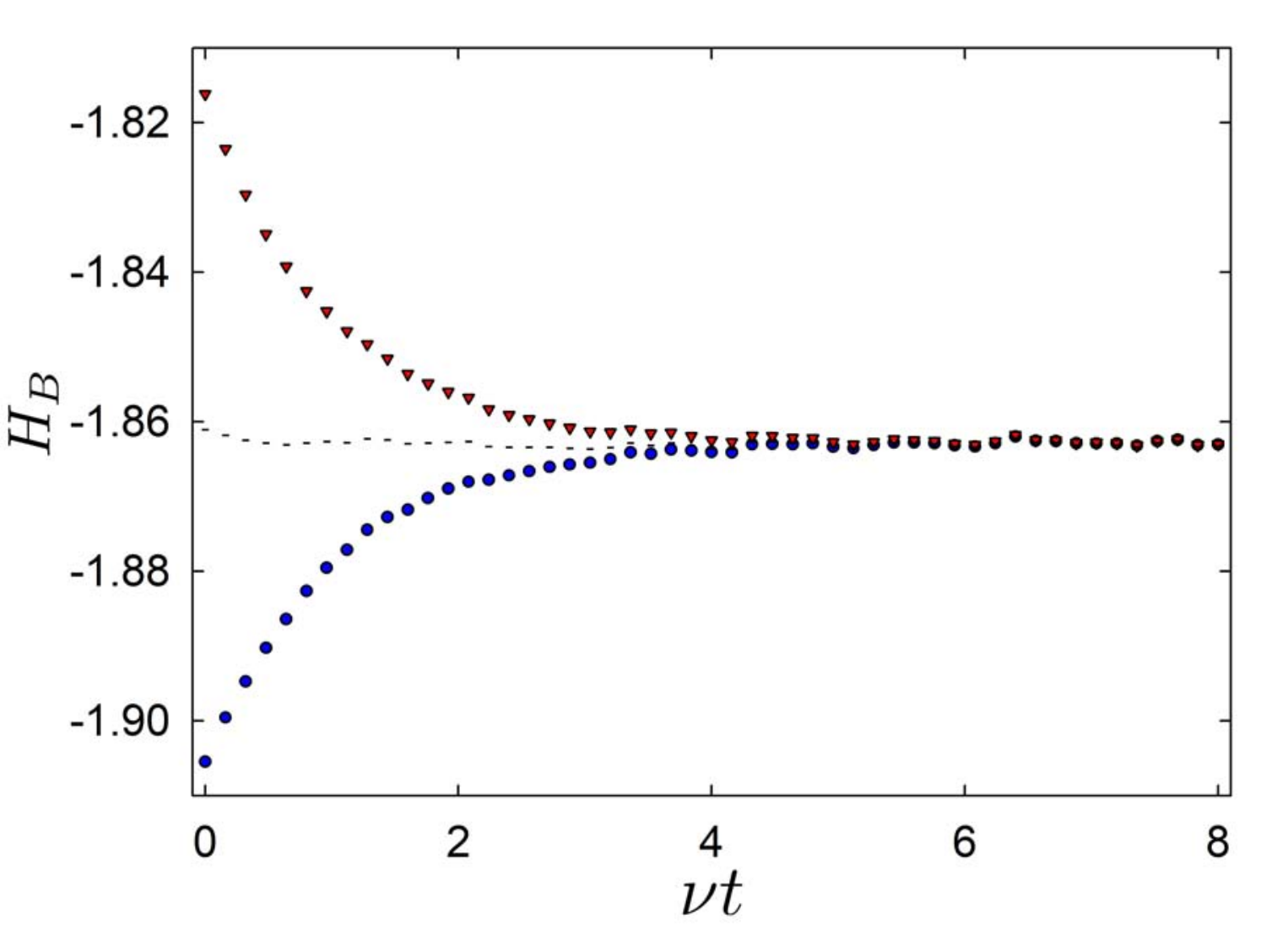}
  \caption{\label{fig:HBreversed} Time evolution of the functional
    $H_{B}$.  Specifically, we consider two different initial
    conditions in a uniformly heated system.  Both simulations start
    from a Gaussian distribution with a homogeneous temperature, which
    slightly differs from the stationary one: (i)
    $T(t=0)=1.1 \, T_{\st}$ (blue circles) and (ii)
    $T(t=0)=0.9 \, T_{\st}$ (red triangles).  Both functionals are
    symmetric with respect to its stationary value in agreement with
    the prediction of the linear approximation.  Consistently, the
    mean value of both curves (dashed line) remains approximately
    constant throughout. We have considered a system with parameters
    $\nu=20$ and $\xi=50$, size $N=330$, and averaged over $3000$
    trajectories.}
\end{figure}

In figure.~\ref{fig:HBreversed}, we show the evolution of $H_{B}$ in our
kinetic model. We consider a uniformly heated system, so that the
system remain homogeneous for all times, as described in
section \ref{NESS}. Two different initial conditions are considered,
corresponding to Gaussian distributions with zero average velocity but
nonsteady values of the temperature, specifically $1.1 \, T_{\st}$ and
$0.9 \, T_{\st}$. We can see how, in agreement with our discussion, not
only is one of the functionals increasing, but also it can be obtained
as the mirror image of the decreasing one through the stationary
value. Simulations, as have been repeatedly stated throughout the thesis, are performed following the recipe given in Appendix
\ref{ch5-6_app-b}.

Taking into account the specific (Gaussian) shape of the steady PDF
for the uniformly heated system, as given by
 \eqref{ch7_eq:P1-Unif-heated}, the time derivative of $H_{B}$ in
 \eqref{ch7_eq:HB-time-ev-linear} reduces to
\begin{equation}
  \label{ch7_eq:dHB/dt-Unif-Heated}
\frac{dH_{B}}{dt}=-\frac{1}{2T_{\st}}\frac{d\langle v^{2}\rangle}{dt}.
\end{equation}
Since the plots in figure~\ref{fig:HBreversed} correspond to
evolutions of the system for which $u(x,t)\equiv 0$ for all times,
therein $\langle v^{2}\rangle =T$ and, consistently, the $H_{B}$-curve
corresponding to an initial value of the temperature that is higher
(lower) than the steady one monotonically increases
(decreases).

The above picture is consistent with the situation found in
\cite{BCDVTyW06}, in which the uniformly heated granular gas described
by the inelastic Boltzmann equation was investigated within the first
Sonine approximation. Therein, the entropy production was shown to
have linear terms in the deviations of the temperature and the excess
kurtosis. Also, our result is consistent with the numerical results in
\cite{MPyV13} for several collision models. In particular, observe the
similarity with the situation reported in panels A and B of figure
\ref{ch1_fig:MPyV13} of the introduction to this thesis. Note that our
argument also proves why $H_{B}$ cannot be nonincreasing for an
elastic system immersed in a heat bath at a temperature different from
the initial temperature of the gas, as also observed in
\cite{MPyV13}. Although $\ln P^{(\st)}_1$ is conserved in collisions,
the evolution operator includes a term coming from the interaction
with the bath that does not conserve the kinetic energy, and again
$dH/dt|_{\text{lin}}\neq 0$, making it impossible for $H_{B}$ to be a
``good'' Lyapunov functional.

\begin{figure}
\centering
  \includegraphics[width=0.8 \textwidth]{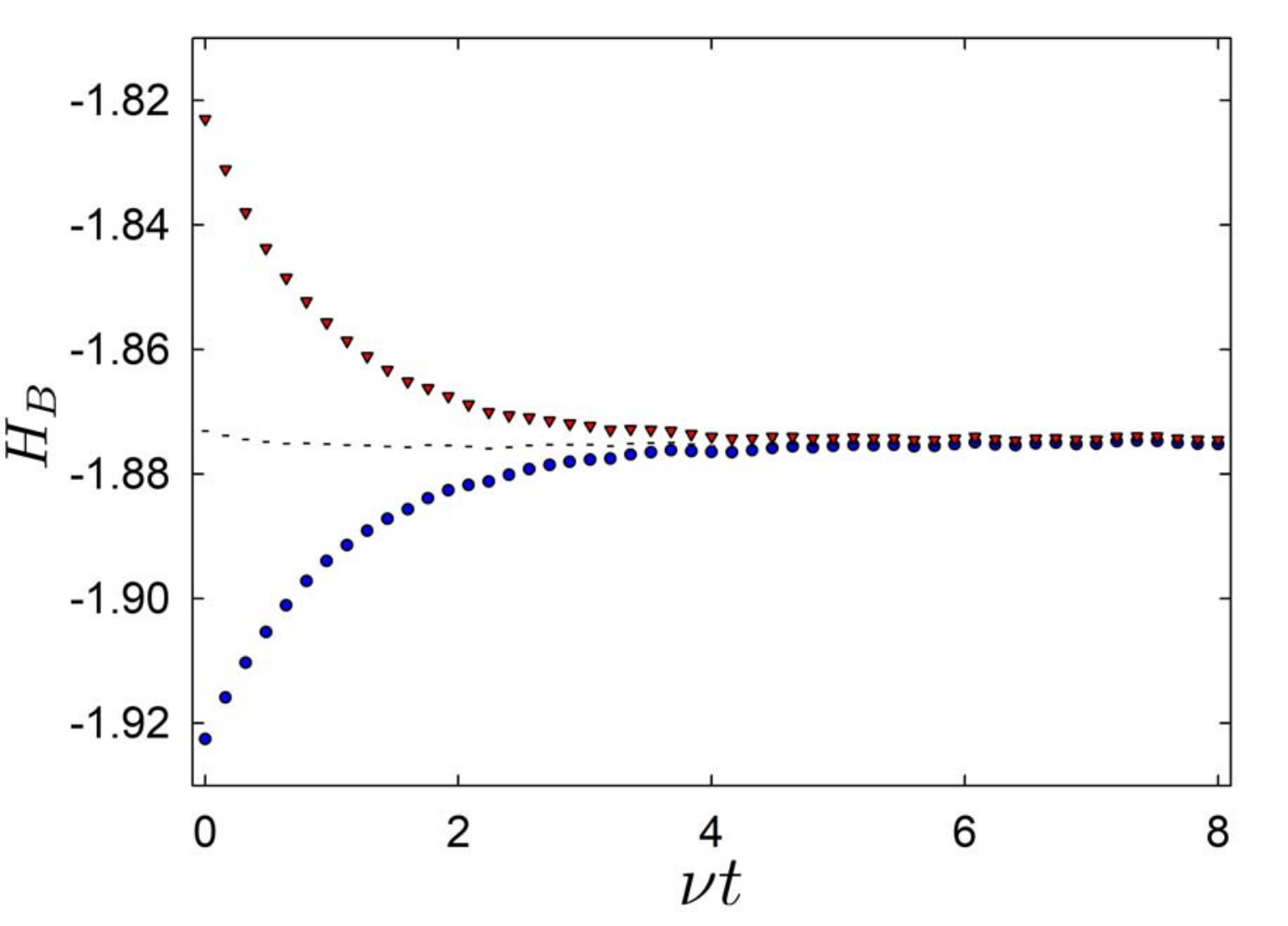}
  \caption{\label{fig:HBreversedv2} Time evolution of the functional $H_{B}$. 
  Analogously to figure \ref{fig:HBreversed}, we consider two different initial 
  conditions, but now for a sheared system.
    Both simulations start from a Gaussian distribution with the stationary
    average velocity profile but
    a homogeneous temperature, which slightly differs from the stationary
    one: (i) $T(t=0)=1.1 \, T_{\st}$ (blue circles) and (ii)
    $T(t=0)=0.9 \, T_{\st}$ (red triangles).     
    We have considered a system with parameters $\nu=20$ and 
    $a=5$, size $N=660$, and averaged over $6000$ trajectories.}
\end{figure}

In spatially inhomogeneous situations, the main difference is
the additional integral over space, both in $H_{B}$ and,
consequently, $dH_{B}/dt$. There is no reason to expect this integral
over space to make $dH/dt|_{\lin}$ vanish, since one still has that
\begin{equation}
\left.\frac{dH_{B}}{dt}\right|_{\lin}=\left.\frac{d}{dt}\la\ln P^{(\st)}_1\ra\right|_{\lin},
\end{equation}
and, in general, $\ln P^{(\st)}_1$ is not a sum of constants of motion. In
fact, again the sign of $\left.dH_{B}/dt\right|_{\lin}$ is reversed
when $\Delta P_1\to-\Delta P_1$, similarly to the homogeneous case. In figure 
\ref{fig:HBreversedv2}, we
have numerically checked this prediction for the sheared system, with
the resulting evolution of $H_{B}$ being completely similar to that
for the uniformly heated case in figure \ref{fig:HBreversed}.

\section{$H$-theorem for some specific NESS}
\label{USF}

Here we prove that the functional $H[f]$ is monotonically decreasing
for all times in some specific physical situations. Our proof applies
both to the sheared and the uniformly heated systems described in
section \ref{NESS}. To be as general as possible, we consider a system
that is both heated and sheared: $a\neq 0$ and
$\xi\neq 0$.  In this situation, the boundary conditions for the
PDF are given by \eqref{ch5_eq:P1-bc-USF}, which lead to the Lees-Edwards boundary 
conditions \eqref{ch5_eq:lees-edwards} for the averages
$u(x,t)$ and $T(x,t)$.

The steady solution of the hydrodynamic equations is
\begin{equation}\label{ch7_eq:USF-st-hydrovar}
u_{\st}(x)=a\left(x-\frac 1 2\right), \qquad T_{\st}=\frac{2a^{2} + \xi}{\nu}.
\end{equation}
On the one hand, the average velocity has a linear profile, similarly
to the situation in the USF state. On the other hand, the  temperature
remains homogeneous but its steady value has two contributions, one
coming from the shear and the other from the stochastic thermostat. The
viscous heating $2(\xder{u})^{2}$ and uniform heating $\xi$ terms
cancel the cooling term $-\nu T$ for all $x$.  The stationary solution
of the kinetic equation is the Gaussian distribution \eqref{USF-P1} corresponding to the hydrodynamic
fields in \eqref{ch7_eq:USF-st-hydrovar}.  Of course, the USF state
and NESS of the uniformly heated system in section \ref{NESS} can be
easily recovered as particular cases of \eqref{USF-P1}:
for $(a \neq 0,\xi=0)$ and $(a=0,\xi >0$), respectively.

Then, we turn now to the question of the existence of an $H$-theorem,
that is, the existence of a nonequilibrium entropy ensuring the
monotonic approach of the one-particle PDF to the steady state. Our
starting point is the following expansion of the one-particle PDF in
Hermite polynomials, 
\begin{equation}
P_1(v;x,t)=\frac{1}{\sqrt{2\pi
            T(x,t)}}\exp\left\{-\frac{[v-u(x,t)]^{2}}{2T(x,t)}\right\}
 \left\{1+\sum_{n=3}^{\infty}
\gamma_{n}(x,t)\,H_{n}\!\!\left[\frac{v-u(x,t)}{\sqrt{T(x,t)}}\right]\right\},
\label{ch7_eq:USF-Hermite-expansion}
\end{equation}
which is known as the Gram-Charlier series
\cite{Ch90,Ch05,Ed05,Wa58}. Therein, $u(x,t)$ and $T(x,t)$ are
the (exact) average velocity and temperature stemming from the
hydrodynamic equations for the considered distribution. Note that the argument of the Hermite polynomials is the scaled velocity $c$ defined in \eqref{P1-scaled-num}. The above
expansion is suggested by the Gaussian shape of the stationary PDF in
 \eqref{USF-P1}.
Making use of the definition \eqref{P1-scaled-num} of the scaled one-particle distribution function $\varphi(c;x,t)$ and the orthogonality relation of the Hermite polynomials
\cite{AyS72}, it is readily obtained that
\begin{equation}
\varphi(c;x,t)=\frac{1}{\sqrt{2\pi
            }}\exp\left(-\frac{c^{2}}{2}\right)
 \left[1+\sum_{n=3}^{\infty}
\gamma_{n}(x,t)\,H_{n}(c)\right],
\label{ch7_eq:USF-Hermite-expansion-phi}
\end{equation}
with
\begin{equation}\label{ch7_eq:kappa-n}
\gamma_{n}(x,t)=\frac{1}{n!}\int\!\! dc\, H_{n}(c) \varphi(c;x,t).
\end{equation}
Also, we could have written $\gamma_n$ as a combination of moments of
the distribution.

Some comments on the Gram-Charlier expansion are pertinent. First,
note that $n\geq 3$ in the sum: $\gamma_{1}=\gamma_{2}=0$ because the
zero-th order Gaussian contribution exactly gives the first two
moments $\la v \ra(x,t)=u(x,t)$ and $\la v^{2}\ra(x,t)=u^{2}(x,t)+T(x,t)$. Second, if
$P_1(v;x,t)$ were symmetric with respect to $v=u$, that is,
$\la(v-u)^{2n+1}\ra=0$ for all $n\in\mathbb{N}$, only even values of
$n$ would be present in the sum and one recovers the usual
expansion in Sonine-Laguerre polynomials of kinetic theory.  Finally,
it is worth stressing that the series \eqref{ch7_eq:USF-Hermite-expansion}
converges for functions such that the tails of $\varphi(c;x,t)$
approach zero faster than $e^{-c^2/4}$  for
$c\to\pm\infty$ \cite{Cr25,Sz39,Wa58}.

Now we substitute the
Gaussian stationary solution \eqref{USF-P1} and the Gram-Charlier 
series \eqref{ch7_eq:USF-Hermite-expansion} into the three
contributions to $dH/dt$, given by \eqref{ch7_eq:H-time-ev-diff},
\eqref{ch7_eq:H-time-ev-inel} and \eqref{ch7_eq:H-time-ev-noise}. 
For the inelastic term, it is readily obtained that
\begin{equation}\label{ch7_eq:H-terms-1}
\left.\frac{dH}{dt}\right|_{\inel} = \frac{\nu}{2 T_{\st}}\int\!\! dx
\left(u-u_{\st}\right)^2.
\end{equation}
For the diffusive and noise terms, the key ideas are a change of the
integration variable from $v$ to $c=(v-u)/\sqrt{T}$ and the use
of the recursion relations and the orthogonality property of the
Hermite polynomials \cite{AyS72}. Working along
these guidelines, we arrive at
\begin{align}
\left.\frac{dH}{dt}\right|_{\diff} =& -\int\!\! dx\, T \left( \frac{u^{\prime}}{T}-\frac{u_{\st}^{\prime}}{T_{\st}}\right)^{2} -\frac{u_{\st}^{\prime 2}}{T_{\st}^2}\int\!\! dx
\left(u-u_{\st}\right)^2\nonumber 
\\
&  -\frac{1}{\sqrt{2\pi}}\int\!\! dx\, dc \, \frac{e^{-c^{2}/2}}{1+\sum_{n=3}^{\infty} \gamma_{n} H_{n}(c)}
\left\{
  \frac{T^{\prime}}{2T}H_{2}(c)+\sum_{n=3}^{\infty}\gamma_{n}^{\prime}H_{n}(c) \right. \nonumber \\
& \qquad \qquad \left. -\sum_{n=3}^{\infty}\frac{\gamma_{n} u^{\prime}}{\sqrt{T}}n H_{n-1}(c) +\sum_{n=3}^{\infty}\frac{\gamma_{n}T^{\prime}}{2T}\left[H_{n+2}(c)+nH_{n}(c)\right] \right\}^2, \label{ch7_eq:H-terms-2-a}
\\
\left.\frac{dH}{dt}\right|_{\noise} =& - \frac{\xi}{2}\int\!\! dx\, T \left( \frac{1}{T}-\frac{1}{T_{\st}}\right)^{2} -\frac{\xi}{2T_{\st}^2}\int\!\! dx
\left(u-u_{\st}\right)^2 \nonumber \\
&-\frac{\xi}{2\sqrt{2\pi}}\int\!\! dx\, dc \, \frac{e^{-c^{2}/2}}{1+\sum_{n=3}^{\infty} \gamma_{n} H_{n}(c)} \frac{1}{T}
\left[  \sum_{n=3}^{\infty} \gamma_{n} n H_{n-1}(c) \right]^2, \label{ch7_eq:H-terms-2-b}
\end{align}
where the prime stands for spatial derivative. In \eqref{ch7_eq:H-terms-1}, \eqref{ch7_eq:H-terms-2-a} and \eqref{ch7_eq:H-terms-2-b}  
there are several terms multiplying
$\int\!\! dx \left(u-u_{\st}\right)^2$: they cancel out when we take into account
the equation for the (spatially homogeneous) stationary
temperature. Therefore, the sum of the remaining terms leads right to
\begin{equation}
\frac{dH}{dt}=A(t)+B(t), \quad \text{with both } A(t),B(t)\leq 0,
\end{equation}
being
\begin{equation}\label{ch7_eq:USF-A(t)}
A(t)=-\int\!\! dx\, T \left[ \left( \frac{u^{\prime}}{T}-\frac{u_{\st}^{\prime}}{T_{\st}}\right)^{2}
+ \frac{\xi}{2} \left( \frac{1}{T}-\frac{1}{T_{\st}}\right)^{2} \right],
\end{equation}
and $B(t)$ the sum of the second and third line in \eqref{ch7_eq:H-terms-2-a} with the second line in \eqref{ch7_eq:H-terms-2-b}, that is,
\begin{align}
B(t)= &  -\frac{1}{\sqrt{2\pi}}\int\!\! dx\, dc \, \frac{e^{-c^{2}/2}}{1+\sum_{n=3}^{\infty} \gamma_{n} H_{n}(c)} \left\{ \frac{\xi}{2T}
\left[  \sum_{n=3}^{\infty} \gamma_{n} n H_{n-1}(c) \right]^2 + \left[
  \frac{T^{\prime}}{2T}H_{2}(c) \right. \right.
 \nonumber \\
&  \qquad +\sum_{n=3}^{\infty}\gamma_{n}^{\prime}H_{n}(c) - \left.  \left. \sum_{n=3}^{\infty}\frac{\gamma_{n} u^{\prime}}{\sqrt{T}}n H_{n-1}(c) +\sum_{n=3}^{\infty}\frac{\gamma_{n}T^{\prime}}{2T}\left\{H_{n+2}(c)+nH_{n}(c)\right\} \right]^2 \right\}, 
\end{align}

In conclusion, $dH/dt\leq 0$ for all times and we have shown that the
$H$-theorem holds for the sheared and heated system. Rigorously, our
proof holds for those PDFs such that the above Hermite expansion
converges. Note that the proof remains valid for the approach to any
NESS, whose PDF is a Gaussian with a homogeneous temperature,
independently of the corresponding boundary conditions.  In
section \ref{global-stability}, we have already demonstrated that $dH/dt$
only vanishes for $P_1(v;x,\infty)=P^{(\st)}_1(v;x)$, but the same result
can be rederived here in a different way. By imposing that both $A(t)$
and $B(t)$ vanish in the long time limit and making use of the
hydrodynamic equations for the averages, it can be shown that
$u(x,\infty)=u_{\st}(x)$, $T(x,\infty)=T_{\st}$ and
$\gamma_{n}(x,\infty)=0$, $\forall n\geq
3$.

\subsection{USF state: simulations}

Here we consider the sheared system  in order to check numerically our
theoretical predictions. Throughout this section, we employ the values of
the parameters $\nu=20$ and $a=5$. Note that $\xi=0$, since there is no stochastic forcing.

\begin{figure}
\centering
  \includegraphics[width=0.8 \textwidth]{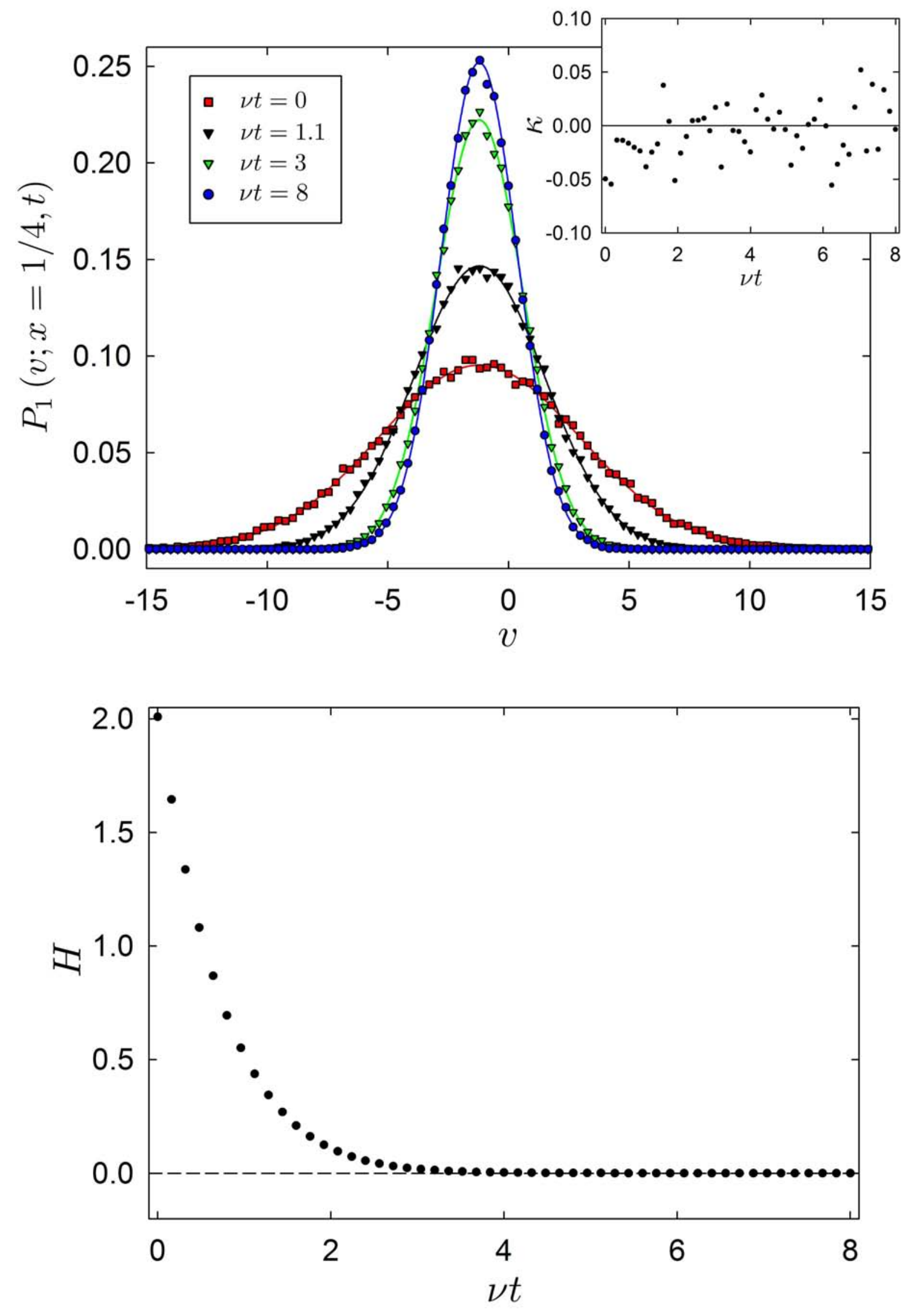}
  \caption{\label{fig:Gauss_hom} Relaxation towards the USF state.
    The initial distribution is Gaussian, with average velocity in
    $u_{\st}(x)$ and temperature $T=7 \, T_{\st}$. (Top) Velocity
    distribution function at $x=1/4$. Four different times are shown,
    as labeled.  In the inset, we present the evolution of the excess
    kurtosis.  Solid lines correspond to the (theoretical) Gaussian
    distributions for the plotted times, except for the longest in
    which it represents the theoretical steady distribution. (Bottom)
    Relaxation of the $H$ functional, which is clearly monotonically
    decreasing to zero.  System size is $N=660$, parameters are
    $\nu=20$ and $a=5$, and curves are averaged over $6000$ runs.  }
\end{figure}

First, in figure~\ref{fig:Gauss_hom}, we show the evolution of the
distribution and the $H$-functional from a Gaussian initial condition
with the steady velocity profile $u(x,0)=u_{\st}(x)$ but a higher
temperature, $T(t=0)=7 \, T_{\st}$. In the top panel, we depict the
velocity distribution at $x=1/4$ for several times. All of them are
Gaussian, which agrees with the theoretical prediction of the kinetic
equation: when the initial velocity profile coincides with the steady
one and only the temperature is perturbed, an initially Gaussian PDF
remains Gaussian for all times. Indeed, we can see in the inset how
the excess kurtosis,
\begin{equation}
\kappa = \langle [v-u(x)]^4 \rangle / \langle [v-u(x)]^2
\rangle^{2}-3,
\end{equation}
only fluctuates around zero at the considered position,
consistently with the Gaussian shape. In
the bottom panel, it is neatly observed that the $H$-functional 
monotonically decreases in time.

\begin{figure}
\centering
  \includegraphics[width=0.8 \textwidth]{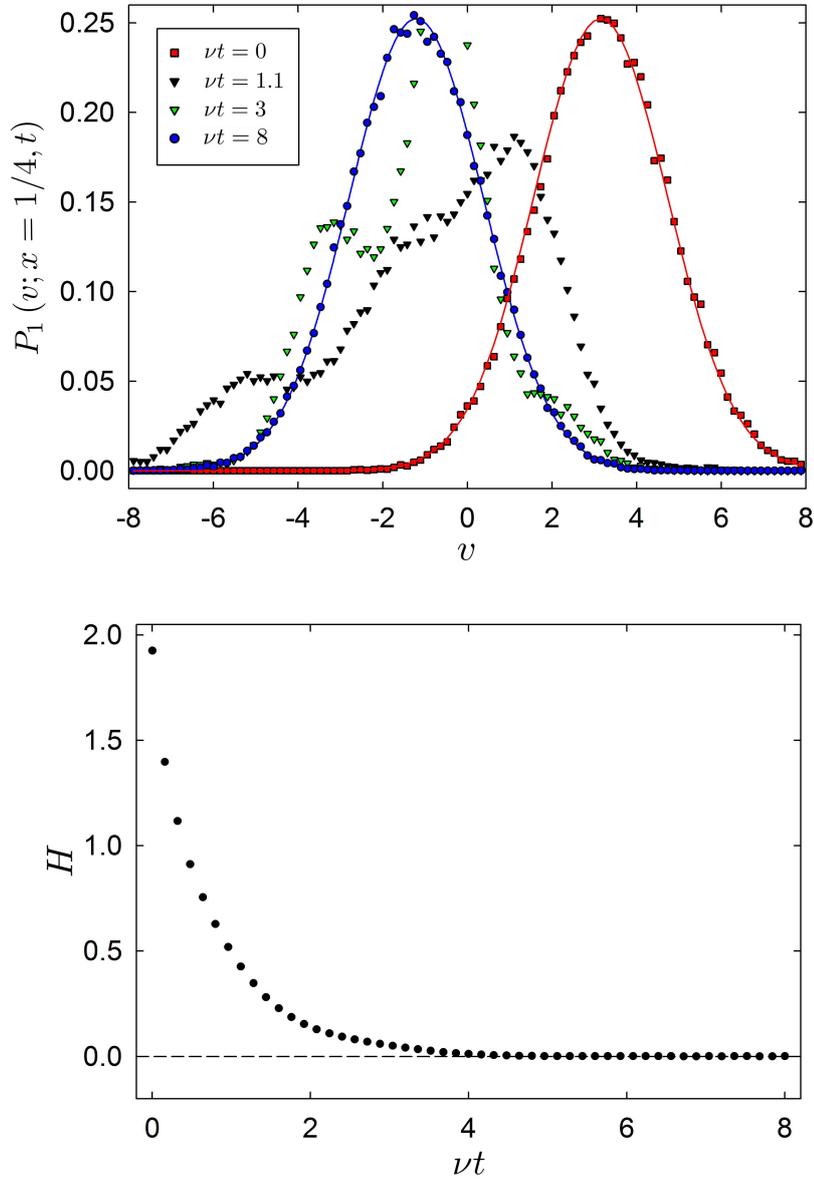}
  \caption{\label{fig:Gauss_Du} The same plots as in
    figure~\ref{fig:Gauss_hom}, but starting from a different initial
    condition. Now, the initial PDF is a Gaussian with average velocity
    $u(x,0)=u_{\st}(x)+ 4.4 \, \sin (2 \pi x)$ and temperature
    $T(t=0)= T_{\st}$. In the top panel, solid lines correspond to
    the theoretical PDFs for the initial time and the steady state. In
    the bottom panel, $H$ decreases again monotonically towards its
    steady value, consistently with our theory.}
\end{figure}

Second, we look into the relaxation to the USF state from another
initial preparation, for which the velocity profile $u(x,0)$ is
different from the stationary but $T(x,0)=T_{\st}$. The numerical
results are shown in figure~\ref{fig:Gauss_Du}, and for the sake of
simplicity we use again an initial Gaussian
distribution. Specifically, we have that
$u(x,0)=u_{\st}(x)+ 4.4 \, \sin (2 \pi x)$. Here, the departure from
the Gaussian shape is evident, and thus we have not plotted the
kurtosis. Consistently with our theoretical prediction, we get again a
monotonous relaxation of $H$ towards its null stationary value.

\begin{figure}
\centering
  \includegraphics[width=0.8 \textwidth]{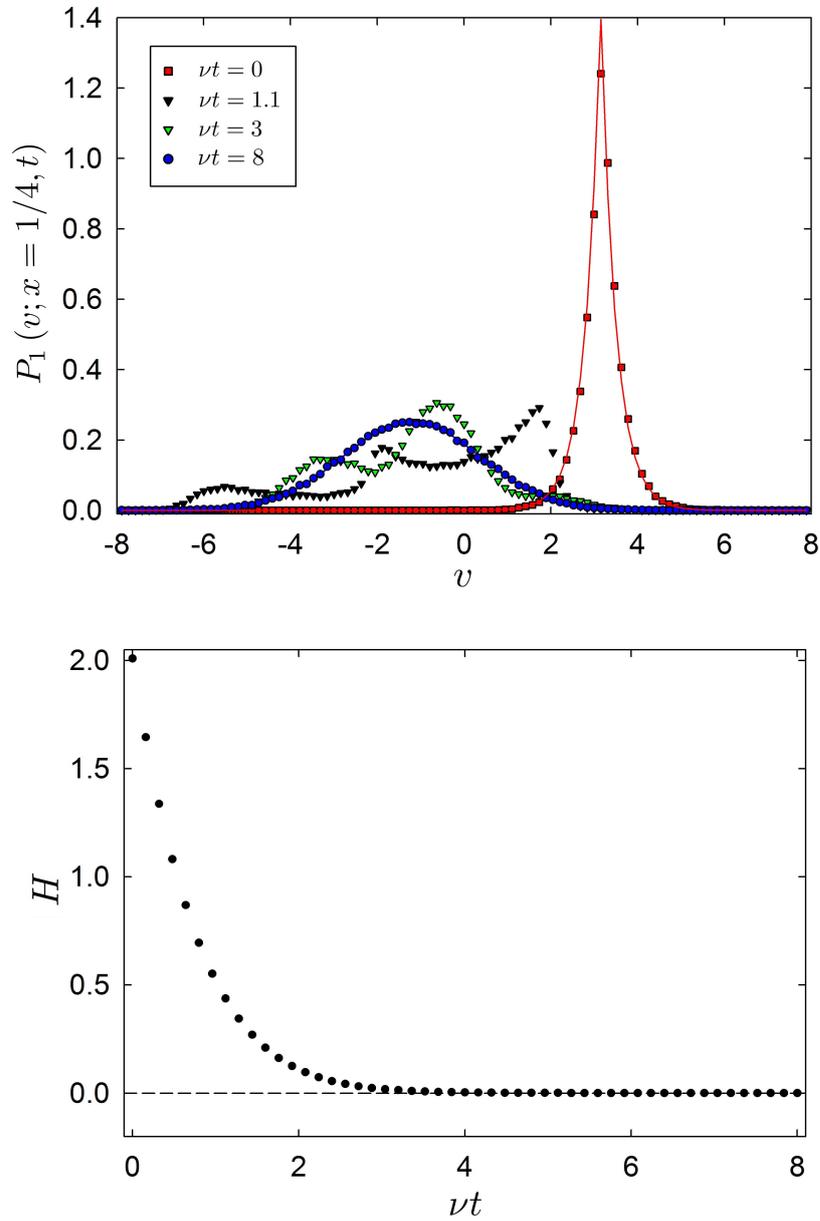}
  \caption{\label{fig:exp} The same plots as in
    figure~\ref{fig:Gauss_hom}, but starting from an initial PDF with a
    divergent Gram-Charlier series. Specifically, they correspond
    to an exponential initial distribution with average velocity 
    $u(x,t=0)=u_{\st}(x)+ 4.4 \, \sin (2 \pi x)$ and temperature
    $T(t=0)=0.1 \,T_{\st}$.  }
\end{figure}

Finally, we consider situations for which the above presented proof is
not rigorously applicable. As stated before, the Gram-Charlier series
does not converge when the tails of the distribution decay to zero
slower than the square root of the Gaussian. Nevertheless, when all
the coefficients $\gamma_{n}$ defined in \eqref{ch7_eq:kappa-n} exist
and are finite, we still expect the $H$-theorem to hold. We illustrate
this situation with an initial exponential distribution; specifically,
we consider
\begin{equation}\label{ch7_eq:exp-PDF}
P_1(v;x,0) = \frac{1}{\sqrt{2 T(t=0)}} \exp \left[ \frac{ \sqrt{2} \left| v-u(x,t=0) \right| }{\sqrt{T(t=0)}} \right],
\end{equation}
with $u(x,t=0)= u_{\st}(x) + 4.4 \sin (2 \pi x)$ and
$T(t=0)=0.1 \,T_{\st}$.  In agreement with our expectation, 
the $H$-functional also
monotonically decreases in figure \ref{fig:exp}.

\subsection{Numerical results in the uniformly heated system}

\begin{figure}
\centering
  \includegraphics[width=0.8 \textwidth]{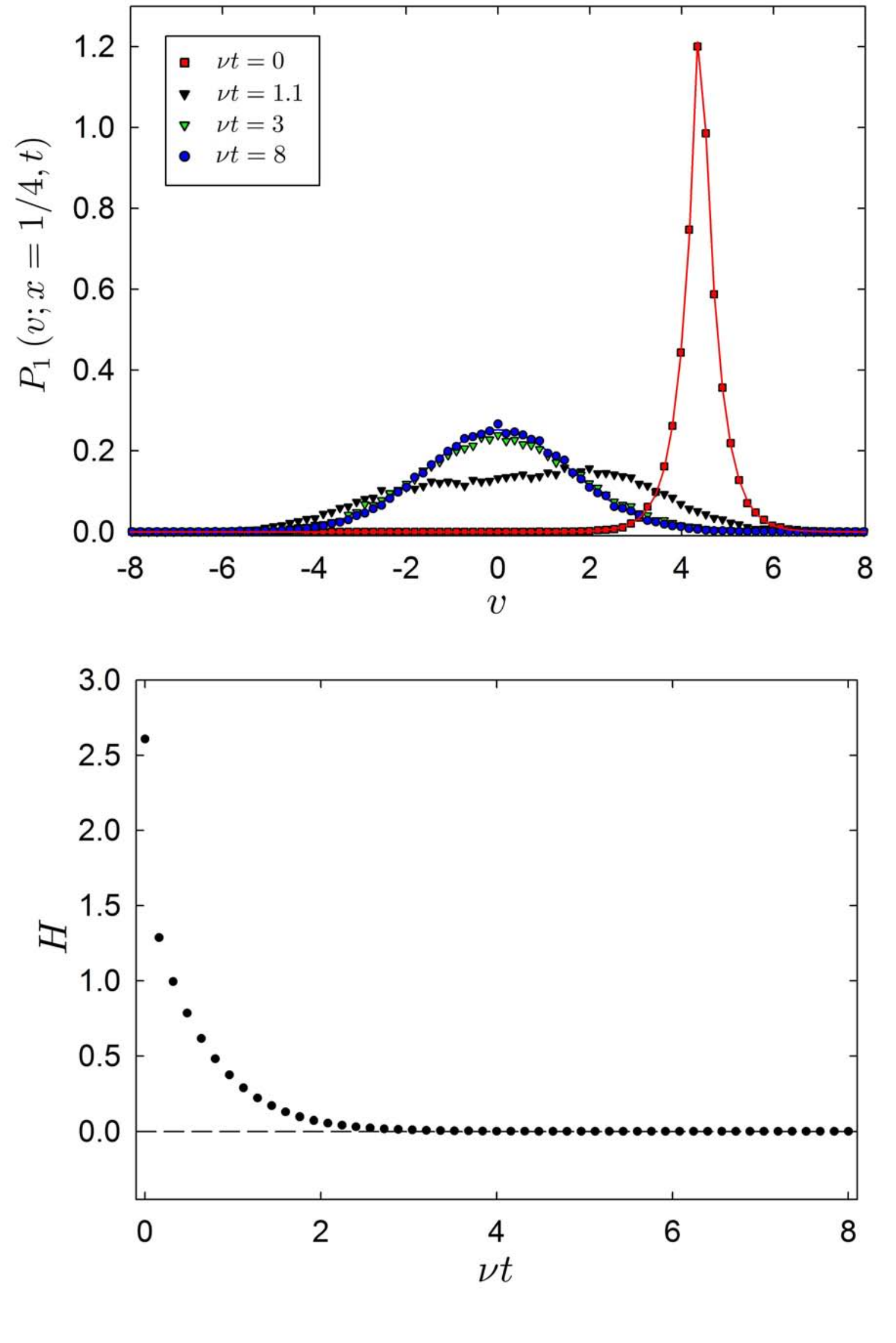}
  \caption{\label{fig:exp-heating} Numerical results for the uniformly
    heated system.  Similar to what we did in 
    figure~\ref{fig:Gauss_hom}, we show the time evolution
    of the PDF (top) and the time evolution of the
    $H$-functional (bottom). The system is initially prepared with an
    exponential PDF with average velocity  $u(x,t=0)= 4.4 \, \sin (2 \pi x)$ and
    temperature $T(t=0)=0.1 \,T_{\st}$.  System size is $N=330$
     and the curves have been averaged over $3000$ runs.  }
\end{figure}

To conclude, we put forward the results of simulations for the
uniformly heated system. Specifically, our simulations have been done
for $\nu=20$, $a=0$ (no shear) and $\xi=50$. In order not to be repetitive,
 we only present the more complex
case in figure~\ref{fig:exp-heating}: the relaxation towards the steady
state from an initial exponential distribution, as given by
 \eqref{ch7_eq:exp-PDF}. In particular, we consider that
$u(x,t=0)= 4.4 \, \sin (2 \pi x)$ and $T(t=0)=0.1 \, T_{\st}$. Note
that the perturbation from the steady values is the same as in
figure~\ref{fig:exp} for the sheared case. Once more, we observe the
monotonic relaxation of $H$ towards its stationary value, in neat agreement
with our theoretical result, even for a initial distribution for which
the Gram-Charlier series does not converge.

\chapter{Memory effects in athermal systems}
\label{ch:Kovacs_pgran}
\newcommand{\ness}{\text{s}}
\newcommand{\calP}{\mathcal{P}}
\newcommand{\calY}{\mathcal{Y}}
\newcommand{\bx}{\bm{x}}
\newcommand{\bxpr}{\bm{x}^{\prime}}
\newcommand{\bxa}{\bm{x}_{\alpha}}
\newcommand{\bxapr}{\bm{x}_{\alpha^{\prime}}}
\newcommand{\bbW}{\mathbb{W}}

The objectives of this chapter are two-fold. First, we present a general, rigorous, 
derivation of the linear response expression for the Kovacs hump, which was 
described in the introduction, specifically in section \ref{ch1_sec:kovacs}. 
This is done for 
systems with a realistic continuous time dynamics, at both the mesoscopic and 
macroscopic levels of description. In the former, our starting point is  the master 
equation for the probability distribution, which is always linear. In the latter, 
we begin by considering the hierarchy of equations for the moments, which in 
general is nonlinear. The presented proof is valid for both molecular and athermal 
systems, since no hypothesis is needed with regard to the form of either the 
stationary probability distribution or the fluctuation-dissipation relation.

The organization of this chapter is as follows. 
To start with, we focus on the linear response regime in both the mesoscopic 
and macroscopic levels of description in sections \ref{ch8_ch:lin-th-1}
and \ref{ch8_ch:lin-th-2}, respectively. Afterwards, 
section  \ref{Section-model} is devoted
to the particularization of this linear theory to an even more generalized version
of our simple model of a granular gas. Here, we consider a  general
collision rate that makes it necessary to introduce the typical 
Sonine expansion of kinetic theory in order to close
the evolution equation of the temperature. Finally, we test our theoretical 
predictions by comparison with Monte Carlo simulations 
in section \ref{ch8_sec:num}.

\section{Linear theory for the Kovacs memory effect}\label{ch8_ch:lin-th}

\subsection{Master equation approach}\label{ch8_ch:lin-th-1}

First, we consider a general system, the state of which is completely
characterized by a vector $\bm{x}$ with $M$ components,
$\bm{x}=\{x_{1},x_{2},\ldots,x_{M}\}$. For example, in a
one-dimensional Ising chain of $N$ spins, $x_{i}=\sigma_{i}=\pm 1$, and
$M=N$; for a gas comprising $N$ particles with positions $\bm{r}_{i}$
and velocities $\bm{v}_{i}$, $M=6N$, and
$\bm{x}=\{\bm{r}_{1},\bm{v}_{1},\ldots,\bm{r}_{N},\bm{v}_{N}\}$. Henceforth,
 for the sake of simplicity,  a
notation suitable for systems in which the states can be labeled with
a discrete index $\alpha$, $1\leq\alpha\leq\Omega$, is employed. For~example, this
is the case of the Ising system above, where $\Omega=2^{N}$. The
generalization for a continuous index is straightforward, by changing
sums into integrals and Kronecker deltas by Dirac deltas
\cite{vK92}.

We assume that $\bm{x}$ is a
Markov process, and its dynamics at the mesoscopic level of description
is thus governed by a master equation for
the probabilities $P(\bxa,t)$, that is,
\begin{equation}\label{ch8_eq:master-eq}
\partial_{t}P(\bxa,t)=\sum_{\alpha^{\prime}}\left[W(\bxa|\bxapr;\xi)P(\bxapr,t)-W(\bxapr|\bxa;\xi)P(\bxa,t)\right].
\end{equation}
Therein, $W(\bxa|\bxapr;\xi)$ are the transition rates from state $\bxapr$
to state $\bxa$, with our notation marking their dependence on
some control parameter $\xi$. Equation \eqref{ch8_eq:master-eq} can be
formally rewritten as
\begin{equation}\label{ch8_eq:master-eq-2}
\partial_{t}\ket{\calP(t)}=\bbW(\xi)\ket{\calP(t)},
\end{equation}
in which $\ket{\mathcal{P}(t)}$ is a vector (column matrix) with 
 the probabilities $P(\bxa,t)$ as its components, and $\bbW(\xi)$ is the linear
operator (square matrix) that generates the dynamical evolution of
$\ket{\mathcal{P}(t)}$,
\begin{equation}\label{bbW}
\bbW(\bxa|\bxapr;\xi)=W(\bxa|\bxapr;\xi)-\delta_{\alpha,\alpha^{\prime}}
\sum_{\alpha^{\prime \prime}}W(\bm{x}_{\alpha^{\prime \prime}}|\bxa;\xi).
\end{equation}

Let us assume that the Markovian dynamics is ergodic---or irreducible
\cite{vK92}---, that is, all the states are
dynamically connected through a chain of transitions with nonzero
probability. Therefore, there is a unique steady solution of the
master equation $\ket{\calP_{\ness}(\xi)}$, which verifies
\begin{equation}\label{ch8_eq:steady-sol-P}
\bbW(\xi)\ket{\calP_{\ness}(\xi)}=0.
\end{equation}
Note that the stationary solution depends 
on the parameter $\xi$ controlling the system
dynamics. Ergodicity does not imply detailed balance, so there can be
nonzero currents in the steady state. Physically, this means
that in general the system approaches a NESS in the long time limit; equilibrium 
is reached only when the currents vanish.

Now, we consider the system evolving from a certain initial state at
time $t_{0}$, characterized by 
$\ket{\calP(t_{0})}$. The formal solution of the master equation is
\begin{equation}\label{ch8_eq:formal-sol}
\ket{\calP(t)-\calP_{\ness}(\xi)}=e^{(t-t_{0})\bbW(\xi)}\ket{\calP(t_{0})-\calP_{\ness}(\xi)}.
\end{equation}
This is the starting point for our derivation of the expression for
the Kovacs effect in the linear response approximation, which is
carried out below.

The time evolution of any physical property $Y$ is obtained right
away. The value of $Y$ for a given configuration $\bx$ is denoted
by $Y(\bx)$ and thus its average value is given by
\begin{equation}\label{ch8_eq:average-def}
\langle Y(t)\rangle=\sum_{\alpha}Y(\bxa)P(\bxa,t)=\braket{\calY}{\calP(t)}.
\end{equation}
Above $\ket{\calY}$ is a ket whose components are $Y(\bxa)$, and
$\bra{\calY}$ its corresponding bra (row matrix with the same
components).  Note that we are assuming that $Y$ is a real
  quantity for all the
  configurations. 
By substituting \eqref{ch8_eq:formal-sol} into \eqref{ch8_eq:average-def},
we get
\begin{equation}\label{ch8_eq:average-time-evol}
\Delta Y(t;\xi)\equiv\langle Y(t)\rangle-\langle Y\rangle_{\ness}(\xi)=\mel{\calY}{e^{(t-t_{0})\bbW(\xi)}}{\calP(t_{0})-\calP_{\ness}(\xi)},
\end{equation}
in which $\langle Y\rangle_{\ness}(\xi)$ is the average value at the
steady state corresponding to $\xi$.
 
Now, we investigate the relaxation of the system from the steady state
for $\xi_{0}=\xi+\Delta\xi$ to the steady state for $\xi$. This is done in
linear response, that is, $\Delta\xi$ is considered to be small and thus
all terms beyond those linear in $\Delta\xi$ are neglected. Hence, at $t=0$
we have that
\begin{equation}\label{ch8_eq:P-linear}
\ket{\calP(t=0)}=\ket{P_{\ness}(\xi+\Delta\xi)} =\ket{P_{\ness}(\xi)}+\Delta\xi \ket{\frac{dP_{\ness}(\xi)}{d\xi}}+\mathcal{O}(\Delta\xi)^{2}.
\end{equation}
Substitution of \eqref{ch8_eq:P-linear} into \eqref{ch8_eq:average-time-evol}
yields the formal expression for the relaxation of $Y$ in linear
response,
\begin{equation}\label{ch8_eq:delta-Y-linear}
\Delta Y(t;\xi)=\Delta \xi \mel{\calY}{e^{(t-t_{0})\bbW(\xi)}}{\frac{dP_{\ness}(\xi)}{d\xi}}.
\end{equation}
In order to have an order of unity function,
one may normalize the relaxation function,
\begin{equation}\label{ch8_eq:phi-Y-linear}
\phi_{Y}(t;\xi)\equiv\lim_{\Delta \xi\to 0}\frac{\Delta
 Y(t;\xi)}{\Delta\xi}=\mel{\calY}{e^{(t-t_{0})\bbW(\xi)}}{\frac{dP_{\ness}(\xi)}{d\xi}}.
\end{equation}
Sometimes, the relaxation function is further normalized by
considering $\phi_{Y}(t)/\phi_{Y}(t=t_{0})$---see~for instance
\cite{ByP93,PyB10}. However, this
is clearly not physically relevant and will not be used here.

Now we introduce a Kovacs-like protocol:
the parameter $\xi$ controlling the dynamics is changed in the
following stepwise manner,
\begin{equation}\label{ch8_eq:stepwise-xi}
\xi(t)=\left\{ 
\begin{array}{ll}
\xi_{0}, & -\infty<t<0, \\ \xi_{1}, & 0<t<t_{w}, \\ \xi, & t>t_{w}.
\end{array}\right.
\end{equation}
Therefore, since $\xi_{0}$ is kept for an infinite time, at $t=0$ the
system is prepared in the corresponding steady state,
$\calP(t=0)=\calP_{\ness}(\xi_{0})$. Our idea is that the
jumps $\xi_{1}-\xi_{0}$ and $\xi-\xi_{1}$ are small, in~the sense that
all expressions can be linearized in them.
This protocol is completely analogous to the Kovacs
protocol described in the introduction (see figure \ref{ch8_fig1}), but
with $\xi$ playing the role of the temperature.

We start by analyzing the relaxation in the first time window,
$0<t<t_{w}$. Therein, we apply \eqref{ch8_eq:formal-sol} with the
substitutions $t_{0}\to 0$ and $\xi\to \xi_{1}$, that is,
\begin{equation}\label{ch8_eq:evol-first-stage}
\ket{\calP(t)-\calP_{\ness}(\xi_{1})}=e^{t\bbW(\xi_{1})}
\ket{\calP_{\ness}(\xi_{0})-\calP_{\ness}(\xi_{1})}, \quad 0\leq
t\leq t_{w}.  
\end{equation} 
The final distribution function, at $t=t_{w}$, is the initial
condition for the next stage, $t>t_{w}$, in which the system relaxes
towards the steady state corresponding to $\xi$. Making use
again of \eqref{ch8_eq:formal-sol} with $t_{0}\to t_{w}$,
\begin{align}
\ket{\calP(t)-\calP_{\ness}(\xi)}&= e^{(t-t_{w})\bbW(\xi)}
\ket{\calP(t_{w})-\calP_{\ness}(\xi)} \nonumber \\
& =  e^{(t-t_{w})\bbW(\xi)} \left[ e^{t_{w}\bbW(\xi_{1})}
\ket{\calP_{\ness}(\xi_{0})-\calP_{\ness}(\xi_{1})}+ \ket{\calP_{\ness}(\xi_{1})-\calP_{\ness}(\xi)}\right],
\label{ch8_eq:evol-second-stage}
\end{align}
with $t\geq t_{w}$. It must be stressed that the expressions above, 
\eqref{ch8_eq:evol-first-stage} and \eqref{ch8_eq:evol-second-stage}, are
exact, no approximation has been made.

The linear response approximation is introduced now:
both jumps $\xi_{0}-\xi_{1}$ and $\xi-\xi_{1}$ are assumed to be
 small. Therefore, we can
expand both $\ket{\calP_{\ness}(\xi_{0})-\calP_{\ness}(\xi_{1})}$ and
$\ket{\calP_{\ness}(\xi_{1})-\calP_{\ness}(\xi)}$ in these jumps,
 similarly to what was
done in  \eqref{ch8_eq:P-linear}. Namely,
\begin{subequations}\label{ch8_eq:lin-approx}
\begin{equation}\label{ch8_eq:lin-approx-a}
\ket{\calP_{\ness}(\xi_{0})-\calP_{\ness}(\xi_{1})}=(\xi_{0}-\xi_{1})\ket{\frac{d\calP_{\ness}(\xi)}{d\xi}}+\mathcal{O}(\xi_{0}-\xi_{1})^{2},
\end{equation}
\begin{equation}\label{ch8_eq:lin-approx-b}
\ket{\calP_{\ness}(\xi_{1})-\calP_{\ness}(\xi)}=(\xi_{1}-\xi)\ket{\frac{d\calP_{\ness}(\xi)}{d\xi}}+\mathcal{O}(\xi_{1}-\xi)^{2}.
\end{equation}
\end{subequations}
In both \eqref{ch8_eq:lin-approx-a} and \eqref{ch8_eq:lin-approx-b},
the derivatives are evaluated at $\xi$; the difference introduced by
evaluating them at either $\xi_{1}$ or $\xi_{0}$ is second order in
the deviations. Then, we have that
\begin{equation}\label{ch8_eq:evol-second-stage-linear-1}
\ket{\calP(t)-\calP_{\ness}(\xi)}= 
(\xi_{0}-\xi_{1}) e^{(t-t_{w})\bbW(\xi)} e^{t_{w}\bbW(\xi_{1})}
\ket{\frac{d\calP_{\ness}(\xi)}{d\xi}}+
(\xi_{1}-\xi)e^{(t-t_{w})\bbW(\xi)}\ket{\frac{d\calP_{\ness}(\xi)}{d\xi}}
,
\end{equation}
which can be simplified as follows. Since the two terms on its
rhs are of first order in the jumps, we can substitute $\bbW(\xi_{1})$
with $\bbW(\xi)$, which yields

\begin{equation}\label{ch8_eq:evol-second-stage-linear-2}
\ket{\calP(t)-\calP_{\ness}(\xi)}= 
(\xi_{0}-\xi_{1}) e^{t\bbW(\xi)}
\ket{\frac{d\calP_{\ness}(\xi)}{d\xi}}-
(\xi-\xi_{1})e^{(t-t_{w})\bbW(\xi)}\ket{\frac{d\calP_{\ness}(\xi)}{d\xi}}.
\end{equation}
This is the superposition of two responses: the first term on the rhs
gives the relaxation from $\xi_{0}$ to $\xi_{1}$, starting at $t=0$,
whereas the second term stands for the relaxation from $\xi_{1}$ to
$\xi$, starting at $t=t_{w}$. We have written
$-(\xi-\xi_{1})$ in the second term because $\xi>\xi_{1}$ in the
Kovacs protocol.

The same structure in \eqref{ch8_eq:evol-second-stage-linear-2}
is transferred to the average values. Taking into account \eqref{ch8_eq:average-time-evol},
\begin{equation}\label{ch8_eq:evol-Yav}
\Delta Y(t)=(\xi_{0}-\xi_{1}) \mel{\calY}{e^{t\bbW(\xi)}}
{\frac{d\calP_{\ness}(\xi)}{d\xi}}-
(\xi-\xi_{1})\mel{\calY}{e^{(t-t_{w})\bbW(\xi)}}{\frac{d\calP_{\ness}(\xi)}{d\xi}},
\quad t\geq t_{w},
\end{equation}
in which we recognize the relaxation function in linear response,
as defined in \eqref{ch8_eq:phi-Y-linear}. We can also normalize the
response in this experiment, by defining a function $K(t)$
as follows, 
\begin{equation}\label{ch8_eq:evol-Yav-phi}
K_{Y}(t)\equiv \lim_{\xi_{0}\to\xi}\frac{\Delta Y(t)}{\xi_{0}-\xi}=\frac{\xi_{0}-\xi_{1}}{\xi_{0}-\xi} \phi_{Y}(t)-
\frac{\xi-\xi_{1}}{\xi_{0}-\xi} \phi_{Y}(t-t_{w}).
\end{equation}
It is understood that, as $\xi_{0}-\xi\to 0$, both
prefactors $\frac{\xi_{0}-\xi_{1}}{\xi_{0}-\xi}$ and
$\frac{\xi-\xi_{1}}{\xi_{0}-\xi}$ are kept of the order of unity.

A few comments on \eqref{ch8_eq:evol-Yav-phi} are in
order. Hitherto, no restriction has been imposed on the state of the
system at $t=t_{w}$; therefore, \eqref{ch8_eq:evol-Yav-phi} is valid for
arbitrary $(\xi_{0},\xi_{1},\xi)$, provided that the jumps are small
enough and the ratio of the jumps is of the order of unity. The
function $K(t)$ corresponds to a Kovacs-like experiment when $\xi$ is
chosen as a function of $t_{w}$ in such a way that $\langle
Y(t_{w})\rangle=\langle Y\rangle_{\ness}(\xi)$ or $K_{Y}(t_{w})=0$, that is,
\begin{equation}\label{ch8_eq:Kovacs-condition}
 \frac{\xi-\xi_{1}}{\xi_{0}-\xi_{1}}=\frac{\phi_{Y}(t_{w})}{\phi_{Y}(0)}.
\end{equation} 
Alternatively, one may consider that \eqref{ch8_eq:Kovacs-condition}
defines $t_{w}$ as a function of $\xi$. 

The complete analogy between \eqref{ch8_eq:evol-Yav-phi} and
\eqref{ch8_eq:Kovacs-condition} and 
\eqref{ch8_eq:kovacs-thermal-linear} and \eqref{ch8_eq:T-tw-relation} is
apparent. Nevertheless, we have made use neither of the explicit form
of the steady state distribution---in general noncanonical---, nor of the
relation between response functions and time correlation functions---fluctuation-dissipation relation---, which were necessary in
\cite{PyB10} to demonstrate
\eqref{ch8_eq:kovacs-thermal-linear}. Therefore, the proof
developed here is more general, being valid for any steady state, either
equilibrium or nonequilibrium. Thus, it specifically holds in
athermal systems. Furthermore, it must be noted that it can be easily
extended to the Fokker--Planck, or the equivalent Langevin, equation.

\subsection{Macroscopic equations approach}\label{ch8_ch:lin-th-2}

In this section, we do not start from the equation for the probability
distribution as before, but from the equations for the relevant physical
properties of the considered system. For example, from the
hydrodynamic equations for a fluid or the law of mass action equations
for chemical reactions. Of~course, these equations can be derived in a
certain ``macroscopic'' approximation
\cite{vK92}, which typically involves neglecting
fluctuations, from the equation for the probability distribution by
taking moments. Although this is not our approach here, we borrow
this term to call them ``equations for the moments''.

We denote the relevant moments by $z_{i}$, $i=1,\ldots,J$,
where $J$ is the number of relevant moments. The equations for the moments have the general form
\begin{equation}\label{ch8_eq:moments-eq}
\frac{d}{dt}z_{i}=f_{i}(z_{1},\ldots,z_{J};\xi),
\end{equation}
where $f_{i}$ are continuous, in general nonlinear, functions of the
moments. This is a key difference between moment equations and the
master (or Fokker--Planck) equation, since the latter is always linear
in the probability distribution. Therefore, unlike the master
equation, \eqref{ch8_eq:moments-eq} cannot be formally solved for
arbitrary initial conditions. However, in the linear response
approximation, we show here that a procedure similar to the one
 in the previous section can be performed and leads to the same expression for
the Kovacs hump.

We assume that there is only one, globally stable, steady solution of 
\eqref{ch8_eq:moments-eq}. The
corresponding values of the moments in this solution are
$z_{i}^{\ness}(\xi)$. Linearization of the dynamical system around the
steady state gives
\begin{equation}\label{ch8_eq:lin-dyn-syst}
\frac{d}{dt} \ket{\Delta z(t)} =\mathbb{M}(\xi) \ket{\Delta z(t)}, \qquad
\ket{\Delta z(t)}\equiv \ket{z(t)-z_{\ness}(\xi)}.
\end{equation}
The notation is completely similar to that in the previous
section: $\ket{z}$ is a vector represented by a column matrix with
components $z_{i}$, and $\mathbb{M}(\xi)$ is a linear operator
represented~by a square matrix with elements
\begin{equation}\label{ch8_eq:M-matrix-elem}
 M_{ij}(\xi)=\left. \partial_{z_{j}}f_{i} \right|_{\ket{z}=\ket{z_{\ness}(\xi)}}.
\end{equation}
The dimensions of these matrices are much smaller than those
for the master equation, since $J$ is of the order of unity and does
not diverge in the thermodynamic limit. In general,
$M_{ij} \neq M_{ji}$, and the operator $\mathbb{M}$ is not
Hermitian. However, we do not need $\mathbb{M}$ to be Hermitian to
solve the linearized system in a formal way, as shown below.

Analogously to what was done for the master equation, the formal
solution of \eqref{ch8_eq:lin-dyn-syst} is
\begin{equation}\label{ch8_eq:lin-dyn-syst-sol1}
\ket{\Delta z(t)}=e^{(t-t_{0})\mathbb{M}(\xi)}\ket{\Delta z(t_{0})}.
\end{equation}
In particular, if the initial condition is chosen to
correspond to the steady state for $\xi_{0}=\xi+\Delta\xi$, one~has
\begin{equation}\label{ch8_eq:lin-dyn-syst-sol2}
\ket{\Delta z(t)}=\Delta\xi\, e^{(t-t_{0})\mathbb{M}(\xi)}\ket{\frac{dz_{\ness}(\xi)}{d\xi}}.
\end{equation}
The response for any of the relevant moments can be extracted by
projecting the above result onto the ``natural'' basis
$\ket{u_{i}}$, whose $j$-th component is $u_{ij}=\delta_{ij}$. Then, the
normalized linear response function for $z_{i}$ can be defined by
\begin{equation}\label{ch8_eq:lin-resp-normalized}
 \phi_{z_{i}}(t)=\lim_{\Delta\xi\to 0}\frac{\braket{u_{i}}{\Delta z(t)}}{\Delta\xi}= \mel{u_{i}}{e^{(t-t_{0})\mathbb{M}(\xi)}}{\frac{dz_{\ness}(\xi)}{d\xi}}.
\end{equation}
Note the utter formal analogy of Expression \eqref{ch8_eq:lin-resp-normalized} with  \eqref{ch8_eq:phi-Y-linear},
which was obtained from the master equation. The proof of the
expression for the Kovacs hump follows exactly the same line of
reasoning, and the result is exactly that in 
\eqref{ch8_eq:evol-Yav-phi} and \eqref{ch8_eq:Kovacs-condition}; thus, it
is not repeated here.

\section{Kovacs hump in the model of granular media}
\label{Section-model}
\subsection{Kinetic approach}

Here, we study the memory effects described above in the generalized lattice model of granular media introduced in chapters \ref{ch:finsize_pgran} and \ref{ch:Hth_pgran}. As anticipated in section \ref{ch5_ev_eq}, we explore in this chapter the model with a more general collision rate. Dynamics is generated by collisions following \eqref{coll_rule}, and the action of the stochastic thermostat given by \eqref{jump-moments-1} and \eqref{jump-moments-2}.

We focus on the one-particle distribution of homogeneous states $P_1(v,\tau)$. 
Following the same line of reasoning we have adopted in previous chapters,
the evolution equation for $P_1(v,\tau)$ is derived. From the master equation \eqref{eq:cma2} with a collision rate proportional to some power $\beta$ of the 
relative velocity, we have to incorporate the Fokker-Planck term 
coming from the heating \eqref{ch7_eq:Fokker-Planck}, 
and integrate over $N-1$ velocities. This leads to the result
\begin{equation}\label{ch8_eq:kinetic-eq-tau}
\partial_{\tau}P_1(v,\tau)=\frac{\omega\epsilon}{2}\partial_{v}\int_{-\infty}^{+\infty}dv^{\prime}
(v-v^{\prime}) |v-v^{\prime}|^{\beta}
P_1(v,\tau)P_1(v^{\prime},\tau)+\frac{\chi}{2} \partial_{v}^{2} P_1(v,\tau),
\end{equation}
where $\epsilon=1-\alpha^2$ and, as usual, we have made use of the Stosszahlansatz.
Note that the first term on the rhs of \eqref{ch8_eq:kinetic-eq-tau} is completely
analogous to the last one in \eqref{eq:P1-hydrobeta}. 
 Here, we define a dimensionless time scale $\tilde{t}=\omega \epsilon \tau$,
 which is slightly 
 different from the time defined in previous sections, namely by a factor 
 $\nu=\epsilon L^{-2}$ \eqref{eq:nu}. However, since we use this time scale along all the 
 chapter and for the sake of a clear notation, we skip the tilde from now on. 
 Introducing this time scale we get 
\begin{equation}\label{ch8_eq:kinetic-eq-t}
\partial_{t}P_1(v,t)=\frac{1}{2}\partial_{v}\int_{-\infty}^{+\infty}dv^{\prime}
(v-v^{\prime}) |v-v^{\prime}|^{\beta}
P_1(v,t)P_1(v^{\prime},t)+\frac{\tilde{\xi}}{2} \partial_{v}^{2} P_1(v,t),
\end{equation}
where $\tilde{\xi}=\frac{\chi}{\omega\epsilon}$ is the rescaled strength of the 
noise, which differs from the $\xi$ parameter of the previous chapter \eqref{ch7_eq:nu} by the same $\nu$ factor as the time scale. Again, 
 we skip the tilde right away.  

Since we are studying homogeneous states, the average velocity is zero and the granular temperature $T$ is simply
\begin{equation}\label{ch8_eq:granular-T}
T = \langle v^{2}\rangle=\int_{-\infty}^{+\infty}dv\, v^{2} P_1(v,t).
\end{equation}
A scaled velocity and its corresponding distribution can be defined as 
\begin{equation}\label{ch8_eq:v-dimensionless}
v=\sqrt{2T} \, \tilde{c}, \qquad P_1(v,t)dv=\varphi(\tilde{c},t)d\tilde{c} \Leftrightarrow \varphi(\tilde{c},t)=\sqrt{2T}P_1(v,t).
\end{equation}
Note that, again, the scaled velocity introduced here is slightly different from 
that defined in previous chapters by a factor $\sqrt{2}$. We do so in order to have 
below the usual expansion in Sonine polynomials of kinetic theory. As with the
previous scalings, we omit the tilde from now on.

Taking moments in \eqref{ch8_eq:kinetic-eq-t} and making the change of
variables above, the (granular) temperature evolves according to
\begin{equation}\label{ch8_eq:evol-T-1}
\frac{d}{d t}T=-\zeta\,T^{1+\frac{\beta}{2}}+\xi, \qquad \zeta=2^{\frac{\beta}{2}}\int_{-\infty}^{+\infty} dc
\int_{-\infty}^{+\infty} dc^{\prime}
|c-c^{\prime}|^{2+\beta}\varphi(c,t)\varphi(c^{\prime},t).
\end{equation}
The first term on the rhs stems from collisions and {cools} the
system, it always makes the granular temperature
decrease. The second term stems from the stochastic thermostat and
{heats} the system. Thus, in the long time limit, a NESS is
attained in which both terms counterbalance each other.

The equation for the granular temperature is not closed in general
because the cooling rate $\zeta$ depends on the whole velocity distribution. Then, 
an expansion in Sonine (or Laguerre) polynomials is typically
introduced,
\begin{equation}\label{ch8_eq:Sonine-expansion}
\varphi(c,t)=\frac{e^{-c^{2}}}{\sqrt{\pi}}\left[
 1+\sum_{k=2}^{\infty}a_{k}(t) L_{k}^{\left(-\frac{1}{2}\right)}(c^{2})\right],
\end{equation}
where $L_{k}^{(m)}(x)$ are the associated Laguerre polynomials
\cite{AyS72}. In kinetic theory, $m=\frac{d}{2}-1$, with $d$ being the
spatial dimension, and often, the notation
$S_{k}(x)\equiv L_{k}^{\left(\frac{d}{2}-1\right)}$ is used. Here, we
mainly use the so-called first Sonine approximation, in which (i) only
the term with $k=2$ is retained and (ii) nonlinear terms in $a_{2}$
are neglected. The coefficient $a_{2}$ is the excess kurtosis,
\begin{equation}
\langle c^{4}\rangle=3(1+a_{2})/4.
\end{equation}

Although the linearization in $a_{2}$ is quite standard in kinetic
theory, we derive firstly the evolution equations considering just
step (i) of the first Sonine approximation, that is we truncate the
expansion for the scaled distribution \eqref{ch8_eq:Sonine-expansion}
after the $k=2$ term. Henceforth, we call this approximation the
nonlinear first Sonine approximation. Afterwards, in the numerical
results, we will discuss how both approximations, nonlinear and
standard, give almost indistinguishable results.

In the nonlinear first Sonine approximation,
 the evolution equation for the temperature is readily obtained 
 \begin{subequations}\label{ch8_eq:evol-Soninenl}
\begin{equation}\label{ch8_eq:evol-T-Soninenl}
\frac{d}{dt}T=- \zeta_{0}\,
T^{1+\frac{\beta}{2}}\left[1+\frac{\beta(2+\beta)}{16}a_{2}+\frac{\beta(2+\beta)(2-\beta)(4-\beta)}{1024}a_{2}^{2}\right]+\xi,
\end{equation}
where
$\zeta_{0}=\pi^{-\nicefrac{1}{2}}\,2^{1+\beta}\,\Gamma\!\left(\frac{3+\beta}{2}\right)$.
Unless $\beta=0$ (the Maxwell molecules we have considered in previous chapters), the equation for the temperature
is not closed. Then, we write down the equation for $a_{2}$: again,
after a lengthy, but straightforward calculation, we derive
\begin{align}
T\frac{d}{dt}a_{2}=&-2\xi a_{2}- \frac{\zeta_{0}}{3}
 \beta\, T^{1+\frac{\beta}{2}} \left[1+\frac{56+\beta(6+\beta)}{16}a_{2} \right. \nonumber\\
& \left. -\frac{(2+\beta)[384+(2-\beta)\beta(4+\beta)]}{1024}a_{2}^{2}-\frac{3(4-\beta)(2-\beta)(2+\beta)}{512}
a_{2}^{3}\right].
 \label{ch8_eq:evol-a2-Soninenl}
\end{align}
\end{subequations}
The evolution equations in the standard first Sonine approximation are
obtained just neglecting nonlinear terms in $a_2$ in 
\eqref{ch8_eq:evol-Soninenl}, that is,
\begin{subequations}\label{ch8_eq:evol-Sonine}
\begin{align}\label{ch8_eq:evol-T-Sonine}
\frac{d}{dt}T&=- \zeta_{0}\,
T^{1+\frac{\beta}{2}}\left[1+\frac{\beta(2+\beta)}{16}a_{2}\right]+\xi,
\\
\label{ch8_eq:evol-a2-Sonine}
 T\frac{d}{dt}a_{2}&=- \frac{\zeta_{0}}{3}
 \beta\, T^{1+\frac{\beta}{2}}
 \left[1+\frac{56+\beta(6+\beta)}{16}a_{2}\right]-2\xi a_{2}.
\end{align}
\end{subequations}
For $\xi\neq 0$, the steady solution of these equations is
\begin{equation}\label{ch8_eq:T-a2-steady}
 T_{s}=\left(\frac{\xi}{\zeta_{0}\left[1+
\frac{\beta(2+\beta)}{16}a_{2}^{\ness}\right]}\right)^{\frac{2}{2+\beta}},\qquad a_{2}^{\ness}=-\frac{16\beta}{96+56\beta+6\beta^{2}+\beta^{3}}.
\end{equation}
Note that (i) $0\leq |a_{2}^{\ness}|\leq 0.133$ for
$0\leq \beta\leq 2$, which makes it reasonable to use the first Sonine
approximation, and (ii) $a_{2}^{\ness}$ is independent of the driving
intensity $\xi$. This will be useful in the linear
response analysis, to be developed below. A sudden change in
the driving only changes the stationary value of the temperature, but
not that of the excess kurtosis. If $\xi=0$, the system evolves
towards the homogeneous cooling state, in which the excess kurtosis
tends to the value
\begin{equation}
\label{ch8_eq:a2HCS}
a_{2}^{\text{HCS}}=-\frac{16}{56+\beta(6+\beta)},
\end{equation}
as predicted by  \eqref{ch8_eq:evol-a2-Sonine}, and the
temperature decays following Haff's law, $dT/dt \propto -T^{1+\frac{\beta}{2}}$.

From now on, we use reduced temperature and time,
\begin{equation}\label{ch8_eq:theta-A2-def}
\theta=\frac{T}{T_{\ness}}, \quad s=\zeta_{0}T_{\st}^{\beta/2}t.
\end{equation}
The steady temperature $T_{\ness}$ plays the role of a natural energy
(or granular temperature) unit. In~reduced variables, the evolution
equations are
\begin{subequations}\label{ch8_eq:reduced-var-evol}
\begin{align}\label{ch8_eq:reduced-var-evol-1}
 \frac{d}{ds}\theta&=1-\theta^{1+\frac{\beta}{2}}+\frac{\beta(2+\beta)}{16}
       \left(a_{2}^{\ness}-a_{2}\theta^{1+\frac{\beta}{2}}\right), \\
 \theta \frac{d}{ds}a_{2}&= \kappa_{1} \left( a_{2} - a_{2}^{\text{HCS}} \right) \left( 1-\theta^{1+\frac{\beta}{2}} \right) -\kappa_{2} \left(a_{2}-a_{2}^{\ness}\right),
\label{ch8_eq:reduced-var-evol-2}
\end{align}
\end{subequations}
where we have introduced two parameters of the order of unity, 
\begin{equation}\label{ch8_eq:B-def}
\kappa_{1}=-\frac{\beta}{3 a_{2}^{\text{HCS}}}, \quad \kappa_{2}=-\frac{\beta}{3 a_{2}^{\ness}},
\end{equation}
$0\leq \kappa_{1}\leq 3$ and $2\leq \kappa_{2}\leq 5$ for
$0\leq\beta\leq 2$. 

The evolution equations in the first Sonine approximation,
\eqref{ch8_eq:evol-Sonine} or \eqref{ch8_eq:reduced-var-evol}, are the
particularization of the equations for the moments
\eqref{ch8_eq:moments-eq} to our model: $J=2$, and $z_{1}=T$ (or
$\theta$), $z_{2}=a_{2}$. Consistently, they~are nonlinear, although
here, due to the simplifications introduced in the first Sonine
approximation, only nonlinear in $\theta$. When the system is close to
the NESS, \eqref{ch8_eq:reduced-var-evol} can be linearized
around it by writing $\theta=1+\Delta\theta$,
$a_{2}=a_{2}^{\ness}+\Delta a_{2}$,
\begin{equation}\label{ch8_eq:linearized-system}
 \frac{d}{ds}\begin{pmatrix}\Delta\theta \\ \Delta a_{2} \end{pmatrix}
 =\bm{M}\cdot\begin{pmatrix}\Delta\theta \\ \Delta a_{2} \end{pmatrix},
 \qquad
 \bm{M}=\begin{pmatrix} -\frac{2(2+\beta)(12+\beta)}{48+4\beta+\beta^{2}} &
 -\frac{\beta(2+\beta)}{16} \\ -\kappa_{1} \left( 1 + \frac{\beta}{2} \right) \left(a_{2}^{\ness}- a_{2}^{\text{HCS}} \right) & -\kappa_{2}
 \end{pmatrix} .
\end{equation}
Of course, the general solution of this linear system for
arbitrary initial conditions $\Delta\theta(0)$ and $\Delta a_{2}(0)$
can be immediately written, but we omit it here.

\subsection{Kovacs hump in linear response}

Now, we look into the Kovacs hump in the linear response
approximation. Following the discussion leading to 
\eqref{ch8_eq:evol-Yav-phi}, first we have to calculate the relaxation
function $\phi_{T}$ for the granular temperature. The system is at the
steady state corresponding to a driving $\xi_{0}$ for
$t<0$; at $t=0$, the driving is instantaneously changed to $\xi$, and
only the linear terms in $\Delta\xi=\xi-\xi_{0}$ are retained. We
choose the normalization of $\phi_{T}(s)$ in such a way that
$\phi_{T}(0)=1$, that is,
\begin{equation}\label{phi-granular-temp-def}
\phi_{T}(s)\equiv\lim_{\Delta T(0)\to 0}\frac{\Delta T(s)}{\Delta
 T(0)}=\lim_{\Delta\theta(0)\to 0}\frac{\Delta\theta(s)}{\Delta\theta(0)}.
\end{equation} 

Since $T_{\ness}$ changes with $\xi$, but $a_{2}$ does not, we have to
solve \eqref{ch8_eq:linearized-system} for $\Delta a_{2}(0)=0$
and arbitrary (small enough) $\Delta\theta(0)$. The solution is
\begin{subequations}\label{phi-granular-temp-exp}
\begin{equation}
\phi_{T}(s)=c_{+}e^{\lambda_{+}s}+c_{-}e^{\lambda_{-}s},
\end{equation}
\begin{equation}
c_{+}=\frac{M_{11}-\lambda_{-}}{\lambda_{+}-\lambda_{-}}, \qquad
c_{-}=\frac{\lambda_{+}-M_{11}}{\lambda_{+}-\lambda_{-}},
\end{equation}
\end{subequations}
where $M_{ij}$ is the $(i,j)$ element of the matrix $\bm{M}$ and
$\lambda_{\pm}$ its eigenvalues,
\begin{equation}\label{ch8_eq:eigenv}
\lambda_{\pm}=\frac{\Tr\bm{M}\pm\sqrt{(\Tr\bm{M})^{2}-4\det\bm{M}}}{2}=\frac{\Tr\bm{M}\pm\sqrt{(M_{11}-M_{22})^{2}+4M_{12}M_{21}}}{2}.
\end{equation}
Both eigenvalues $\lambda_{\pm}$ are negative, since $\Tr\bm{M}<0$ and
$\det\bm{M}>0$ for all $\beta>0$. Therefore,
$|\lambda_{+}|<|\lambda_{-}|$, and it is $\lambda_{+}$ that dominates
the relaxation of the granular temperature for long times. Moreover,
$c_{\pm}>0$, and thus, the linear relaxation function $\phi_{T}(s)$ is
always positive and decays monotonically to~zero.

Next, we consider a Kovacs-like experiment: the system was at the NESS
corresponding to a driving $\xi_{0}$, with granular temperature
$T_{\ness,0}$ for $t<0$; the driving is suddenly changed to $\xi_{1}$
at $t=0$ so that the system starts to relax towards a new steady
temperature $T_{\ness,1}$ for $0\leq t\leq t_{w}$, and this relaxation
is interrupted at $t=t_{w}$, because the driving is again suddenly
changed to the value $\xi$ such that the stationary granular
temperature $T_{\ness}$ equals its instantaneous value at $t_{w}$. The
time evolution of the granular temperature for $t\geq t_{w}$ is given
by the particularization of \eqref{ch8_eq:evol-Yav-phi} and
\eqref{ch8_eq:Kovacs-condition} to our situation, that is,
\begin{equation}\label{ch8_eq:Kovacs-Sonine}
K_{T}(s)=\frac{\xi_{0}-\xi_{1}}{\xi_{0}-\xi} \phi_{T}(s)-
\frac{\xi-\xi_{1}}{\xi_{0}-\xi} \phi_{T}(s-s_{w}), \qquad
 \frac{\xi-\xi_{1}}{\xi_{0}-\xi_{1}}=\phi_{T}(s_{w}),
\end{equation} 
where we have made use of the normalization $\phi_{T}(0)=1$. In the
linear response approximation, the~jumps in the driving values can be
substituted by the corresponding jumps in the stationary values of the
granular temperature.

{The structure of the linear relaxation function
 $ \phi_{T}(s)$, as a linear combination of decreasing exponentials
 $\exp(\lambda_{\pm}t)$, $\lambda_{\pm}<0$, with positive weights
 $c_{\pm}$,} assures that the Kovacs behavior is {normal}:
(i) $K_{T}(s)$ is always positive and bounded from above by
$\phi_{T}(s)$ and (ii) there is only one maximum at a certain time
$s_{k}>s_{w}$ \cite{PyB10}. The {anomalous}
behavior found in the uniformly heated hard-sphere granular for large
enough inelasticity is thus not present here. This is consistent with
the quasi-elastic limit we have introduced to simplify the collision
operator.

\subsection{Nonlinear Kovacs hump}

Here, we consider the Kovacs hump for arbitrary large driving jumps. In
our model, we can make use of the smallness of $a_{2}$, which is
assumed in the first Sonine approximation, in order to introduce a
perturbative expansion of \eqref{ch8_eq:reduced-var-evol} in powers
of $a_{2}^{\ness}$. The procedure is completely analogous to that
performed in \cite{PyT14,TyP14} for a
dilute gas of inelastic hard spheres, and thus, we omit the details
here. 

We start by writing $a_{2}=a_{2}^{\ness}A_{2}$, with $A_{2}$ of the
order of unity, and
\begin{equation}\label{ch8_eq:generic-expansion-in-a2}
\theta(s)=\theta_{0}(s)+a_{2}^{\ness}\theta_{1}(s)+\ldots, \qquad A_{2}(s)=A_{20}(s)+a_{2}^{\ness}A_{21}(s)+\ldots.
\end{equation}
These expansions are inserted into 
\eqref{ch8_eq:reduced-var-evol}, which have to be solved with initial
conditions $\theta(s_{w})=1$, $A_{2}(s_{w})=A_{2}^{\text{ini}}$. To
the lowest order, $\theta_{0}(s)=1$, whereas $A_{20}(s)$ decays
exponentially to one,
\begin{equation}\label{ch8_eq:expansion_a2s-kurtosis}
 A_{20}(s) - 1 \sim \left( A_{2}^{\text{ini}} - 1 \right) e^{-\kappa_{2} \left(s-s_{w} \right)}.
\end{equation} 
In order to describe the Kovacs hump, we compute 
$\theta_{1}(s)$ that verifies the evolution equation
\begin{equation}\label{ch8_eq:theta1-evol}
\frac{d\theta_{1}}{ds}=-\left(1+\frac{\beta}{2}\right)\theta_{1}+\frac{\beta(2+\beta)}{16}\left(A_{20}-1\right),
\end{equation}
which gives
\begin{equation}\label{ch8_eq:expansion_a2s-theta}
\theta(s)-1 \sim \left(a_{2}^{\text{ini}}-a_{2}^{\ness}\right)
\frac{ \beta (2+\beta)}{8(2+\beta-2\kappa_{2})} \left[
e^{-\kappa_{2}\left(s-s_{w} \right)}-e^{-\left(1+\frac{\beta}{2}
 \right) \left(s-s_{w} \right)} \right], \qquad s\geq s_{w}.
\end{equation}
The structure of this result is completely analogous to those in
\cite{PyT14,TyP14}, and thus, the
conclusions can also be drawn in a similar way. In particular, we want
to highlight that (i) the factor that controls the size of the hump is
proportional to $a_{2}^{\text{ini}}-a_{2}^{\ness}$ and (ii) the
shape of the hump is codified in the factor between brackets that only
depends on $\beta$. Note that $(a_{2}^{\text{ini}}-a_{2}^{\ness})>0$
for the considered cooling protocols ($\xi_{1}<\xi<\xi_{0}$).
Thus, no anomalous Kovacs hump is expected in the nonlinear
regime either.

\section{Numerical results} \label{ch8_sec:num}

Here, we compare the theoretical approach above to numerical results.
 Specifically, we focus on the case $\beta=1$ that gives
a collision rate linear in the relative velocity and thus
similar to that of hard-spheres. All simulations have
been carried out with a restitution coefficient $\alpha=0.999$, which
corresponds to the quasi-elastic limit in which our simplified kinetic
description holds. Furthermore, we have set $\omega=1$ without loss of~generality.

\subsection{Validation of the first Sonine approximation}

\begin{figure}
\centering
\includegraphics[width=0.8 \textwidth]{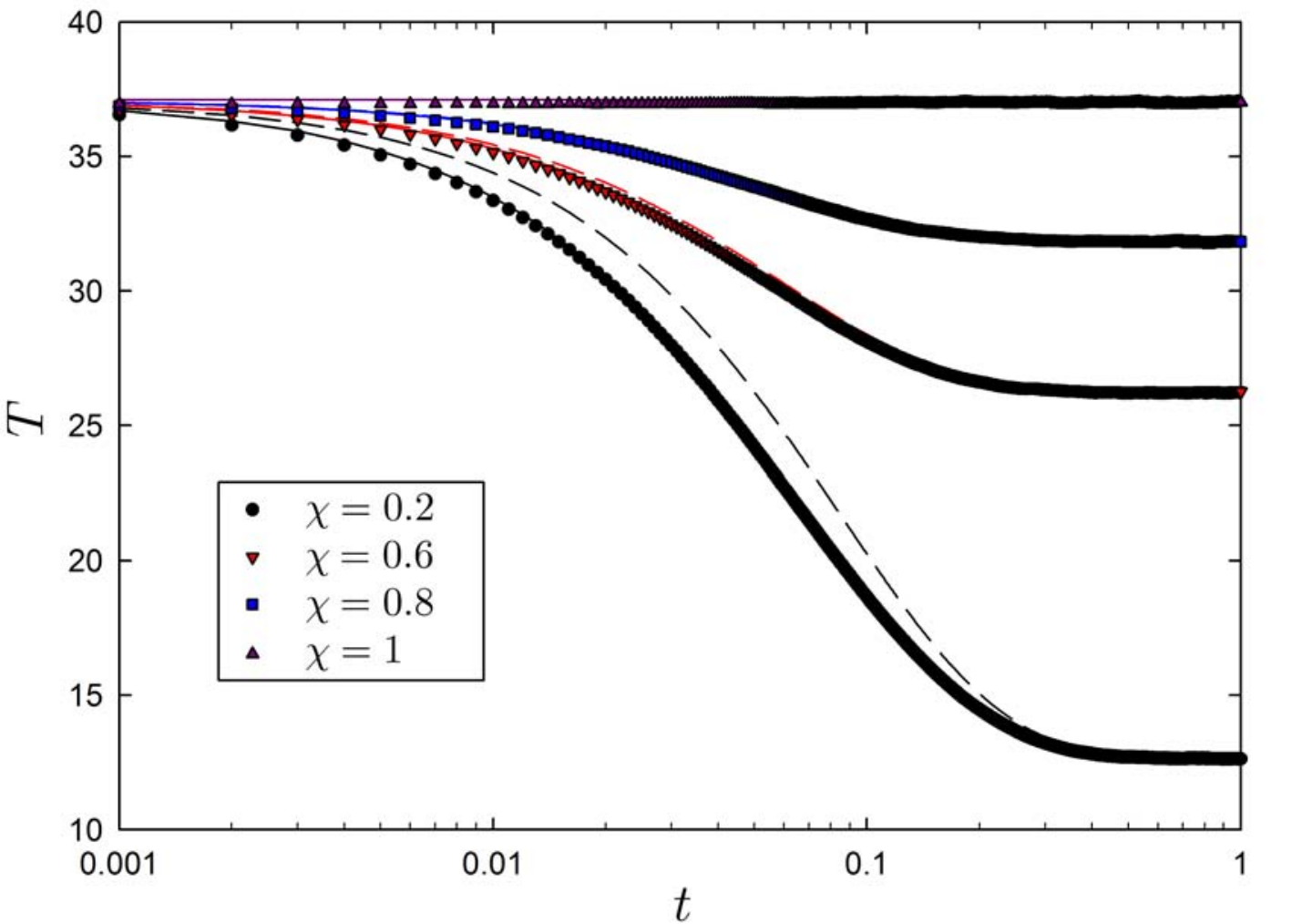}
\caption{Direct relaxation of the granular
 temperature $T$ for different final noise amplitudes. All curves
 start from the stationary state corresponding to $\chi_{0}=1$. 
 Monte Carlo simulations for a system of $N=100$ sites
 (symbols) are compared with the numerical solution of the first Sonine
 approximation  \eqref{ch8_eq:evol-Sonine} (solid lines), and
 the analytic solution of the linear response system
 \eqref{ch8_eq:linearized-system} (dashed lines). \label{ch8_fig2}}
\end{figure}

First of all, we check the validity of the first Sonine approximation,
as given by \eqref{ch8_eq:evol-Sonine}, to describe the time
evolution of our system. In order to do so, we compare several
relaxation curves between the NESS corresponding to two different
noise strengths. In particular, we always depart from the stationary
state corresponding to $\chi_{0}=1$ and afterwards let the system
evolve with $\chi=\{ 0.2, 0.6, 0.8, 1 \}$ for $t>0$. In Figure
\ref{ch8_fig2}, we compare the Monte Carlo simulations---see once more appendix
\ref{ch5-6_app-b} for details---with the numerical solution of
the evolution equations in the first Sonine approximation
\eqref{ch8_eq:evol-Sonine}. In addition,
we also have plotted the analytical solution of the linear response
system, \eqref{ch8_eq:linearized-system}.~The agreement is
complete between simulation and theory, and it can be observed how the
linear response result becomes more accurate as the temperature jump
decreases.

In order not to clutter the plot in Figure \ref{ch8_fig2}, we have not
shown the results for the nonlinear first Sonine approximation \eqref{ch8_eq:evol-Soninenl}. The relative error between their
numerical solution and that of the standard first Sonine approximation
\eqref{ch8_eq:evol-Sonine} is at most of order $0.1 \%$, for all the cases
we have considered. Henceforth, we always use the latter, which is the
usual approach in kinetic theory.

\subsection{Kovacs hump}

Since the numerical integration of the first Sonine approximation
perfectly agrees with Monte Carlo simulations, we compare the
analytical results for the Kovacs hump with the former. Specifically,
we work in reduced variables, and therefore, we integrate numerically
\eqref{ch8_eq:reduced-var-evol}.

\subsubsection{Linear response}

It is convenient to rewrite the expression for the Kovacs hump in an
alternative form to compare our theory with numerical results. We take
advantage of the simple structure of the relaxation function in the
first Sonine approximation, which is the sum of two exponentials, to
introduce the factorization \cite{PyB10}
\begin{subequations}\label{ch8_eq:Kovacs-factorization-and-factors}
\begin{equation}\label{ch8_eq:Kovacs-factorization}
K_{T}(s)=K_{0}(s_{w})K_{1}(s-s_{w}),
\end{equation}
where
\begin{equation}\label{ch8_eq:Kovacs-factors}
K_{0}(s_{w})=c_{+}c_{-}
\frac{e^{\lambda_{+}s_{w}}-e^{\lambda_{-}s_{w}}}{1-\phi_{T}(s_{w})},
\qquad
K_{1}(s-s_{w})=e^{\lambda_{+}(s-s_{w})}-e^{\lambda_{-}(s-s_{w})}.
\end{equation}
\end{subequations}
Firstly, this factorization property shows that the position $s_{k}$
of the maximum relative to the waiting time $s_{w}$, that is,
$s_{k}-s_{w}$, is controlled by the function $K_{1}$. Thus,
$s_{k}-s_{w}$ does not depend on the waiting time, but only on the two
eigenvalues $\lambda_{\pm}$. Namely,
\begin{equation}\label{ch8_eq:max-pos}
s_{k}-s_{w}=\frac{1}{\lambda_{+}-\lambda_{-}}
\ln\left(\frac{\lambda_{-}}{\lambda_{+}}\right) \underset{\beta=1}{\simeq} 0.442.
\end{equation}
Secondly, the height of the maximum $K_{\max}$ does depend on the
waiting time $s_{w}$ due to the factor $K_{0}(s_{w})$. Specifically,
it can be shown that $K_{\max}$ is a monotonically decreasing function
of the waiting time $s_{w}$ that vanishes in the limit as
$s_{w}\to\infty$.

\begin{figure}
\centering
\includegraphics[width=0.8\textwidth]{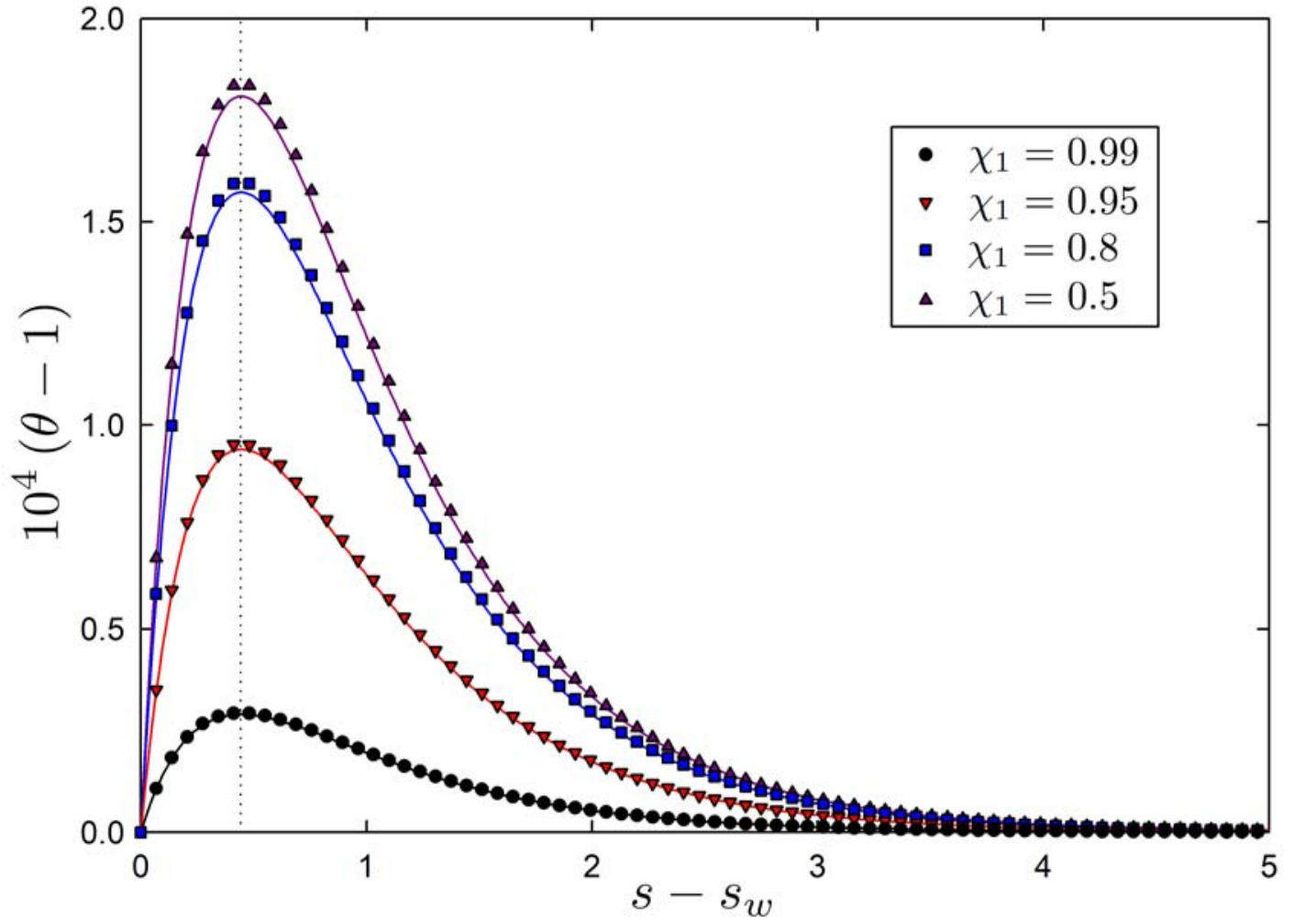}
\caption{Kovacs hump in linear response. 
Initial and final drivings, $\chi_{0}=1.05$ and $\chi=1$ are fixed.
We have considered four values for the intermediate driving, namely
 $\chi_1=\{0.5,0.8,0.95,0.99\}$. The linear response result
 \eqref{ch8_eq:Kovacs-factorization-and-factors} (solid line) perfectly
 agrees with the numerical solution of the first Sonine approximation
 \eqref{ch8_eq:reduced-var-evol} (symbols). In addition, we plot the
 theoretical prediction for the maximum~\eqref{ch8_eq:max-pos}, which
 again agrees with the numerics (dotted
 line). \label{ch8_fig3}}
\end{figure}

In order to check the above results, we have fixed the initial and
final drivings in the Kovacs protocol $\chi_{0}$ and $\chi$ and
changed the intermediate driving value $\chi_{1}$. We do so to
simplify the comparison, because the time scale $s$ involves the
steady value of the temperature; see \eqref{ch8_eq:theta-A2-def}.
Note that the smaller $\chi_{1}$ is, the shorter the waiting time
becomes. Therefore, one expects to get a Kovacs hump whose maximum
remains at $s-s_w\simeq 0.44$, but raises as $\chi_{1}$ decreases. This
is shown in figure \ref{ch8_fig3}, where~the numerical solution of the
first Sonine approximation \eqref{ch8_eq:reduced-var-evol} and
the analytical result \eqref{ch8_eq:Kovacs-factorization-and-factors} are
compared. Their agreement is almost perfect for the two lowest
curves, corresponding to $\chi_{1}=0.99$ and $\chi_{1}=0.95$, as
expected, but is still remarkably good for the two topmost ones,
corresponding to the not-so-small jumps for $\chi_{1}=0.8$ and
$\chi_{1}=0.5$.

\subsubsection{Nonlinear regime}

\begin{figure}
\centering
\includegraphics[width=0.8\textwidth]{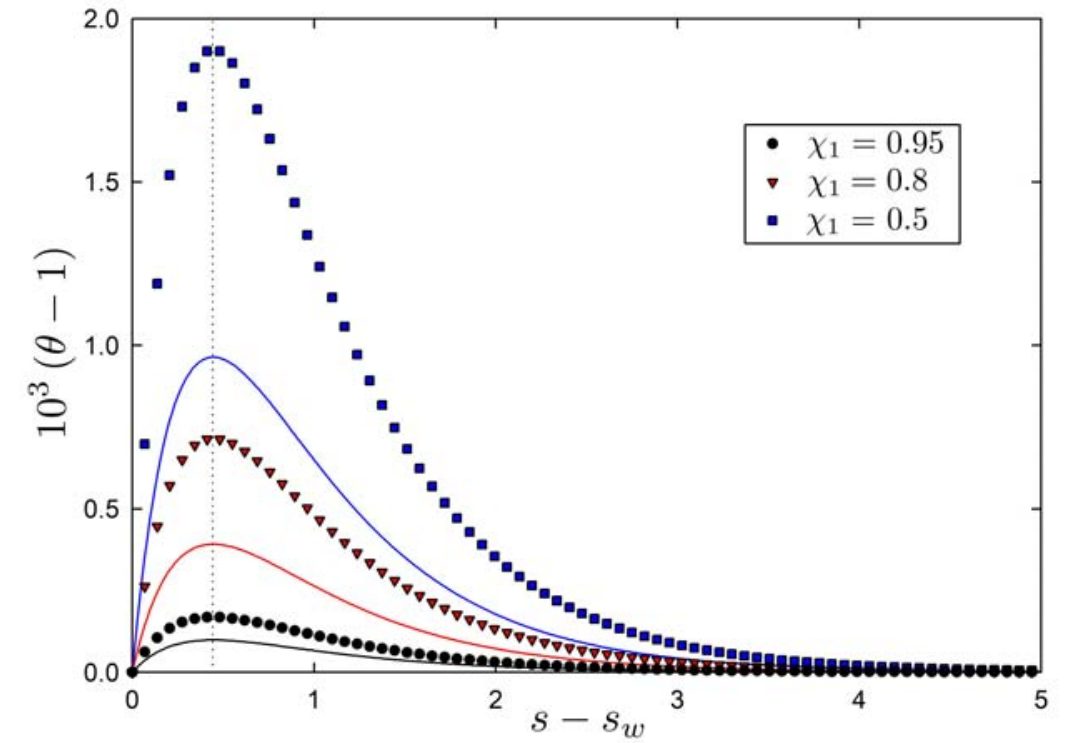}
\caption{Kovacs hump in the nonlinear regime. The
 initial driving is much higher than that in figure \ref{ch8_fig3},
 $\chi_{0}=10$, whereas the final and intermediate values of the
 driving are the same, $\chi=1$ and $\chi_1=\{0.5,0.8,0.95\}$. The
 linear response theoretical expression \eqref{ch8_eq:Kovacs-Sonine}
 (solid line) seriously underestimates the actual humps, given by
  the numerical solutions of the
 first Sonine approximation \eqref{ch8_eq:reduced-var-evol}
 (symbols). However, the theoretical prediction for the maximum 
 position in linear
 response \eqref{ch8_eq:max-pos} (dotted line) 
is still quite accurate, 
  see also figure \ref{ch8_fig5}. \label{ch8_fig4}}
\end{figure}

Furthermore, we explore the Kovacs effect out of the linear
regime. Figure~\ref{ch8_fig4} is similar to figure~\ref{ch8_fig3}, but for
larger temperature (or driving) jumps. We have also fixed the initial
and final values of the driving, $\chi_{0}=10$ and $\chi=1$. The
intermediate values of the driving are the same as in the linear case
except for the largest one, $\chi_{1}=0.99$, which we have omitted for
the sake of clarity---its hump is too small in the scale of the
figure. Now, the linear response theory results just provide the
qualitative behavior of the hump, correctly predicting the position
of the maximum, but not its height. On the one hand, and consistently
with the numerical results in an active matter model
\cite{KSyI17}, the Kovacs hump out of the linear response
regime is larger than the prediction of linear response theory. On
the other hand, the position of the maximum remains basically
unchanged, and its height still increases as $\chi_{1}$~decreases.

We also compare our analytical expansion in $a_{2}^{\ness}$ with the
numerical solutions of  \eqref{ch8_eq:reduced-var-evol} for large
jumps. Specifically, in order to make things as simple as possible, we
choose $\chi_{1}=0$. If the waiting time is long enough, the system
reaches the homogeneous cooling state and
$a_{2}(s_{w})=a_{2}^{\text{HCS}}$, which is then the initial condition
for the final stage of the Kovacs protocol. Moreover, we can compute
the location of the maximum of the hump from 
\eqref{ch8_eq:expansion_a2s-theta}, obtaining
\begin{equation}
\label{maximium_expansion}
s_{k}-s_{w}=\frac{2}{\kappa_{2}-2-\beta} \ln \left( \frac{2\kappa_{2}}{2+\beta}\right) \underset{\beta=1}{\simeq} 0.437.
\end{equation}
This result is numerically indistinguishable from that of linear
response, as given by \eqref{ch8_eq:max-pos}, since the relative
error is around $1 \%$.

\begin{figure}
\centering
\includegraphics[width=0.8\textwidth]{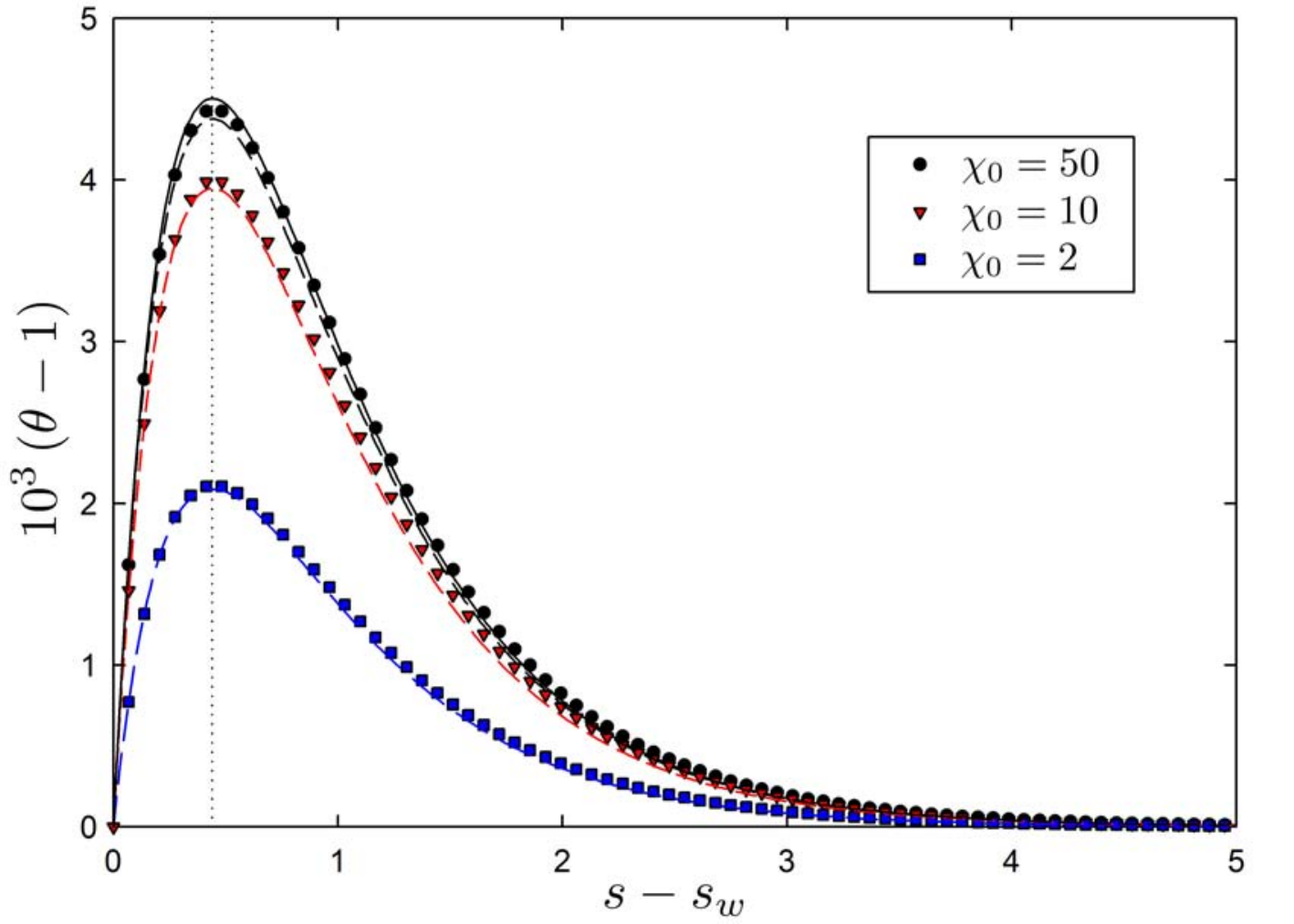}
\caption{Kovacs hump in the nonlinear regime.
 We have
 considered the following values of the drivings:
 $\chi_0=\{2,10,50\} $, $\chi_1=0$ and $\chi=1$. Symbols stand for
 the numerical solutions of the first Sonine approximation
 \eqref{ch8_eq:reduced-var-evol}, whereas lines correspond to the
 theoretical expression stemming from a perturbative expansion in $a_2^s$, 
 as given by  \eqref{ch8_eq:expansion_a2s-theta}. For the solid
 line, $a_{2}^{\text{ini}}=a_{2}^{\text{HCS}}$, whereas
  we have used the value of $a_{2}^{\text{ini}}$ in the
 numerical solution  for the dashed lines. 
 An almost perfect agreement is observed. Finally, we~
 also mark the theoretical prediction for the maximum position in
 nonlinear response \eqref{maximium_expansion} (dotted line),
 which also shows an excellent agreement
 with the numerics. \label{ch8_fig5}}
\end{figure}

In figure \ref{ch8_fig5}, we put forward a comparison between the Kovacs
hump obtained from the numerical solution of the first Sonine
approximation equations and our theoretical expression for the
nonlinear regime, as given by \eqref{ch8_eq:expansion_a2s-theta}. Fixing
$\xi_{1}=0$ and $\xi=1$, as $\xi_{0}$ increases (as the waiting time
is increased), the~hump approaches  \eqref{ch8_eq:expansion_a2s-theta}
with $a_{2}^{\text{ini}}=a_{2}^{\text{HCS}}$. Moreover, our theory
perfectly reproduces all the simulations when we substitute the
actual values of $a_{2}^{\text{ini}}$ into 
\eqref{ch8_eq:expansion_a2s-theta}.

\subsection{Time evolution of the $H$-functional}

The nonmonotonicity in the relaxation of the granular temperature
that is brought about by the Kovacs protocol is not automatically
transferred to other relevant physical magnitudes. Specifically, here,
we deal with the $H$-functional
\begin{equation}\label{ch8_eq:H-func}
H(t)=\int_{-\infty}^{+\infty}dv P_1(v,t) \log \left[\frac{P_1(v,t)}{P_1^{(\st)}(v)} \right],
\end{equation}
which we have analyzed in chapter \ref{ch:Hth_pgran}. Therein, 
we have analytically proven that 
$H(t)$ is a Lyapunov functional in
our system for the Maxwell collision rule, which corresponds to the case $\beta=0$.
However, it is precisely for this case that the Kovacs effect vanishes, since 
the evolution equation for the temperature is closed. Therefore, we look into the 
time evolution of $H(t)$ for the hard particle case $\beta=1$ in this section. Note 
that, in addition to our rigorous proof in the model for the Maxwell case, there is 
strong numerical evidence of $H(t)$ being a Lyapunov functional for granular 
systems \cite{MPyV13,GMMMRyT15}.

\begin{figure}
\centering
\includegraphics[width=0.8\textwidth]{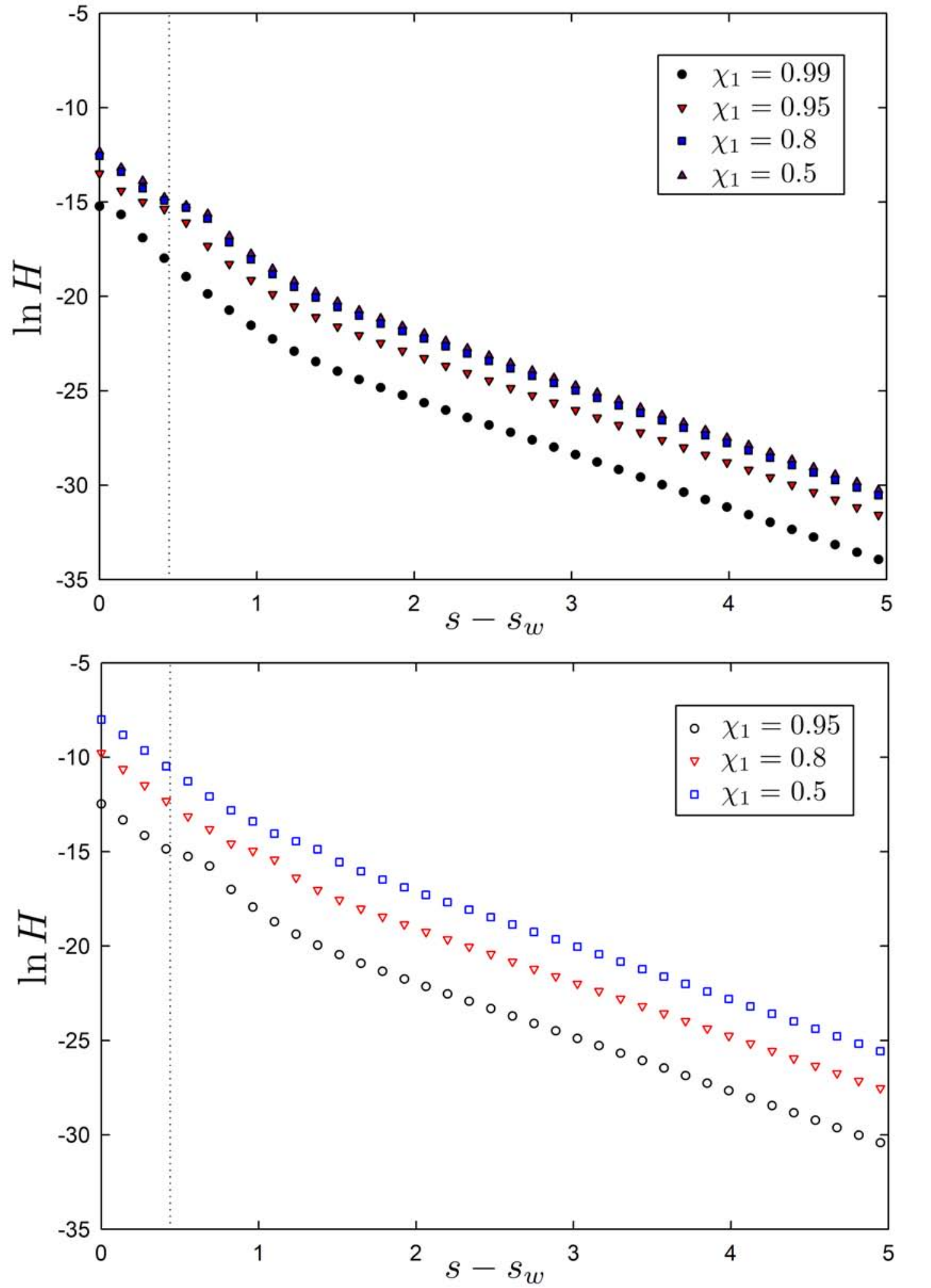}
\caption{Time evolution of the $H$-functional in the Kovacs experiments. 
The top and bottom panels correspond, respectively, to the protocols
 in figures \ref{ch8_fig3} (filled symbols) and \ref{ch8_fig4} (open symbols),
 that is, to the linear and nonlinear regimes. The~vertical dotted
 line marks the maximum of the hump in the
 corresponding regime. \label{ch8_fig6}}
\end{figure}

We have computed $H(t)$ numerically from \eqref{ch8_eq:H-func}
within the first Sonine approximation, that is, we have
 substituted both $P_1(v,t)$ and $P_1^{(\st)}(v)$ by their expressions in
 the first Sonine approximation and calculated the integral
 numerically. This has been done for the Kovacs protocols considered in figures~
\ref{ch8_fig3}~and~\ref{ch8_fig4}.
The results are shown in figure \ref{ch8_fig6} and make it clear
that $H(t)$ still monotonically decreases for the Kovacs-like
protocols, in both the linear (top panel) and nonlinear (bottom panel)
regimes. At the time of the maximum in the hump, $s-s_{w}\simeq 0.44$,
no special signature is observed in the ``entropy'', which would be given 
by $-k_B H(t)$.

\bookmarksetup{startatroot}

\chapter{Conclusions}
\pagestyle{mystyle4}
\label{ch:conclusions}
This thesis intends to be a showcase for the relevance of simple models as 
useful tools for improving our understanding of complex systems. 
We have analyzed, mainly,
two models of very different nature with the fundamental tools of the statistical 
mechanics. On the one hand, we have studied a model capable of predicting the
deterministic unfolding pathway in modular biomolecules. On the other hand, 
we have thoroughly
explored a model with nonconservative interactions, 
motivated within the field of soft matter and granular media.

In this final chapter we enumerate the major conclusions of this thesis
as well as some future perspectives. 
Regarding the first part devoted to the biophysical model: 
\begin{enumerate}
\item We have put forward a 1d model for modular biomolecules that comprises several consecutive nonharmonic units, which is submitted to a mechanical pull that tries to control the total length of the chain. Langevin dynamics is assumed for the endpoints of the different units.
\item Dynamical equations are solved, in the limit of negligible thermal noise, by means of a perturbative approach in both the pulling velocity and the asymmetry between units. Within this quite drastic approach, the dynamics becomes deterministic. With no thermal activation being allowed, the only mechanism  for unfolding is reaching the stability threshold, where the folded state ceases to exist.  
\item We derive the critical values of the pulling velocity at which the unfolding pathway changes. For slow enough pulling, it is the weakest unit the first to unfold. By contrast, for fast enough pulling, the first unit to unfold is the closest to the pulled end. Intermediate situations are also possible, depending on the details of the system. Comparison between numerical solutions of specific systems and theoretical predictions shows a remarkably good agreement. 
\item The applicability of the model has been carefully looked into. The pulling velocity has to lie within a very specific range: low enough to make our perturbative approach possible, and, at the same time, high enough to prevent thermal activation and  make the system follow the maximum hysteresis path. These conditions make it difficult to choose the ``right'' molecule to test our theory.
\item The robustness of the theoretical model has been tested by varying the model in different ways. The described phenomenology does not depend on the details of the pulling device. Specifically, we have analyzed both the impact of the location of the elastic reaction of the pulling device and the finite value of its stiffness. The latter leads to an imperfect control of the length.
\item Steered molecular dynamics of a model system composed of two consecutive coiled-coil structures have been performed. The statistical study of the trajectories reveals (i) an excellent agreement with the theoretical prediction for high enough pulling velocities and (ii) anisotropic behavior of the molecule.
\item We have limited ourselves to the investigation of the first unfolding event. However, our argument can be easily generalized to the next unfolding event: the difference is that the zero-th order approximation is no longer given by all the units sweeping the all-units-folded branch, but by the sweeping of the branch with one module unfolded and the remainder folded. Then, a similar perturbative expansion around this zero-th order solution, in powers of the asymmetry and the pulling speed, would give the next unit that opens.
\item Although we focus on the biophysical application of this model, it is worth highlighting that similar models have been employed in other fields. Many physical systems are also ``modular'', since they comprise several units \cite{BZDyG16}, and thus a similar phenomenology may emerge. Some examples can be found in studies of plasticity \cite{MyV77,PyT02}, lithium-ion batteries \cite{DJGHMyG10,DGyH11}, and ferromagnetic alloys \cite{BFSyG13}.
\item Finally, we briefly comment on some prospects for future work:
\begin{enumerate}
\item As stressed throughout the thesis, we have considered
the deterministic approximation of the Langevin dynamics. 
Including temperature effects,
in a way similar to that of Kramers theory \cite{Kr40,HTyB90,Be78,EyR97}, 
would complete our theoretical approach.
\item We assume that the dynamical stability threshold equals that of
  the static case.  We have discussed how seeking a dynamical
  criterion could lead to an improvement of our theory, reducing its
  discrepancies with the numerical results.
\item After analyzing in detail the unfolding scenario in a simple molecule
such as the two coiled-coil construct, the next natural  step is exploring more 
complex molecules, in which the modules are not identical. 
Therein, we expect the emergence of the predicted critical
velocities.
\item Once simulations have supported our theoretical prediction, the 
desirable continuation would be going to real experiments. In the current state
of the theory, where thermal activation is neglected, the difficulties of this
stage are neat: the test molecule should be carefully chosen. 
\end{enumerate}
\end{enumerate}

\newpage 

Concerning the second part devoted to the granular model: 
\begin{enumerate}
\item We have analyzed a simple model for the ``transversal'' component of the velocity of a granular fluid. The model captures many of the essential features of granular fluids, for example the existence of some characteristic ``basic'' states like the Homogeneous Cooling State and the Uniform Shear Flow steady state.
\item Specifically, we have focused  on the continuum
  limit of the one-particle distribution function and its
  solutions. In this regard, we would like to stress that we have completely characterized the Homogeneous Cooling
  State and the Uniform Shear Flow steady state. Also, we have compared, with excellent results, our theoretical predictions with numerical simulations.   
\item We have looked into the relevant fluctuating fields of the
  model. Some of them need a proper regularization since their
  definitions in the continuum limit lead to singularities. This
  issue, although  known in the literature \cite{BSGJyL15}, is neatly understood within our theoretical framework.
\item We have exactly solved the average two-particle correlations in a system of arbitrary size, by reducing the solution of the evolution equations to an eigenproblem. The obtained spectrum is particularly interesting, since it allows us to explain the shear instability of the system as the crossing of the two first eigenvalues. Moreover, it improves our understanding of  the situation beyond the instability ($\nu>\nu_c$) in the model: we find that both the energy and the correlations still decay to zero but with a rate that is independent of the inelasticity. 
\item The $H$-theorem we have derived here is, to the best of our knowledge, the first rigorous proof of its kind in systems  with nonconservative interactions. Not only have we proven the global stability of a family of nonequilibrium steady solutions in our simplified model, but also understood why the ``Boltzmann entropy'' $H_B$ cannot be a good candidate for a Lyapunov functional in inelastic systems. In addition, our work strongly supports previous numerical evidence \cite{GMMMRyT15,MPyV13} in favor of the master-equation-like $H$-functional being the ``correct'' Lyapunov functional in the granular case.
\item The ideas introduced in our proof of the $H$-theorem (expansion of the distribution and splitting of the evolution operator) could be useful in more complex systems, closer to the real granular gas. If this extension is not possible, one would like, at least, to show that the long-time solutions are globally stable by establishing that $H$ is asymptotically nonincreasing, similarly to what has been done in this thesis.  
\item Also, we have analyzed Kovacs-like memory effects in our model. We have done so by introducing a kinetic approach---specifically, a Sonine expansion---in our model. This Sonine expansion makes it possible to close the temperature equation  together with that for the excess kurtosis.  Theoretical predictions for the linear response have been successfully validated by the numerical results.  We have also checked that the nonmonotonic behavior of the temperature in the Kovacs experiment is utterly compatible with the monotonic decrease of the $H$-functional to its steady value. 
\item The theoretical framework developed for the study of memory effects here goes beyond its application to our model. In particular, it establishes  the basis for a general approach to this kind of memory effects, within linear response, in athermal systems. In this regard, it is worth emphasizing that our approach is very general, since it works both at the mesoscopic and macroscopic levels of description.
\item Also for the granular part of the thesis, we discuss some possibilities for further development:
\begin{enumerate}
\item Throughout the thesis, we have focused on 
the study of the ``Maxwellian'' case $\beta=0$  of our
granular model. The investigation of the general case for $\beta \neq 0$
is worthy of further consideration. In particular, the case $\beta=1$ leads to a collision rate analogous to that of hard particles. 
For instance, the nonequilibrium stationary
states in the general case are more involved and deserve to be
analyzed in depth.
\item Considering other physically relevant energy injection mechanisms
would extend the applicability of the model. This generalization could lead us
closer to a ``master'' lattice model, which merge the results stemming from 
this thesis with other works, as those in \cite{PSyD13}. 
\item One of the most appealing future prospects is the possibility of
  exploring a derivation of global stability or even an $H$-theorem
  for the inelastic Boltzmann equation, by using ideas similar to those presented
  in chapter \ref{ch:Hth_pgran}.
\item A complete theoretical description of the Kovacs effect out of
  the linear response regime is lacking. Extending our results to the
  nonlinear regime would be, undoubtedly, of utmost relevance in
  nonequilibrium physics.
\end{enumerate}
\end{enumerate}

\clearpage

\appendix

\chapter{Stability threshold} \label{ch2:appA}

\pagestyle{mystyle3}

To first order in $\xi$,  the extension $x_{i,b}$ such that
$a''_{i}(x_{i,b})=0$ verifies
\begin{equation}
  \label{ch2_eq:a0}
  a'''(\ell_{b})(x_{i,b}-\ell_{b})+\xi \deltaf'_{i}(\ell_{b})=0,
\end{equation}
that is,
\begin{equation}
  \label{ch2_eq:a1}
  x_{i,b}=\ell_{b}-\xi \frac{\deltaf'_{i}(\ell_{b})}{a'''(\ell_{b})}.
\end{equation}
The corresponding force at the stability threshold is obtained from
\eqref{ch2_eq:10}. To the lowest order in the deviations,
\begin{equation}
  \label{ch2_eq:28}
  F_{i,b}\equiv a'_{i}(x_{i,b})\sim a'_{i}(\ell_{b})=F_{b}+\xi\, \deltaf_{i}(\ell_{b}),
\end{equation}
because the next term, $a'''(\ell_{b})(x_{i,b}-\ell_{b})^{2}/2$, is of
the order of $\xi^{2}$. Therefore, the $i$-th module reaches its limit
of stability at the time for which $x_{i}=x_{i}^{(0)}+\xi\delta
x_{i}=x_{i,b}$, that is, when the length per unit $\ell$ has the value
$\ell_{i}$ verifying
\begin{equation}
  \label{ch2_eq:a2}
  \ell_{i}+\xi  
\frac{\overline{\deltaf}(\ell_{i})-\deltaf_{i}(\ell_{i})}{a''(\ell_{i})}=
\ell_{b}-\xi \frac{\deltaf'_{i}(\ell_{b})}{a'''(\ell_{b})},
\end{equation}
or, equivalently,
\begin{equation}
  \label{ch2_eq:a3}
  \ell_{i}-\ell_{b}=\xi \frac{\deltaf_{i}(\ell_{i})-\overline{\deltaf}(\ell_{i})}{a''(\ell_{i})}-\xi \frac{\deltaf'_{i}(\ell_{b})}{a'''(\ell_{b})}.
\end{equation}
We know that $\ell\to\ell_{b}$ when $\xi\to 0$. But $a''(\ell_{b})=0$
and thus we cannot substitute $\ell_{b}$ on the rhs of \eqref{ch2_eq:a3}. 
On the other hand, this means that the dominant balance for $\xi\to 0$ involves the 
lhs and the first term on the rhs of \eqref{ch2_eq:a3}. 
Therefore, making use of $a''(\ell)\sim
a'''(\ell_{b})(\ell-\ell_{b})$, we get 
\begin{equation}
  \label{ch2_eq:a4}
  (\ell_{i}-\ell_{b})^{2} \sim \xi \frac{\deltaf_{i}(\ell_{b})-\overline{\deltaf}(\ell_{b})}{a'''(\ell_{b})}.
\end{equation}
Since $a'''(\ell_{b})<0$ (see figure \ref{fig:2ch2}), this means that only
the units with $\deltaf_{i}(\ell_{b})$ smaller than the average (that
is, weaker than average) reach the limit of stability in the limit as
$v_p \to 0$. 
In fact, it is the weakest unit, that is, the unit with
smallest $\deltaf_{i}(\ell_{b})$, that unfolds first.

It is interesting to note that, in order to obtain \eqref{ch2_eq:a4},
we have completely neglected the last term on the rhs of
\eqref{ch2_eq:a3}. 
Since, in turn, this term stems from the last term
on the rhs of \eqref{ch2_eq:a1}, to the lowest order we are solving
the equation $x_{i}=\ell_{b}$. In other words, to the lowest order the
stability threshold can be considered to be given by the
non-disordered, zero-asymmetry case, free energy $a(x)$. For the
sake of concreteness and simplicity, we have stuck to the asymmetry
contribution $\delta x_{i}$ in this appendix, but the same condition
$x_{i}=\ell_{b}$ would still be valid, had we taken into account the
kinetic contribution $\Delta x_{i}$ derived in
section \ref{sec:pulling_speed}. The reason is that there is also a
factor $a''(\ell)$ in the denominator of $\Delta x_{i}$, see
\eqref{ch2_eq:22a}, and thus both the terms coming from $\delta x_{i}$ and
$\Delta x_{i}$ are dominant against the last term on the rhs of
\eqref{ch2_eq:a1}.

\chapter{Fluctuating expression for the dissipation}\label{ch5-6_app-a}

Let us consider the dissipation $d_{l,p}$ at site $l$ and at time $p$.
Its main part is obtained by averaging \eqref{micro-ener-dis} with
respect to the fast variables $y_{p}$, i.e.
\begin{equation}
  \label{eq:d-main}
  \bar{d}_{l,p}= \frac{\alpha^2-1}{4L}\left(\Delta_{l,p}^2+\Delta_{l-1,p}^2 \right)<0.
\end{equation}
This is the expression that we have to analyze in the fluctuating
hydrodynamic description, since the amplitude of the dissipation
noise scales as $L^{-3}$ \cite{MPLPyP16,AM_thesis}.  If we consider the average
  of the dissipation field, it is readily obtained that
  $d^{\av}_{l,p}=(\alpha^{2}-1)T_{l,p}/L$, which gives
  \eqref{eq:av-d-nu} in the continuum limit by using
  $d(x,t)=L^{3}d_{l,p}$ and the definition of $\nu$. Therefore, it is
  consistent to write at the fluctuating level that
\begin{equation}
  \label{eq:d-fluct}  
\bar{d}_{l,p}=\frac{\alpha^{2}-1}{L} \theta_{l,p},
\end{equation}
by defining the fluctuating temperature as
\begin{equation}\label{eq:temp-fluct}
\theta_{l,p}=\frac{1}{4}\left(\Delta_{l,p}^2+\Delta_{l-1,p}^{2}\right)=\frac{v_{l-1,p}^{2}+2v_{l,p}^{2}+v_{l+1,p}^{2}}{4}-v_{l,p}\frac{v_{l+1,p}+v_{l-1,p}}{2}.
\end{equation}
The first term on the rhs,
$(v_{l-1,p}^{2}+2v_{lp}^{2}+v_{l+1,p}^{2})/4$, reduces to $e_{l,p}$
plus terms of the order of $L^{-2}$, which are neglected.

Our main goal is to obtain a correct expression for $v_{l,p}v_{l\pm1,p}$ at
the fluctuating level. In general, we have for the average correlations
\begin{equation}
  \label{eq:discrete-corr}
  \langle v_{l,p}
  v_{l',p}\rangle=E_{l,p}\delta_{ll'}+C_{l,l'-l;p}\left(1-\delta_{ll'} \right),
\end{equation}
with the definition 
\begin{equation}
  \label{eq:def-C-dos-l}
  C_{l,l'-l;p}=\langle v_{l,p} v_{l',p}\rangle \quad \text{for $l'\neq
  l$}.
\end{equation}
The functions $C_{k,p}$ defined in section~\ref{sec:HCS-exact}
are the particularisation of $C_{l,l'-l;p}$ to a homogeneous
situation ($k=l'-l$). Consistently with \eqref{eq:discrete-corr}, we
write
\begin{equation}\label{eq:def-gamma-dos-l}
v_{l,p}v_{l',p}=e_{l,p}\delta_{ll'}+\gamma_{l,l'-l,p}\left(1-\delta_{ll'}\right)=\gamma_{l,l'-l,p}+\left(e_{l,p}-\gamma_{l,l'-l,p}\right)\delta_{ll'},
\end{equation}
at the fluctuating level. We have introduced the fluctuating
correlations $\gamma_{l,l'-l,p}$, such that
$\langle\gamma_{l,l'-l;p}\rangle=C_{l,l'-l,;p}$. In the continuum
limit, $x=(l-1)/L$ and $x'=(l'-1)/L$ and~\eqref{eq:def-gamma-dos-l} is transformed into
\begin{equation}\label{eq:def-gamma-continuum}
v(x,t)v(x',t)=\gamma(x,x'-x;t)+L^{-1}\delta(x-x')\left[e(x,t)-\gamma(x,x'-x;t)\right],
\end{equation}
because $\delta_{l,l'}\sim L^{-1} \, \delta(x-x')$, see note at the
end of the appendix. 

Taking into account~\eqref{eq:temp-fluct} and
the above definitions, the fluctuating temperature in the continuum
limit is
\begin{equation}
\theta(x,t)=e(x,t)-\gamma(x,0;t),
\end{equation}
where we have neglected terms of the order of $L^{-2}$. Since we are interested in the limit of $\gamma(x,\Delta x;t)$ when
$\Delta x\to 0$, we use~\eqref{eq:def-gamma-continuum} with
$\Delta x=x'-x=\pm L^{-1}$ to obtain
\begin{equation}
  \label{eq:v2-continuum}
  v^{2}(x,t)=\gamma(x,0;t)+L^{-1}\left[ e(x,t)-\gamma(x,0;t)
  \right]\lim_{x'\to x}\delta(x'-x).
\end{equation}
 Thus, we have that
\begin{equation}
  \label{eq:gamma-x-0-def}
 \gamma(x,0;t)=v^{2}(x,t)-L^{-1}\theta(x,t)
  \lim_{x'\to x}\delta(x'-x).
\end{equation}
Note that $v^{2}(x,t)$ always has a singular part that stems from the
$\delta(\Delta x)$ factors on the rhs
of~\eqref{eq:v2-continuum}. Therefore, $\gamma(x,0;t)$ can be
considered as the ``regular'' part of $v^{2}(x,t)$, and  we
introduce the notation
\begin{equation}
  \label{eq:vR2-def-app}
  v_{R}^{2}(x,t)\equiv \gamma(x,0;t)=v^{2}(x,t)-L^{-1}\theta(x,t)
  \lim_{x'\to x}\delta(x'-x).
\end{equation}

By combining the previous results, and recalling that
$d(x,t)=L^{3}d_{l,p}$, we finally conclude
\begin{equation}
  \label{eq:d-cont}
  d(x,t)=-\nu\theta(x,t), \quad \theta(x,t)= e(x,t)-v_{R}^{2}(x,t).
\end{equation}
This tells us that the fluctuations of the dissipation field are
enslaved to those of the temperature. Moreover, the appearance of
$v_{R}^{2}$ in \eqref{eq:d-cont} is easy to understand on a physical
basis, since
$\langle v_{R}^{2}(x,t)\rangle=\langle\gamma(x,0;t)\rangle=
u^{2}(x,t)+O(L^{-1})$.
Equations~\eqref{eq:vR2-def-app} and \eqref{eq:d-cont} make it
possible to write a closed expression for the fluctuating temperature,
  \begin{equation}
    \label{eq:theta-closed}
    \theta(x,t)=\beta \left[e(x,t)-v^{2}(x,t)\right], \quad \beta=\left[1-L^{-1}
  \lim_{x'\to x}\delta(x'-x)\right]^{-1},
  \end{equation}
in which $\beta$ is a regularisation factor, which ``heals'' the
singularity of $v^{2}(x,t)$ in the large system size limit.

\underline{Note:} The appearance of $\delta(0)$---more accurately,
$\lim_{x'\to x}\delta(x'-x)$---can be avoided in the
following way: for discrete $(l,l')$ we may write 
\begin{equation}
  \delta_{ll'}=\Theta(l-l'+1/2)\Theta(l'-l+1/2),
\end{equation}
in which $\Theta(x)$ is the Heaviside step function. Therefore, in the
continuum limit,  we have that
\begin{equation}
  \delta_{ll'}\sim\Theta\left(x-x'+\frac{1}{2L}\right)\Theta\left(x'-x+\frac{1}{2L}\right).
\end{equation}
When used inside an integral, the relative error introduced by using
the expression above is of the order of $L^{-2}$, since
\begin{align}
   \sum_{l=1}^{L} f_{l} \, \delta_{ll'}&=f_{l'},       \\
  L \int_{0}^{1}dx f(x) \,\Theta\left(x-x'+\frac{1}{2L}\right)\Theta\left(x'-x+\frac{1}{2L}\right)&= \int_{x'-\frac{1}{2L}}^{{x'+\frac{1}{2L}}}dx \,
  f(x)  \nonumber \\
  &= f(x')+O(L^{-2}).
\end{align}
Therefore, both expressions, (i)
$L^{-1}\delta(x-x')$ and (ii) the product of Heaviside functions, can
be used indistinctly within the mesoscopic fluctuation framework.

Consistently with the above discussion, the Fourier components of the
product of Heaviside functions are the same as those of
$L^{-1}\delta(x-x')$, with a relative error of the order of $L^{-2}$,
\begin{align}
  \int_{0}^{1} dx\, \Theta\left(x-x'+\frac{1}{2L}\right)\Theta\left(x'-x+\frac{1}{2L}\right) e^{-i k_{n}x}&=\int_{x'-\frac{1}{2L}}^{{x'+\frac{1}{2L}}}dx
  e^{-ik_{n}x} \nonumber \\
  &=L^{-1}e^{-i k_{n}x'}+O(L^{-3}).
\end{align}
Therefore,
\begin{align}  \Theta\left(x-x'+\frac{1}{2L}\right)\Theta\left(x'-x+\frac{1}{2L}\right)&=L^{-1}\sum_{n}e^{i
    k_{n} (x-x')}+O(L^{-3}) \nonumber \\
&= L^{-1}\delta(x-x')+O(L^{-3}).
\end{align}

{\clearpage \thispagestyle{empty}}

\chapter{Simulation method}\label{ch5-6_app-b}

Simulations have been made reproducing $M$ times the phase-space
trajectory of a system of $N$ particles, each one carrying a velocity
$v_l$ and being at a definite position $l=1,\ldots,L$, with $L=N$ for
periodic or Lees-Edwards boundaries.  For each trajectory, the system starts with a random
extraction of velocities $v_l$ normally distributed with
$\left\langle v_l \right\rangle =0$ and
$\left\langle v^2_l \right\rangle = T_0$, unless otherwise
specified. Afterwards, we move to the centre of mass frame making the
transformation
$v_l \Rightarrow v'_l = v_l - \frac{1}{L} \sum^L_{l=1} v_l$, so that
the total momentum of the system is zero.

We carry out the Monte Carlo simulation of the system time-evolution
through the residence time algorithm, which gives the
numerical integration of a master equation in the limit of
infinite trajectories \cite{BKyL75,PByS97}. The
basic numerical recipe is as follows:
\begin{enumerate}

\item\label{stepone} At time $\tau$, a random ``free time''
 $\tau_{f} > 0$ is extracted with an exponential probability density
 $\Omega(\bm{v})\exp[-\Omega(\bm{v})\tau_{f}]$, where
 $\Omega(\bm{v})=\sum_l \omega |v_l-v_{l+1}|^\beta$ depends on the
 state of the system $\bm{v}$;

\item Time is advanced by such a free time
$\tau \to \tau+\tau_f$;

\item A pair $(l,l+1)$ is chosen to collide with probability
 $\omega|v_l-v_{l+1}|^\beta /\Omega(\bm{v})$;
 
\item The chosen pair collide following the collision rule \eqref{coll_rule}.

\item In the case of the thermostated system, all particles are heated by the stochastic thermostat, by adding
 independent Gaussian random numbers of zero mean and variance $\chi\tau_{f}$ to their velocities;

 \item In the case of the thermostated system, the mean value of the random
 numbers generated in the previous step is subtracted from the
 velocities of all particles to conserve momentum;

\item The process is repeated from Step~\ref{stepone}.
\end{enumerate}

Regarding the measurement of $P_1(v;x,t)$, we sample both position and
velocity spaces by defining $N_x$ bins of width $\Delta x$ and $N_v$
bins of with $\Delta v$. Of course, the product $N_x \Delta x = 1$,
covering the whole lattice, whereas $N_v \Delta v$ gives the range of
velocities bounded by the cutoffs $v_\text{\text{min}}$ and
$v_{\text{max}}$. In our simulations, we control that the contribution
to the PDFs coming from velocities outside the considered interval
$[v_{\text{min}},v_{\text{max}}]$ is negligible. With such a binning,
we build up an histogram and therefrom the distribution function
$P_1(v;x,t)$, which is represented by a $N_x \times N_v $ matrix for
each time $t$. Functionals ($H$ and $H_{B}$) are computed by numerically
replacing the integral over $x$ and $v$ with sums over the prescribed
bins.

\chapter{Lees-Edwards boundary conditions} \label{app-lees}

Here, we derive the Lees-Edwards boundary conditions in the continuum limit 
\eqref{ch5_eq:lees-edwards}, starting from the collision rules for the
boundary sites of our lattice.
As stated in section 
\ref{ch5_sec:haveUSF}, the collision rule for the pair $(1,L)$ is
\begin{subequations}\label{coll_rule_esp}
\begin{align}
v_{L,p+1} &= v_{L,p}-\frac{1+\alpha}{2}(v_{L,p}-v_{1,p}-a), \\
v_{1,p+1} &= v_{1,p}+\frac{1+\alpha}{2}(v_{L,p}-a-v_{1,p}),
\end{align}
\end{subequations}
which, as the bulk rule \eqref{coll_rule}, conserves momentum. The evolution equation for
$v_1$ is readily obtained,
\begin{equation}
\label{eq:v-app-LE}
v_{1,p+1}-v_{1,p}=-\frac{1+\alpha}{2}\delta_{y_p,1}(v_{1,p}-v_{2,p})+ \frac{1+\alpha}{2} \delta_{y_p,L}(v_{L,p}-a-v_{1,p}),
\end{equation}
whereas for the energy we have
\begin{align}
\label{eq:E-app-LE}
e_{1,p+1}-e_{1,p}=&-(1+\alpha)v_{1,p}(v_{1,p}-v_{2,p})\delta_{y_p,1}+(1+\alpha)v_{1,p}(v_{L,p}-a-v_{1,p})\delta_{y_p,L} \nonumber \\
&+ \left( \frac{1+\alpha}{2} \right)^2 \left( v_{1,p}- v_{2,p} \right)^2 \delta_{y_p,1}
+ \left( \frac{1+\alpha}{2} \right)^2 \left( v_{L,p}- a-v_{1,p} \right)^2 \delta_{y_p,L}.
\end{align}
Taking averages in \eqref{eq:v-app-LE} and \eqref{eq:E-app-LE}, we get
\begin{subequations}
\label{eq:uyE-app-LE}
\begin{align}
u_{1,p+1}-u_{1,p}&=\frac{1+\alpha}{2L} (u_{L,p}-a-2u_{1,p}+u_{2,p}), \\
E_{1,p+1}-E_{1,p}&= \frac{1+\alpha}{L} 
\left( \left\langle v_{L,p} v_{1,p} \right\rangle -a u_{1,p} -2 E_{1,p} 
+ \left\langle v_{1,p} v_{2,p} \right\rangle \right) \nonumber \\
&+ \frac{(1+\alpha)^2}{4L} \left( 2 E_{1,p}+2 E_{2,p}+a^2
-2\left\langle v_{L,p} v_{1,p} \right\rangle \right. \nonumber \\
& \qquad \qquad \qquad \left. -2\left\langle v_{1,p} v_{2,p} \right\rangle-2au_{1,p}-2au_{L,p}\right) .
\end{align}
\end{subequations}

Now, we turn our attention to the continuum limit defined in \eqref{eq:hydro-scales-x-t}. Specifically, we identify $f_{l,p} = f \left[x=(l-1)/L, t=p/L^3\right]$, being $f (x, t)$ a “smooth” function. Therefore,
\begin{subequations}
\label{eq:ordenes-app-LE}
\begin{align}
f_{1,p}&=f(0,t), \\
f_{1,p+1}&=f(0,t+1/L^3)=f(0,t)+L^{-3} \partial_t f(0,t)+ \mathcal{O}\left(L^{-6}\right) ,\\
f_{2,p}&=f(1/L,t)=f(0,t)+L^{-1}\left.\partial_x f(x,t)\right|_{x=0}+\mathcal{O}\left(L^{-2}\right),\\
f_{L,p}&=f(1-1/L,t)=f(1,t)-L^{-1}\left.\partial_x f(x,t)\right|_{x=1}+\mathcal{O}\left(L^{-2}\right).
\end{align}
\end{subequations}
Substituting \eqref{eq:ordenes-app-LE} into \eqref{eq:uyE-app-LE}, and using that
\begin{equation}
\frac{1+\alpha}{2}=1+\mathcal{O} \left( L^{-2}\right),
\end{equation} 
one can obtain, after some algebra,
\begin{subequations}
\begin{align}
0&=u(1,t)-u(0,t)-a+L^{-1}\left[ \left.\partial_x u(x,t)\right|_{x=1} - \left.\partial_x u(x,t)\right|_{x=1} \right] + \mathcal{O}\left(L^{-2} \right), \\
0&=E(1,t)-E(0,t)-2au(1,t)+a^2 \nonumber \\
&\, \quad + L^{-1} \left[ \left.\partial_x E(x,t)\right|_{x=0}- \left.\partial_x E(x,t)\right|_{x=1} +2a \left.\partial_x u(x,t)\right|_{x=1}  \right] + \mathcal{O}\left(L^{-2} \right).
\end{align}
\end{subequations}
Above, as explicitly notated, we have only retained the two lowest orders in the expansion in powers of $L^{-1}$. Imposing that the considered orders vanish 
separately and taking into account the definition of the temperature, 
as given by \eqref{eq:definitio_energy},
finally lead us to the boundary conditions in \eqref{ch5_eq:lees-edwards}.

\chapter{Derivation of the expression for $\lowercase{d}H/\lowercase{dt}$ in a general driven
  state}\label{app_gran-dHdt-general}

Let us consider the three contributions to $dH/dt$ in
\eqref{ch7_eq:H-time-ev-total}. We start with the diffusive
one, 
\begin{align}\label{ch7_eq:dH/dt-diff-start}
\left.\frac{dH}{dt}\right|_{\diff}=\int\!\! dx\, dv\,\left( \mathcal{L}_{\diff}P_1 \right)
  \,\ln\left(\frac{P_1}{P_1^{(s)}}\right)-\int\!\! dx\,dv\,\frac{P_1}{P_1^{(s)}}\,  \mathcal{L}_{\diff}P_1^{(s)} ,
\end{align}
where $\mathcal{L}_{\diff}P_1=\xxder{P_1}$ and we have used that
$\int\!\! dx\, dv\, \xxder{P_1^{(s)}}$ vanishes identically. Integrating
by parts the first term on the rhs of \eqref{ch7_eq:dH/dt-diff-start},
the result is
\begin{equation}
\int\!\! dv\, \xder{P_1}
\left.\ln\left(\frac{P_1}{P_1^{(s)}}\right)\right|_{0}^{1}-\int\!\! dx\,
  dv\, P_1 \xder{\ln P_1} \left( \xder{\ln P_1}-\xder{\ln P_1^{(s)}}\right).
\end{equation}
Also integrating by parts the second term, one obtains
\begin{equation}
-\int\!\! dv\, \left.\frac{P_1}{P_1^{(s)}} \xder{P_1^{(s)}}\right|_{0}^{1}
+\int\!\! dx\,
  dv\, P_1 \xder{\ln P_1^{(s)}} \left( \xder{\ln P_1}-\xder{\ln P_1^{(s)}}\right).
\end{equation}
We assume that the boundary terms are equal to zero, that is,
\begin{equation}
\int\!\! dv\,\left[\xder{P_1} \ln\left(\frac{P_1}{P_1^{(s)}}\right)-\frac{P_1}{P_1^{(s)}}
    \xder{P_1^{(s)}}\right]_{0}^{1}=0.
\end{equation}
This is obviously true for Lees-Edwards and periodic boundary conditions. For the Couette state,
  in which the PDF at the boundaries is Gaussian with zero average
  velocity and a given temperature $T_{B}$ for all times, the first
  term is identically zero and the second vanishes because $\int\!\!
  dv\, P_1^{(s)}(v;x)=1$ for all $x$. Summing the two contributions to
the diffusive term above, we have 
\begin{align}\label{ch7_eq:dH/dt-diff-final}
\left.\frac{dH}{dt}\right|_{\diff}=\int\!\! dx\, dv\, P_1 \left( \xder{\ln P_1}-\xder{\ln P_1^{(s)}}\right)^{2},
\end{align}
which is \eqref{ch7_eq:H-time-ev-diff} of the main text. 

The noise term is treated along the same lines as above, but
integrating by parts in $v$ instead of $x$, since
$\mathcal{L}_{\noise}P_1=\frac{\xi}{2}\vvder{P_1}$. Therein, the boundary
terms vanish if $P_1$ and $P_1^{(s)}$ tend to zero fast
enough for $v\to\pm\infty$, and
\begin{align}\label{ch7_eq:dH/dt-noise-final}
\left.\frac{dH}{dt}\right|_{\noise}=\int\!\! dx\, dv\, P_1 \left( \vder{\ln P_1}-\vder{\ln P_1^{(s)}}\right)^{2},
\end{align}
which is \eqref{ch7_eq:H-time-ev-noise}.

Now we focus on the inelastic contribution,
\begin{align}\label{ch7_eq:dH/dt-inel-start}
\left.\frac{dH}{dt}\right|_{\inel}=\int\!\! dx\, dv\, \left( \mathcal{L}_{\inel}P_1\right)
  \,\ln\left(\frac{P_1}{P_1^{(s)}}\right)-\int\!\! dx\,dv\, \frac{P_1}{P_1^{(s)}}\, \mathcal{L}_{\inel}P_1^{(s)} ,
\end{align}
in which $\mathcal{L}_{\inel}P_1=\frac{\nu}{2}\vder{[(v-u)P_1]}$. Then,
\begin{align}
\left.\frac{dH}{dt}\right|_{\inel}&=\;\frac{\nu}{2}\int\!\! dx\, dv\,\left\{ \vder{[(v-u)P_1]}\right\}
  \,\ln\left(\frac{P_1}{P_1^{(s)}}\right)
\nonumber \\
& \, \quad -\frac{\nu}{2}\int\!\! dx\,dv\,\left\{ \vder{[(v-u_{s})P_1^{(s)}]} \right\}\, \frac{P_1-P_1^{(s)}}{P_1^{(s)}}.
\label{ch7_eq:dH/dt-inel-develop-2}
\end{align}
Again, integrating by parts in $v$ (here we do not write the boundary
terms at $v\to\pm\infty$), the first term on the rhs of
\eqref{ch7_eq:dH/dt-inel-develop-2} is
\begin{align}
-\frac{\nu}{2}\int\!\! dx\, dv\, (v-u) P_1 
  \, \left(\vder{\ln P_1}-\vder{\ln P_1^{(s)}}\right),
\label{ch7_eq:dH/dt-inel-develop-21}
\end{align}
whereas the second term gives
\begin{align}
\frac{\nu}{2}\int\!\! dx\, dv\, (v-u_{s}) P_1
  \, \left(\vder{\ln P_1}-\vder{\ln P_1^{(s)}}\right).
\label{ch7_eq:dH/dt-inel-develop-22}
\end{align}
Summing up these two contributions, and taking into account that both
$u$ and $u_{s}$ do not depend on $v$,
\begin{align}
\left.\frac{dH}{dt}\right|_{\inel}=\frac{\nu}{2} \int\!\! dx \, (u-u_{s}) \int\!\! dv\, P_1 \left(\vder{\ln P_1}-\vder{\ln P_1^{(s)}}\right)
\label{ch7_eq:dH/dt-inel-final}
\end{align}
Since $\int\! dv\, P_1 \,\vder{\ln P_1}\equiv \int\! dv\, \vder{P_1}=0$,
this leads to \eqref{ch7_eq:H-time-ev-inel}.

\chapter{Completion of the proof of global stability}\label{app_final_glob_stab}

Here, we complete the proof of the global stability for $\xi=0$.
In that case, \eqref{ch7_eq:H-time-ev-noise} identically
vanishes and the proof presented in the main text for deriving
$\phi(v;t)=1$ in \eqref{ch7_eq:phi_stab} does not hold.

For $\xi=0$, one can use the kinetic equation in the limit as
$t\to\infty$, that is,
\begin{equation} \label{app_eq:kin_infty}
\lim_{t \to \infty} \frac{\partial \phi(v;t)}{\partial t} = \frac{\nu}{2} [v-
u_{\st}(x)] \lim_{t \to \infty} \frac{\partial \phi(v;t)}{\partial v}.
\end{equation}
Since $\phi(v;t)$ has no spatial dependence, the limit of its time
derivative must vanish and, consistently, the same applies for its
velocity derivative. In this way, $\phi(v;t)=1$ is reobtained, by
employing \eqref{ch7_eq:phi_stab} and the normalization of the
distributions.

The above result can be proven by a \textit{reductio ad absurdum}
argument. If the long time limit of $\partial_v \phi(v,t)$ were
nonzero, we would show $u_s(x)$ to depend only on $v$ by solving
\eqref{app_eq:kin_infty} for $u_s(x)$. This is contradictory, thus
$\lim_{t\rightarrow\infty} \partial_v \phi(v,t)$ must vanish and so
must $\lim_{t\rightarrow\infty} \partial_t \phi(v,t)$, by taking into
account again \eqref{app_eq:kin_infty}. 

Note that the argument in the previous paragraph ceases to be valid in
case of a homogeneous $u_{\st}$. However, the only possible flat
profile is $u_{\st}=0$, and this implies a null shear rate. Were that
the case, both $a$ and $\xi$ would vanish and no energy injection
mechanism would be present: in a few words, no stationary state would
be attained.

{\clearpage \thispagestyle{empty}}

\clearpage

\pagestyle{mystyle2} \bibliography{Bib-tesis}

\begin{thebibliography}{100}

\bibitem{fig-structure-url}
\url{https://commons.wikimedia.org/wiki/File:Main_protein_structure_levels_en.svg}.

\bibitem{So03}
C.~Soto, Nat. Rev. Neurosci. \textbf{4}, 49 (2003).

\bibitem{Ha17}
F.~U. Hartl, Annu. Rev. Biochem. \textbf{86}, 21 (2017).

\bibitem{TOMyH10}
D.~Thirumalai, E.~P. O'Brien, G.~Morrison, and C.~Hyeon, Annu. Rev. Biophys.
  \textbf{39}, 159 (2010).

\bibitem{fig-origami-url}
\url{https://phys.org/news/2015-05-tale-roads-protein-unfolding.html}.

\bibitem{R06}
F.~Ritort, J. Phys. Condens. Matter \textbf{18}, R531 (2006).

\bibitem{KyL10}
S.~Kumar and M.~S. Li, Phys. Rep. \textbf{486}, 1  (2010).

\bibitem{MyD12}
P.~E. Marszalek and Y.~F. Dufr\^{e}ne, Chem. Soc. Rev. \textbf{41}, 3523
  (2012).

\bibitem{HyD12}
T.~Hoffmann and L.~Dougan, Chem. Soc. Rev. \textbf{41}, 4781 (2012).

\bibitem{COFMBCyF99}
M.~Carrion-Vazquez, A.~F. Oberhauser, S.~B. Fowler, P.~E. Marszalek, S.~E.
  Broedel, J.~Clarke, and J.~M. Fernandez, Proc. Natl. Acad. Sci. U.S.A.
  \textbf{96}, 3694 (1999).

\bibitem{FMyF00}
T.~E. Fisher, P.~E. Marszalek, and J.~M. Fernandez, Nat. Struct. Biol.
  \textbf{7}, 719  (2000).

\bibitem{HDyT06}
C.~Hyeon, R.~I. Dima, and D.~Thirumalai, Structure \textbf{14}, 1633  (2006).

\bibitem{AyA06}
J.~L.~R. Arrondo and A.~Alonso (editors), \emph{Advanced Techniques in
  Biophysics},  (Springer, Berlin, 2007).

\bibitem{CSMGCyB15}
T.~Cr\'epin, C.~Swale, A.~Monod, F.~Garzoni, M.~Chaillet, and I.~Berger, Curr.
  Opin. Struct. Biol. \textbf{32}, 139  (2015).

\bibitem{MyB17}
V.~E.~T. Maervoet and Y.~Briers, Bioengineered \textbf{8}, 196 (2017).

\bibitem{ByR08}
M.~Bertz and M.~Rief, J. Mol. Biol. \textbf{378}, 447 (2008).

\bibitem{GMTCyC14}
C.~Guardiani, D.~D. Marino, A.~Tramontano, M.~Chinappi, and F.~Cecconi, J.
  Chem. Theory Comput. \textbf{10}, 3589 (2014).

\bibitem{BBTBSRyC03}
R.~B. Best, D.~J. Brockwell, J.~L. Toca-Herrera, A.~W. Blake, D.~Smith, S.~E.
  Radford, and J.~Clarke, Anal. Chim. Acta \textbf{479}, 87  (2003).

\bibitem{STyC09}
A.~Steward, J.~L. Toca-Herrera, and J.~Clarke, Protein Sci. \textbf{11}, 2179
  (2009).

\bibitem{KyL11}
M.~Kr\"uger and W.~A. Linke, J. Biol. Chem. \textbf{286}, 9905 (2011).

\bibitem{Hi02}
T.~Hill, \emph{Thermodynamics of Small Systems},  (Dover Publications, Mineola,
  NY, 2002).

\bibitem{Ja97}
C.~Jarzynski, Phys. Rev. E \textbf{56}, 5018 (1997).

\bibitem{Ja97b}
C.~Jarzynski, Phys. Rev. Lett. \textbf{78}, 2690 (1997).

\bibitem{Cr98}
G.~E. Crooks, J. Stat. Phys. \textbf{90}, 1481 (1998).

\bibitem{Cr99}
G.~E. Crooks, Phys. Rev. E \textbf{60}, 2721 (1999).

\bibitem{CRJSTyB05}
D.~Collin, F.~Ritort, C.~Jarzynski, S.~B. Smith, I.~Tinoco~Jr, and
  C.~Bustamante, Nature \textbf{437}, 231 (2005).

\bibitem{Fl89}
P.~Flory and J.~Jackson, \emph{Statistical Mechanics of Chain Molecules},
  (Hanser Publishers, Munich, 1989).

\bibitem{Ru03}
M.~Rubinstein and R.~Colby, \emph{Polymer Physics},  (Oxford University Press,
  Oxford, 2003).

\bibitem{MyS95}
J.~F. Marko and E.~D. Siggia, Macromolecules \textbf{28}, 8759 (1995).

\bibitem{SFyB92}
S.~Smith, L.~Finzi, and C.~Bustamante, Science \textbf{258}, 1122 (1992).

\bibitem{BMSyS94}
C.~Bustamante, J.~Marko, E.~Siggia, and S.~Smith, Science \textbf{265}, 1599
  (1994).

\bibitem{RFyG98}
M.~Rief, J.~M. Fernandez, and H.~E. Gaub, Phys. Rev. Lett. \textbf{81}, 4764
  (1998).

\bibitem{ByS05}
O.~Braun and U.~Seifert, Eur. Phys. J. E \textbf{18}, 1 (2005).

\bibitem{MHyM01}
D.~E. Makarov, P.~K. Hansma, and H.~Metiu, J. Chem. Phys. \textbf{114}, 9663
  (2001).

\bibitem{Kr40}
H.~Kramers, Physica \textbf{7}, 284  (1940).

\bibitem{HTyB90}
P.~H\"anggi, P.~Talkner, and M.~Borkovec, Rev. Mod. Phys. \textbf{62}, 251
  (1990).

\bibitem{Be78}
G.~Bell, Science \textbf{200}, 618 (1978).

\bibitem{EyR97}
E.~Evans and K.~Ritchie, Biophys. J. \textbf{72}, 1541  (1997).

\bibitem{PCyB13}
A.~Prados, A.~Carpio, and L.~L. Bonilla, Phys. Rev. E \textbf{88}, 012704
  (2013).

\bibitem{BCyP14}
L.~L. Bonilla, A.~Carpio, and A.~Prados, EPL \textbf{108}, 28002 (2014).

\bibitem{BCyP15}
L.~L. Bonilla, A.~Carpio, and A.~Prados, Phys. Rev. E \textbf{91}, 052712
  (2015).

\bibitem{BZDyG16}
I.~Benichou, Y.~Zhang, O.~K. Dudko, and S.~Givli, J. Mech. Phys. Solids
  \textbf{95}, 44  (2016).

\bibitem{LyK09}
M.~S. Li and M.~Kouza, J. Chem. Phys. \textbf{130}, 145102 (2009).

\bibitem{KHLyK13}
M.~Kouza, C.-K. Hu, M.~S. Li, and A.~Kolinski, J. Chem. Phys. \textbf{139},
  065103 (2013).

\bibitem{JNyB96}
H.~M. Jaeger, S.~R. Nagel, and R.~P. Behringer, Rev. Mod. Phys. \textbf{68},
  1259 (1996).

\bibitem{PyL01}
T.~P\"oschel and S.~Luding (editors), \emph{Granular Gases},  (Springer-Verlag,
  Berlin, 2001).

\bibitem{fig-granular-url}
\url{https://commons.wikimedia.org/wiki/File:Granular_matter_examples.PNG}.

\bibitem{OLLyN02}
C.~S. O'Hern, S.~A. Langer, A.~J. Liu, and S.~R. Nagel, Phys. Rev. Lett.
  \textbf{88}, 075507 (2002).

\bibitem{Pu14}
A.~Puglisi, \emph{Transport and Fluctuations in Granular Fluids},  (Springer,
  Berlin, 2014).

\bibitem{ByP04}
N.~Brilliantov and T.~P\"oschel (editors), \emph{Kinetic Theory of Granular
  Gases},  (Oxford University Press, New York, 2004).

\bibitem{LHMyZ98}
S.~Luding, M.~Huthmann, S.~McNamara, and A.~Zippelius, Phys. Rev. E
  \textbf{58}, 3416 (1998).

\bibitem{SKyS11}
A.~Santos, G.~M. Kremer, and M.~dos Santos, Phys. Fluids \textbf{23}, 030604
  (2011).

\bibitem{RSyK14}
F.~V. Reyes, A.~Santos, and G.~M. Kremer, Phys. Rev. E \textbf{89}, 020202
  (2014).

\bibitem{GSyK18}
V.~Garz\'o, A.~Santos, and G.~M. Kremer, Phys. Rev. E \textbf{97}, 052901
  (2018).

\bibitem{Sa18}
A.~Santos, Phys. Rev. E \textbf{98}, 012904 (2018).

\bibitem{BSHyP96}
N.~V. Brilliantov, F.~Spahn, J.~M. Hertzsch, and T.~P\"oschel, Phys. Rev. E
  \textbf{53}, 5382 (1996).

\bibitem{NyE98}
T.~P.~C. van Noije and M.~Ernst, Granul. Matter \textbf{1}, 57 (1998).

\bibitem{LSJyC84}
C.~K.~K. Lun, S.~B. Savage, D.~J. Jeffrey, and N.~Chepurniy, J. Fluid Mech.
  \textbf{140}, 223–256 (1984).

\bibitem{BDKyS98}
J.~J. Brey, J.~W. Dufty, C.~S. Kim, and A.~Santos, Phys. Rev. E \textbf{58},
  4638 (1998).

\bibitem{Go99}
I.~Goldhirsch, Chaos \textbf{9}, 659 (1999).

\bibitem{Ka99}
L.~P. Kadanoff, Rev. Mod. Phys. \textbf{71}, 435 (1999).

\bibitem{NyE00}
T.~P.~C. van Noije and M.~H. Ernst, Phys. Rev. E \textbf{61}, 1765 (2000).

\bibitem{Ei04}
A.~Einstein, Ann. Phys. (Berl.) \textbf{319}, 354 (1904).

\bibitem{OyM53}
L.~Onsager and S.~Machlup, Phys. Rev. \textbf{91}, 1505 (1953).

\bibitem{LyL80}
L.~D. Landau and E.~M. Lifshitz, \emph{Statistical Physics (3rd edition Course
  of Theoretical Physics Vol. 5)},  (Pergamon Press, Oxford, 1980).

\bibitem{BMyG09}
J.~J. Brey, P.~Maynar, and M.~I. {Garc\'{\i}a de Soria}, Phys. Rev. E
  \textbf{79}, 051305 (2009).

\bibitem{BSGJyL01}
L.~Bertini, A.~De~Sole, D.~Gabrielli, G.~Jona-Lasinio, and C.~Landim, Phys.
  Rev. Lett. \textbf{87}, 040601 (2001).

\bibitem{KyL99}
C.~Kipnis and C.~Landim, \emph{Scaling Limits of Interacting Particle Systems},
   (Springer-Verlag, Berlin, 1999).

\bibitem{KMyP82}
C.~Kipnis, C.~Marchioro, and E.~Presutti, J. Stat. Phys. \textbf{27}, 65
  (1982).

\bibitem{HyG09}
P.~I. Hurtado and P.~L. Garrido, Phys. Rev. Lett. \textbf{102}, 250601 (2009).

\bibitem{HyG10}
P.~I. Hurtado and P.~L. Garrido, Phys. Rev. E \textbf{81}, 041102 (2010).

\bibitem{HyG09b}
P.~I. Hurtado and P.~L. Garrido, J. Stat. Mech. Theory Exp. \textbf{2009},
  P02032 (2009).

\bibitem{HyK11}
P.~I. Hurtado and P.~L. Krapivsky, Phys. Rev. E \textbf{85}, 060103 (2012).

\bibitem{SyL04}
Y.~Srebro and D.~Levine, Phys. Rev. Lett. \textbf{93}, 240601 (2004).

\bibitem{PLyH12a}
A.~Prados, A.~Lasanta, and P.~I. Hurtado, Phys. Rev. E \textbf{86}, 031134
  (2012).

\bibitem{PLyH11a}
A.~Prados, A.~Lasanta, and P.~I. Hurtado, Phys. Rev. Lett. \textbf{107}, 140601
  (2011).

\bibitem{PLyH13}
P.~I. Hurtado, A.~Lasanta, and A.~Prados, Phys. Rev. E \textbf{88}, 022110
  (2013).

\bibitem{PLyH16}
A.~Lasanta, P.~I. Hurtado, and A.~Prados, Eur. Phys. J. E \textbf{39} (2016).

\bibitem{LMPyP15}
A.~Lasanta, A.~Manacorda, A.~Prados, and A.~Puglisi, New J. Phys. \textbf{17},
  083039 (2015).

\bibitem{GyS13}
V.~Garz{\'o} and A.~Santos, \emph{Kinetic Theory of Gases in Shear Flows:
  Nonlinear Transport},  (Springer, Netherlands, 2003).

\bibitem{Ha83}
P.~K. Haff, J. Fluid Mech. \textbf{134}, 401–430 (1983).

\bibitem{Er81}
M.~Ernst, Phys. Rep. \textbf{78}, 1  (1981).

\bibitem{BRyC96}
J.~J. Brey, M.~J. Ruiz-Montero, and D.~Cubero, Phys. Rev. E \textbf{54}, 3664
  (1996).

\bibitem{BPGyM07}
J.~J. Brey, A.~Prados, M.~I. {Garc\'ia de Soria}, and P.~Maynar, J. Phys. A
  \textbf{40}, 14331 (2007).

\bibitem{MN93}
S.~McNamara, Phys. Fluid. Fluid Dynam. \textbf{5}, 3056 (1993).

\bibitem{Ga06}
V.~Garz\'o, Phys. Rev. E \textbf{73}, 021304 (2006).

\bibitem{Ly92}
A.~M. Lyapunov, Int. J. Control \textbf{55}, 531 (1992).

\bibitem{Bo95}
L.~Boltzmann, \emph{Lectures on Gas Theory},  (Dover Publications, New York,
  1995).

\bibitem{Le93}
J.~L. Lebowitz, Physica A \textbf{194}, 1  (1993).

\bibitem{Le93b}
J.~L. Lebowitz, Phys. Today \textbf{46}, 32 (1993).

\bibitem{Le99}
J.~L. Lebowitz, Physica A \textbf{263}, 516  (1999).

\bibitem{Pr99}
I.~Prigogine, Physica A \textbf{263}, 528  (1999).

\bibitem{Ru99}
D.~Ruelle, Physica A \textbf{263}, 540  (1999).

\bibitem{CyC90}
S.~Chapman, T.~Cowling, D.~Burnett, and C.~Cercignani, \emph{The Mathematical
  Theory of Non-uniform Gases: An Account of the Kinetic Theory of Viscosity,
  Thermal Conduction and Diffusion in Gases},  (Cambridge University Press,
  Cambridge, New York, 1990).

\bibitem{vK92}
N.~G. Van~Kampen, \emph{Stochastic Processes in Physics and Chemistry},
  (North-Holland, Amsterdam, 1992).

\bibitem{KyL51}
S.~Kullback and R.~A. Leibler, Ann. Math. Statist. \textbf{22}, 79 (1951).

\bibitem{MPyV13}
U.~M.~B. Marconi, A.~Puglisi, and A.~Vulpiani, J. Stat. Mech. Theory Exp.
  \textbf{2013}, P08003 (2013).

\bibitem{GMMMRyT15}
M.~I. {Garc\'ia de Soria}, P.~Maynar, S.~Mischler, C.~Mouhot, T.~Rey, and
  E.~Trizac, J. Stat. Mech. Theory Exp. \textbf{2015}, P11009 (2015).

\bibitem{GSyB90}
V.~Garz\'o, A.~Santos, and J.~Brey, Physica A \textbf{163}, 651  (1990).

\bibitem{Vi06}
C.~Villani, J. Stat. Phys. \textbf{124}, 781 (2006).

\bibitem{BCDVTyW06}
I.~Bena, F.~Coppex, M.~Droz, P.~Visco, E.~Trizac, and F.~van Wijland, Physica A
  \textbf{370}, 179  (2006).

\bibitem{Bi63}
G.~A. Bird, Phys. Fluids \textbf{6}, 1518 (1963).

\bibitem{Bi13}
G.~Bird, \emph{The DSMC Method},  (CreateSpace Independent Publishing Platform,
  USA, 2013).

\bibitem{KAHyR79}
A.~J. Kovacs, J.~J. Aklonis, J.~M. Hutchinson, and A.~R. Ramos, J. Polym. Sci.,
  Polym. Phys. Ed. \textbf{17}, 1097 (1979).

\bibitem{BBDyG03}
E.~M. Bertin, J.~P. Bouchaud, J.~M. Drouffe, and C.~Godrèche, J. Phys. A
  \textbf{36}, 10701 (2003).

\bibitem{Bu03}
A.~Buhot, J. Phys. A \textbf{36}, 12367 (2003).

\bibitem{MyS04}
S.~Mossa and F.~Sciortino, Phys. Rev. Lett. \textbf{92}, 045504 (2004).

\bibitem{ALyN06}
G.~Aquino, L.~Leuzzi, and T.~M. Nieuwenhuizen, Phys. Rev. B \textbf{73}, 094205
  (2006).

\bibitem{PyB10}
A.~Prados and J.~J. Brey, J. Stat. Mech. Theory Exp. \textbf{2010}, P02009
  (2010).

\bibitem{DyH11}
G.~Diezemann and A.~Heuer, Phys. Rev. E \textbf{83}, 031505 (2011).

\bibitem{RyP14}
M.~Ruiz-Garc\'{\i}a and A.~Prados, Phys. Rev. E \textbf{89}, 012140 (2014).

\bibitem{PyT14}
A.~Prados and E.~Trizac, Phys. Rev. Lett. \textbf{112}, 198001 (2014).

\bibitem{TyP14}
E.~Trizac and A.~Prados, Phys. Rev. E \textbf{90}, 012204 (2014).

\bibitem{Wi96}
D.~Williams, Physica A: Statistical Mechanics and its Applications
  \textbf{233}, 718  (1996).

\bibitem{WyM96}
D.~R.~M. Williams and F.~C. MacKintosh, Phys. Rev. E \textbf{54}, R9 (1996).

\bibitem{SBCyM98}
M.~R. Swift, M.~Boamf\ifmmode~\check{a}\else \v{a}\fi{}, S.~J. Cornell, and
  A.~Maritan, Phys. Rev. Lett. \textbf{80}, 4410 (1998).

\bibitem{BGMyB14}
J.~J. Brey, M.~I. {Garc\'{\i}a de Soria}, P.~Maynar, and V.~Buz\'on, Phys. Rev.
  E \textbf{90}, 032207 (2014).

\bibitem{LGAyR17}
Y.~Lahini, O.~Gottesman, A.~Amir, and S.~M. Rubinstein, Phys. Rev. Lett.
  \textbf{118}, 085501 (2017).

\bibitem{KSyI17}
R.~K\"ursten, V.~Sushkov, and T.~Ihle, Phys. Rev. Lett. \textbf{119}, 188001
  (2017).

\bibitem{HyS03}
G.~Hummer and A.~Szabo, Biophys. J. \textbf{85}, 5—15 (2003).

\bibitem{ACDNKMNLyR10}
G.~Arad-Haase, S.~G. Chuartzman, S.~Dagan, R.~Nevo, M.~Kouza, B.~K. Mai, H.~T.
  Nguyen, M.~S. Li, and Z.~Reich, Biophys. J. \textbf{99}, 238  (2010).

\bibitem{ByO99}
C.~M. Bender and S.~A. Orszag, \emph{Advanced Mathematical Methods for
  Scientists and Engineers {I}: {Asymptotic} Methods and Perturbation Theory},
  (Springer, New York, 1999).

\bibitem{RGPCyS13}
F.~Rico, L.~Gonzalez, I.~Casuso, M.~Puig-Vidal, and S.~Scheuring, Science
  \textbf{342}, 741 (2013).

\bibitem{DHyS06}
O.~K. Dudko, G.~Hummer, and A.~Szabo, Phys. Rev. Lett. \textbf{96}, 108101
  (2006).

\bibitem{PCCyP15}
C.~A. Plata, F.~Cecconi, M.~Chinappi, and A.~Prados, J. Stat. Mech. Theory Exp.
  \textbf{2015}, P08003 (2015).

\bibitem{BGUKyF10}
R.~Berkovich, S.~Garcia-Manyes, M.~Urbakh, J.~Klafter, and J.~M. Fernandez,
  Biophys. J. \textbf{98}, 2692 (2010).

\bibitem{BHPSGByF12}
R.~Berkovich, R.~I. Hermans, I.~Popa, G.~Stirnemann, S.~Garcia-Manyes, B.~J.
  Berne, and J.~M. Fernandez, Proc. Natl. Acad. Sci. U.S.A. \textbf{109}, 14416
  (2012).

\bibitem{ber10}
R.~Berkovich, S.~Garcia-Manyes, M.~Urbakh, J.~Klafter, and J.~M. Fernandez,
  Biophys. J. \textbf{98}, 2692  (2010).

\bibitem{LSyM14}
Q.~Li, Z.~N. Scholl, and P.~E. Marszalek, Angew. Chem. Int. Ed. \textbf{53},
  13429 (2014).

\bibitem{SKSNyR03}
I.~Schwaiger, A.~Kardinal, M.~Schleicher, A.~A. Noegel, and M.~Rief, Nat.
  Struct. Mol. Biol. \textbf{11}, 81 (2004).

\bibitem{GHyT96}
H.~Grubm{\"u}ller, B.~Heymann, and P.~Tavan, Science \textbf{271}, 997 (1996).

\bibitem{LIKVyS98}
H.~Lu, B.~Isralewitz, A.~Krammer, V.~Vogel, and K.~Schulten, Biophys. J.
  \textbf{75}, 662  (1998).

\bibitem{Is11}
J.~N. Israelachvili, \emph{Intermolecular and Surface Forces (Third Edition)},
  (Academic Press, San Diego, 2011).

\bibitem{Zu10}
D.~Zuckerman, \emph{Statistical Physics of Biomolecules: An Introduction},
  (CRC Press, Boca Raton, Florida, 2010).

\bibitem{FyS01}
D.~Frenkel and B.~Smit, \emph{Understanding Molecular Simulation: From
  Algorithms to Applications},  (Academic Press, San Diego, 2001).

\bibitem{PByS08}
D.~A. Parry, R.~B. Fraser, and J.~M. Squire, J. Struct. Biol. \textbf{163}, 258
   (2008).

\bibitem{RMAVWyG10}
O.~J. Rackham, M.~Madera, C.~T. Armstrong, T.~L. Vincent, D.~N. Woolfson, and
  J.~Gough, J. Mol. Biol. \textbf{403}, 480  (2010).

\bibitem{TyL16}
L.~Truebestein and T.~A. Leonard, BioEssays \textbf{38}, 903 (2016).

\bibitem{ARCPDyL03}
M.~Aittaleb, R.~Rashid, Q.~Chen, J.~R. Palmer, C.~J. Daniels, and H.~Li, Nat.
  Struct. Biol. \textbf{10}, 256 (2003).

\bibitem{HDyS96}
W.~Humphrey, A.~Dalke, and K.~Schulten, J. Mol. Graph. \textbf{14}, 33  (1996).

\bibitem{PBWGTVCSKyS05}
J.~C. Phillips, R.~Braun, W.~Wang, J.~Gumbart, E.~Tajkhorshid, E.~Villa,
  C.~Chipot, R.~D. Skeel, L.~Kal\'e, and K.~Schulten, J. Comput. Chem.
  \textbf{26}, 1781 (2005).

\bibitem{BHyE13}
R.~B. Best, G.~Hummer, and W.~A. Eaton, Proc. Natl. Acad. Sci. U.S.A.
  \textbf{110}, 17874 (2013).

\bibitem{PSMyP18}
C.~A. Plata, Z.~N. Scholl, P.~E. Marszalek, and A.~Prados, J. Chem. Theory
  Comput. \textbf{14}, 2910 (2018).

\bibitem{DBByR06}
H.~Dietz, F.~Berkemeier, M.~Bertz, and M.~Rief, Proc. Natl. Acad. Sci. U.S.A.
  \textbf{103}, 12724 (2006).

\bibitem{GSyZ11}
Y.~Gao, G.~Sirinakis, and Y.~Zhang, J. Am. Chem. Soc. \textbf{133}, 12749
  (2011).

\bibitem{BByP02}
A.~Baldassarri, U.~M.~B. Marconi, and A.~Puglisi, EPL \textbf{58}, 14 (2002).

\bibitem{GLyS90}
G.~Grinstein, D.-H. Lee, and S.~Sachdev, Phys. Rev. Lett. \textbf{64}, 1927
  (1990).

\bibitem{GLMyS90}
P.~L. Garrido, J.~L. Lebowitz, C.~Maes, and H.~Spohn, Phys. Rev. A \textbf{42},
  1954 (1990).

\bibitem{Sp80}
H.~Spohn, Rev. Mod. Phys. \textbf{52}, 569 (1980).

\bibitem{MPLPyP16}
A.~Manacorda, C.~A. Plata, A.~Lasanta, A.~Puglisi, and A.~Prados, J. Stat.
  Phys. \textbf{164}, 810 (2016).

\bibitem{AM_thesis}
A.~Manacorda, \emph{Lattice Models for Fluctuating Hydrodynamics in Granular
  and Active Matter}, Springer Theses,  (Springer, 2018).

\bibitem{ByK03}
E.~Ben-Naim and P.~L. Krapivsky, in T.~P\"oschel and N.~Brilliantov (editors),
  \emph{Granular Gas Dynamics}, volume 624 of \emph{Lecture Notes in Physics},
  65--94,  (Springer, Berlin, 2003).

\bibitem{ETyB06}
M.~H. Ernst, E.~Trizac, and A.~Barrat, EPL \textbf{76}, 56 (2006).

\bibitem{BPyP18}
A.~Baldassarri, A.~Puglisi, and A.~Prados, Phys. Rev. E \textbf{97}, 062905
  (2018).

\bibitem{PMLPyP16}
C.~A. Plata, A.~Manacorda, A.~Lasanta, A.~Puglisi, and A.~Prados, J. Stat.
  Mech. Theory Exp. \textbf{2016}, 093203 (2016).

\bibitem{NEByO97}
T.~P.~C. van Noije, M.~H. Ernst, R.~Brito, and J.~A.~G. Orza, Phys. Rev. Lett.
  \textbf{79}, 411 (1997).

\bibitem{GMSByT08}
M.~I. {Garc\'{\i}a de Soria}, P.~Maynar, G.~Schehr, A.~Barrat, and E.~Trizac,
  Phys. Rev. E \textbf{77}, 051127 (2008).

\bibitem{LyE72}
A.~W. Lees and S.~F. Edwards, J. Phys. C \textbf{5}, 1921 (1972).

\bibitem{SyG07}
A.~Santos and V.~Garz\'o, J. Stat. Mech. Theory Exp. \textbf{2007}, P08021
  (2007).

\bibitem{SGyD04}
A.~Santos, V.~Garz\'o, and J.~W. Dufty, Phys. Rev. E \textbf{69}, 061303
  (2004).

\bibitem{Re77}
P.~Resibois and M.~de~Leener, \emph{Classical Kinetic Theory of Fluids},
  (Wiley, New York, 1977).

\bibitem{GMyT12}
M.~I. {Garc\'{\i}a de Soria}, P.~Maynar, and E.~Trizac, Phys. Rev. E
  \textbf{85}, 051301 (2012).

\bibitem{GMyT13}
M.~I. {Garc\'{\i}a de Soria}, P.~Maynar, and E.~Trizac, Phys. Rev. E
  \textbf{87}, 022201 (2013).

\bibitem{BDGyM06}
J.~J. Brey, A.~Dom\'{\i}nguez, M.~I. {Garc\'{\i}a de Soria}, and P.~Maynar,
  Phys. Rev. Lett. \textbf{96}, 158002 (2006).

\bibitem{PSDyN17}
V.~V. Prasad, S.~Sabhapandit, A.~Dhar, and O.~Narayan, Phys. Rev. E
  \textbf{95}, 022115 (2017).

\bibitem{ByP93b}
J.~J. Brey and A.~Prados, Phys. Rev. E \textbf{47}, 1541 (1993).

\bibitem{ByP94}
J.~J. Brey and A.~Prados, Phys. Rev. B \textbf{49}, 984 (1994).

\bibitem{BPyR94}
J.~Brey, A.~Prados, and M.~Ruiz-Montero, J. Non-Cryst. Solids \textbf{172-174},
  371  (1994).

\bibitem{VyR97}
M.~O. Vlad and J.~Ross, J. Phys. Chem. B \textbf{101}, 8756 (1997).

\bibitem{VMyR98}
M.~O. Vlad, F.~Moran, and J.~Ross, J. Phys. Chem. B \textbf{102}, 4598 (1998).

\bibitem{PByS00}
A.~Prados, J.~Brey, and B.~S\'anchez-Rey, Physica A \textbf{284}, 277  (2000).

\bibitem{EyK10}
B.~Earnshaw and J.~Keener, SIAM J. Appl. Dyn. Syst. \textbf{9}, 220 (2010).

\bibitem{MyS00}
J.~M. Montanero and A.~Santos, Granul. Matter \textbf{2}, 53 (2000).

\bibitem{GMyT09}
M.~I. {Garc\'ia de Soria}, P.~Maynar, and E.~Trizac, Mol. Phys. \textbf{107},
  383 (2009).

\bibitem{PTNyE01}
I.~Pagonabarraga, E.~Trizac, T.~P.~C. van Noije, and M.~H. Ernst, Phys. Rev. E
  \textbf{65}, 011303 (2001).

\bibitem{MGyT09}
P.~Maynar, M.~I. {Garc\'ia de Soria}, and E.~Trizac, Eur. Phys. J. Spec. Top.
  \textbf{179}, 123 (2009).

\bibitem{PSyD13}
V.~V. Prasad, S.~Sabhapandit, and A.~Dhar, EPL \textbf{104}, 54003 (2013).

\bibitem{Ch90}
P.~L. Chebyshev, Acta Math. \textbf{14}, 305 (1890).

\bibitem{Ch05}
C.~V.~L. Charlier, Ark. Math. Astr. och Phys. \textbf{2}, 1 (1905-06).

\bibitem{Ed05}
F.~Y. Edgeworth, Math. Proc. Cambridge Philos. Soc. \textbf{20}, 36 (1905).

\bibitem{Wa58}
D.~L. Wallace, Ann. Math. Stat. \textbf{29}, 635 (1958).

\bibitem{AyS72}
M.~Abramowitz and I.~A. Stegun (editors), \emph{Handbook of Mathematical
  Functions with Formulas, Graphs, and Mathematical Tables},  (Dover,
  Washington DC, 1972).

\bibitem{Cr25}
H.~Cram\'er, in \emph{Proceedings of the Sixth Scandinavian Congress of
  Mathematicians, Copenhagen}, 399--425 (1925).

\bibitem{Sz39}
G.~Szeg\"o, \emph{Orthogonal Polynomials},  (American Mathematical Society,
  New York, 1939).

\bibitem{ByP93}
J.~Brey and A.~Prados, Physica A \textbf{197}, 569  (1993).

\bibitem{MyV77}
I.~M{\"u}ller and P.~Villaggio, Arch. Ration. Mech. Anal. \textbf{65}, 25
  (1977).

\bibitem{PyT02}
G.~Puglisi and L.~Truskinovsky, Cont. Mech. Thermodyn. \textbf{14}, 437 (2002).

\bibitem{DJGHMyG10}
W.~Dreyer, J.~Jamnik, C.~Guhlke, R.~Huth, J.~Moškon, and M.~Gaberšček, Nat.
  Mater. \textbf{9}, 448 (2010).

\bibitem{DGyH11}
W.~Dreyer, C.~Guhlke, and M.~Herrmann, Cont. Mech. Thermodyn. \textbf{23}, 211
  (2011).

\bibitem{BFSyG13}
I.~Benichou, E.~Faran, D.~Shilo, and S.~Givli, Appl. Phys. Lett. \textbf{102},
  011912 (2013).

\bibitem{BSGJyL15}
L.~Bertini, A.~De~Sole, D.~Gabrielli, G.~Jona-Lasinio, and C.~Landim, Rev. Mod.
  Phys. \textbf{87}, 593 (2015).

\bibitem{BKyL75}
A.~Bortz, M.~Kalos, and J.~Lebowitz, J. Comput. Phys. \textbf{17}, 10  (1975).

\bibitem{PByS97}
A.~Prados, J.~J. Brey, and B.~S\'anchez-Rey, J. Stat. Phys. \textbf{89}, 709
  (1997).

\end{thebibliography}

{\clearpage \thispagestyle{empty}}
\chapter*{List of acronyms}
\addcontentsline{toc}{chapter}{List of acronyms}
\begin{description}
\item[LOT] Laser optical tweezers
\item[AFM] Atomic force microscopy
\item[FJC] Freely-jointed chain
\item[WLC] Worm-like chain
\item[MBP] Maltose binding protein
\item[PDB] Protein data bank
\item[HCS] Homogeneous cooling state
\item[USF] Uniform shear flow
\item[PDF] Probability density function
\item[NESS] Nonequilibrium steady state
\item[SMD] Steered molecular dynamics
\item[CC] Coiled-coil
\item[ARG] Arginine
\item[ILE] Isoleucine
\end{description}

{\clearpage \thispagestyle{empty}}

\renewcommand{\listfigurename}{List of figures}
\listoffigures

\end{document}